\documentclass{article}
\usepackage{amssymb}
\usepackage{graphicx}

\input{tcilatex}

\begin{document}

\begin{center}
\bigskip

{\LARGE QUANTUM\ ENTANGLEMENT\ FOR\bigskip }

{\LARGE SYSTEMS\ OF\ IDENTICAL\ BOSONS.\bigskip }

{\LARGE II. SPIN\ SQUEEZING\bigskip\ }

{\LARGE AND\ OTHER\ ENTANGLEMENT\ TESTS\bigskip\ \ }

\bigskip

B. J. Dalton$^{1,2}$, J. Goold$^{1,3}$, B. M. Garraway$^{4}$ and M. D. Reid$%
^{2}${\LARGE \bigskip }

$^{1}$\textit{Physics Department, University College Cork, Cork City, Ireland%
}

$^{2}$\textit{Centre for Quantum and Optical Science}$^{\ast }$\textit{,
Swinburne University of Technology, Melbourne, Victoria 3122, Australia}

$^{3}$\textit{The Abdus Salam International Centre for Theoretical Physics
(ICTP), Trieste 34151, Italy}

$^{4}$\textit{Department of Physics and Astronomy, University of Sussex,
Falmer, Brighton BN19QH, United Kingdom}

\textit{\bigskip }

\textit{\bigskip \bigskip \bigskip }
\end{center}

Email: bdalton@swin.edu.au

$\ast $Formerly \textit{Centre for Atom Optics and Ultrafast Spectroscopy}

\pagebreak

$\mathbf{Abstract}$

These two accompanying papers are concerned with entanglement for systems of
identical massive bosons and the relationship to spin squeezing and other
quantum correlation effects. The main focus is on two mode entanglement, but
multi-mode entanglement is also considered. The bosons may be atoms or
molecules as in cold quantum gases. The previous paper I\textbf{\ }dealt
with the general features of quantum entanglement and its specific
definition in the case of systems of identical bosons. Entanglement is a
property shared between two (or more) quantum sub-systems. In defining
entanglement for systems of identical massive particles, it was concluded
that the single particle states or modes are the most appropriate choice for
sub-systems that are distinguishable, that the general quantum states must
comply both with the symmetrisation principle and the super-selection rules
(SSR) that forbid quantum superpositions of states with differing total
particle number (global SSR compliance). Further, it was concluded that (in
the separable states) quantum superpositions of sub-system states with
differing sub-system particle number (local SSR compliance) also do not
occur. The present paper II determines possible tests for entanglement based
on the treatment of entanglement set out in paper I.

Several inequalities involving variances and mean values of operators have
been previously proposed as tests for entanglement between two sub-systems.
These inequalities generally involve mode annihilation and creation
operators and include the inequalities that define spin squeezing. In this
paper, spin squeezing criteria for two mode systems are examined, and spin
squeezing is also considered for principle spin operator components where
the covariance matrix is diagonal. The proof, which is based on our SSR
compliant approach shows that the presence of spin squeezing in any one of
the spin components requires entanglement of the relevant pair of modes. A
simple Bloch vector test for entanglement is also derived. Thus we show that
spin squeezing bdecomes a rigorous test for entanglement in a system of
massive bosons, when viewed as a test for entanglement between two modes.

In addition, other previously proposed tests for entanglement involving spin
operators are considered, including those based on the sum of the variances
for two spin components. All of the tests are still valid when the present
concept of entanglement based on the symmetrisation and super-selection rule
criteria is applied. These tests also apply in cases of multi-mode
entanglement, though with restrictions in the case of sub-systems each
consisting of pairs of modes. Tests involving quantum correlation functions
are also considered and for global SSR compliant states these are shown to
be equivalent to tests involving spin operators. A new weak correlation test
is derived for entanglement based on local SSR compliance for separable
states, complementing the stronger correlation test obtained previously when
this is ignored. The Bloch vector test is equivalent to one case of this
weak correlation test. Quadrature squeezing for single modes is also
examined but not found to yield a useful entanglement test, whereas two mode
quadrature squeezing proves to be a valid entanglement test, though not as
useful as the Bloch vector test. The various entanglement tests are
considered for well-known entangled states, such as binomial states,
relative phase eigenstates and NOON states - sometimes the new tests are
satisfied whilst than those obtained in other papers are not.

The present paper II then outlines the theory for a simple\emph{\ }two mode
interferometer showing that such an interferometer can be used to measure
the mean values and covariance matrix for the spin operators involved in
entanglement tests for the two mode bosonic system. The treatment is also
generalised to cover multi-mode interferometry. The interferometer involves
a pulsed classical field characterised by a phase variable and an area
variable defined by the time integral of the field amplitude, and leads to a
coupling between the two modes. For simplicity the centre frequency was
chosen to be resonant with the inter-mode transition frequency. Measuring
the mean and variance of the\emph{\ }population difference between the two
modes for the output state of the interferometer for various choices of
interferometer variables is shown to enable the mean values and covariance
matrix for the spin operators for the input quantum state of the two mode
system to be determined. The paper concludes with a discussion of several
key experimental papers on spin squeezing.

\bigskip

\bigskip


\begin{center}
\pagebreak
\end{center}

{\Large Contents\bigskip }

{\Large 1. Introduction\medskip }

{\Large 2. Spin Squeezing}

\qquad \textbf{2.1 Spin Operators, Bloch Vector and Covariance Matrix}

\qquad \qquad \textit{2.1.1 Spin Operators}

\qquad \qquad \textit{2.1.2 Bloch Vector and Covariance Matrix}

\qquad \textbf{2.2 Spin Operators - Multimode Case}

\qquad \textbf{2.3 New Spin Operators and Principle Spin Fluctuations}

\qquad \textbf{2.4 Spin Squeezing Definition }

\qquad \qquad \textit{2.4.1 Heisenberg Uncertainty Principle - Spin Squeezing%
}

\qquad \qquad \textit{2.4.2 Alternative Spin Squeezing Criteria}

\qquad \qquad \textit{2.4.3 Planar Spin Squeezing}

\qquad \qquad \textit{2.4.4 Spin Squeezing in Multi-Mode Case}

\qquad \textbf{2.5 Rotation Operators and New Modes}

\qquad \qquad \textit{2.5.1 Rotation Operators}

\qquad \qquad \textit{2.5.2 New Mode Operators}

\qquad \qquad \textit{2.5.3 New Modes}

\qquad \textbf{2.6 Old and New Modes - Coherence Terms}

\qquad \textbf{2.7 Quantum Correlation Functions and Spin
Measurements\medskip }

{\Large 3. Spin Squeezing Test for Entanglement}

\qquad \textbf{3.1 Spin Squeezing and Entanglement - Original Modes}

\qquad \qquad \textit{3.1.1 Mean values for }$\widehat{\mathit{S}}_{x},$%
\textit{\ }$\widehat{\mathit{S}}_{y}$\textit{\ and }$\widehat{\mathit{S}}%
_{z} $

\qquad \qquad \textit{3.1.2 Variances for }$\widehat{\mathit{S}}_{x}$ 
\textit{and }$\widehat{\mathit{S}}_{y}$

\qquad \qquad \textit{3.1.3 Significance of Spin Squeezing Test}

\qquad \qquad \textit{3.1.4 Variance for }$\widehat{\mathit{S}}_{z}$

\qquad \qquad \textit{3.1.5 No Spin Squeezing for Separable States}

\qquad \qquad \textit{3.1.6 Spin Squeezing Tests for Entanglement}

\qquad \qquad \textit{3.1.7 Inequality for }$|\left\langle \widehat{\mathit{S%
}}_{z}\right\rangle |$

\qquad \textbf{3.2 Spin Squeezing and Entanglement - New Modes}

\qquad \textbf{3.3 Spin Squeezing and Entanglement - Multi-Mode Case}

\qquad \textbf{3.4 Bloch Vector Entanglement Test}

\qquad \textbf{3.5 Entanglement Test for Number Difference and Sum}

\qquad \textbf{3.6 Entangled States that are Non Spin Squeezed - NOON State}

\qquad \textbf{3.7 Non-Entangled States that are Non Spin Squeezed -
Binomial State}

\qquad \textbf{3.8 Entangled States that are Spin Squeezed - Relative Phase
Eigenstate}

\qquad \qquad \textit{3.8.1 New Spin Operators}

\qquad \qquad \textit{3.8.2 Bloch Vector and Covariance Matrix}

\qquad \qquad \textit{3.8.3 New Mode Operators}\textbf{\medskip }

{\Large 4. Other Spin Operator Tests for Entanglement}

\qquad \textbf{4.1 Hillery et al. 2006}

\qquad \qquad \textit{4.1.1 Hillery Spin Variance Entanglement Test}

\qquad \qquad \textit{4.1.2 Validity of Hillery Test for Local SSR Compliant
States}

\qquad \qquad \textit{4.1.3 Non-Applicable Entanglement Test Involving }$%
|\left\langle \widehat{\mathit{S}}_{z}\right\rangle |$

\qquad \qquad \textit{4.1.4 Hillery Entanglement Test - Multi-Mode Case}

\qquad \textbf{4.2 He et al 2012}

\qquad \qquad \textit{4.6.1 Spin Operator Tests for Entanglement}

\qquad \textbf{4.3 Raymer et al 2003}

\qquad \textbf{4.4 Sorensen et al 2001}

\qquad \qquad \textit{4.4.1 Sorensen Spin Squeezing Entanglement Test}

\qquad \qquad \textit{4.4.2 Revising Sorensen Spin Squeezing Test -
Localised Modes}

\qquad \qquad \textit{4.4.3 Revising Sorensen Spin Squeezing Test -
Separable State of Single Modes}

\qquad \qquad \textit{4.4.4 Revising Sorensen Spin Squeezing Test -
Separable State of Pairs of Modes with Single Boson Occupancy}

\qquad \textbf{4.5 Sorensen and Molmer 2001\medskip }

{\Large 5. Correlation Tests for Entanglement}

\qquad \textbf{5.1 Dalton et al 2014}

\qquad \qquad \textit{5.1.1 Weak Correlation Test for Local SSR Compatible
States}

\qquad \textbf{5.2 Hillery et al 2009}

\qquad \qquad \textit{5.2.1 Hillery Strong Correlation Entanglement Test}

\qquad \qquad \textit{5.2.2 Applications of Correlation Tests for
Entanglement}

\qquad \textbf{5.3 He et al 2012}

\qquad \qquad \textit{5.3.1 Correlation Tests for Entanglement\medskip }

{\Large 6. Quadrature Squeezing Tests for Entanglement}

\qquad \textbf{6.1 Duan et al 2000}

\qquad \qquad \textit{6.1.1 Two Distinguishable Particles}

\qquad \qquad \textit{6.1.2 Two Mode Systems of Identical Bosons}

\qquad \qquad \textit{6.1.3 Non SSR Compliant States}

\qquad \textbf{6.2 Reid 1989}\textit{\medskip }

\qquad \textbf{6.3 Two Mode Quadrature Squeezing}

{\Large 7. Interferometry in Bosonic Systems}

\qquad \textbf{7.1 Simple Two Mode Interferometer}

\qquad \textbf{7.2 General Two Mode Interferometers}

\qquad \textbf{7.3 Measurements in Simple Two Mode Interferometer}

\qquad \qquad \textit{7.3.1 Tomography in xy Plane - Beam Splitter}

\qquad \qquad \textit{7.3.2 Tomography in yz Plane}

\qquad \qquad \textit{7.3.3 Phase Changer}

\qquad \qquad \textit{7.3.4 Other Measurements in Simple Two Mode
Interferometer}

\qquad \textbf{7.4 Multi-Mode Interferometers}

\qquad \textbf{7.5 Application to Squeezing Tests for Entanglement}

\qquad \qquad \textit{7.5.1 Spin Squeezing in }$\widehat{S}_{x}$, \textit{\ }%
$\widehat{S}_{y}$

\qquad \qquad \textit{7.5.2 Spin Squeezing in }$xy$ \textit{Plane}

\qquad \qquad \textit{7.5.3 Measurement of }$\left\langle \widehat{\mathit{S}%
}_{z}\right\rangle $

\qquad \qquad \textit{7.5.4 Spin Squeezing in }$\widehat{S}_{z}$, \textit{\ }%
$\widehat{S}_{y}$

\qquad \textbf{7.6 Application to Correlation Tests for Entanglement}

\qquad \qquad \textit{7.6.1 First Order Correlation Test}

\qquad \qquad \textit{7.6.2 Second Order Correlation Test}

\qquad \textbf{7.7 Application to Quadrature Tests for Entanglement}{\Large %
\medskip }

{\Large 8. Experiments on Spin Squeezing}

\qquad \textbf{8.1 Esteve et al (2008)}

\qquad \textbf{8.2 Reidel et al (2010)}

\qquad \textbf{8.3 Gross et al (2010)}

\qquad \textbf{8.4 Gross et al (2011)}{\Large \medskip }

{\Large 9. Discussion and Summary of Key Results\bigskip }

\textbf{References}

\textbf{Acknowledgements\bigskip }

{\Large Appendices\bigskip }

{\Large A. Spin Squeezing Test for Bipartite Multi-Mode Systems}

\qquad \textbf{A.1 Mode Expansions}

\qquad \textbf{A.2 Positive Definiteness}{\Large \medskip }

{\Large B. Spin Squeezing Test for Other Multi-Mode Systems}

\qquad \textbf{B.1 Single Mode Sub-Systems}

\qquad \textbf{B.2 Variance }$\left\langle \Delta \widehat{S}%
_{x}^{2}\right\rangle $\textbf{\ Two Mode Sub-Systems{\Large \medskip }}

{\Large C. Hillery Spin Variance - Multi\_Mode}\textbf{\Large \medskip }

{\Large D. Derivation of Sorensen et al Results\medskip }

{\Large E. Heisenberg Uncertainty Principle Results\medskip }

{\Large F. "Separable but Non-Local" States\medskip }

{\Large G. Derivation of Interferometer Results}

\qquad \textbf{H.1 General Theory - Two Mode Interferometer}

\qquad \textbf{H.2 Beam Splitter and Phase Changer}

\qquad \textbf{H.3 Other Measurables}

\qquad \textbf{H.4 General Theory - Multi-Mode Interferometer}{\Large %
\medskip }

{\Large H. Limits on Interferometer Tests\medskip }

{\Large I. Relative Phase State\medskip \bigskip }

\pagebreak

\section{Introduction}

\label{Section- - Intrduction}

The previous paper I\textbf{\ }dealt with the general features of quantum
entanglement and its specific definition in the case of systems of identical
bosons. In defining entanglement for systems of identical massive particles,
it was concluded that the single particle states or modes are the most
appropriate choice for sub-systems that are distinguishable. Further, it was
conclude that the general quantum states must comply both with the
symmetrisation principle and the super-selection rules (SSR) forbidding
quantum superpositions of states with differing total particle number
(global SSR compliance). As a consequence, it was then reasoned that in the
separable states quantum superpositions of sub-system states with differing
sub-system particle number (local SSR compliance) do not occur \cite%
{Dalton14a}. Other approaches - such as sub-systems consisting of labelled
indistinguishable particles and entanglement due to symmetrisation \cite%
{Peres93a} or allowing for non-entangled separable but non-local states \cite%
{Verstraete03a}- were found to be unsuitable. The local (and global) SSR
compliant definition of entanglement used here was justified on the basis of
there being no non-relativistic quantum processes available to create SSR
non-compliant states and alternatively on the absence of a phase reference 
\cite{Bartlett06a}.

Paper I can be summarised as follows. Section \textbf{2 }covered the key
definitions of entangled states, the relationship to hidden variable theory
and some of the key paradoxes associated with quantum entanglement such as
EPR and Bell inequalities. A detailed discussion on why the symmetrisation
principle and the super-selection rule is invoked for entanglement in
identical particle systems was discussed in Section \textbf{3}. Challenges
to the necessity of the super-selection rule were outlined, with arguments
against such challenges dealt with in Appendices \textbf{D} and \textbf{E}.
Two key mathematical inequalities were derived in Appendix \textbf{B }and
details about the spin EPR paradox set out in Appendix \textbf{C}. The final
Section \textbf{4} summarised the key features of quantum entanglement
discussed in the paper.

The present paper II focuses on tests for entanglement in two mode systems
of identical bosons, with particular emphasis on spin squeezing and
correlation tests and how the quantities involved in these tests can be
measured via two mode interferometry. Two mode bosonic systems are of
particular interest because cold atomic gases cooled well below the
Bose-Einstein condensation (BEC) transition temperature can be prepared
where essentially only two modes are occupied (\cite{Leggett01a}, \cite%
{Dalton12a}). This can be achieved for cases involving a single hyperfine
components using a double well trap potential or for two hyperfine
components using a single well. At higher temperatures more than two modes
may be occupied, so multi-mode systems are also of importance and the two
mode treatment is extended to this situation.

As well as their relevance for entanglement tests, states that are spin
squeezed have important applications in \emph{quantum metrology}\textbf{. }%
That squeezed states can improve interferometry via the quantum noise in
quadrature variables being reduced to below the standard quantum limit has
been known since the pioneeering work of Caves \cite{Caves81a} on optical
systems. The extension to spin squeezing in systems of massive bosons
originates with the work of Kitagawa and Ueda \cite{Kitagawa93a}, who
considered systems of two state atoms. As this review is focused on spin
squeezing as an entanglement test rather than the use of spin squeezing in
quantum metrology, the latter subject will not be covered here. In quantum
metrology involving spin operators the quantity $\sqrt{\left\langle \Delta 
\widehat{S}\,_{x}^{2}\right\rangle }/|\left\langle \widehat{S}%
\,_{z}\right\rangle |$ (which involves the variance and mean value of
orthogonal spin operators) is a measure of the uncertainty $\Delta \theta $
in measuring the interferometer\textbf{\ }\emph{phase}\textbf{. }The
interest in spin squeezing lies in the possibility of improvement over the 
\emph{standard quantum limit} where $\Delta \theta =1/\sqrt{N}$ (see
Subsection \ref{SubSection - Binomial State}). As we will see, for squeezed
states $\sqrt{\left\langle \Delta \widehat{S}\,_{x}^{2}\right\rangle }<\sqrt{%
|\left\langle \widehat{S}\,_{z}\right\rangle |/2}$ so we could have $\Delta
\theta <\sim 1/\sqrt{|\left\langle \widehat{S}\,_{z}\right\rangle |}\,\sim
\,1/\sqrt{N}$ which is less than the standard quantum limit. In SubSection %
\ref{SubSection - Ent States that are Spin Sq} we give an example of a
highly squeezed state where $\Delta \theta \sim \sqrt{\ln N}/N$ which is
near the \emph{Heisenberg limit}. Suffice to say that increasing the number
of particles in the squeezed state has the effect of improving the
sensitivity of the interferometer. Aspects of quantum metrology are covered
in a number of papers (see \cite{Pezze09a}, \cite{Berry00a}), based on
concepts such as quantum Fisher information, Cramers-Rao bound \cite%
{Braunstein94a}, \cite{Helstrom76a}, quantum phase eigenstates.

The proof of the key conclusion that spin squeezing in any spin component is
a sufficiency test for entanglement \cite{Dalton14a} is set out in this
paper, as is that for a new Bloch vector test. A previous proof \cite%
{Sorensen01a} that spin squeezing in the $z$ spin component $\sqrt{%
\left\langle \Delta \widehat{S}\,_{z}^{2}\right\rangle }<\sqrt{|\left\langle 
\widehat{S}\,_{x}\right\rangle |/2}$demonstrates entanglement based on
treating identical bosonic atoms as distinguishable sub-systems has
therefore now been superceded. It is seen that correlation tests for
entanglement of quantum states complying with the global particle number SSR
can be expressed in terms of inequalities involving powers of spin
operators. Section \ref{Section - Spin Squeezing} sets out the definitions
of spin squeezing and in the following Section \ref{Section - Relationship
Spin Squeezing & Entanglement} it is shown that spin squeezing is a
signature of entanglement, both for the original spin operators with
entanglement of the original modes, for the principle spin operators with
entanglement of the two new modes and finally for several multi-mode cases.
Details of the latter are set out in Appendices \ref{Appendix = MultiMo Spin
Sq Choice 1} and \ref{Appendix - Revised Sorensen}. A number of other
correlation, spin operator and quadrature operator tests for entanglement
proposed by other authors are considered in Sections \ref{Section - Criteria
for Spin Squeezing Based on Non-Physical States}\textbf{,} \ref{Section -
Correlation Tests for Entanglement} and \ref{Section - Quadrature Tesrs for
Entanglement}, with details of these treatments set out in Appendices \ref%
{Appendix - Sorensen Results} and \ref{Appendix - Heisenberg Uncertainty
Principle Results}. Some tests also apply in cases of multi-mode
entanglement, though with restrictions in the case of sub-systems each
consisting of pairs of modes. A new weak correlation test is derived and for
one case is equivalent to the Bloch vector test.

In Section\textbf{\ }\ref{section - Interferometry in Two Mode BEC}\textbf{\ 
}it is shown that a simple two mode interferometer can be used to measure
the mean values and covariance matrix for the spin operators involved in
entanglement tests, with details covered in Appendices \ref{Appendix -
Derivatio n of Interferometer Result}\textbf{\ }and \ref{Appendix - Limits
on Interferometry Tests}. The treatment is also generalised to cover
multi-mode interferometry. Actual experiments aimed at detecting
entanglement via spin squeezing tests are examined in Section \ref{Section -
Experiments}\textbf{. }The final Section \ref{Section - Discussion & Summary
of Key Results} summarises and discusses the key results regarding
entanglement tests. Appendices \ref{Appendix - Separable but Non Local
States}\textbf{\ }and \ref{Appendix - Relative Phase State}\textbf{\ }%
provide details regarding certain important states whose features are
discussed in this paper - the "separable but non-local " states and the
relative phase eigenstate.

\pagebreak

\section{Spin Squeezing}

\label{Section - Spin Squeezing}

The basic concept of spin squeezing was first introduced by Kitagawa and
Ueda \cite{Kitagawa93a} for general spin systems. These include cases based
on two mode systems, such as may occur both for optical fields and for
Bose-Einstein condensates. Though focused on systems of massive identical
bosons, the treatment in this paper also applies to photons though details
will differ.

\subsection{Spin Operators, Bloch Vector and Covariance Matrix}

\label{SubSection - Spin Operators}

\subsubsection{Spin Operators}

For two mode systems with mode annihilation operators $\widehat{a}$, $%
\widehat{b}$ associated with the two single particle states $\left\vert \phi
_{a}\right\rangle $, $\left\vert \phi _{b}\right\rangle $, and where the
non-zero bosonic commutation rules are $[\widehat{e},\widehat{e}^{\dag }]=%
\widehat{1}$ ($\widehat{e}=\widehat{a}$ or $\widehat{b}$), Schwinger \emph{%
spin angular momentum operators} $\widehat{S}_{\xi }$ ($\xi =x,y,z$) are
defined as 
\begin{equation}
\widehat{S}_{x}=(\widehat{b}^{\dag }\widehat{a}+\widehat{a}^{\dag }\widehat{b%
})/2\qquad \widehat{S}_{y}=(\widehat{b}^{\dag }\widehat{a}-\widehat{a}^{\dag
}\widehat{b})/2i\qquad \widehat{S}_{z}=(\widehat{b}^{\dag }\widehat{b}-%
\widehat{a}^{\dag }\widehat{a})/2  \label{Eq.OldSpinOprs}
\end{equation}%
and which satisfy the commutation rules $[\widehat{S}_{\xi }$ $,\widehat{S}%
_{\mu }$ $]=i\epsilon _{\xi \mu \lambda }\widehat{S}_{\lambda }$ for angular
momentum operators. For bosons the square of the angular momentum operators
is given by $\widehat{S}_{x}^{2}+\widehat{S}_{y}^{2}+\widehat{S}_{z}^{2}=(%
\widehat{N}/2)(\widehat{N}/2+1)$, where $\widehat{N}=(\widehat{b}^{\dag }%
\widehat{b}+\widehat{a}^{\dag }\widehat{a})$ is the boson total number
operator, those for the separate modes being $\widehat{n}_{e}=\widehat{e}%
^{\dag }\widehat{e}$ ($\widehat{e}=\widehat{a}$ or $\widehat{b}$). The
Schwinger spin operators are the second quantization form of symmetrized one
body operators $\widehat{S}_{x}=\sum_{i}(\left\vert \phi
_{b}(i)\right\rangle \left\langle \phi _{a}(i)\right\vert +\left\vert \phi
_{a}(i)\right\rangle \left\langle \phi _{b}(i)\right\vert )/2$ ; $\widehat{S}%
_{y}=\sum_{i}(\left\vert \phi _{b}(i)\right\rangle \left\langle \phi
_{a}(i)\right\vert -\left\vert \phi _{a}(i)\right\rangle \left\langle \phi
_{b}(i)\right\vert )/2i$ ; $\widehat{S}_{z}=\sum_{i}(\left\vert \phi
_{b}(i)\right\rangle \left\langle \phi _{b}(i)\right\vert -\left\vert \phi
_{a}(i)\right\rangle \left\langle \phi _{a}(i)\right\vert )/2$ , where the
sum $i$ is over the identical bosonic particles. In the case of the two mode
EM\ field the spin angular momentum operators are related to the Stokes
parameters.

\subsubsection{Bloch Vector and Covariance Matrix}

If the density operator for the overall system is $\widehat{\rho }$ then
expectation values of the three spin operators $\left\langle \widehat{S}%
_{\xi }\right\rangle =Tr(\widehat{\rho }\widehat{S}_{\xi })$ ($\xi =x,y,z$)
define the \emph{Bloch vector}. Spin squeezing is related to the fluctuation
operators $\Delta \widehat{S}_{\xi }=\widehat{S}_{\xi }-\left\langle 
\widehat{S}_{\xi }\right\rangle $, in terms of which a real, symmetric \emph{%
covariance matrix} $C(\widehat{S}_{\xi },\widehat{S}_{\mu })$ ($\xi ,\mu
=x,y,z$) is defined \cite{Jaaskelainen06a}, \cite{Dalton12a} via%
\begin{eqnarray}
C(\widehat{S}_{\xi },\widehat{S}_{\mu }) &=&(\left\langle \Delta \widehat{S}%
_{\xi }\,\Delta \widehat{S}_{\mu }\right\rangle +\left\langle \Delta 
\widehat{S}_{\mu }\,\Delta \widehat{S}_{\xi }\right\rangle )/2  \nonumber \\
&=&\left\langle \widehat{S}_{\xi }\,\widehat{S}_{\mu }+\widehat{S}_{\mu }\,%
\widehat{S}_{\xi }\right\rangle /2-\left\langle \widehat{S}_{\xi
}\right\rangle \left\langle \widehat{S}_{\mu }\right\rangle
\label{Eq.CovMatrix}
\end{eqnarray}%
and whose diagonal elements $C(\widehat{S}_{\xi },\widehat{S}_{\xi
})=\left\langle \Delta \widehat{S}_{\xi }{}^{2}\right\rangle $ gives the
variance for the fluctuation operators. The covariance matrix is also \emph{%
positive definite}. The variances for the spin operators satisfy the three
Heisenberg uncertainty principle reations $\left\langle \Delta \widehat{S}%
_{x}{}^{2}\right\rangle \left\langle \Delta \widehat{S}_{y}{}^{2}\right%
\rangle \geq \frac{1}{4}|\left\langle \widehat{S}_{z}\right\rangle |^{2}$; $%
\left\langle \Delta \widehat{S}_{y}{}^{2}\right\rangle \left\langle \Delta 
\widehat{S}_{z}{}^{2}\right\rangle \geq \frac{1}{4}|\left\langle \widehat{S}%
_{x}\right\rangle |^{2}$; $\left\langle \Delta \widehat{S}%
_{z}{}^{2}\right\rangle \left\langle \Delta \widehat{S}_{x}{}^{2}\right%
\rangle \geq \frac{1}{4}|\left\langle \widehat{S}_{y}\right\rangle |^{2}$,
and spin squeezing is defined via conditions such as $\left\langle \Delta 
\widehat{S}_{x}{}^{2}\right\rangle <\frac{1}{2}|\left\langle \widehat{S}%
_{z}\right\rangle |$ with $\left\langle \Delta \widehat{S}%
_{y}{}^{2}\right\rangle >\frac{1}{2}|\left\langle \widehat{S}%
_{z}\right\rangle |,$ for $\widehat{S}_{x}{}$ being squeezed compared to $%
\widehat{S}_{y}$ and so on. Spin squeezing in these components is relevant
to tests for entanglement of the modes $\widehat{a}$ and $\widehat{b}$, as
will be shown later. Spin squeezing in rotated components is also important,
in particular in the so-called \emph{principal} components for which the
covariance matrix is diagonal.

\subsection{Spin Operators - Multi-Mode Case}

\label{SubSection - Spin Operators Multimode Case}

As well as spin operators for the simple case of two modes we can also
define spin operators in multimode cases involving two sub-systems $A$ and $%
B $. For example, there may be two types of bosonic particle involved, each 
\emph{component} distinguished from the other by having different hyperfine
internal states $\left\vert A\right\rangle ,\left\vert B\right\rangle $.
Each component may be associated with a complete orthonormal set of \emph{%
spatial mode functions} $\phi _{ai}(\mathbf{r})$ and $\phi _{bi}(\mathbf{r)}$%
, so there will be two sets of modes $\left\vert \phi _{ai}\right\rangle $, $%
\left\vert \phi _{bi}\right\rangle $, where in the $\left\vert \mathbf{r}%
\right\rangle $ representation we have $\left\langle \mathbf{r\,|}\phi
_{ai}\right\rangle =\phi _{ai}(\mathbf{r})\left\vert A\right\rangle $ and $%
\left\langle \mathbf{r\,|}\phi _{bi}\right\rangle =\phi _{bi}(\mathbf{r}%
)\left\vert B\right\rangle $. Mode orthogonality between $A$ and $B$ modes
arises from $\left\langle A|B\right\rangle =0$ rather from the spatial mode
functions being orthogonal.

We can define \emph{spin operators} for the combined \emph{multimode} $A$
and $B$ sub-systems \cite{Li09a}\textbf{\ }via 
\begin{eqnarray}
\widehat{S}_{x} &=&\frac{1}{2}\tint d\mathbf{r\,}\left( \widehat{\Psi }%
_{b}^{\dag }(\mathbf{r})\widehat{\Psi }_{a}(\mathbf{r})+\widehat{\Psi }%
_{a}^{\dag }(\mathbf{r})\widehat{\Psi }_{b}(\mathbf{r})\right)  \nonumber \\
\widehat{S}_{y} &=&\frac{1}{2i}\tint d\mathbf{r\,}\left( \widehat{\Psi }%
_{b}^{\dag }(\mathbf{r})\widehat{\Psi }_{a}(\mathbf{r})-\widehat{\Psi }%
_{a}^{\dag }(\mathbf{r})\widehat{\Psi }_{b}(\mathbf{r})\right)  \nonumber \\
\widehat{S}_{z} &=&\frac{1}{2}\tint d\mathbf{r\,}\left( \widehat{\Psi }%
_{b}^{\dag }(\mathbf{r})\widehat{\Psi }_{b}(\mathbf{r})-\widehat{\Psi }%
_{a}^{\dag }(\mathbf{r})\widehat{\Psi }_{a}(\mathbf{r})\right)
\label{Eq.SpinFieldOprs}
\end{eqnarray}%
where the field operators satisfy the non-zero commutation rules%
\begin{equation}
\lbrack \widehat{\Psi }_{c}(\mathbf{r}),\widehat{\Psi }_{d}^{\dag }(\mathbf{r%
}^{\prime })]=\delta _{cd}\,\delta (\mathbf{r}-\mathbf{r}^{\prime })\qquad
c,d=a,b  \label{Eq.FldOprCommRules}
\end{equation}%
It is then easy to show that the standard spin angular momentum conmmutation
rules are satisfied. $[\widehat{S}_{\xi }$ $,\widehat{S}_{\mu }$ $%
]=i\epsilon _{\xi \mu \lambda }\widehat{S}_{\lambda }$.

For convenience we can expand the field operators in terms of an orthonormal
set of spatial mode functions $\phi _{i}(\mathbf{r})$. We can choose the
spatial mode functions to be the same $\phi _{ai}(\mathbf{r})=\phi _{bi}(%
\mathbf{r)=}\,\phi _{i}(\mathbf{r})$ (these might be momentum
eigenfunctions) and then the field annihilation operators for each component
are 
\begin{equation}
\widehat{\Psi }_{a}(\mathbf{r})=\tsum\limits_{i}\widehat{a}_{i}\,\,\phi _{i}(%
\mathbf{r})\qquad \widehat{\Psi }_{b}(\mathbf{r})=\tsum\limits_{i}\widehat{b}%
_{i}\,\,\phi _{i}(\mathbf{r})  \label{Eq.FieldOprs}
\end{equation}%
These expansions are consistent with the field operator commutation rules (%
\ref{Eq.FldOprCommRules}) based on the usual non-zero mode operator
commutation rules $[\widehat{a}_{i},\widehat{a}_{j}^{\dag }]=[\widehat{b}%
_{i},\widehat{b}_{j}^{\dag }]=\delta _{ij}$.

By substituting for the field operators we can then express the spin
operators in terms of mode operators as 
\begin{eqnarray}
\widehat{S}_{x} &=&\frac{1}{2}\tsum\limits_{i}\left( \widehat{b}_{i}^{\dag }%
\widehat{a}_{i}+\widehat{a}_{i}^{\dag }\widehat{b}_{i}\right) \quad \widehat{%
S}_{y}=\frac{1}{2i}\tsum\limits_{i}\left( \widehat{b}_{i}^{\dag }\widehat{a}%
_{i}-\widehat{a}_{i}^{\dag }\widehat{b}_{i}\right) \quad \widehat{S}_{z}=%
\frac{1}{2}\tsum\limits_{i}\left( \widehat{b}_{i}^{\dag }\widehat{b}_{i}-%
\widehat{a}_{i}^{\dag }\widehat{a}_{i}\right)  \nonumber \\
&&  \label{Eq.SpinOprs2}
\end{eqnarray}%
and it is then easy to confirm that the standard spin angular momentum
conmmutation rules are satisfied. $[\widehat{S}_{\xi }$ $,\widehat{S}_{\mu }$
$]=i\epsilon _{\xi \mu \lambda }\widehat{S}_{\lambda }$. We now have both
field and mode expressions for spin operators in multimode cases involving
two sub-systems $A$ and $B$.

Finally, the\emph{\ total number of particles} is given by 
\begin{eqnarray}
\widehat{N} &=&\tint d\mathbf{r\,}\left( \widehat{\Psi }_{b}^{\dag }(\mathbf{%
r})\widehat{\Psi }_{b}(\mathbf{r})+\widehat{\Psi }_{a}^{\dag }(\mathbf{r})%
\widehat{\Psi }_{a}(\mathbf{r})\right)  \nonumber \\
&=&\tsum\limits_{i}\left( \widehat{b}_{i}^{\dag }\widehat{b}_{i}+\widehat{a}%
_{i}^{\dag }\widehat{a}_{i}\right)  \nonumber \\
&=&\widehat{N}_{b}+\widehat{N}_{a}  \label{Eq.NumberOprsFields}
\end{eqnarray}%
in an obvious notation.

\subsection{New Spin Operators and Principal Spin Fluctuations}

The covariance matrix has real, non-negative eigenvalues and can be
diagonalised via an orthogonal \emph{rotation matrix} $M(-\alpha ,-\beta
,-\gamma )$ that defines \emph{new spin angular momentum operators} $%
\widehat{J}_{\xi }$ ($\xi =x,y,z$) via 
\begin{equation}
\widehat{J}_{\xi }=\sum_{\mu }M_{\xi \mu }(-\alpha ,-\beta ,-\gamma )%
\widehat{S}_{\mu }  \label{Eq.NewSpinOprsRotnMatrix}
\end{equation}%
and where 
\begin{eqnarray}
C(\widehat{J}_{\xi },\widehat{J}_{\mu }) &=&\sum_{\lambda \theta }M_{\xi
\lambda }(-\alpha ,-\beta ,-\gamma )C(\widehat{S}_{\lambda },\widehat{S}%
_{\theta })M_{\mu \theta }(-\alpha ,-\beta ,-\gamma )  \nonumber \\
&=&\delta _{\xi \mu }\left\langle \Delta \widehat{J}_{\xi
}{}^{2}\right\rangle  \label{Eq.DiagnCovMatrix}
\end{eqnarray}%
is the covariance matrix for the new spin angular momentum operators $%
\widehat{J}_{\xi }$ ($\xi =x,y,z$), and which is \emph{diagonal} with the
diagonal elements $\left\langle \Delta \widehat{J}_{x}{}^{2}\right\rangle
,\left\langle \Delta \widehat{J}_{y}{}^{2}\right\rangle $ and $\left\langle
\Delta \widehat{J}_{z}{}^{2}\right\rangle $ giving the so-called \emph{%
principal spin fluctuations}. The matrix $M(\alpha ,\beta ,\gamma )$ is
parameterised in terms of three Euler angles $\alpha ,\beta ,\gamma $ and is
given in \cite{Rose57a} (see Eq. (4.43)).

The Bloch vector and spin fluctuations are illustrated in Figure 1. In Fig 1
the Bloch vector and spin fluctuation ellipsoid is shown in terms of the
original spin operators $\widehat{S}_{\xi }$ ($\xi =x,y,z$)

\bigskip

\includegraphics[width=4.1442in]{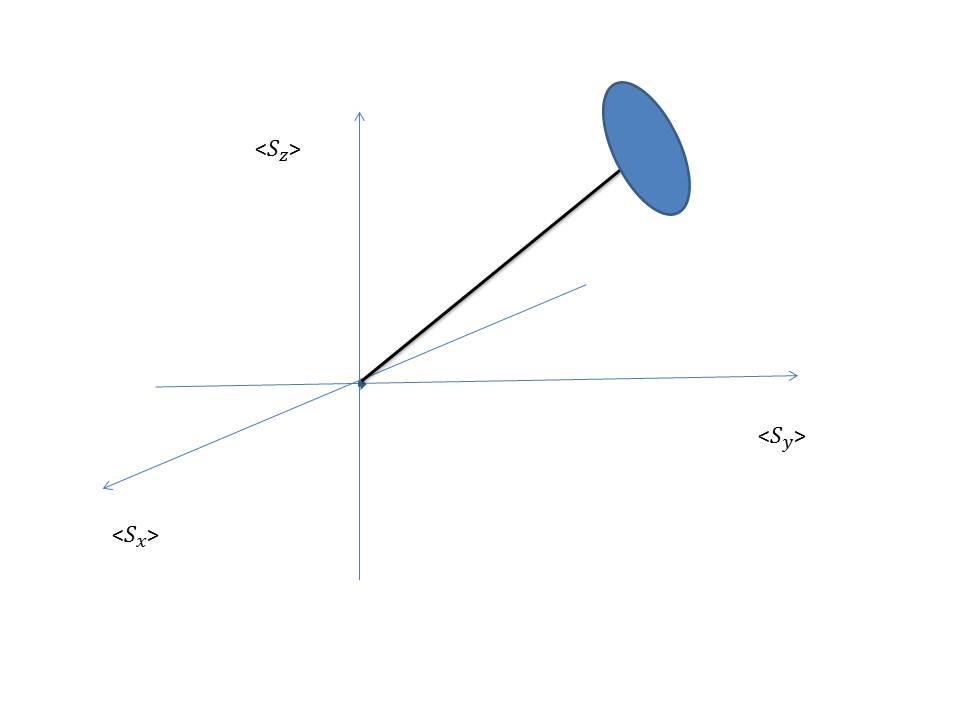}
%
\medskip

\begin{center}
Figure 1. Bloch vector and spin fluctuations shown for original spin
operators.

\bigskip
\end{center}

These rules also apply to multimode spin operators as defined in SubSection %
\ref{SubSection - Spin Operators Multimode Case}.

\begin{center}
\pagebreak
\end{center}

\subsection{Spin Squeezing Definitions}

\label{SubSection - Spin Squeezing New Spin Oprs}

We will begin by considering the case of the spin operators in the most
general case. We will also specifically consider spin squeezing for the two
new modes and for the multi-mode situation.

\subsubsection{Heisenberg Uncertainty Principle and Spin Squeezing}

Since the spin operators also satisfy \emph{Heisenberg uncertainty principle}
relationships 
\begin{eqnarray}
\left\langle \Delta \widehat{S}_{x}{}^{2}\right\rangle \left\langle \Delta 
\widehat{S}_{y}{}^{2}\right\rangle &\geq &\frac{1}{4}|\left\langle \widehat{S%
}_{z}\right\rangle |^{2}  \nonumber \\
\left\langle \Delta \widehat{S}_{y}{}^{2}\right\rangle \left\langle \Delta 
\widehat{S}_{z}{}^{2}\right\rangle &\geq &\frac{1}{4}|\left\langle \widehat{S%
}_{x}\right\rangle |^{2}  \nonumber \\
\left\langle \Delta \widehat{S}_{z}{}^{2}\right\rangle \left\langle \Delta 
\widehat{S}_{x}{}^{2}\right\rangle &\geq &\frac{1}{4}|\left\langle \widehat{S%
}_{y}\right\rangle |^{2}  \label{Eq.HeisenbergUncert}
\end{eqnarray}%
\emph{spin squeezing} will now be defined for the\emph{\ spin operators} via
condtions such as 
\begin{eqnarray}
\left\langle \Delta \widehat{S}_{x}{}^{2}\right\rangle &<&\frac{1}{2}%
|\left\langle \widehat{S}_{z}\right\rangle |\;and\;\left\langle \Delta 
\widehat{S}_{y}{}^{2}\right\rangle >\frac{1}{2}|\left\langle \widehat{S}%
_{z}\right\rangle |  \nonumber \\
\left\langle \Delta \widehat{S}_{y}{}^{2}\right\rangle &<&\frac{1}{2}%
|\left\langle \widehat{S}_{x}\right\rangle |\;and\;\left\langle \Delta 
\widehat{S}_{z}{}^{2}\right\rangle >\frac{1}{2}|\left\langle \widehat{S}%
_{x}\right\rangle |  \nonumber \\
\left\langle \Delta \widehat{S}_{z}{}^{2}\right\rangle &<&\frac{1}{2}%
|\left\langle \widehat{S}_{y}\right\rangle |\;and\;\left\langle \Delta 
\widehat{S}_{x}{}^{2}\right\rangle >\frac{1}{2}|\left\langle \widehat{S}%
_{y}\right\rangle |  \label{Eq.SpinSqueezingJXJY}
\end{eqnarray}%
for $\widehat{S}_{x}{}$being squeezed compared to $\widehat{S}_{y}$, and so
on.

Note also that the Heisenberg uncertainty principle proof (based on

$\left\langle \left( \Delta \widehat{S}_{\alpha }+i\lambda \Delta \widehat{S}%
_{\beta }\right) \left( \Delta \widehat{S}_{\alpha }-i\lambda \Delta 
\widehat{S}_{\beta }\right) \right\rangle \geq 0$ for all real $\lambda $)
also establishes the general result for all quantum states 
\begin{equation}
\left\langle \Delta \widehat{S}_{\alpha }{}^{2}\right\rangle +\left\langle
\Delta \widehat{S}_{\beta }{}^{2}\right\rangle \geq |\left\langle \widehat{S}%
_{\gamma }\right\rangle |  \label{Eq.GeneralVarianceSumResult}
\end{equation}%
where $\alpha $, $\beta $ and $\gamma $ are $x$, $y$ and $z$ in cyclic order.

Since the two new mode spin operators defined in Eq. (\ref%
{Eq.NewSpinOprsRotnMatrix}) satisy the standard angular momentum operator
commutation rules, the usual Heisenberg Uncertainty rules analogous to (\ref%
{Eq.HeisenbergUncert}) apply, so that spin squeezing can also exist in the
two mode cases involving the\emph{\ new spin operators} $\widehat{J}_{x},%
\widehat{J}_{y}$ and $\widehat{J}_{z}$ as well. These uncertainty principle
features also apply to multimode spin operators as defined in SubSection \ref%
{SubSection - Spin Operators Multimode Case}.

It should be noted that finding spin squeezing for one principal spin
operator $\widehat{J}_{y}$ with respect to another $\widehat{J}_{x}$ does 
\emph{not} mean that there is spin squeezing for \emph{any} of the\emph{\ }%
old spin operators $\widehat{S}_{x},\widehat{S}_{y}$ and $\widehat{S}_{z}$ \
In the case of the relative phase eigenstate (see SubSection \ref{SubSection
- Ent States that are Spin Sq}) $\widehat{J}_{y}$ is squeezed with respect
to $\widehat{J}_{x}$ - however none of the old spin components are spin
squeezed.

\subsubsection{Alternative Spin Squeezing Criteria}

\emph{Other criteria} for spin squeezing are also used, for example in the
article by Wineland et al \cite{Wineland94a} . To focus on spin squeezing
for $\widehat{S}_{z}{}$compared to \emph{any} orthogonal spin operators we
can combine the second and third Heisenberg uncertainty principle
relationships to give%
\begin{equation}
\left\langle \Delta \widehat{S}_{z}{}^{2}\right\rangle \left( \left\langle
\Delta \widehat{S}_{x}{}^{2}\right\rangle +\left\langle \Delta \widehat{S}%
_{y}{}^{2}\right\rangle \right) \geq \frac{1}{4}\left( |\left\langle 
\widehat{S}_{x}\right\rangle |^{2}+|\left\langle \widehat{S}%
_{y}\right\rangle |^{2}\right)  \label{Eq.AdditionalHeisebergUncertainty}
\end{equation}%
Then we may define two new spin operators via%
\begin{equation}
\widehat{S}_{\perp \,1}=\cos \theta \;\widehat{S}_{x}+\sin \theta \;\widehat{%
S}_{y}\qquad \widehat{S}_{\perp \,2}=-\sin \theta \;\widehat{S}_{x}+\cos
\theta \;\widehat{S}_{y}  \label{Eq.NewOrthogSpinOprs}
\end{equation}%
where $\theta $ corresponds to a rotation angle in the $xy$ plane, and which
satisfy the standard angular momentun commutation rules $[\widehat{S}_{\perp
\,1}$ $,\widehat{S}_{\perp \,2}]=i\widehat{S}_{z}$, $[\widehat{S}_{\perp
\,2} $ $,\widehat{S}_{z}$ $]=i\widehat{S}_{\perp \,1}$, $[\widehat{S}_{z}$ $,%
\widehat{S}_{\perp \,1}]=i\widehat{S}_{\perp \,2}$. It is straightforward to
show that $\left\langle \Delta \widehat{S}_{x}{}^{2}\right\rangle
+\left\langle \Delta \widehat{S}_{y}{}^{2}\right\rangle =\left\langle \Delta 
\widehat{S}_{\perp \,1}{}^{2}\right\rangle +\left\langle \Delta \widehat{S}%
_{\perp \,2}{}^{2}\right\rangle $ and $|\left\langle \widehat{S}_{\perp
\,1}\right\rangle |^{2}+|\left\langle \widehat{S}_{\perp \,2}\right\rangle
|^{2}=$ $|\left\langle \widehat{S}_{x}\right\rangle |^{2}+|\left\langle 
\widehat{S}_{y}\right\rangle |^{2}$ so that 
\begin{equation}
\left\langle \Delta \widehat{S}_{z}{}^{2}\right\rangle \left( \left\langle
\Delta \widehat{S}_{\perp \,1}{}^{2}\right\rangle +\left\langle \Delta 
\widehat{S}_{\perp \,2}{}^{2}\right\rangle \right) \geq \frac{1}{4}\left(
|\left\langle \widehat{S}_{\perp \,1}\right\rangle |^{2}+|\left\langle 
\widehat{S}_{\perp \,2}\right\rangle |^{2}\right)
\label{Eq.NewHeisenbergUncertaintyPpl}
\end{equation}%
so that \emph{spin squeezing} for $\widehat{S}_{z}{}$compared to \emph{any
two }orthogonal spin operators such as $\widehat{S}_{\perp \,1}$ or $%
\widehat{S}_{\perp \,2}$ would be defined as 
\begin{eqnarray}
\left\langle \Delta \widehat{S}_{z}{}^{2}\right\rangle &<&\frac{1}{2}\sqrt{%
\left( |\left\langle \widehat{S}_{\perp \,1}\right\rangle
|^{2}+|\left\langle \widehat{S}_{\perp \,2}\right\rangle |^{2}\right) } 
\nonumber \\
&&and  \nonumber \\
\left\langle \Delta \widehat{S}_{\perp \,1}{}^{2}\right\rangle +\left\langle
\Delta \widehat{S}_{\perp \,2}{}^{2}\right\rangle &>&\frac{1}{2}\sqrt{\left(
|\left\langle \widehat{S}_{\perp \,1}\right\rangle |^{2}+|\left\langle 
\widehat{S}_{\perp \,2}\right\rangle |^{2}\right) }
\label{Eq.NewCriterionSpinSqueezing}
\end{eqnarray}%
For spin squeezing in $\left\langle \Delta \widehat{S}_{z}{}^{2}\right%
\rangle $ we require the \emph{spin squeezing parameter }$\xi $ to satisfy
an inequality 
\begin{equation}
\xi ^{2}=\frac{\left\langle \Delta \widehat{S}_{z}{}^{2}\right\rangle }{%
\left( |\left\langle \widehat{S}_{\perp \,1}\right\rangle
|^{2}+|\left\langle \widehat{S}_{\perp \,2}\right\rangle |^{2}\right) }<%
\frac{1}{2\sqrt{\left( |\left\langle \widehat{S}_{\perp \,1}\right\rangle
|^{2}+|\left\langle \widehat{S}_{\perp \,2}\right\rangle |^{2}\right) }}\sim 
\frac{1}{N}  \label{Eq.SpinSqueezingMeasure}
\end{equation}%
The last step is an approximation for an $N$ particle state based on the
assumption that the Bloch vector lies in the $xy$ plane and close to the
Bloch sphere, this situation being the most conducive to detecting the
fluctuation $\left\langle \Delta \widehat{S}_{z}{}^{2}\right\rangle $. In \
this situation $\sqrt{\left( |\left\langle \widehat{S}_{\perp
\,1}\right\rangle |^{2}+|\left\langle \widehat{S}_{\perp \,2}\right\rangle
|^{2}\right) }$ is approximately $N/2$. The condition $N\xi ^{2}<1$ is
sometimes taken as the condition for spin squeezing \cite{Toth09a}, but it
should be noted that this is approximate and Eq. (\ref%
{Eq.NewCriterionSpinSqueezing}) gives the correct expression.

\subsubsection{Planar Spin Squeezing}

A special case of recent interest is that referred to as \emph{planar
squeezing} \cite{He11b} in which the Bloch vector for a suitable choice of
spin operators lies in a \emph{plane} and along one of the \emph{axes}. If
this plane is chosen to be the $xy$ plane and the $x$ axis is chosen then $%
\left\langle \widehat{S}_{z}\right\rangle =0$ and $\left\langle \widehat{S}%
_{y}\right\rangle =0$, resulting in only one Heisenberg uncertainty
principle relationship where the right side is non-zero, namely $%
\left\langle \Delta \widehat{S}_{y}{}^{2}\right\rangle \left\langle \Delta 
\widehat{S}_{z}{}^{2}\right\rangle \geq \frac{1}{4}|\left\langle \widehat{S}%
_{x}\right\rangle |^{2}$. Combining this with $\left\langle \Delta \widehat{S%
}_{x}{}^{2}\right\rangle \left\langle \Delta \widehat{S}_{y}{}^{2}\right%
\rangle \geq 0$ gives $\left( \left\langle \Delta \widehat{S}%
_{y}{}^{2}\right\rangle +\left\langle \Delta \widehat{S}_{x}{}^{2}\right%
\rangle \right) \left\langle \Delta \widehat{S}_{z}{}^{2}\right\rangle \geq 
\frac{1}{4}|\left\langle \widehat{S}_{x}\right\rangle |^{2}$. So the total
spin fluctuation \emph{in} the $xy$ plane defined as $\left\langle \Delta 
\widehat{S}_{{\LARGE para}}{}^{2}\right\rangle =\left\langle \Delta \widehat{%
S}_{y}{}^{2}\right\rangle +\left\langle \Delta \widehat{S}%
_{x}{}^{2}\right\rangle $ will be squeezed compared to the spin fluctuation
perpendicular to the $xy$ plane given by $\left\langle \Delta \widehat{J}_{%
{\LARGE perp}}{}^{2}\right\rangle =\left\langle \Delta \widehat{S}%
_{z}{}^{2}\right\rangle $ if 
\begin{equation}
\left\langle \Delta \widehat{S}_{{\LARGE para}}{}^{2}\right\rangle <\frac{1}{%
2}|\left\langle \widehat{S}_{x}\right\rangle |\;and\;\left\langle \Delta 
\widehat{S}_{{\LARGE perp}}{}^{2}\right\rangle >\frac{1}{2}|\left\langle 
\widehat{S}_{x}\right\rangle |  \label{Eq.PlanarSpinSqg}
\end{equation}%
By minimising $\left\langle \Delta \widehat{S}_{{\LARGE para}%
}{}^{2}\right\rangle $ whilst satisfying the constraints $\left\langle 
\widehat{S}_{z}\right\rangle =\left\langle \widehat{S}_{y}\right\rangle =0$
a spin squeezed state is found that satisfies (\ref{Eq.PlanarSpinSqg}) with $%
\left\langle \Delta \widehat{S}_{{\LARGE para}}{}^{2}\right\rangle \;\symbol{%
126}\;J^{2/3}$, $\left\langle \Delta \widehat{S}_{{\LARGE perp}%
}{}^{2}\right\rangle \;\sim \;J^{4/3}$, $|\left\langle \widehat{S}%
_{x}\right\rangle |\;\sim \;J$ for large $J=N/2$ \cite{He11b}. The Bloch
vector is on the Bloch sphere.

\subsubsection{Spin Squeezing in Multi-Mode Cases}

Since the multi-mode spin operators defined in Eq.(\ref{Eq.SpinFieldOprs})
satisy the standard angular momentum operator commutation rules, the usual
Heisenberg Uncertainty rules analogous to (\ref{Eq.HeisenbergUncert}) apply,
so that spin squeezing can also exist in the multi-mode case as well. Thus

\begin{eqnarray}
\left\langle \Delta \widehat{S}_{x}{}^{2}\right\rangle &<&\frac{1}{2}%
|\left\langle \widehat{S}_{z}\right\rangle |\;and\;\left\langle \Delta 
\widehat{S}_{y}{}^{2}\right\rangle >\frac{1}{2}|\left\langle \widehat{S}%
_{z}\right\rangle |  \nonumber \\
\left\langle \Delta \widehat{S}_{y}{}^{2}\right\rangle &<&\frac{1}{2}%
|\left\langle \widehat{S}_{x}\right\rangle |\;and\;\left\langle \Delta 
\widehat{S}_{z}{}^{2}\right\rangle >\frac{1}{2}|\left\langle \widehat{S}%
_{x}\right\rangle |  \nonumber \\
\left\langle \Delta \widehat{S}_{z}{}^{2}\right\rangle &<&\frac{1}{2}%
|\left\langle \widehat{S}_{y}\right\rangle |\;and\;\left\langle \Delta 
\widehat{S}_{x}{}^{2}\right\rangle >\frac{1}{2}|\left\langle \widehat{S}%
_{y}\right\rangle |  \label{Eq.SpinSqgMultiMode}
\end{eqnarray}%
for $\widehat{S}_{x}{}$being squeezed compared to $\widehat{S}_{y}$, and so
on.

Sinmilar alternative criteria to (\ref{Eq.NewCriterionSpinSqueezing}) can
also be obtained, for example for $\widehat{S}_{z}$ being squeezed compared
to \emph{any two }orthogonal spin operators such as $\widehat{S}_{\perp \,1}$
or $\widehat{S}_{\perp \,2}$ defined similarly to (\ref{Eq.NewOrthogSpinOprs}%
).

\subsection{Rotation Operators and New Modes}

\label{SubSection - Rotation Operators}

\subsubsection{Rotation Operators}

The new spin operators are also related to the original spin operators via a 
\emph{unitary rotation operator }$\widehat{R}(\alpha ,\beta ,\gamma )$
parameterised in terms of Euler angles so that 
\begin{equation}
\widehat{J}_{\xi }=\widehat{R}(\alpha ,\beta ,\gamma )\,\widehat{S}_{\xi }\,%
\widehat{R}(\alpha ,\beta ,\gamma )^{-1}  \label{Eq.UnitaryTransfnSpinOprs}
\end{equation}%
where 
\begin{equation}
\widehat{R}(\alpha ,\beta ,\gamma )=\widehat{R}_{z}(\alpha )\widehat{R}%
_{y}(\beta )\widehat{R}_{z}(\gamma )  \label{Eq.UnitaryRotnOpr}
\end{equation}%
with $\widehat{R}_{\xi }(\phi )=\exp (i\phi \widehat{S}_{\xi })$ describing
a rotation about the $\xi $ axis anticlockwise through an angle $\phi $.
Details for the rotation operators and matrices are set out in \cite%
{Dalton12a}. Note that Eq. (\ref{Eq.UnitaryTransfnSpinOprs}) specifies a
rotation of the vector spin operator rather than a rotation of the axes, so $%
\widehat{J}_{\xi }$ ($\xi =x,y,z$) are the components of the rotated vector
spin operator with respect to the original axes.

\subsubsection{New Mode Operators}

We can also see that the new spin operators are related to \emph{new mode
operators} $\widehat{c}$ and $\widehat{d}$ via 
\begin{equation}
\widehat{J}_{x}=(\widehat{d}^{\dag }\widehat{c}+\widehat{c}^{\dag }\widehat{d%
})/2\qquad \widehat{J}_{y}=(\widehat{d}^{\dag }\widehat{c}-\widehat{c}^{\dag
}\widehat{d})/2i\qquad \widehat{J}_{z}=(\widehat{d}^{\dag }\widehat{d}-%
\widehat{c}^{\dag }\widehat{c})/2  \label{Eq.NewSpinOprs}
\end{equation}%
where 
\begin{equation}
\widehat{c}=\widehat{R}(\alpha ,\beta ,\gamma )\,\widehat{a}\,\widehat{R}%
(\alpha ,\beta ,\gamma )^{-1}\qquad \widehat{d}=\widehat{R}(\alpha ,\beta
,\gamma )\,\widehat{b}\,\widehat{R}(\alpha ,\beta ,\gamma )^{-1}
\label{Eq.UnitaryTransfnModeOprs}
\end{equation}

For the bosonic case a straight-forward calculation gives the new mode
operators as 
\begin{eqnarray}
\widehat{c} &=&\exp (\frac{1}{2}i\gamma )\left( \cos (\frac{\beta }{2}%
)\,\exp (\frac{1}{2}i\alpha )\,\widehat{a}+\sin (\frac{\beta }{2})\,\exp (-%
\frac{1}{2}i\alpha )\,\widehat{b}\right)  \nonumber \\
\widehat{d} &=&\exp (-\frac{1}{2}i\gamma )\left( -\sin (\frac{\beta }{2}%
)\,\exp (\frac{1}{2}i\alpha )\,\widehat{a}+\cos (\frac{\beta }{2})\,\exp (-%
\frac{1}{2}i\alpha )\,\widehat{b}\right)  \nonumber \\
&&  \label{Eq.NewModeOprs}
\end{eqnarray}%
and it is easy to then check that $\widehat{c}$ and $\widehat{d}$ satisfy
the expected non-zero bosonic commutation rules are $[\widehat{e},\widehat{e}%
^{\dag }]=\widehat{1}$ ($\widehat{e}=\widehat{c}$ or $\widehat{d}$) and that
the \emph{total boson number operator} is $\widehat{N}=(\widehat{d}^{\dag }%
\widehat{d}+\widehat{c}^{\dag }\widehat{c})$. As $\widehat{N}$ is invariant
under unitary rotation operators it follows that $\widehat{J}_{x}^{2}+%
\widehat{J}_{y}^{2}+\widehat{J}_{z}^{2}=(\widehat{N}/2)(\widehat{N}/2+1)$.

\subsubsection{New Modes}

The new mode operators correspond to \emph{new single particle states} $%
\left\vert \phi _{c}\right\rangle $, $\left\vert \phi _{d}\right\rangle $
where 
\begin{eqnarray}
\left\vert \phi _{c}\right\rangle &=&\exp (-\frac{1}{2}i\gamma )\left( \cos (%
\frac{\beta }{2})\,\exp (-\frac{1}{2}i\alpha )\,\left\vert \phi
_{a}\right\rangle +\sin (\frac{\beta }{2})\,\exp (\frac{1}{2}i\alpha
)\,\left\vert \phi _{b}\right\rangle \right)  \nonumber \\
\left\vert \phi _{d}\right\rangle &=&\exp (\frac{1}{2}i\gamma )\left( -\sin (%
\frac{\beta }{2})\,\exp (-\frac{1}{2}i\alpha )\,\left\vert \phi
_{a}\right\rangle +\cos (\frac{\beta }{2})\,\exp (\frac{1}{2}i\alpha
)\,\left\vert \phi _{b}\right\rangle \right)  \nonumber \\
&&  \label{Eq.NewModes}
\end{eqnarray}%
These are two orthonormal quantum superpositions of the original single
particle states $\left\vert \phi _{a}\right\rangle $, $\left\vert \phi
_{b}\right\rangle $, and as such represent an \emph{alternative choice} of
modes that could be realised experimentally.

Eqs. (\ref{Eq.NewModeOprs}) can be inverted to give the old mode operators
via 
\begin{eqnarray}
\widehat{a} &=&\exp (-\frac{1}{2}i\alpha )\left( \cos (\frac{\beta }{2}%
)\,\exp (-\frac{1}{2}i\gamma )\,\widehat{c}-\sin (\frac{\beta }{2})\,\exp (+%
\frac{1}{2}i\gamma )\,\widehat{d}\right)  \nonumber \\
\widehat{b} &=&\exp (+\frac{1}{2}i\alpha )\left( \sin (\frac{\beta }{2}%
)\,\exp (\frac{1}{2}i\gamma )\,\widehat{c}+\cos (\frac{\beta }{2})\,\exp (-%
\frac{1}{2}i\gamma )\,\widehat{d}\right)  \nonumber \\
&&  \label{Eq.OldModeOperators}
\end{eqnarray}

\subsection{Old and New Modes - Coherence Terms}

The general non-entangled state for modes $\widehat{a}$ and $\widehat{b}$ is
given by 
\begin{equation}
\widehat{\rho }=\sum_{R}P_{R}\,\widehat{\rho }_{R}^{A}\otimes \widehat{\rho }%
_{R}^{B}  \label{Eq.NonEntStateModesCD}
\end{equation}%
and as a consequence of the requirement that $\widehat{\rho }_{R}^{A}$ and $%
\widehat{\rho }_{R}^{B}$ are physical states for modes $\widehat{a}$ and $%
\widehat{b}$ satisying the super-selection rule, it follows that 
\begin{eqnarray}
\left\langle (\widehat{a})^{n}\right\rangle _{c} &=&Tr(\widehat{\rho }%
_{R}^{A}(\widehat{a})^{n})=0\qquad \left\langle (\widehat{a}^{\dag
})^{n}\right\rangle _{c}=Tr(\widehat{\rho }_{R}^{A}(\widehat{a}^{\dag
})^{n})=0  \nonumber \\
\left\langle (\widehat{b})^{m}\right\rangle _{d} &=&Tr(\widehat{\rho }%
_{R}^{B}(\widehat{b})^{m})=0\qquad \left\langle (\widehat{b}^{\dag
})^{m}\right\rangle _{d}=Tr(\widehat{\rho }_{R}^{B}(\widehat{b}^{\dag
})^{m})=0  \nonumber \\
&&  \label{Eq.CondNonEntStateCD}
\end{eqnarray}%
Thus \emph{coherence} terms are zero.

For our two-mode case we have also seen that the original choice of modes
with annihilation operators $\widehat{a}$ and $\widehat{b}$ may be replaced
by new modes with annihilation operators $\widehat{c}$ and $\widehat{d}$.
Since the new modes are associated with new spin operators $\widehat{J}_{\xi
}$ ($\xi =x,y,z$) for which the covariance matrix is diagonal and where the
diagonal elements give the variances, it is therefore also relevant to
consider entanglement for the case where the sub-systems are modes $\widehat{%
c}$ and $\widehat{d}$, rather than $\widehat{a}$ and $\widehat{b}$.
Consequently we also consider general non-entangled states for modes $%
\widehat{c}$ and $\widehat{d}$ in which the density operator is of the same
form as (\ref{Eq.NonEntStateModesCD}), but with $\widehat{\rho }%
_{R}^{A}\rightarrow \widehat{\rho }_{R}^{C}$ and $\widehat{\rho }%
_{R}^{B}\rightarrow \widehat{\rho }_{R}^{D}$. Results analogous to (\ref%
{Eq.CondNonEntStateCD}) apply in this case, but with $\widehat{a}\rightarrow 
\widehat{c}$ and $\widehat{b}\rightarrow \widehat{d}$.

\subsection{Quantum Correlation Functions and Spin Measurements}

Finally, we note that the principal spin fluctuations can be related to 
\emph{quantum correlation functions}. For example, it is easy to show that%
\begin{eqnarray}
\left\langle \Delta \widehat{S}_{x}{}^{2}\right\rangle &=&\frac{1}{4}\left(
\left\langle (\widehat{b}^{\dag })^{2}(\widehat{a})^{2}\right\rangle
+\left\langle (\widehat{a}^{\dag })^{2}(\widehat{b})^{2}\right\rangle
+2\left\langle \widehat{b}^{\dag }\widehat{a}^{\dag }\widehat{a}\widehat{b}%
\right\rangle +\left\langle \widehat{b}^{\dag }\widehat{b}\right\rangle
+\left\langle \widehat{a}^{\dag }\widehat{a}\right\rangle \right)  \nonumber
\\
&&-\frac{1}{4}\left( \left\langle (\widehat{b}^{\dag }\widehat{a}%
\right\rangle ^{2}+\left\langle (\widehat{a}^{\dag }\widehat{b}\right\rangle
^{2}+2\left\langle (\widehat{b}^{\dag }\widehat{a}\right\rangle \left\langle
(\widehat{a}^{\dag }\widehat{b}\right\rangle \right)  \label{Eq.QCFReln}
\end{eqnarray}%
showing that $\left\langle \Delta \widehat{S}_{x}{}^{2}\right\rangle $ is
related to various first and second order quantum correlation functions.
These can be measured experimentally and are given theoretically in terms of
phase space integrals involving distribution functions to represent the
density operator and phase space variables to represent the mode
annihilation, creation operators.\pagebreak

\section{Spin Squeezing Test for Entanglement}

\label{Section - Relationship Spin Squeezing & Entanglement}

With the general non-entangled state now required to be such that the
density operators for the individual sub-systems must represent quantum
states that conform to the super-selection rule, the consequential link
between entanglement in two mode bosonic systems and spin squeezing can now
be established. We first consider spin squeezing for the original spin
operators $\widehat{S}_{x}$, $\widehat{S}_{y},$ $\widehat{S}_{z}$ and
entangled states of the original modes $\widehat{a}$, $\widehat{b}$. and
then for the principal spin operators $\widehat{J}_{x}$, $\widehat{J}_{y},$ $%
\widehat{J}_{z}$ and entangled states of the related new modes $\widehat{c}$%
, $\widehat{d}$. We show \cite{Dalton14a} that spin squeezing in \emph{any}
spin component is a \emph{sufficiency test} for entanglement of the two
modes involved. Examples of entangled states that are not spin squeezed and
states that are not entangled nor spin squeezed for one choice of mode
sub-systems, but are entangled and spin squeezed for another choice are then
presented.

\subsection{Spin Squeezing and Entanglement - Old Modes}

\label{SubSection - SpinSqueezingEntangNewModes}

Firstly, the \emph{mean} for a Hermitian operator $\widehat{\Omega }$ in a
mixed state 
\begin{equation}
\widehat{\rho }=\sum_{R}P_{R}\,\widehat{\rho }_{R}  \label{Eq.MixedState}
\end{equation}%
is the \emph{average} of means for separate components%
\begin{equation}
\left\langle \widehat{\Omega }\right\rangle =\sum_{R}P_{R}\,\left\langle 
\widehat{\Omega }\right\rangle _{R}  \label{Eq.MeanResult}
\end{equation}%
where $\left\langle \widehat{\Omega }\,\right\rangle _{R}=Tr(\widehat{\rho }%
_{R}\,\widehat{\Omega })$.

Secondly, the \emph{variance} for a Hermitian operator $\widehat{\Omega }$
in a mixed state is always \emph{never less} than the the \emph{average} of
the variances for the separate components (see \cite{Hoffmann03a}) 
\begin{equation}
\left\langle \Delta \widehat{\Omega }\,^{2}\right\rangle \geq
\sum_{R}P_{R}\,\left\langle \Delta \widehat{\Omega }_{R}{}^{2}\right\rangle
_{R}  \label{Eq.VarianceResult}
\end{equation}%
where $\left\langle \Delta \widehat{\Omega }\,^{2}\right\rangle =Tr(\widehat{%
\rho }\,\Delta \widehat{\Omega }\,^{2})$ with $\Delta \widehat{\Omega }=%
\widehat{\Omega }-\left\langle \widehat{\Omega }\right\rangle $ and $%
\left\langle \Delta \widehat{\Omega }\,^{2}\right\rangle _{R}=Tr(\widehat{%
\rho }_{R}\,\Delta \widehat{\Omega }_{R}\,^{2})$ with $\Delta \widehat{%
\Omega }_{R}=\widehat{\Omega }-\left\langle \widehat{\Omega }\right\rangle
_{R}$ . To prove this result we have using (\ref{Eq.MeanResult}) both for $%
\widehat{\Omega }$ and $\widehat{\Omega }^{2}$%
\begin{eqnarray}
\left\langle \Delta \widehat{\Omega }\,^{2}\right\rangle &=&\left\langle 
\widehat{\Omega }\,^{2}\right\rangle -\left\langle \widehat{\Omega }%
\right\rangle ^{2}  \nonumber \\
&=&\sum_{R}P_{R}\,\left( \left\langle \widehat{\Omega }{}^{2}\right\rangle
_{R}-\left\langle \widehat{\Omega }\right\rangle _{R}^{2}\right)
+\sum_{R}P_{R}\,\left\langle \widehat{\Omega }\right\rangle _{R}^{2}-\left(
\sum_{R}P_{R}\,\left\langle \widehat{\Omega }\right\rangle _{R}\right) ^{2} 
\nonumber \\
&=&\sum_{R}P_{R}\,\left\langle \Delta \widehat{\Omega }{}_{R}^{2}\right%
\rangle _{R}+\sum_{R}P_{R}\,\left\langle \widehat{\Omega }\right\rangle
_{R}^{2}-\left( \sum_{R}P_{R}\,|\left\langle \widehat{\Omega }\right\rangle
_{R}|\right) ^{2}
\end{eqnarray}%
The variance result (\ref{Eq.VarianceResult}) follows because the sum of the
last two terms is always $\geq 0$ using the result (\textbf{175})\ \textbf{%
in Appendix 2 of paper 1}, with $C_{R}=\left\langle \widehat{\Omega }%
\right\rangle _{R}^{2}$, $\sqrt{C_{R}}=\,|\left\langle \widehat{\Omega }%
\right\rangle _{R}|$- which are real and positive.

\subsubsection{Mean Values for $\widehat{S}_{x},$ $\widehat{S}_{y}$ and $%
\widehat{S}_{z}$}

Next, we find the \emph{mean values} of the spin operators for the product
state $\widehat{\rho }_{R}=$ $\widehat{\rho }_{R}^{A}\otimes \widehat{\rho }%
_{R}^{B}$%
\begin{eqnarray}
\left\langle \widehat{S}\,_{x}\right\rangle _{R} &=&\frac{1}{2}(\left\langle 
\widehat{b}^{\dag }\right\rangle _{R}\left\langle \widehat{a}\right\rangle
_{R}+\left\langle \widehat{a}^{\dag }\right\rangle _{R}\left\langle \widehat{%
b}\right\rangle _{R})=0  \nonumber \\
\left\langle \widehat{S}\,_{y}\right\rangle _{R} &=&\frac{1}{2i}%
(\left\langle \widehat{b}^{\dag }\right\rangle _{R}\left\langle \widehat{a}%
\right\rangle _{R}-\left\langle \widehat{a}^{\dag }\right\rangle
_{R}\left\langle \widehat{b}\right\rangle _{R})=0
\label{Eq.MeanNewSpinXYProdState}
\end{eqnarray}%
and%
\begin{equation}
\left\langle \widehat{S}\,_{z}\right\rangle _{R}=\frac{1}{2}(\left\langle 
\widehat{b}^{\dag }\widehat{b}\right\rangle _{R}-\left\langle \widehat{a}%
^{\dag }\widehat{a}\right\rangle _{R})  \label{Eq.MeanNewSpinZProdState}
\end{equation}%
for SSR compliant sub-system states using (\ref{Eq.CondNonEntStateCD}), and
thus using (\ref{Eq.MeanResult}) the \emph{overall mean} values for the 
\emph{separable} state is%
\begin{equation}
\left\langle \widehat{S}\,_{x}\right\rangle =0\qquad \left\langle \widehat{S}%
\,_{y}\right\rangle =0  \label{Eq.MeanNewSpinXY}
\end{equation}%
and 
\begin{equation}
\left\langle \widehat{S}\,_{z}\right\rangle =\sum_{R}P_{R}\,\frac{1}{2}%
(\left\langle \widehat{b}^{\dag }\widehat{b}\right\rangle _{R}-\left\langle 
\widehat{a}^{\dag }\widehat{a}\right\rangle _{R}))  \label{Eq.MeanNewSpinZ}
\end{equation}%
Hence if either $\left\langle \widehat{S}\,_{x}\right\rangle \neq 0$ or $%
\left\langle \widehat{S}\,_{y}\right\rangle \neq 0$ the state must be
entangled. This may be called the \emph{Bloch vector} test. This result will
also have later significance.

\subsubsection{Variances for $\widehat{S}_{x}$ and $\widehat{S}_{y}$}

Next we calculate $\left\langle \Delta \widehat{S}\,_{x}^{2}\right\rangle
_{R}$, $\left\langle \Delta \widehat{S}\,_{y}^{2}\right\rangle _{R}$ and $%
\left\langle \widehat{S}_{x}\right\rangle _{R}$, $\left\langle \widehat{S}%
_{y}\right\rangle _{R}$, $\left\langle \widehat{S}_{z}\right\rangle _{R}$
for the case of the \emph{separable state} (\ref{Eq.NonEntStateModesCD})
where $\widehat{\rho }_{R}=$ $\widehat{\rho }_{R}^{A}\otimes \widehat{\rho }%
_{R}^{B}$. From Eqs. (\ref{Eq.OldSpinOprs}) we find that%
\begin{eqnarray}
\widehat{S}\,_{x}^{2} &=&\frac{1}{4}((\widehat{b}^{\dag })^{2}(\widehat{a}%
)^{2}+\widehat{b}^{\dag }\widehat{b}\widehat{a}\widehat{a}^{\dag }+\widehat{a%
}^{\dag }\widehat{a}\widehat{b}\widehat{b}^{\dag }+(\widehat{b})^{2}(%
\widehat{a}^{\dag })^{2})  \nonumber \\
\widehat{S}\,_{y}^{2} &=&-\frac{1}{4}((\widehat{b}^{\dag })^{2}(\widehat{a}%
)^{2}-\widehat{b}^{\dag }\widehat{b}\widehat{a}\widehat{a}^{\dag }-\widehat{a%
}^{\dag }\widehat{a}\widehat{b}\widehat{b}^{\dag }+(\widehat{b})^{2}(%
\widehat{a}^{\dag })^{2})  \label{Eq.SquareNewSpinOprs}
\end{eqnarray}%
so that on taking the trace with $\widehat{\rho }_{R}$ and using Eqs. (\ref%
{Eq.CondNonEntStateCD}) we get after applying the commutation rules $[%
\widehat{e},\widehat{e}^{\dag }]=\widehat{1}$ ($\widehat{e}=\widehat{a}$ or $%
\widehat{b}$) 
\begin{eqnarray}
\left\langle \widehat{S}\,_{x}^{2}\right\rangle _{R} &=&\frac{1}{4}%
(\left\langle \widehat{b}^{\dag }\widehat{b}\right\rangle _{R}+\left\langle 
\widehat{a}^{\dag }\widehat{a}\right\rangle _{R})+\frac{1}{2}(\left\langle 
\widehat{a}^{\dag }\widehat{a}\right\rangle _{R}\left\langle \widehat{b}%
^{\dag }\widehat{b}\right\rangle _{R})  \nonumber \\
\left\langle \widehat{S}\,_{y}^{2}\right\rangle _{R} &=&\frac{1}{4}%
(\left\langle \widehat{b}^{\dag }\widehat{b}\right\rangle _{R}+\left\langle 
\widehat{a}^{\dag }\widehat{a}\right\rangle _{R})+\frac{1}{2}(\left\langle 
\widehat{a}^{\dag }\widehat{a}\right\rangle _{R}\left\langle \widehat{b}%
^{\dag }\widehat{b}\right\rangle _{R})
\label{Eq.MeanSquareNewSpinXYProdState}
\end{eqnarray}

Hence using (\ref{Eq.MeanNewSpinXYProdState}) for $\left\langle \widehat{S}%
\,_{x}\right\rangle _{R}$ and $\left\langle \widehat{S}\,_{y}\right\rangle
_{R}$ we see finally that the variances are%
\begin{eqnarray}
\left\langle \Delta \widehat{S}\,_{x}^{2}\right\rangle _{R} &=&\frac{1}{4}%
(\left\langle \widehat{b}^{\dag }\widehat{b}\right\rangle _{R}+\left\langle 
\widehat{a}^{\dag }\widehat{a}\right\rangle _{R})+\frac{1}{2}(\left\langle 
\widehat{a}^{\dag }\widehat{a}\right\rangle _{R}\left\langle \widehat{b}%
^{\dag }\widehat{b}\right\rangle _{R})  \nonumber \\
\left\langle \Delta \widehat{S}\,_{y}^{2}\right\rangle _{R} &=&\frac{1}{4}%
(\left\langle \widehat{b}^{\dag }\widehat{b}\right\rangle _{R}+\left\langle 
\widehat{a}^{\dag }\widehat{a}\right\rangle _{R})+\frac{1}{2}(\left\langle 
\widehat{a}^{\dag }\widehat{a}\right\rangle _{R}\left\langle \widehat{b}%
^{\dag }\widehat{b}\right\rangle _{R})
\label{Eq.VariancesNewSpinXYProdState}
\end{eqnarray}%
and therefore from Eq. (\ref{Eq.VarianceResult}) 
\begin{eqnarray}
\left\langle \Delta \widehat{S}\,_{x}^{2}\right\rangle &\geq &\sum_{R}P_{R}\,%
\frac{1}{4}(\left\langle \widehat{b}^{\dag }\widehat{b}\right\rangle
_{R}+\left\langle \widehat{a}^{\dag }\widehat{a}\right\rangle _{R})+\frac{1}{%
2}(\left\langle \widehat{a}^{\dag }\widehat{a}\right\rangle _{R}\left\langle 
\widehat{b}^{\dag }\widehat{b}\right\rangle _{R})  \nonumber \\
\left\langle \Delta \widehat{S}\,_{y}^{2}\right\rangle &\geq &\sum_{R}P_{R}\,%
\frac{1}{4}(\left\langle \widehat{b}^{\dag }\widehat{b}\right\rangle
_{R}+\left\langle \widehat{a}^{\dag }\widehat{a}\right\rangle _{R})+\frac{1}{%
2}(\left\langle \widehat{a}^{\dag }\widehat{a}\right\rangle _{R}\left\langle 
\widehat{b}^{\dag }\widehat{b}\right\rangle _{R})
\label{Eq.InequalityXYVariancesNonEntState}
\end{eqnarray}

Now using (\ref{Eq.MeanNewSpinZ}) for $\left\langle \widehat{S}%
\,_{z}\right\rangle $ we see that 
\begin{equation}
\frac{1}{2}|\left\langle \widehat{S}\,_{z}\right\rangle |\leq \sum_{R}P_{R}\,%
\frac{1}{4}|(\left\langle \widehat{b}^{\dag }\widehat{b}\right\rangle
_{R}-\left\langle \widehat{a}^{\dag }\widehat{a}\right\rangle _{R}))|\leq
\sum_{R}P_{R}\,\frac{1}{4}(\left\langle \widehat{b}^{\dag }\widehat{b}%
\right\rangle _{R}+\left\langle \widehat{a}^{\dag }\widehat{a}\right\rangle
_{R}))  \label{Eq.InequalityMeanNewSpinZEntState}
\end{equation}%
and thus for any non-entangled state of modes $\widehat{a}$ and $\widehat{b}$
\begin{eqnarray}
&&\left\langle \Delta \widehat{S}\,_{x}^{2}\right\rangle -\frac{1}{2}%
|\left\langle \widehat{S}\,_{z}\right\rangle |  \nonumber \\
&\geq &\sum_{R}P_{R}\,\frac{1}{4}(\left\langle \widehat{b}^{\dag }\widehat{b}%
\right\rangle _{R}+\left\langle \widehat{a}^{\dag }\widehat{a}\right\rangle
_{R})+\frac{1}{2}(\left\langle \widehat{a}^{\dag }\widehat{a}\right\rangle
_{R}\left\langle \widehat{b}^{\dag }\widehat{b}\right\rangle
_{R})-\sum_{R}P_{R}\,\frac{1}{4}(\left\langle \widehat{b}^{\dag }\widehat{b}%
\right\rangle _{R}+\left\langle \widehat{a}^{\dag }\widehat{a}\right\rangle
_{R}))  \nonumber \\
&\geq &\sum_{R}P_{R}\,\frac{1}{2}(\left\langle \widehat{a}^{\dag }\widehat{a}%
\right\rangle _{R}\left\langle \widehat{b}^{\dag }\widehat{b}\right\rangle
_{R})  \nonumber \\
&\geq &0  \label{Eq.NonSpinSqResult}
\end{eqnarray}%
Similar final steps show that $\left\langle \Delta \widehat{S}%
\,_{y}^{2}\right\rangle -\frac{1}{2}|\left\langle \widehat{S}%
\,_{z}\right\rangle |\geq 0$ for all non-entangled state of modes $\widehat{a%
}$ and $\widehat{b}$.

This shows that for the general non-entangled state with modes $\widehat{a}$
and $\widehat{b}$ as the sub-systems, the variances for two of the principal
spin fluctuations $\left\langle \Delta \widehat{S}\,_{x}^{2}\right\rangle $
and $\left\langle \Delta \widehat{S}\,_{y}^{2}\right\rangle $ are \emph{both}
greater than $\frac{1}{2}|\left\langle \widehat{S}\,_{z}\right\rangle |$,
and hence there is no spin squeezing for $\widehat{S}_{x}$ compared to $%
\widehat{S}_{y}$ (or vice versa). Note that as $|\left\langle \widehat{S}%
\,_{y}\right\rangle |=0$, the quantity $\sqrt{\left( |\left\langle \widehat{S%
}_{\perp \,1}\right\rangle |^{2}+|\left\langle \widehat{S}_{\perp
\,2}\right\rangle |^{2}\right) }$ is the same as $|\left\langle \widehat{S}%
\,_{z}\right\rangle |$, so the alternative criterion in Eq. (\ref%
{Eq.NewCriterionSpinSqueezing}) is the same as that in Eq. (\ref%
{Eq.SpinSqueezingJXJY}) which is used here.

It is easy to see from (\ref{Eq.MeanNewSpinXY}) that 
\begin{equation}
\left\langle \Delta \widehat{S}\,_{x}^{2}\right\rangle -\frac{1}{2}%
|\left\langle \widehat{S}\,_{y}\right\rangle |\geq 0\qquad \left\langle
\Delta \widehat{S}\,_{y}^{2}\right\rangle -\frac{1}{2}|\left\langle \widehat{%
S}\,_{x}\right\rangle |\geq 0  \label{Eq.NonSpinSqResultB}
\end{equation}%
for any non-entangled state of modes $\widehat{a}$ and $\widehat{b}$. This
completes the set of inequalities for the variances of $\widehat{S}_{x}$ and 
$\widehat{S}_{y}$. These last inequalities are of course trivially true and
amount to no more than showing that the variances $\left\langle \Delta 
\widehat{S}\,_{x}^{2}\right\rangle $ and $\left\langle \Delta \widehat{S}%
\,_{y}^{2}\right\rangle $ are not negative.

\subsubsection{Variance for $\widehat{S}_{z}$}

For the other principal spin fluctuation we find that for separable states 
\begin{equation}
\left\langle \Delta \widehat{S}\,_{z}^{2}\right\rangle _{R}=\frac{1}{4}%
(\left\langle \left( \widehat{b}^{\dag }\widehat{b}-\left\langle \widehat{b}%
^{\dag }\widehat{b}\right\rangle _{R}\right) \left( \widehat{b}^{\dag }%
\widehat{b}-\left\langle \widehat{b}^{\dag }\widehat{b}\right\rangle
_{R}\right) \right\rangle _{R}+\left\langle \left( \widehat{a}^{\dag }%
\widehat{a}-\left\langle \widehat{a}^{\dag }\widehat{a}\right\rangle
_{R}\right) \left( \widehat{a}^{\dag }\widehat{a}-\left\langle \widehat{a}%
^{\dag }\widehat{a}\right\rangle _{R}\right) \right\rangle _{R}
\label{Eq.VariancesNewSpinZProdState}
\end{equation}%
so that using (\ref{Eq.VarianceResult}) 
\begin{equation}
\left\langle \Delta \widehat{S}\,_{z}^{2}\right\rangle \geq \sum_{R}P_{R}\,%
\frac{1}{4}(\left\langle \left( \widehat{b}^{\dag }\widehat{b}-\left\langle 
\widehat{b}^{\dag }\widehat{b}\right\rangle _{R}\right) ^{2}\right\rangle
_{R}+\left\langle \left( \widehat{a}^{\dag }\widehat{a}-\left\langle 
\widehat{a}^{\dag }\widehat{a}\right\rangle _{R}\right) ^{2}\right\rangle
_{R}  \label{Eq.VariancesNewSpinZNonEntState}
\end{equation}%
From Eq. (\ref{Eq.MeanNewSpinXY}) it follows that 
\begin{eqnarray}
&&\left\langle \Delta \widehat{S}\,_{z}^{2}\right\rangle -\frac{1}{2}%
|\left\langle \widehat{S}\,_{x}\right\rangle |\,  \nonumber \\
&\geq &\sum_{R}P_{R}\,\frac{1}{4}(\left\langle \left( \widehat{b}^{\dag }%
\widehat{b}-\left\langle \widehat{b}^{\dag }\widehat{b}\right\rangle
_{R}\right) ^{2}\right\rangle _{R}+\left\langle \left( \widehat{a}^{\dag }%
\widehat{a}-\left\langle \widehat{a}^{\dag }\widehat{a}\right\rangle
_{R}\right) ^{2}\right\rangle _{R}  \nonumber \\
&\geq &0  \label{Eq.NonSpinSqResultC}
\end{eqnarray}%
Similarly $\left\langle \Delta \widehat{S}\,_{z}^{2}\right\rangle -\frac{1}{2%
}|\left\langle \widehat{S}\,_{y}\right\rangle |\,\geq 0$. Again, these
results are trivial and just show that the variances are non-negative.

\subsubsection{No Spin Squeezing for Separable States}

So overall, we have for the general non-entangled state of modes $\widehat{a}
$ and $\widehat{b}$%
\begin{eqnarray}
\left\langle \Delta \widehat{S}_{x}{}^{2}\right\rangle &\geq &\frac{1}{2}%
|\left\langle \widehat{S}_{z}\right\rangle |\;and\;\left\langle \Delta 
\widehat{S}_{y}{}^{2}\right\rangle \geq \frac{1}{2}|\left\langle \widehat{S}%
_{z}\right\rangle |  \nonumber \\
\left\langle \Delta \widehat{S}_{y}{}^{2}\right\rangle &\geq &\frac{1}{2}%
|\left\langle \widehat{S}_{x}\right\rangle |\;and\;\left\langle \Delta 
\widehat{S}_{z}{}^{2}\right\rangle \geq \frac{1}{2}|\left\langle \widehat{S}%
_{x}\right\rangle |  \nonumber \\
\left\langle \Delta \widehat{S}_{z}{}^{2}\right\rangle &\geq &\frac{1}{2}%
|\left\langle \widehat{S}_{y}\right\rangle |\;and\;\left\langle \Delta 
\widehat{S}_{x}{}^{2}\right\rangle \geq \frac{1}{2}|\left\langle \widehat{S}%
_{y}\right\rangle |  \label{Eq.CombinedResultNonEntState}
\end{eqnarray}%
The first result tells us that for \emph{any} non-entangled state of modes $%
\widehat{a}$ and $\widehat{b}$ the spin operator $\widehat{S}_{x}$ is \emph{%
not} squeezed compared to $\widehat{S}_{y}$ (or vice-versa). The same is
also true for the other pairs of spin operators, as we will now see.

\subsubsection{Spin Squeezing Tests for Entanglement}

The key value of these results is the \emph{spin squeezing test} for \emph{%
entanglement}. We see that from the first inequality in (\ref%
{Eq.CombinedResultNonEntState}) for separable states, that \emph{if} for a
quantum state we find that 
\begin{equation}
\left\langle \Delta \widehat{S}\,_{x}^{2}\right\rangle <\frac{1}{2}%
|\left\langle \widehat{S}\,_{z}\right\rangle |\qquad \emph{or\quad }%
\left\langle \Delta \widehat{S}\,_{y}^{2}\right\rangle <\frac{1}{2}%
|\left\langle \widehat{S}\,_{z}\right\rangle |
\label{Eq.SpinSqueezeEntangleTest}
\end{equation}%
then the state \emph{must} be entangled for modes $\widehat{a}$ and $%
\widehat{b}$. Thus we only need to have spin squeezing in \emph{either} of $%
\widehat{S}_{x}$ with respect to $\widehat{S}_{y}$ \emph{or} vice-versa to
demonstrate entanglement. Note that one cannot have \emph{both} $%
\left\langle \Delta \widehat{S}\,_{x}^{2}\right\rangle <\frac{1}{2}%
|\left\langle \widehat{S}\,_{z}\right\rangle |$ and$\left\langle \Delta 
\widehat{S}\,_{y}^{2}\right\rangle <\frac{1}{2}|\left\langle \widehat{S}%
\,_{z}\right\rangle |$ etc. due to the Heisenberg uncertainty principle.

Because $\left\langle \widehat{S}_{x}\right\rangle _{\rho }=\left\langle 
\widehat{S}_{y}\right\rangle _{\rho }=0$ the second and third results in (%
\ref{Eq.CombinedResultNonEntState}) merely show that $\left\langle \Delta 
\widehat{S}_{x}{}^{2}\right\rangle \geq 0$, $\left\langle \Delta \widehat{S}%
_{y}{}^{2}\right\rangle \geq 0$ and $\left\langle \Delta \widehat{S}%
_{z}{}^{2}\right\rangle \geq 0\;$ for SSR compliant non-entangled states, it
may be thought that no conclusion follows regarding the spin squeezing
involving $\widehat{S}_{z}$ for entangled states. This is not the case. 
\emph{If} for a given state we find that 
\begin{equation}
\left\langle \Delta \widehat{S}\,_{y}^{2}\right\rangle <\frac{1}{2}%
|\left\langle \widehat{S}\,_{x}\right\rangle |\qquad \emph{or\quad }%
\left\langle \Delta \widehat{S}\,_{z}^{2}\right\rangle <\frac{1}{2}%
|\left\langle \widehat{S}\,_{x}\right\rangle |
\end{equation}%
or 
\begin{equation}
\left\langle \Delta \widehat{S}\,_{z}^{2}\right\rangle <\frac{1}{2}%
|\left\langle \widehat{S}\,_{y}\right\rangle |\qquad \emph{or\quad }%
\left\langle \Delta \widehat{S}\,_{x}^{2}\right\rangle <\frac{1}{2}%
|\left\langle \widehat{S}\,_{y}\right\rangle |
\end{equation}%
then the state \emph{must} be entangled. This is because if any of these
situations apply then \emph{at least one} of $\left\langle \widehat{S}%
_{x}\right\rangle _{\rho }$ or $\left\langle \widehat{S}_{y}\right\rangle
_{\rho }$ must be non-zero. But as we have seen from (\ref{Eq.MeanNewSpinXY}%
) both of the quantities are zero in non-entangled states. Thus we only need
to have spin squeezing in \emph{either} of $\widehat{S}_{z}$ with respect to 
$\widehat{S}_{y}$ \emph{or} vice-versa or spin squeezing in \emph{either} of 
$\widehat{S}_{z}$ with respect to $\widehat{S}_{x}$ \emph{or} vice-versa to
demonstrate entanglement.

Hence the general conclusion stated in \cite{Dalton14a}, that spin squeezing
in \emph{any} spin operator $\widehat{S}_{x}$, $\widehat{S}_{y},$ $\widehat{S%
}_{z}$ shows that the state must be entangled for modes $\widehat{a}$ and $%
\widehat{b}$.. The presence of spin squeezing is a conclusive test for
entanglement. Note that the reverse is not true - there are many entangled
states that are \emph{not} spin squeezed. A notable example is the\emph{\ }%
particular \emph{binomial} state $\left\vert \Phi \right\rangle =((\,%
\widehat{a}+\,\widehat{b})^{\dag }/\sqrt{2})^{N}/\sqrt{N!}\;\left\vert
0\right\rangle $ for which $\left\langle \widehat{S}_{x}\right\rangle _{\rho
}=N/2$, $\left\langle \widehat{S}_{y}\right\rangle _{\rho }=\left\langle 
\widehat{S}_{z}\right\rangle _{\rho }=0$ and $\left\langle \Delta \widehat{S}%
_{y}{}^{2}\right\rangle _{\rho }=\left\langle \Delta \widehat{S}%
_{z}{}^{2}\right\rangle _{\rho }=N/4$, $\left\langle \Delta \widehat{S}%
_{x}{}^{2}\right\rangle _{\rho }=0$ (see \cite{Dalton12a}) The spin
fluctuations in $\widehat{S}_{y}$ and $\widehat{S}_{z}$ correspond to the 
\emph{standard quantum limit}.

This is a \emph{key result} for two mode entanglement. \emph{All }spin
squeezed states are\emph{\ entangled}. We emphasise again that the converse
is not true. \emph{Not all} entangled two mode states are \emph{spin squeezed%
}. This important distinction is not always recognised - entanglement and
spin squeezing are two \emph{distinct} features of a two mode quantum state
that do not always occur together.

For the two orthogonal spin operator components (\ref{Eq.NewOrthogSpinOprs}%
)\ in the $xy$ plane $\widehat{S}\,_{\perp 1}$ and $\widehat{S}\,_{\perp 1}$
it is then straightforward to show that 
\begin{equation}
If\emph{\quad }\left\langle \Delta \widehat{S}\,_{\perp 1}^{2}\right\rangle <%
\frac{1}{2}|\left\langle \widehat{S}\,_{z}\right\rangle |
\label{Eq.SpinSqEntangleTestD}
\end{equation}%
or%
\begin{equation}
If\emph{\quad }\left\langle \Delta \widehat{S}\,_{\perp 2}^{2}\right\rangle <%
\frac{1}{2}|\left\langle \widehat{S}\,_{z}\right\rangle |
\label{Eq.SpinSqEntangleTestE}
\end{equation}%
that is, if $\widehat{S}\,_{\perp 1}$ is squeezed compared to $\widehat{S}%
\,_{\perp 2}$ or vice versa - then the state must be entangled. Spin
squeezing in \emph{any} of the spin operator component in the $xy$ plane
will demonstrate entanglement.

\subsubsection{Significance of Spin Squeezing Test}

\label{SubSubSection - Significance of Spin Squeezing Test}

The spin squeezing test for two mode systems was based on the general form (%
\ref{Eq.NonEntStateModesCD}) for all \emph{separable} states together with
the requirement that the sub-system density operators $\widehat{\rho }%
_{R}^{A}$ and $\widehat{\rho }_{R}^{B}$ were compliant with the \emph{local}
particle number \emph{SSR}. From the point of view of a \emph{supporter} for
applying the local particle number SSR if the result of an experiment is
that spin squeezing has occurred, the immediate conclusion is that the state
is entangled. On the other hand from the point of view of a\emph{\ sceptic}
about being required to apply the local particle number SSR for the
sub-system states, such a sceptic would draw different conclusions from an
experiment that demonstrated spin squeezing. They would immediately point
out that in this case spin squeezing is\textbf{\ }\emph{not}\textbf{\ }a test%
\textbf{\ }for entanglement.\textbf{\ }However, as we will now see these
conclusions are still of some interest.

To discuss this it is convenient to divide possible\emph{\ mathematical}
forms for the density operator into categories. Considering \emph{all}
general two mode quantum states that are compliant with the \emph{global}
particle number SSR, we may first divide such quantum states into three
categories, as set out in Table 1.\medskip

\begin{tabular}{|l|l|l|l|}
\hline
\underline{\textbf{REGION}} & \underline{\textbf{OVERALL}} & \underline{%
\textbf{SUB-SYSTEM}} & \underline{\textbf{CATEGORY}} \\ \hline
& \underline{\textbf{QUANTUM\ STATE}} & \underline{\textbf{QUANTUM\ STATE}}
&  \\ \hline
&  &  &  \\ \hline
\textbf{A} & $\widehat{\rho }=\dsum\limits_{R}P_{R}\,\widehat{\rho }%
_{R}^{A}\otimes \widehat{\rho }_{R}^{B}$ & $\,$Both $\widehat{\rho }%
_{R}^{A}\;$and\ $\,\widehat{\rho }_{R}^{B}$ are local & * \textbf{Separable}
\\ \hline
&  & particle\ number\ SSR\ compliant &  \\ \hline
&  &  &  \\ \hline
\textbf{B} & $\widehat{\rho }=\dsum\limits_{R}P_{R}\,\widehat{\rho }%
_{R}^{A}\otimes \widehat{\rho }_{R}^{B}$ & Neither $\widehat{\rho }%
_{R}^{A}\; $nor$\,\widehat{\rho }_{R}^{B}$ is local & * \textbf{Separable
but } \\ \hline
&  & particle\ number\ SSR\ compliant & \ \textbf{non-local \cite%
{Verstraete03a}; } \\ \hline
&  &  & * \textbf{Entangled \cite{Dalton14a}} \\ \hline
&  &  &  \\ \hline
\textbf{C} & $\widehat{\rho }\neq \dsum\limits_{R}P_{R}\,\widehat{\rho }%
_{R}^{A}\otimes \widehat{\rho }_{R}^{B}$ & N/A & * \textbf{Entangled} \\ 
\hline
\end{tabular}

\medskip

\begin{center}
\bigskip Table I. Categories of two mode quantum states.
\end{center}

The regions referred to are shown in Figure 2. All authors would regard the
quantum states in Region A as being separable and those in Region C as being
entangled - it is only those in Region B where the category is disputed.
Those such as \cite{Dalton14a}\textbf{\ }(local SSR supporters)\textbf{\ }%
who require local particle number SSR compliance for each sub-system state
would classify the overall state as entangled, those who do not require this
(local SSR sceptics) such as \cite{Verstraete03a} would classify the overall
state as separable but non-local. Note that no further sub-classification is
needed.

In \textbf{SubSection 3.2}$\partial $\textbf{\ of paper I} we show that if
states of the form (\ref{Eq.NonEntangStateModesAB}) are globally SSR
compliant, then \emph{both }the sub-system states $\,\widehat{\rho }%
_{R}^{A}\;$and\ $\,\widehat{\rho }_{R}^{B}\;$are\ local\ particle\ number\
SSR\ compliant \emph{in general}. However, we point out that there are
special matched choices for both $\widehat{\rho }_{R}^{A}\;$and\ $\,\widehat{%
\rho }_{R}^{B}\;$along with the $P_{R}$, where neither $\widehat{\rho }%
_{R}^{A}\;$nor\ $\,\widehat{\rho }_{R}^{B}$ is local particle number SSR
compliant even though $\widehat{\rho }$ is global particle number SSR
compliant. But the case where just \emph{one} of $\widehat{\rho }_{R}^{A}\;$%
or\ $\,\widehat{\rho }_{R}^{B}$ is non SSR compliant does not occur, so
Region B does not need to be sub-divided along these lines.

\includegraphics[height=3.7507in]{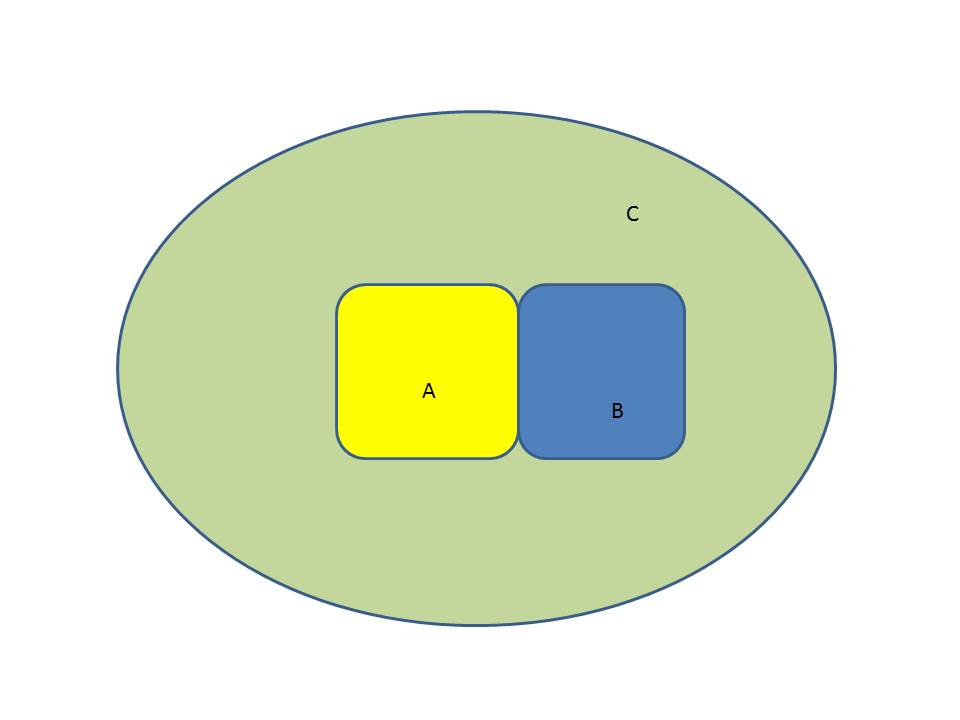}
%
%
%
\newline

\begin{center}
\bigskip

Figure 2. Categories of two mode quantum states that are global particle
number SSR compliant. The regions A, B and C are described in Table I.%
\textbf{\ } \medskip\ 
\end{center}

Now let us consider quantum states for which $\left\langle \Delta \widehat{S}%
_{x}{}^{2}\right\rangle _{\rho }\geq \frac{1}{2}|\left\langle \widehat{S}%
_{z}\right\rangle _{\rho }|$ and $\left\langle \Delta \widehat{S}%
_{y}{}^{2}\right\rangle _{\rho }\geq \frac{1}{2}|\left\langle \widehat{S}%
_{z}\right\rangle _{\rho }|$. Such states are clearly not spin squeezed.
Firstly, we know that \emph{all} states in Region A satisfy these
inequalities. However, \emph{some} states in Region B and \emph{some} states
in Region C may also satisfy these inequalities. In Figure 3 the quantum
states in Region B that satisfy these inequalities are depicted as lying in
Region D, those in Region C that do so are depicted as lying in Region F.
Hence, if we find that the quantum state is such that spin squeezing \emph{%
does} occur (as in the test of (\ref{Eq.SpinSqEntTestOrigModes})) we can
definitely say that it does \emph{not} lie in Regions A, D or F. It must
therefore be located in Regions E or L. The question is - Does this
determine whether the state is entangled or not according to the \emph{%
supporters} of applying the local SSR as in the definition of entanglement
used in the present paper? The answer is that it does. This is because the
quantum state must be located within either of Regions B or C, since these
regions include E and L respectively. In both cases it would be \emph{%
entangled} according to the definition used here \cite{Dalton14a} (see Table
1).

\includegraphics[height=3.8597in]{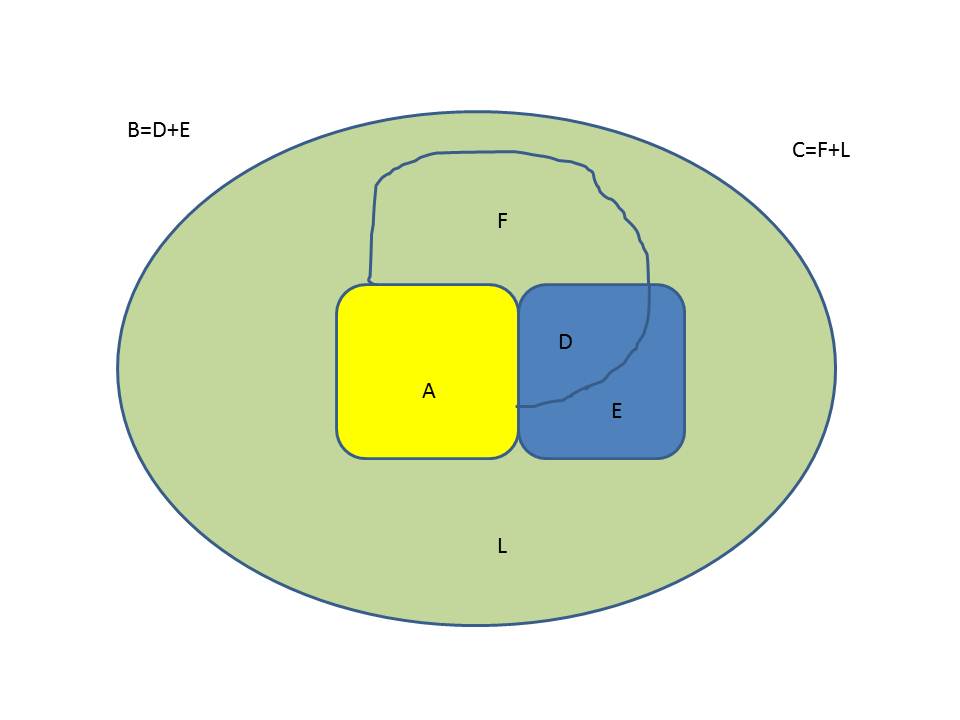}


\begin{center}
\medskip \medskip

Figure 3. Categories of two mode quantum states satisfying inequalities $%
\left\langle \Delta \widehat{S}_{x}{}^{2}\right\rangle _{\rho }\geq \frac{1}{%
2}|\left\langle \widehat{S}_{z}\right\rangle _{\rho }|$ and $\left\langle
\Delta \widehat{S}_{y}{}^{2}\right\rangle _{\rho }\geq \frac{1}{2}%
|\left\langle \widehat{S}_{z}\right\rangle _{\rho }|$ \ These regions are A,
D, F. Set unions are denoted +. Referring to Figure 2,\textbf{\ }$B=D\oplus
E $ and\textbf{\ }$C=F\oplus L$\textbf{.}
\end{center}

However, the \emph{sceptics} of applying the local SSR would draw a
different conclusion from the experiment that demonstrated spin squeezing
(as in the test of (\ref{Eq.SpinSqEntTestOrigModes})). They would agree that
the mathematics shows that a state in Region A could not demonstrate spin
squeezing. Nor by assumption could states in Regions D or F. This means that
the state must lie in either Region L or Region E. So from the point of view
of the sceptic, \emph{either} the state is \emph{entangled} (if it lies in
Region L) \emph{or} the sub-system states in \emph{all} separable states
(Region E) do not comply with the local particle number \emph{SSR}. The
sceptic's conclusion is clearly interesting - in the first case the quantum
state is entangled, and hence may demonstrate other non-classical features,
and in the second case the possibility exists of finding sub-systems in
states that have the unexpected feature in non-relativistic many body
physics of having coherences between states with differing particle number.
If there was a \emph{second} experimental test that could show that the
state was not entangled, then this would demonstrate the existence of
quantum states (sub-systems are themselves possible quantum systems) in
which the particle number SSR breaks down.

The second experiment would seem to require a test for entanglement which is 
\emph{necessary} as well as being sufficient - the latter alone being
usually the case for entanglement tests.\textbf{\ }Such criteria and
measurements are a challenge, but not impossible even though we have not met
this challenge in these two papers. Thus, \emph{in principle} there could be
a\emph{\ pair} of experiments that give evidence of entanglement, \emph{or}
failure of the Super Selection Rule. For such investigations to be possible,
the use of entanglement criteria that \emph{do} invoke the local
super-selection rules is \emph{also} required. Such tests are the focus of
these two papers, though here our primary reason is because we consider
applying the local particle number SSR is required by the physics of
non-relativistic quantum many body systems involving massive particles.

\subsubsection{Inequality for $|\left\langle \widehat{S}\,_{z}\right\rangle
|\,$}

Of the results for a \emph{non-entangled} physical state for modes $\widehat{%
a}$ and $\widehat{b}$ we will later find it particularly important to
consider the first of (\ref{Eq.CombinedResultNonEntState})%
\begin{equation}
\left\langle \Delta \widehat{S}\,_{x}^{2}\right\rangle \geq \frac{1}{2}%
|\left\langle \widehat{S}\,_{z}\right\rangle |\quad and\quad \left\langle
\Delta \widehat{S}\,_{y}^{2}\right\rangle \geq \frac{1}{2}|\left\langle 
\widehat{S}\,_{z}\right\rangle |  \label{Eq.NonEntStateSpinSqCondn}
\end{equation}%
This is because we can show that for any quantum state%
\begin{equation}
|\left\langle \widehat{S}\,_{z}\right\rangle |\,=|\left\langle \frac{1}{2}(%
\widehat{n}_{b}-\widehat{n}_{a})\right\rangle |\,\leq \frac{1}{2}%
(|\left\langle \widehat{n}_{b}\right\rangle |\,+|\left\langle \widehat{n}%
_{a}\right\rangle |)\,=\frac{1}{2}\left\langle \widehat{N}\right\rangle
\label{Eq.GenInequalNJZ}
\end{equation}%
there is an inequality involving $|\left\langle \widehat{S}%
\,_{z}\right\rangle |\,$\ and the mean number of bosons $\left\langle 
\widehat{N}\right\rangle $ in the two mode system. Note that there \emph{are}
some entangled states (see SubSection \ref{SubSection - Ent State Non
Squeezed}) for which $\left\langle \Delta \widehat{S}\,_{x}^{2}\right\rangle 
$ and $\left\langle \Delta \widehat{S}\,_{y}^{2}\right\rangle $ are both
greater than $\frac{1}{2}|\left\langle \widehat{S}\,_{z}\right\rangle |$,
since all that has been proven is that for \emph{all}\textbf{\ }%
non-entangled states we must have \emph{both} $\left\langle \Delta \widehat{S%
}\,_{x}^{2}\right\rangle \geq \frac{1}{2}|\left\langle \widehat{S}%
\,_{z}\right\rangle |$ \emph{and} $\left\langle \Delta \widehat{S}%
\,_{y}^{2}\right\rangle \geq \frac{1}{2}|\left\langle \widehat{S}%
\,_{z}\right\rangle |$.

Hence we may conclude that spin squeezing in any of the original spin
fluctuations $\widehat{S}_{x}$ , $\widehat{S}_{y}$ or $\widehat{S}_{z}$
requires the quantum state to be entangled for the modes $\widehat{a}$ and $%
\widehat{b}$ as the sub-systems. Similarly, we may conclude that spin
squeezing in any of the principal spin fluctuations $\widehat{J}_{x}$ , $%
\widehat{J}_{y}$ or $\widehat{J}_{z}$ requires the quantum state to be
entangled for the modes $\widehat{c}$ and $\widehat{d}$ as the sub-systems,
these modes being associated with the principal spin fluctuations via Eq. (%
\ref{Eq.NewSpinOprs}). Although finding spin squeezing tells us that the
state is entangled, there are however no simple relationships between the
measures of entanglement and those of spin squeezing, so the linkage is
essentially a qualitative one. For general quantum states, measures of
entanglement for the specific situation of two sub-systems (bi-partite
entanglement) are reviewed in \cite{Amico08a}. \ 

\subsection{Spin Squeezing and Entanglement - New Modes}

\label{SubSection - Spin Squeezing Original Modes}

It is also of some interest to consider spin squeezing for the new spin
operators $\widehat{J}_{x}$, $\widehat{J}_{y}$, $\widehat{J}_{z}$ with the
new modes $\widehat{c}$ and $\widehat{d}$ as the sub-systems, where these
spin operators are associated with a diagonal covariance matrix. The
definition of spin squeezing in this case is set out analogous to that in
Eq.(\ref{Eq.SpinSqgMultiMode}). In this case the general non-entangled state
for the \emph{new} modes is given by 
\begin{equation}
\widehat{\rho }=\sum_{R}P_{R}\,\widehat{\rho }_{R}^{C}\otimes \widehat{\rho }%
_{R}^{D}  \label{Eq.NonEntangStateModesAB}
\end{equation}%
with the $\widehat{\rho }_{R}^{C}$ and $\widehat{\rho }_{R}^{D}$
representing physical states for modes $\widehat{c}$ and $\widehat{d}$, and
where results analogous to Eqs. (\ref{Eq.CondNonEntStateCD}) apply. The same
treatment applies as for spin operators $\widehat{S}_{x}$, $\widehat{S}_{y}$%
, $\widehat{S}_{z}$ with the modes $\widehat{a}$ and $\widehat{b}$ as the
sub-systems and leads to the result for a \emph{non-entangled} state of
modes $\widehat{c}$ and $\widehat{d}$ 
\begin{equation}
\left\langle \Delta \widehat{J}\,_{x}^{2}\right\rangle \geq \frac{1}{2}%
|\left\langle \widehat{J}\,_{z}\right\rangle |\quad and\quad \left\langle
\Delta \widehat{J}\,_{y}^{2}\right\rangle \geq \frac{1}{2}|\left\langle 
\widehat{J}\,_{z}\right\rangle  \label{Eq.NonSpinSqResultSXSY}
\end{equation}%
showing that neither $\widehat{J}_{x}$ or $\widehat{J}_{y}$ is spin squeezed
for the general non-entangled state for modes $\widehat{c}$ and $\widehat{d}$
given in Eq. (\ref{Eq.NewModeOprs}). We also have 
\begin{equation}
\left\langle \widehat{J}\,_{x}\right\rangle =\sum_{R}P_{R}\left\langle 
\widehat{J}\,_{x}\right\rangle _{R}=0\qquad \left\langle \widehat{J}%
\,_{y}\right\rangle =\sum_{R}P_{R}\left\langle \widehat{J}%
\,_{y}\right\rangle _{R}=0  \label{Eq.MeanOldSpinXY}
\end{equation}%
so all the results analogous to Eqs. (\ref{Eq.CombinedResultNonEntState})
also follow. Following similar arguements as in SubSection \ref{SubSection -
SpinSqueezingEntangNewModes} we may also conclude that spin squeezing in 
\emph{any} of the original spin fluctuations requires the quantum state to
be entangled when the original modes $\widehat{c}$ and $\widehat{d}$ are the
sub-systems. Thus the \emph{entanglement test} is 
\begin{equation}
If\emph{\quad }\left\langle \Delta \widehat{J}\,_{x}^{2}\right\rangle <\frac{%
1}{2}|\left\langle \widehat{J}\,_{z}\right\rangle |\qquad \emph{or\quad }%
\left\langle \Delta \widehat{J}\,_{y}^{2}\right\rangle <\frac{1}{2}%
|\left\langle \widehat{J}\,_{z}\right\rangle |
\label{Eq.SpinSqEntTestOrigModes}
\end{equation}%
or%
\begin{equation}
If\emph{\quad }\left\langle \Delta \widehat{J}\,_{y}^{2}\right\rangle <\frac{%
1}{2}|\left\langle \widehat{J}\,_{x}\right\rangle |\qquad \emph{or\quad }%
\left\langle \Delta \widehat{J}\,_{z}^{2}\right\rangle <\frac{1}{2}%
|\left\langle \widehat{J}\,_{x}\right\rangle |
\label{Eq.SpinSqEntTestOrigModesB}
\end{equation}%
or%
\begin{equation}
If\emph{\quad }\left\langle \Delta \widehat{J}\,_{z}^{2}\right\rangle <\frac{%
1}{2}|\left\langle \widehat{J}\,_{y}\right\rangle |\qquad \emph{or\quad }%
\left\langle \Delta \widehat{J}\,_{x}^{2}\right\rangle <\frac{1}{2}%
|\left\langle \widehat{J}\,_{y}\right\rangle |
\label{Eq.SpinSqEntTestOrigModesC}
\end{equation}%
then we have an entangled state for the original modes $\widehat{c}$ and $%
\widehat{d}$.

The result (\ref{Eq.MeanOldSpinXY}) also provides a \emph{Bloch vector}
entanglement test - if either $\left\langle \widehat{S}\,_{x}\right\rangle
\neq 0$ or $\left\langle \widehat{S}\,_{y}\right\rangle \neq 0$ the state
must be entangled.

Hence we have seen that spin squeezing - either of the new or original spin
operators requires entanglement of the new or original modes. Which spin
operators to consider depends on which pairs of modes are being tested for
entanglement.

\subsection{Spin Squeezing and Entanglement - Multi-Mode Case}

\label{SubSection - SpinSqgEnt MultiMode}

As we have seen the multi-mode case involves a set of $n$ modes with
annihilation operators $\widehat{a}_{i}$ for bosons with hyperfine component 
$A$, and another set of $n$ modes with annihilation operators $\widehat{b}%
_{i}$ for bosons with hyperfine component $B$. Since entanglement implies a
clear choice of what sub-systems are to be entangled, there are numerous
choices possible here for the present multi-mode case. \emph{Case 1}
involves two sub-systems, one consisting of all the $\widehat{a}_{i}$ modes
as sub-system $A$ and the other consisting of all the $\widehat{b}_{i}$
modes as sub-system $B$. \emph{Case 2} involves $2n$ sub-systems, the $Ai$
th containing the mode $\widehat{a}_{i}$ and the $Bi$ th containing the mode 
$\widehat{b}_{i}$. \emph{Case 3} involves $n$ sub-systems, the $i$th
containiing the two modes $\widehat{a}_{i}$ and $\widehat{b}_{i}$. These
three cases relate to entanglement causing interactions in differing
circumstances. Case 1 might apply to cases where separable states can be
created with all the $\widehat{a}_{i}$ modes coupled together to produce
states $\widehat{\rho }_{R}^{A}$ and the $\widehat{b}_{i}$ modes coupled
together to produce states $\widehat{\rho }_{R}^{B}$ . Case 2 might apply
cases where separable states can be created with the $\widehat{a}_{i}$ and
all the $\widehat{b}_{i}$ modes independent of each together to produce
states $\widehat{\rho }_{R}^{a(i)}\otimes \widehat{\rho }_{R}^{b(i)}$.Case 3
might apply cases where separable states can be created with the $\widehat{a}%
_{i}$ and the matching $\widehat{b}_{i}$ modes coupled together to produce
states $\widehat{\rho }_{R}^{ab(i)}$. Cases 2 and 3 will be discussed
further in SubSection \ref{SubSection - Sorensen 2001} dealing with the
entanglement test introduced by Sorensen et al \cite{Sorensen01a}.

The density operators for \emph{separable} states in the three cases will be
of the form%
\begin{eqnarray}
\widehat{\rho }_{sep} &=&\sum_{R}P_{R}\,\widehat{\rho }_{R}  \nonumber \\
\widehat{\rho }_{R} &=&\widehat{\rho }_{R}^{A}\otimes \widehat{\rho }%
_{R}^{B}\qquad Case\quad 1  \label{Eq.SepStatesMultiModeCase1} \\
\widehat{\rho }_{R} &=&\widehat{\rho }_{R}^{a(1)}\otimes ..\otimes \widehat{%
\rho }_{R}^{a(i)}..\otimes \widehat{\rho }_{R}^{a(n)}\otimes \widehat{\rho }%
_{R}^{b(1)}\otimes ..\otimes \widehat{\rho }_{R}^{b(n)}\qquad Case\quad 2
\label{Eq.SepStatesMultiModeCase2} \\
\widehat{\rho }_{R} &=&\widehat{\rho }_{R}^{ab(1)}\otimes \widehat{\rho }%
_{R}^{ab(2)}\otimes ..\otimes \widehat{\rho }_{R}^{ab(i)}..\otimes \widehat{%
\rho }_{R}^{ab(n)}\qquad Case\quad 3  \label{Eq.SepStatesMultiModeCase3} \\
&&  \label{Eq.SepStatesMultiMode}
\end{eqnarray}

Discussion of whether there is a spin squeezing test for Case 1 in the
multi-mode case involves a generalisation of the theory set out in
SubSection \ref{SubSection - SpinSqueezingEntangNewModes}. There is a \emph{%
Bloch vector} entanglement test, in that if either of $\left\langle \widehat{%
S}\,_{x}\right\rangle $ or $\left\langle \widehat{S}\,_{y}\right\rangle $ is
non-zero, then the state is entangled. We also find that spin squeezing in 
\emph{any} spin component requires the state to be entangled, thus
generalising the spin squeezing test to the\emph{\ multi-mode} case, for%
\emph{\ two} sub-systems consisting of all the modes $\widehat{a}_{i}$ and
al the modes $\widehat{b}_{i}$ The details are covered in Appendix \ref%
{Appendix = MultiMo Spin Sq Choice 1}.

For Case 2 a spin squeezing test for entanglement also be obtained. The test
is again that \emph{spin squeezing} in \emph{any} spin component $\widehat{S}%
_{x},\widehat{S}_{y}$ or $\widehat{S}_{z}$ confirms entanglement of the $2n$
sub-systems consisting of \emph{single} modes $\widehat{a}_{i}$ and $%
\widehat{b}_{i}$. Furthermore, there is also a \emph{Bloch vector}
entanglement test, in that if either of $\left\langle \widehat{S}%
\,_{x}\right\rangle $ or $\left\langle \widehat{S}\,_{y}\right\rangle $ is
non-zero, then the state is entangled. As these systems can have quantum 
\emph{states} with large numbers $N$ of bosonic particles, it can be said
that entanglement in an $N$ particle \emph{system} has occurred if spin
squeezing is found. The proof of these tests is set out in SubSection \ref%
{SubSection - Var Sz One Mode Subsystems} of Appendix \ref{Appendix -
Revised Sorensen}.

For Case 3 there is also a spin squeezing test for entanglement, but it is
restricted. Here the test is that\emph{\ spin squeezing} in $\widehat{S}_{z}$
confirms entanglement of the $n$ sub-systems consisting of \emph{pairs} of
modes $\widehat{a}_{i}$ and $\widehat{b}_{i}$, but the test is \emph{%
restricted} to the situation where exactly \emph{one boson} occupies each
mode pair. No spin squeezing test was found for the other spin operators,
nor was a Bloch vector entanglement test obtained. The proof of this result
is set out in SubSection \ref{SubSection - Var Sz Two Mode SubSystems} of
Appendix \ref{Appendix - Revised Sorensen}. That no general spin squeezing
test for entanglement exists can be shown by a counter-example. If all the $%
N $ bosons occupied one mode pair $\widehat{a}_{i}$ and $\widehat{b}_{i}$,
and the quantum state $\widehat{\rho }_{R}^{ab(i)}$ for this pair
corresponded to the \emph{relative phase eigenstate} with phase $\theta
_{p}=0$ (see SubSection \ref{SubSection - Ent States that are Spin Sq}) then
although the overall state is separable, spin squeezing in $\widehat{S}_{y}$
compared to $\widehat{S}_{z}$ occurs (with $\left\langle \Delta \widehat{S}%
_{y}^{2}\right\rangle =\frac{1}{4}+\frac{1}{8}\ln N$, $\left\langle \Delta 
\widehat{S}_{z}^{2}\right\rangle =\left( \frac{{\LARGE 1}}{{\LARGE 6}}-\frac{%
{\LARGE \pi }^{2}}{{\LARGE 64}}\right) N^{2}$ and $\left\langle \widehat{S}%
\,_{x}\right\rangle =N\,\frac{\pi }{8}$. Thus there is a situation where a 
\emph{non-entangled} state for sub-systems consisting of \emph{mode pairs}
is spin squeezed, so spin squeezing does \emph{not} always confirm
entanglement.

As in the previous two mode cases, having established in multi-mode cases
that spin squeezing requires entanglement a further question then is: Does
entanglement automatically lead to spin squeezing? The answer is no, since
cases where the quantum state is entangled but not spin squeezed can be
found - an example is given in the previous paragraph . Thus in general,
spin squeezing and entanglement are \emph{not equivalent}.- they do not
occur \emph{together} for all states. Spin squeezing is a \emph{sufficient}
condition for entanglement, it is not a \emph{necessary} condition.

\subsection{Bloch Vector Entanglement Test}

We have seen for the general non-entangled states of modes $\widehat{c}$ and 
$\widehat{d}$ or of modes\ $\widehat{a}$ and $\widehat{b}$ that 
\begin{eqnarray}
\left\langle \widehat{J}\,_{x}\right\rangle &=&0\qquad \left\langle \widehat{%
J}\,_{y}\right\rangle =0  \label{Eq.BlochVectNonEntStateCD} \\
\left\langle \widehat{S}\,_{x}\right\rangle &=&0\qquad \left\langle \widehat{%
S}\,_{y}\right\rangle =0  \label{Eq.BlochVectNonEntStateAB}
\end{eqnarray}%
Hence the two mode \emph{Bloch vector entanglement} tests%
\begin{eqnarray}
\left\langle \widehat{J}\,_{x}\right\rangle &\neq &0\qquad or\qquad
\left\langle \widehat{J}\,_{y}\right\rangle \neq 0  \nonumber \\
\left\langle \widehat{S}\,_{x}\right\rangle &\neq &0\qquad or\qquad
\left\langle \widehat{S}\,_{y}\right\rangle \neq 0
\label{Eq.BlochVectorEntTest}
\end{eqnarray}%
for modes $\widehat{c}$ and $\widehat{d}$ or of modes\ $\widehat{a}$ and $%
\widehat{b}.$ The same Bloch vector test also applies in the \emph{%
multi-mode case} for Case 1, where there are just two sub-systems each
consisting of all the modes $\widehat{a}_{i}$ or all the modes $\widehat{b}%
_{i}$ and in Case 2, where there are $2n$ sub-systems consisting of all the
modes $\widehat{a}_{i}$ and all the modes $\widehat{b}_{i}$.

From Eqs. (\ref{Eq.NewSpinOprs}) and (\ref{Eq.OldSpinOprs}) these results
are equivalent to 
\begin{eqnarray}
\left\langle \widehat{d}\,\widehat{c}^{\dag }\right\rangle &=&0\qquad
\left\langle \widehat{c}\,\widehat{d}^{\dag }\right\rangle =0
\label{Eq.CondNESCD} \\
\left\langle \widehat{b}\,\widehat{a}^{\dag }\right\rangle &=&0\qquad
\left\langle \widehat{a}\,\widehat{b}^{\dag }\right\rangle =0
\label{Eq.CondNESAB}
\end{eqnarray}

Hence we find further \emph{tests} for \emph{entangled states} of modes $%
\widehat{c}$ and $\widehat{d}$ or of modes \ $\widehat{a}$ and $\widehat{b}$%
\begin{eqnarray}
|\left\langle \widehat{d}\,\widehat{c}^{\dag }\right\rangle |^{2}\,
&>&0\qquad |\left\langle \widehat{c}\,\widehat{d}^{\dag }\right\rangle
|^{2}\,>0  \label{Eq.EntangTestModesCD} \\
|\left\langle \widehat{b}\,\widehat{a}^{\dag }\right\rangle |^{2}\,
&>&0\qquad |\left\langle \widehat{a}\,\widehat{b}^{\dag }\right\rangle
|^{2}\,>0  \label{Eq.EntTestsModesAB}
\end{eqnarray}%
As we will see in Section \ref{Section - Criteria for Spin Squeezing Based
on Non-Physical States}, these tests are particular cases with $m=n=1$ of
the simpler entanglement test in Eq. (\ref{Eq.EntangTest}) that applies for
the situation in the present paper where non-entangled states are required
to satisfy the super-selection rule.

\subsection{Entanglement Test for Number Difference and Sum}

\label{SubSection - Test for Number Difference}

There is also a further spin squeezing test involving the operator $\widehat{%
S}_{z}$, which is equal to half the \emph{number difference} $\frac{1}{2}(%
\widehat{n}_{b}-\widehat{n}_{a})$. We note that simultaneous eigenststes of $%
\widehat{n}_{a}$ and $\widehat{n}_{b}$ exist, which are also eigenstates of
the total number operator. For such states the variances $\left\langle
\Delta \widehat{n}_{a}^{2}\right\rangle $, $\left\langle \Delta \widehat{n}%
_{b}^{2}\right\rangle $, $\left\langle \Delta \widehat{S}_{z}^{2}\right%
\rangle $ and $\left\langle \Delta \widehat{N}^{2}\right\rangle $ are all
zero, which does not suggest that useful general inequalities for these
variances would be found. However, a\ useful entanglement test - which does
not require SSR compliance can be found. For the variance of $\widehat{S}%
_{z} $ in a separable state we have%
\begin{eqnarray}
\left\langle \Delta \widehat{S}_{z}^{2}\right\rangle &\geq
&\dsum\limits_{R}P_{R}\left\langle \Delta \widehat{S}_{z}^{2}\right\rangle
_{R}=\dsum\limits_{R}P_{R}(\left\langle \widehat{S}_{z}^{2}\right\rangle
_{R}-\left\langle \widehat{S}_{z}\right\rangle _{R}^{2})  \nonumber \\
&=&\frac{1}{4}\dsum\limits_{R}P_{R}(\left\langle \widehat{n}%
_{b}^{2}\right\rangle _{R}+\left\langle \widehat{n}_{a}^{2}\right\rangle
_{R}-2\left\langle \widehat{n}_{b}\right\rangle _{R}\left\langle \widehat{n}%
_{a}\right\rangle _{R}-\left\langle \widehat{n}_{b}\right\rangle
_{R}^{2}-\left\langle \widehat{n}_{a}\right\rangle _{R}^{2}+2\left\langle 
\widehat{n}_{b}\right\rangle _{R}\left\langle \widehat{n}_{a}\right\rangle
_{R})  \nonumber \\
&=&\frac{1}{4}\dsum\limits_{R}P_{R}(\left\langle \Delta \widehat{n}%
_{b}^{2}\right\rangle _{R}+\left\langle \Delta \widehat{n}%
_{a}^{2}\right\rangle _{R})  \label{Eq.SzIneqality}
\end{eqnarray}%
For such a separable state we also find%
\begin{equation}
\left\langle \Delta \widehat{N}^{2}\right\rangle \geq
\dsum\limits_{R}P_{R}(\left\langle \Delta \widehat{n}_{b}^{2}\right\rangle
_{R}+\left\langle \Delta \widehat{n}_{a}^{2}\right\rangle _{R})
\label{Eq.NumberInequal}
\end{equation}%
This leads to the useful if somewhat qualitative test that if we have a
state with a \emph{large} fluctuation in the total boson number and a \emph{%
small} fluctuation in the number difference, then it must be an entangled
state. If it was separable and $\left\langle \Delta \widehat{N}%
^{2}\right\rangle $ is large, then $\left\langle \Delta \widehat{S}%
_{z}^{2}\right\rangle $ must also be large. There is also the converse test
- if we have a state with a small fluctuation in the total boson number and
a large fluctuation in the number difference, then it must be an entangled
state.

\subsection{Entangled States that are Non Spin-Squeezed - NOON State}

\label{SubSection - Ent State Non Squeezed}

One such example is the generalised $N$ boson \emph{NOON\ state} defined as%
\begin{eqnarray}
\widehat{\rho } &=&\left\vert \Phi \right\rangle \left\langle \Phi
\right\vert  \nonumber \\
\left\vert \Phi \right\rangle &=&\cos \theta \,\frac{(\widehat{a}^{\dag
})^{N}}{\sqrt{N!}}\left\vert 0\right\rangle +\sin \theta \,\frac{(\widehat{b}%
^{\dag })^{N}}{\sqrt{N!}}\left\vert 0\right\rangle  \nonumber \\
&=&\cos \theta \,\left\vert \frac{N}{2},-\frac{N}{2}\right\rangle +\sin
\theta \,\left\vert \frac{N}{2},+\frac{N}{2}\right\rangle
\label{Eq.GeneralNOONState}
\end{eqnarray}%
which is an entangled state for modes $\widehat{a}$ and $\widehat{b}$ in all
cases except where $\cos \theta \,$or $\sin \theta \,$is zero. In the last
form the state is expressed in terms of the eigenstates for $(%
\underrightarrow{\widehat{S}})^{2}\,$and $\widehat{S}\,_{z}$, as detailed in 
\cite{Dalton12a}.

A straight-forward calculation gives 
\begin{eqnarray}
\left\langle \widehat{S}\,_{x}\right\rangle &=&0\qquad \left\langle \widehat{%
S}\,_{y}\right\rangle =0\qquad \left\langle \widehat{S}\,_{z}\right\rangle =-%
\frac{N}{2}\cos 2\theta  \nonumber \\
\left\langle \Delta \widehat{S}\,_{x}^{2}\right\rangle &=&\frac{N}{4}\qquad
\left\langle \Delta \widehat{S}\,_{y}^{2}\right\rangle =\frac{N}{4}\qquad
\left\langle \Delta \widehat{S}\,_{z}^{2}\right\rangle =\frac{N^{2}}{4}%
(1-\cos ^{2}2\theta )  \label{Eq.MenVarianceNOONState}
\end{eqnarray}%
for $N>1$, so that using the criteria for spin squeezing given in Eq. (\ref%
{Eq.SpinSqueezingJXJY}) we see that as $\left\langle \Delta \widehat{S}%
\,_{x}^{2}\right\rangle -\frac{1}{2}|\left\langle \widehat{S}%
\,_{z}\right\rangle |\,\geq 0$, etc, and hence spin squeezing does not occur
for this entangled state.

\subsection{Non-Entangled States that are Non Spin Squeezed - Binomial State}

\label{SubSection - Binomial State}

Of course from the previous section \emph{any} non entangled state is
definitely not spin squeezed. A specific example illustrating this is the $N$
boson binomial state given by 
\begin{eqnarray}
\widehat{\rho } &=&\left\vert \Phi \right\rangle \left\langle \Phi
\right\vert  \nonumber \\
\left\vert \Phi \right\rangle &=&\frac{(-\widehat{c}^{\dag })^{N}}{\sqrt{N!}}%
\left\vert 0\right\rangle  \label{Eq.BinomialState}
\end{eqnarray}%
where $\widehat{c}$ and $\widehat{d}$ are given by Eqs. (\ref{Eq.NewModeOprs}%
) with Euler angles $\alpha =-\pi +\chi $, $\beta =-2\theta $ and $\gamma
=-\pi $, we find that 
\begin{eqnarray}
\widehat{c} &=&-\cos \theta \,\exp (\frac{1}{2}i\chi )\,\widehat{a}-\sin
\theta \,\exp (-\frac{1}{2}i\chi )\,\widehat{b}=-\widehat{a}_{1}  \nonumber
\\
\widehat{d} &=&\sin \theta \,\exp (\frac{1}{2}i\chi )\,\widehat{a}-\cos
\theta \,\exp (-\frac{1}{2}i\chi )\,\widehat{b}=-\widehat{a}_{2}
\label{Eq.SpecialNewModeOprs}
\end{eqnarray}%
where the mode operators $\widehat{a}_{1}$ and $\widehat{a}_{2}$ are as
defined in \cite{Dalton12a} (see Eqs. (53) therein). The new spin angular
momentum operators $\widehat{J}_{\xi }$ ($\xi =x,y,z$) are the same as those
defined in \cite{Dalton12a} (see Eqs. (64) therein) and in \cite{Dalton12a}
it has been shown (see Eq. (60) therein) for the same binomial state as in (%
\ref{Eq.BinomialState}) that 
\begin{eqnarray}
\left\langle \widehat{J}\,_{x}\right\rangle &=&0\qquad \left\langle \widehat{%
J}\,_{y}\right\rangle =0\qquad \left\langle \widehat{J}\,_{z}\right\rangle =-%
\frac{N}{2}  \nonumber \\
\left\langle \Delta \widehat{J}\,_{x}^{2}\right\rangle &=&\frac{N}{4}\qquad
\left\langle \Delta \widehat{J}\,_{y}^{2}\right\rangle =\frac{N}{4}\qquad
\left\langle \Delta \widehat{J}\,_{z}^{2}\right\rangle =0
\label{Eq.MeanVarianceBinomialState}
\end{eqnarray}%
(see Eqs. (162) and (176) therein). Hence the binomial state is not spin
squeezed since $\left\langle \Delta \widehat{J}\,_{x}^{2}\right\rangle
=\left\langle \Delta \widehat{J}\,_{y}^{2}\right\rangle =\frac{1}{2}%
|\left\langle \widehat{J}\,_{z}\right\rangle |.$ It is of course a \emph{%
minimum uncertainty state} with spin fluctuations at the \emph{standard
quantum limit}. Here\textbf{\ }$\sqrt{\left\langle \Delta \widehat{J}%
\,_{x,y}^{2}\right\rangle }/|\left\langle \widehat{J}\,_{z}\right\rangle |=1/%
\sqrt{N}$. Clearly, it is a non-entangled state for modes $\widehat{c}$ and $%
\widehat{d}$ , being the product of a number state for mode $\widehat{c}$
with the vacuum state for mode $\widehat{d}$.

Note that from the point of view of the original modes $\widehat{a}$ and $%
\widehat{b}$, this is an entangled state. so the question is: Is it a spin
squeezed state with respect to the original spin operators $\widehat{S}_{\xi
}$ ($\xi =x,y,z$) ? The Bloch vector and variances for this binomial state
are given in \cite{Dalton12a} (see Eq. (163) in the main paper and Eq. (410)
in the Appendix). The results include:

\begin{eqnarray}
\left\langle \widehat{S}\,_{z}\right\rangle &=&-\frac{N}{2}\cos 2\theta 
\nonumber \\
\left\langle \Delta \widehat{S}\,_{x}^{2}\right\rangle &=&\frac{N}{4}(\cos
^{2}2\theta \,\cos ^{2}\chi +\sin ^{2}\chi )\qquad \left\langle \Delta 
\widehat{S}\,_{y}^{2}\right\rangle =\frac{N}{4}(\cos ^{2}2\theta \,\sin
^{2}\chi +\cos ^{2}\chi )  \nonumber \\
&&  \label{Eq.BinomStateOrigSpinOprs}
\end{eqnarray}%
This gives $\left\langle \Delta \widehat{S}\,_{x}^{2}\right\rangle
\left\langle \Delta \widehat{S}\,_{y}^{2}\right\rangle -\frac{1}{4}%
|\left\langle \widehat{S}\,_{z}\right\rangle |^{2}=\frac{1}{16}N^{2}(\cos
^{2}2\theta -1)^{2}\cos ^{2}\chi \sin ^{2}\chi \geq 0$ as required for the
Heisenberg uncertainty principle. With $\chi =0$ we have $\left\langle
\Delta \widehat{S}\,_{x}^{2}\right\rangle =\frac{N}{4}\cos ^{2}2\theta \,$\
and $\left\langle \Delta \widehat{S}\,_{y}^{2}\right\rangle =\frac{N}{4}$,
whilst $\frac{1}{2}|\left\langle \widehat{S}\,_{z}\right\rangle |=\frac{N}{4}%
|\cos 2\theta |$. As $\left\langle \Delta \widehat{S}\,_{x}^{2}\right\rangle
<\frac{1}{2}|\left\langle \widehat{S}\,_{z}\right\rangle |$ there is spin
squeezing in $\widehat{S}\,_{x}$ for this entangled state of modes $\widehat{%
a}$ and $\widehat{b}$, though not of course for the new spin operator $%
\widehat{J}\,_{x}$ since this state is non-entangled for modes $\widehat{c}$
and $\widehat{d}$. This example illustrates the need to carefully define
spin squeezing and entanglement in terms of related sets of spin operators
and modes. The same state is entangled with respect to one choice of modes -
and spin squeezing occurs, whilst it is non-entangled with respect to
another set of modes - and no spin squeezing occurs.

To summarise - with a physically based definition of non-entangled states
for bosonic systems with two modes (related to the principal spin operators
that have a diagonal covariance matrix) being the sub-systems and with a
criterion for spin squeezing that focuses on these principal spin
fluctuations, it seen that whilst non-entangled states are never spin
squeezed and therefore although entanglement is a necessary condition for
spin squeezing, it is not a sufficient one. There are entangled states that
are not spin squeezed. Furthermore, as there is no simple quantitative links
between measures of spin squeezing and those for entanglement, the mere
presence of spin squeezing only demonstrates the qualitative result that the
quantum state is entangled. Nevertheless, for high precision measurements
based on spin operators where the primary emphasis is on how much spin
squeezing can be achieved, knowing that entangled states are needed provides
an impetus for studying such states and how they might be produced.

\subsection{Entangled States that are Spin Squeezed - Relative Phase
Eigenstate}

\label{SubSection - Ent States that are Spin Sq}

As an example of an entangled state that is spin squeezed we consider the
relative phase eigenstate $\left\vert \frac{N}{2},\theta _{p}\right\rangle $
for a two mode system in which there are $N$ bosons. For modes with
annihilation operators $\widehat{a}$, $\widehat{b}$ the \emph{relative phase
eigenstate} is defined as 
\begin{equation}
\left\vert \frac{N}{2},\theta _{p}\right\rangle =\frac{1}{\sqrt{N+1}}%
\dsum\limits_{k=\,-N/2}^{N/2}\exp (ik\theta _{p})\,\frac{(\widehat{a}^{\dag
})^{N/2-k}}{\sqrt{(N/2-k)!}}\frac{(\widehat{b}^{\dag })^{N/2+k}}{\sqrt{%
(N/2+k)!}}\left\vert 0\right\rangle  \label{Eq.RelativePhaseState}
\end{equation}%
where the relative phase $\theta _{p}=p(2\pi /(N+1))$ with $%
p=-N/2,-N/2+1,...,+N/2$, is an eigenvalue of the relative phase Hermitian
operator of the type introduced by Barnett and Pegg \cite{Barnett89a} (see 
\cite{Dalton12a} and references therein). Note that the eigenvalues form a
quasi-continuum over the range $-\pi $ to $+\pi $, with a small separation
between neighboring phases of $O(1/N)$. The relative phase state is
consistent with the super-selection rule and is an entangled state for modes 
$\widehat{a}$, $\widehat{b}$. The Bloch vector for spin operators $\widehat{S%
}_{x}$, $\widehat{S}_{y}$, $\widehat{S}_{z}$ is given by (see \cite%
{Dalton12a}) 
\begin{equation}
\left\langle \widehat{S}\,_{x}\right\rangle =N\,\frac{\pi }{8}\cos \theta
_{p}\qquad \left\langle \widehat{S}\,_{y}\right\rangle =-N\,\frac{\pi }{8}%
\sin \theta _{p}\qquad \left\langle \widehat{S}\,_{z}\right\rangle =0
\end{equation}%
but the covariance matrix (see Eq. (178) in \cite{Dalton12a}) is
non-diagonal.

\subsubsection{New Spin Operators}

It is more instructive to consider spin squeezing in terms of new spin
operators $\widehat{J}_{x}$, $\widehat{J}_{y}$, $\widehat{J}_{z}$ for which
the covariance matrix is diagonal. The new spin operators are related to the
original spin operators via 
\begin{eqnarray}
\widehat{J}_{x} &=&\widehat{S}_{z}  \nonumber \\
\widehat{J}_{y} &=&\sin \theta _{p}\,\widehat{S}_{x}+\cos \theta _{p}\,%
\widehat{S}_{y}  \nonumber \\
\widehat{J}_{z} &=&-\cos \theta _{p}\,\widehat{S}_{x}+\sin \theta _{p}\,%
\widehat{S}_{y}  \label{Eq.NewSpinOprsPhaseState}
\end{eqnarray}%
corresponding to the transformation in Eq. (\ref{Eq.NewSpinOprsRotnMatrix})
with Euler angles $\alpha =-\pi +\theta _{p}$, $\beta =-\pi /2$ and $\gamma
=-\pi $.

\subsubsection{Bloch Vector and Covariance Matrix}

The Bloch vector and covariance matrix for spin operators $\widehat{J}_{x}$, 
$\widehat{J}_{y}$, $\widehat{J}_{z}$ are given by (see Eqs. (180), (181) in 
\cite{Dalton12a} - note that the $C(\widehat{J}_{y},\widehat{J}_{y})$
element is incorrect in Eq. (181)) 
\begin{equation}
\left\langle \widehat{J}\,_{x}\right\rangle =0\qquad \left\langle \widehat{J}%
\,_{y}\right\rangle =0\qquad \left\langle \widehat{J}\,_{z}\right\rangle
=-N\,\frac{\pi }{8}  \label{Eq.BlochVectorPhaseState}
\end{equation}%
and%
\begin{eqnarray}
&&\left[ 
\begin{tabular}{lll}
$C(\widehat{J}_{x},\widehat{J}_{x})$ & $C(\widehat{J}_{x},\widehat{J}_{y})$
& $C(\widehat{J}_{x},\widehat{J}_{z})$ \\ 
$C(\widehat{J}_{y},\widehat{J}_{x})$ & $C(\widehat{J}_{y},\widehat{J}_{y})$
& $C(\widehat{J}_{y},\widehat{J}_{z})$ \\ 
$C(\widehat{J}_{z},\widehat{J}_{y})$ & $C(\widehat{J}_{z},\widehat{J}_{y})$
& $C(\widehat{J}_{z},\widehat{J}_{z})$%
\end{tabular}%
\right]  \nonumber \\
&\doteqdot &\left[ 
\begin{tabular}{lll}
$\frac{{\LARGE 1}}{{\LARGE 12}}N^{2}$ & $0$ & $0$ \\ 
$0$ & $\frac{1}{4}+\frac{1}{8}\ln N$ & $0$ \\ 
$0$ & $0$ & $\left( \frac{{\LARGE 1}}{{\LARGE 6}}-\frac{{\LARGE \pi }^{2}}{%
{\LARGE 64}}\right) N^{2}$%
\end{tabular}%
\right] \qquad N\gg 1  \label{Eq.CovMatrixPhaseStateNewSpins}
\end{eqnarray}%
With \ $\left\langle \Delta \widehat{J}\,_{x}^{2}\right\rangle =\frac{%
{\LARGE 1}}{{\LARGE 12}}N^{2}$, $\left\langle \Delta \widehat{J}%
\,_{y}^{2}\right\rangle =\frac{1}{4}+\frac{1}{8}\ln N$ and $\left\langle
\Delta \widehat{J}\,_{z}^{2}\right\rangle =\left( \frac{{\LARGE 1}}{{\LARGE 6%
}}-\frac{{\LARGE \pi }^{2}}{{\LARGE 64}}\right) N^{2}$ and the only non-zero
Bloch vector component being $\left\langle \widehat{J}\,_{z}\right\rangle
=-N\,\frac{\pi }{8}$ it is easy to see that $\left\langle \Delta \widehat{J}%
\,_{x}^{2}\right\rangle \left\langle \Delta \widehat{J}\,_{y}^{2}\right%
\rangle \geq \frac{1}{4}|\left\langle \widehat{J}\,_{z}\right\rangle |^{2}$
as required by the Heisenberg Uncertainty Principle. The principal spin
fluctuations in both $\widehat{J}_{x}$ and $\widehat{J}_{z}$ are comparable
to the length of the Bloch vector and no spin squeezing occurs in either of
these components. However, spin squeezing occurs in that $\widehat{J}_{y}$
is squeezed with respect to $\widehat{J}_{x}$ - $\left\langle \Delta 
\widehat{J}\,_{y}^{2}\right\rangle $ only increases as $\frac{1}{8}\ln N$
whilst $\frac{1}{2}|\left\langle \widehat{J}\,_{z}\right\rangle |$ increases
as $\frac{\pi }{16}N\,$\ for large $N$. Hence the relative phase state
satisfies the test in Eq. (\ref{Eq.SpinSqueezeEntangleTest}) to demonstate
entanglement for modes $\widehat{c}$, $\widehat{d}$. Here $\sqrt{%
\left\langle \Delta \widehat{J}\,_{,y}^{2}\right\rangle }/|\left\langle 
\widehat{J}\,_{z}\right\rangle |\sim \sqrt{\ln N}/N$ which indicates that
the Heisenberg limit is being reached.

Note that \emph{none} of the old spin components are spin squeezed. As shown
in Ref. \cite{Dalton12a} $\left\langle \Delta \widehat{S}\,_{x}^{2}\right%
\rangle =\left( \frac{{\LARGE 1}}{{\LARGE 6}}-\frac{{\LARGE \pi }^{2}}{%
{\LARGE 64}}\right) \cos ^{2}\theta _{p}\,N^{2}$, $\left\langle \Delta 
\widehat{S}\,_{y}^{2}\right\rangle =\left( \frac{{\LARGE 1}}{{\LARGE 6}}-%
\frac{{\LARGE \pi }^{2}}{{\LARGE 64}}\right) \sin ^{2}\theta _{p}\,N^{2}$
and $\left\langle \Delta \widehat{S}\,_{z}^{2}\right\rangle =\frac{{\LARGE 1}%
}{{\LARGE 12}}N^{2}$ , along with $\left\langle \widehat{S}%
\,_{x}\right\rangle =N\,\frac{\pi }{8}\cos \theta _{p}\qquad \left\langle 
\widehat{S}\,_{y}\right\rangle =-N\,\frac{\pi }{8}\sin \theta _{p}\qquad
\left\langle \widehat{S}\,_{z}\right\rangle =0$. All variances are of order $%
N^{2}$ whilst the non-zero means are only of order $N$. Hence spin squeezing
in one of the principal spin operators does \emph{not} imply spin squeezing
in any of the original spin operators. This is relevant to spin squeezing
tests for entanglement of the \emph{original} modes.

\subsubsection{New Modes Operators}

To confirm that the relative phase state is in fact an entangled state for
modes $\widehat{c}$, $\widehat{d}$ as well as for the original modes $%
\widehat{a}$, $\widehat{b}$, we note that the new mode operators $\widehat{c}
$, $\widehat{d}$ are given in in Eq. (\ref{Eq.NewModeOprs}) with Euler
angles $\alpha =-\pi +\theta _{p}$, $\beta =-\pi /2$ and $\gamma =-\pi $.
The old mode operators are given in Eq. (\ref{Eq.OldModeOperators}) and with
these Euler angles we have 
\begin{eqnarray}
\widehat{a} &=&-\exp (\frac{1}{2}i\theta _{p})\frac{1}{\sqrt{2}}\left( 
\widehat{c}-\widehat{d}\right)  \nonumber \\
\widehat{b} &=&-\exp (-\frac{1}{2}i\theta _{p})\frac{1}{\sqrt{2}}\left( 
\widehat{c}+\widehat{d}\right)  \nonumber \\
&&  \label{Eq.OldModeOprsPhaseState}
\end{eqnarray}%
This enables us to write the phase state in terms of new mode operators $%
\widehat{c}$, $\widehat{d}$ as 
\begin{eqnarray}
\left\vert \frac{N}{2},\theta _{p}\right\rangle &=&\frac{1}{\sqrt{N+1}}%
\left( \frac{-1}{\sqrt{2}}\right)
^{N}\dsum\limits_{k=\,-N/2}^{N/2}\dsum\limits_{r=\,-N/4+k/2}^{\,N/4-k/2}%
\dsum\limits_{s=\,-N/4-k/2}^{N/4+k/2}  \nonumber \\
&&\times \frac{1}{\sqrt{(N/2-k)!}}\frac{1}{\sqrt{(N/2+k)!}}(-1)^{N/4-k/2+r} 
\nonumber \\
&&\times \frac{(N/2-k)!}{(N/4-k/2-r)!(N/4-k/2+r)!}\frac{(N/2+k)!}{%
(N/4+k/2-s)!(N/4+k/2+s)!}  \nonumber \\
&&\times (\widehat{c}^{\dag })^{N/2-(r+s)}\,(\widehat{d}^{\dag
})^{N/2+(r+s)}\left\vert 0\right\rangle  \nonumber \\
&&  \label{Eq.PhaseStateNewModes}
\end{eqnarray}%
We see that the expression does not depend explicitly on the relative phase $%
\theta _{p}$ when written in terms of the new mode unnormalised Fock states $%
(\widehat{c}^{\dag })^{N/2-(r+s)}\,(\widehat{d}^{\dag
})^{N/2+(r+s)}\left\vert 0\right\rangle $. This pure state is a linear
combination of product states of the form $\left\vert N/2-m\right\rangle
_{c}\otimes \left\vert N/2+m\right\rangle _{d}$ for various $m$ - each of
which is an $N$ boson state and an eigenstate for $\widehat{J}\,_{z}$ with
eigenvalue $m$, and therefore is an entangled state for modes $\widehat{c}$, 
$\widehat{d}$ which is compatible with the global super-selection rule. Note
that there cannot just be a single term $m$ involved, otherwise the variance
for $\widehat{J}\,_{z}$ would be zero instead of $\left( \frac{{\LARGE 1}}{%
{\LARGE 6}}-\frac{{\LARGE \pi }^{2}}{{\LARGE 64}}\right) N^{2}$. We will
return to the relative phase state again in SubSection \ref{SubSection -
Hillery 2006}.

\pagebreak

\section{Other Spin Operator Tests for Entanglement}

\label{Section - Criteria for Spin Squeezing Based on Non-Physical States}

In this Section we examine a number of previously stated entanglement tests
involving spin operators. It turns out that \emph{many} of the tests do
confirm entanglement for massive bosons according to the SSR and
symmetrisation principle compliant definition as it is defined here, though
not always for the reasons given in their original proofs. Importantly, in
some cases for massive bosons the tests can be made more general.

There are a number of inequalities involving the \emph{spin operators} that
have previously been derived for testing whether a state for a system of
identical bosons is entangled. These are \emph{not} always associated with
criteria for spin squeezing - which involve the variances and mean values of
the spin operators. Also, some of these inequalities are \emph{not} based on
the requirement that the density operators $\widehat{\rho }_{R}^{A}$, $%
\widehat{\rho }_{R}^{B}$ in the expression for a non-entangled state conform
to the \emph{super-selection rule} that prohibits quantum superpositions of
single mode states with differing numbers of bosons (which was invoked
because they represent possible quantum states for the separate modes - 
\emph{local particle number SSR compliance}). Only generic quantum
properties of the sub-system density operators $\widehat{\rho }_{R}^{A}$, $%
\widehat{\rho }_{R}^{B}$ were used in the derivations. In contrast, our
results are based in effect on a \emph{stricter criterion} as to what
constitutes a \emph{separable state}, so of course we obtain \emph{new}
entanglement tests. However, entanglement tests which were based on \emph{%
not requiring} SSR compliance for $\widehat{\rho }_{R}^{A}$, $\widehat{\rho }%
_{R}^{B}$ will \emph{also} confirm entanglement when SSR compliance is \emph{%
required}. This outcome occurs in the SubSection \ref{SubSection - Hillery
2006} in the case of the Hillery spin variance entanglement test. It also
occurs in SubSection \ref{SubSection - He 2012} for the entanglement test in
(\ref{Eq.SpinEntTestPairsModes}) involving spin operators for two mode
sub-systems, in SubSection \ref{SubSection - Raymer 2003} for the
entanglement test in (\ref{Eq.RaymerTest}) involving mean values of powers
of local spin operators, and in two entanglement tests (\ref%
{Eq.BennattiTest1}), (\ref{Eq.BennattiTest2}) in SubSection \ref{SubSection
- Benatti et al 2011} that involve variances of two mode spin operators.

Other entanglement tests have been proposed whose proofs were based on forms
of the density operator for non-entangled states that are \emph{not}
consistent with the \emph{symmetrisation principle}. The sub-systems were
regarded as labelled individual particles, and strictly speaking, this
should only apply to systems of \emph{distinguishable} particles. These
include the spin squeezing in the total spin operator $\widehat{S}_{z}$ test
(\ref{Eq.SorensenSqg})\ in SubSection \ref{SubSection - Sorensen 2001}. In
that SubSection we show that the original proof in \cite{Sorensen01a} can be
modified to treat \emph{identical} particles but now with distinguishable
pairs of modes as the sub-systems, but the proof requires that the separable
states are \emph{restricted} to one boson per mode pair. However, in
SubSection \ref{SubSection - SpinSqueezingEntangNewModes} we have already
shown that for two mode systems in which SSR compliance applies spin
squeezing in $\widehat{S}_{z}$ demonstrates two mode entanglement. Also, in
SubSection \ref{SubSection - SpinSqgEnt MultiMode} we showed that in
multi-mode cases modes associated with two different internal (hyperfine)
components, that spin squeezing in $\widehat{S}_{z}$ also shows entanglement
occurs in two situations - one where there are \emph{two} sub-systems each
just consisting of modes associated with the same internal component (Case
1), the second where \emph{each} mode counts as a separate sub-system (Case
2). Thus the spin squeezing in $\widehat{S}_{z}$ test does demonstrate
entanglement for identical massive bosons, though not for the reasons given
in the original proof. These new SSR compliant proofs now confirm the spin
squeezing in $\widehat{S}_{z}$ as a valid test for entanglement in two
component or two mode BECs.

\subsection{Hillery et al 2006}

\label{SubSection - Hillery 2006}

\subsubsection{Hillery Spin Variance Entanglement Test}

An entanglement test in which local particle number SSR compliance is
ignored is presented in the paper by Hillery and Zubairy \cite{Hillery06a}
entitled "Entanglement conditions for two-mode states". The paper actually
dealt with EM\ field modes, and the density operators $\widehat{\rho }%
_{R}^{A}$, $\widehat{\rho }_{R}^{B}$ for photon modes allowed for coherences
between states with differing photon numbers. A discussion of SSR for the
case of photons is presented in Paper I, \textbf{in SubSection 3.2}. Hence
the conditions in Eq. (\ref{Eq.CondNonEntStateCD}) were not applied.

We will now derive the Hillery spin variance inequalities involving $%
\left\langle \Delta \widehat{S}\,_{x}\right\rangle ^{2}$, $\left\langle
\Delta \widehat{S}\,_{y}\right\rangle ^{2}$ by applying a similar treatment
to that in SubSection \ref{SubSection - SpinSqueezingEntangNewModes}, but
now ignoring local particle number SSR compliance. It is found that for the
original spin operators $\widehat{S}_{x}$, $\widehat{S}_{y}$, $\widehat{S}%
_{z}$ and modes $\widehat{a}$ and $\widehat{b}$ 
\begin{eqnarray}
\left\langle \widehat{S}\,_{x}^{2}\right\rangle _{R} &=&\frac{1}{4}%
(\left\langle \widehat{b}^{\dag }\widehat{b}\right\rangle _{R}+\left\langle 
\widehat{a}^{\dag }\widehat{a}\right\rangle _{R})+\frac{1}{2}(\left\langle 
\widehat{a}^{\dag }\widehat{a}\right\rangle _{R}\left\langle \widehat{b}%
^{\dag }\widehat{b}\right\rangle _{R})  \nonumber \\
&&+\frac{1}{4}(\left\langle (\widehat{b}^{\dag })^{2}\right\rangle
_{R}\left\langle (\widehat{a})^{2}\right\rangle _{R}+\left\langle (\widehat{b%
})^{2}\right\rangle _{R}\left\langle (\widehat{a}^{\dag })^{2}\right\rangle )
\nonumber \\
\left\langle \widehat{S}\,_{y}^{2}\right\rangle _{R} &=&\frac{1}{4}%
(\left\langle \widehat{b}^{\dag }\widehat{b}\right\rangle _{R}+\left\langle 
\widehat{a}^{\dag }\widehat{a}\right\rangle _{R})+\frac{1}{2}(\left\langle 
\widehat{a}^{\dag }\widehat{a}\right\rangle _{R}\left\langle \widehat{b}%
^{\dag }\widehat{b}\right\rangle _{R})  \nonumber \\
&&-\frac{1}{4}(\left\langle (\widehat{b}^{\dag })^{2}\right\rangle
_{R}\left\langle (\widehat{a})^{2}\right\rangle _{R}+\left\langle (\widehat{b%
})^{2}\right\rangle _{R}\left\langle (\widehat{a}^{\dag })^{2}\right\rangle )
\end{eqnarray}%
where terms such as $\left\langle (\widehat{b}^{\dag })^{2}\right\rangle
_{R} $ and $\left\langle (\widehat{a})^{2}\right\rangle _{R}$ previously
shown to be zero have been retained. Note that in \cite{Hillery06a} the
operators $\widehat{S}_{x}$, $\widehat{S}_{y}$, $\widehat{S}_{z}$
constructed from the EM field mode operators as in Eq. (\ref{Eq.OldSpinOprs}%
) would be related to Stokes parameters Hence 
\begin{eqnarray}
&&\left\langle \widehat{S}\,_{x}^{2}\right\rangle _{R}+\left\langle \widehat{%
S}\,_{y}^{2}\right\rangle _{R}  \nonumber \\
&=&\frac{1}{2}(\left\langle \widehat{b}^{\dag }\widehat{b}\right\rangle
_{R}+\left\langle \widehat{a}^{\dag }\widehat{a}\right\rangle
_{R})+(\left\langle \widehat{a}^{\dag }\widehat{a}\right\rangle
_{R}\left\langle \widehat{b}^{\dag }\widehat{b}\right\rangle _{R})  \nonumber
\\
&=&\frac{1}{2}(\left\langle \widehat{b}^{\dag }\widehat{b}\right\rangle
_{R}(\left\langle \widehat{a}^{\dag }\widehat{a}\right\rangle _{R}+1)+\frac{1%
}{2}(\left\langle \widehat{a}^{\dag }\widehat{a}\right\rangle
_{R}(\left\langle \widehat{b}^{\dag }\widehat{b}\right\rangle _{R}+1))
\end{eqnarray}%
where the terms $\left\langle (\widehat{b}^{\dag })^{2}\right\rangle _{R}$,
..., $\left\langle (\widehat{a}^{\dag })^{2}\right\rangle $ cancel out. This
is the same as before.

However, 
\begin{eqnarray}
\left\langle \widehat{S}\,_{x}\right\rangle _{R} &=&\frac{1}{2}(\left\langle 
\widehat{b}^{\dag }\right\rangle _{R}\left\langle \widehat{a}\right\rangle
_{R}+\left\langle \widehat{a}^{\dag }\right\rangle _{R}\left\langle \widehat{%
b}\right\rangle _{R})  \nonumber \\
\left\langle \widehat{S}\,_{y}\right\rangle _{R} &=&\frac{1}{2i}%
(\left\langle \widehat{b}^{\dag }\right\rangle _{R}\left\langle \widehat{a}%
\right\rangle _{R}-\left\langle \widehat{a}^{\dag }\right\rangle
_{R}\left\langle \widehat{b}\right\rangle _{R})
\end{eqnarray}%
is now non-zero, since the previously zero terms have again been retained.
Hence 
\begin{equation}
\left\langle \widehat{S}\,_{x}\right\rangle _{R}^{2}+\left\langle \widehat{S}%
\,_{y}\right\rangle _{R}^{2}=\left\langle \widehat{b}^{\dag }\right\rangle
_{R}\left\langle \widehat{b}\right\rangle _{R}\left\langle \widehat{a}^{\dag
}\right\rangle _{R}\left\langle \widehat{a}\right\rangle _{R}
\end{equation}%
so that we now have%
\begin{eqnarray}
&&\left\langle \Delta \widehat{S}\,_{x}^{2}\right\rangle _{R}+\left\langle
\Delta \widehat{S}\,_{y}^{2}\right\rangle _{R}  \nonumber \\
&=&\frac{1}{2}(\left\langle \widehat{b}^{\dag }\widehat{b}\right\rangle
_{R}(\left\langle \widehat{a}^{\dag }\widehat{a}\right\rangle _{R}+1)+\frac{1%
}{2}(\left\langle \widehat{a}^{\dag }\widehat{a}\right\rangle
_{R}(\left\langle \widehat{b}^{\dag }\widehat{b}\right\rangle
_{R})+1)-\left\langle \widehat{b}^{\dag }\right\rangle _{R}\left\langle 
\widehat{b}\right\rangle _{R}\left\langle \widehat{a}\right\rangle
_{R}\left\langle \widehat{a}^{\dag }\right\rangle _{R}\left\langle \widehat{a%
}\right\rangle _{R}  \nonumber \\
&=&\frac{1}{2}\left( \left\langle \widehat{b}^{\dag }\widehat{b}%
\right\rangle _{R}+\left\langle \widehat{a}^{\dag }\widehat{a}\right\rangle
_{R}\right) +\left( \left\langle \widehat{b}^{\dag }\widehat{b}\right\rangle
_{R}(\left\langle \widehat{a}^{\dag }\widehat{a}\right\rangle
_{R}-|\left\langle \widehat{a}\right\rangle _{R}|^{2}|\left\langle \widehat{b%
}^{\dag }\right\rangle _{R}|^{2}\right)  \label{Eq.ProductStateInequality}
\end{eqnarray}%
But from the Schwarz inequality - which is based on $\left\langle (\widehat{a%
}^{\dag }-\left\langle \widehat{a}^{\dag }\right\rangle )(\widehat{a}%
-\left\langle \widehat{a}\right\rangle )\right\rangle \geq 0$ for any state 
\begin{equation}
|\left\langle \widehat{a}\right\rangle _{R}|^{2}\leq \left\langle \widehat{a}%
^{\dag }\widehat{a}\right\rangle _{R}\qquad |\left\langle \widehat{b}%
\right\rangle _{R}|^{2}\leq \left\langle \widehat{b}^{\dag }\widehat{b}%
\right\rangle _{R}  \label{Eq.Schwarz}
\end{equation}%
so that 
\begin{equation}
\left\langle \Delta \widehat{S}\,_{x}^{2}\right\rangle _{R}+\left\langle
\Delta \widehat{S}\,_{y}^{2}\right\rangle _{R}\geq \frac{1}{2}(\left\langle 
\widehat{b}^{\dag }\widehat{b}\right\rangle _{R}+\left\langle \widehat{a}%
^{\dag }\widehat{a}\right\rangle _{R})
\end{equation}%
and thus from Eq. (\ref{Eq.VarianceResult}) it follows that for a general
non entangled state 
\begin{equation}
\left\langle \Delta \widehat{S}\,_{x}^{2}\right\rangle +\left\langle \Delta 
\widehat{S}\,_{y}^{2}\right\rangle \geq \sum_{R}P_{R}\frac{1}{2}\left(
\left\langle \widehat{n}_{b}\right\rangle _{R}+\left\langle \widehat{n}%
_{a}\right\rangle _{R}\right)
\end{equation}%
However, half the expectation value of the number operator is 
\begin{equation}
\frac{1}{2}\left\langle \widehat{N}\right\rangle =\frac{1}{2}\left\langle (%
\widehat{n}_{a}+\widehat{n}_{b})\right\rangle =\sum_{R}P_{R}\frac{1}{2}%
\left( \left\langle \widehat{n}_{b}\right\rangle _{R}+\left\langle \widehat{n%
}_{a}\right\rangle _{R}\right)
\end{equation}%
so for a non-entangled state%
\begin{equation}
\left\langle \Delta \widehat{S}\,_{x}^{2}\right\rangle +\left\langle \Delta 
\widehat{S}\,_{y}^{2}\right\rangle \geq \frac{1}{2}\left\langle \widehat{N}%
\right\rangle  \label{Eq.InvalidInequalityVariance}
\end{equation}%
This inequality for non-entangled states is given in \cite{Hillery06a} (see
their Eq. (3)). The above proof was based on \emph{not} invoking the SSR
requirements for separable states that we apply in this paper.

\subsubsection{Validity of Hillery Test for Local SSR Compliant
Non-Entangled States}

However, it is interesting that the inequality (\ref%
{Eq.InvalidInequalityVariance}) can be more readily derived from the
definition of entangled states used in the present paper - which is based on
local particle number SSR compliance for separable states. We would then
find that $\left\langle \widehat{S}\,_{x}\right\rangle _{R}=\left\langle 
\widehat{S}\,_{y}\right\rangle _{R}=0$ and hence 
\begin{equation}
\left\langle \Delta \widehat{S}\,_{x}^{2}\right\rangle _{R}+\left\langle
\Delta \widehat{S}\,_{y}^{2}\right\rangle _{R}=\frac{1}{2}\left(
\left\langle \widehat{b}^{\dag }\widehat{b}\right\rangle _{R}+\left\langle 
\widehat{a}^{\dag }\widehat{a}\right\rangle _{R}\right) +\left( \left\langle 
\widehat{b}^{\dag }\widehat{b}\right\rangle _{R}(\left\langle \widehat{a}%
^{\dag }\widehat{a}\right\rangle _{R}\right)  \label{Eq.NewProductInequality}
\end{equation}%
instead of Eq.(\ref{Eq.ProductStateInequality}). Since the term $%
\left\langle \widehat{b}^{\dag }\widehat{b}\right\rangle _{R}(\left\langle 
\widehat{a}^{\dag }\widehat{a}\right\rangle _{R}$ is always positive we find
after applying Eq. (\ref{Eq.VarianceResult}) that 
\begin{equation}
\left\langle \Delta \widehat{S}\,_{x}^{2}\right\rangle +\left\langle \Delta 
\widehat{S}\,_{y}^{2}\right\rangle \geq \frac{1}{2}\left\langle \widehat{N}%
\right\rangle  \label{Eq.InequalityVariance2}
\end{equation}%
which is the same as in Eq. (\ref{Eq.InvalidInequalityVariance}). Hence,
finding that $\left\langle \Delta \widehat{S}\,_{x}^{2}\right\rangle
+\left\langle \Delta \widehat{S}\,_{y}^{2}\right\rangle <\frac{1}{2}%
\left\langle \widehat{N}\right\rangle $ would show that the state was
entangled, irrespective of whether or not entanglement is defined in terms
of non-physical unentangled states.

Thus, the \emph{Hillery} \emph{spin variance} entanglement test \cite%
{Hillery06a} is that \emph{if} 
\begin{equation}
\left\langle \Delta \widehat{S}\,_{x}^{2}\right\rangle +\left\langle \Delta 
\widehat{S}\,_{y}^{2}\right\rangle <\frac{1}{2}\left\langle \widehat{N}%
\right\rangle  \label{Eq.HillerySpinEntTest}
\end{equation}%
then the state is an entangled state of modes $\widehat{a}$ and $\widehat{b}$%
. \ This test is still used in recent papers, for example \cite{He12a}, \cite%
{He12b} which deal with the entanglement of sub-systems each consisting of
single modes $\widehat{a}$, $\widehat{b}$ for a double well situation (in
these papers $\widehat{S}\,_{x}\rightarrow \widehat{J}_{AB}^{X}$, $\widehat{S%
}\,_{y}\rightarrow -\widehat{J}_{AB}^{Y}$, $\widehat{S}\,_{z}\rightarrow -%
\widehat{J}_{AB}^{Z}$).

\subsubsection{Non-Applicable Entanglement Test Involving $|\left\langle 
\widehat{S}\,_{z}\right\rangle |$}

\label{SubSubSection - Non Applicable Test}

Previously we had found for a general non-entangled state that is based on
physically valid density operators $\widehat{\rho }_{R}^{A}$, $\widehat{\rho 
}_{R}^{B}$ 
\begin{eqnarray}
\left\langle \Delta \widehat{S}\,_{x}^{2}\right\rangle -\frac{1}{2}%
|\left\langle \widehat{S}\,_{z}\right\rangle | &\geq &0  \nonumber \\
\left\langle \Delta \widehat{S}\,_{y}^{2}\right\rangle -\frac{1}{2}%
|\left\langle \widehat{S}\,_{z}\right\rangle | &\geq &0
\end{eqnarray}%
so that the sum of the variances satisfies the inequality 
\begin{equation}
\left\langle \Delta \widehat{S}\,_{x}^{2}\right\rangle +\left\langle \Delta 
\widehat{S}\,_{y}^{2}\right\rangle \geq |\left\langle \widehat{S}%
\,_{z}\right\rangle |\,  \label{Eq.CorrectInequalityVariances}
\end{equation}%
This is another correct inequality required for a non-entangled state as
defined in the present paper. It follows that if only physical states $%
\widehat{\rho }_{R}^{A}$, $\widehat{\rho }_{R}^{B}$ are allowed, the related 
\emph{entanglement test} involving $\left\langle \Delta \widehat{S}%
\,_{x}^{2}\right\rangle +\left\langle \Delta \widehat{S}\,_{y}^{2}\right%
\rangle $ would be 
\begin{equation}
\left\langle \Delta \widehat{S}\,_{x}^{2}\right\rangle +\left\langle \Delta 
\widehat{S}\,_{y}^{2}\right\rangle <|\left\langle \widehat{S}%
\,_{z}\right\rangle |\,  \label{Eq.TrueSpinEntTest}
\end{equation}%
For \emph{any} quantum state we have 
\begin{equation}
|\left\langle \widehat{S}\,_{z}\right\rangle |\,=\frac{1}{2}|\left(
\left\langle \widehat{n}_{b}\right\rangle -\left\langle \widehat{n}%
_{a}\right\rangle \right) |\,\leq \frac{1}{2}\left( \left\langle \widehat{n}%
_{b}\right\rangle +\left\langle \widehat{n}_{a}\right\rangle \right) =\frac{1%
}{2}\left\langle \widehat{N}\right\rangle  \label{Eq.GeneralInequalSZN}
\end{equation}%
which means that it is now required that $\left\langle \Delta \widehat{S}%
\,_{x}^{2}\right\rangle +\left\langle \Delta \widehat{S}\,_{y}^{2}\right%
\rangle $ be less than a quantity that is \emph{smaller} than in the
criterion in (\ref{Eq.InvalidInequalityVariance}).

However, it should be noted (see (\ref{Eq.GeneralVarianceSumResult})) that 
\emph{all} states, entangled or otherwise, satisfy the inequality 
\begin{equation}
\left\langle \Delta \widehat{S}\,_{x}^{2}\right\rangle +\left\langle \Delta 
\widehat{S}\,_{y}^{2}\right\rangle \geq |\left\langle \widehat{S}%
\,_{z}\right\rangle |  \label{Eq.GeneralInequality}
\end{equation}%
so the inequality in (\ref{Eq.CorrectInequalityVariances}) - though true, is
of no use in establishing whether a state is entangled in the terms of the
meaning of entanglement in the present paper. There are \emph{no} quantum
states, entangled or otherwise that satisfy the proposed entanglement test
given in Eq. (\ref{Eq.TrueSpinEntTest}). This general result was stated by
Hillery et al \cite{Hillery06a}. To show this we have

\begin{eqnarray}
\left\langle \left( \Delta \widehat{S}\,_{x}-i\lambda \Delta \widehat{S}%
\,_{y}\right) ^{\dag }\left( \Delta \widehat{S}\,_{x}+i\lambda \Delta 
\widehat{S}\,_{y}\right) \right\rangle &\geq &0 \\
\left\langle \Delta \widehat{S}\,_{x}^{2}\right\rangle +\lambda \left\langle 
\widehat{S}_{z}\right\rangle +\lambda ^{2}\left\langle \Delta \widehat{S}%
\,_{y}^{2}\right\rangle &\geq &0
\end{eqnarray}%
for all real $\lambda $. The condition that this function of $\lambda $ is
never negative gives the Heisenberg uncertainty principle $\left\langle
\Delta \widehat{S}\,_{x}^{2}\right\rangle \left\langle \Delta \widehat{S}%
\,_{y}^{2}\right\rangle \geq \frac{1}{4}|\left\langle \widehat{S}%
\,_{z}\right\rangle |^{2}$ and (\ref{Eq.GeneralInequality}) follows from
taking $\lambda =1$ and $\lambda =-1$. Even spin squeezed states with $%
\left\langle \Delta \widehat{S}\,_{x}^{2}\right\rangle <\frac{1}{2}%
|\left\langle \widehat{S}\,_{z}\right\rangle |$ still have $\left\langle
\Delta \widehat{S}\,_{x}^{2}\right\rangle +\left\langle \Delta \widehat{S}%
\,_{y}^{2}\right\rangle \geq |\left\langle \widehat{S}\,_{z}\right\rangle
|\, $, so it is \emph{never} found that $\left\langle \Delta \widehat{S}%
\,_{x}^{2}\right\rangle +\left\langle \Delta \widehat{S}\,_{y}^{2}\right%
\rangle <|\left\langle \widehat{S}\,_{z}\right\rangle |$ and hence this
latter inequality \emph{cannot} used as a test for entanglement.

Fortunately - as we have seen, showing that spin squeezing occurs via \emph{%
either} $\left\langle \Delta \widehat{S}\,_{x}^{2}\right\rangle <\frac{1}{2}%
|\left\langle \widehat{S}\,_{z}\right\rangle |$ \emph{or} $\left\langle
\Delta \widehat{S}\,_{y}^{2}\right\rangle <\frac{1}{2}|\left\langle \widehat{%
S}\,_{z}\right\rangle |$ is sufficient to establish that the state is an
entangled state for modes $\widehat{a},\widehat{b}$, with analogous results
if principle spin operators are considered. Applying the Hillery et al
entanglement test in Eq. (\ref{Eq.HillerySpinEntTest}) involving $\frac{1}{2}%
\left\langle \widehat{N}\right\rangle $ is also a valid entanglement test,
but is usually \emph{less stringent} than the spin squeezing test involving
either $\left\langle \Delta \widehat{S}\,_{x}^{2}\right\rangle <\frac{1}{2}%
|\left\langle \widehat{S}\,_{z}\right\rangle |$ \emph{or} $\left\langle
\Delta \widehat{S}\,_{y}^{2}\right\rangle <\frac{1}{2}|\left\langle \widehat{%
S}\,_{z}\right\rangle |$. For the Hillery et al entanglement test to be
satisfied at least one of $\left\langle \Delta \widehat{S}%
\,_{x}^{2}\right\rangle $ or $\left\langle \Delta \widehat{S}%
\,_{y}^{2}\right\rangle $ is required to be less than $\frac{1}{2}%
\left\langle \widehat{N}\right\rangle $, whereas for the spin squeezing test
to apply at least one of $\left\langle \Delta \widehat{S}\,_{x}^{2}\right%
\rangle $ or $\left\langle \Delta \widehat{S}\,_{y}^{2}\right\rangle $ must
be less than $\frac{1}{2}|\left\langle \widehat{S}\,_{z}\right\rangle |$.
The quantity $\frac{1}{2}|\left\langle \widehat{S}\,_{z}\right\rangle |$ is
likely to be smaller than $\frac{1}{2}\left\langle \widehat{N}\right\rangle $
- for example the Bloch vector may lie close to the $xy$ plane, so a greater
degree of reduction in spin fluctuation is needed to satisfy the spin
squeezing test for entanglement.

However, this is not always the case as the example of the \emph{relative
phase state} discussed in SubSection \ref{SubSection - Ent States that are
Spin Sq} shows. The results in the current SubSection can easily be modified
to apply to new spin operators $\widehat{J}_{x}$, $\widehat{J}_{y}$, $%
\widehat{J}_{z}$ , with entanglement being considered for new modes $%
\widehat{c}$ and $\widehat{d}$. The Hillery et al \cite{Hillery06a}
entanglement test then becomes 
\begin{equation}
\left\langle \Delta \widehat{J}\,_{x}^{2}\right\rangle +\left\langle \Delta 
\widehat{J}\,_{y}^{2}\right\rangle <\frac{1}{2}\left\langle \widehat{N}%
\right\rangle  \label{Eq.HilleryTestB}
\end{equation}%
In the case of the relative phase eigenstate we have from Eq. (\ref%
{Eq.CovMatrixPhaseStateNewSpins}) that $\left\langle \Delta \widehat{J}%
\,_{x}^{2}\right\rangle +\left\langle \Delta \widehat{J}\,_{y}^{2}\right%
\rangle =\frac{{\LARGE 1}}{{\LARGE 12}}N^{2}+\frac{1}{4}+\frac{1}{8}\ln
N\approx \frac{{\LARGE 1}}{{\LARGE 12}}N^{2}$ for large $N$. This clearly
exceeds $\frac{1}{2}\left\langle \widehat{N}\right\rangle =\frac{{\LARGE 1}}{%
{\LARGE 2}}N$, so the Hillery et al \cite{Hillery06a} test for entanglement
fails. On the other hand, as we have seen in SubSection \ref{SubSection -
Ent States that are Spin Sq} $\left\langle \Delta \widehat{J}%
\,_{y}^{2}\right\rangle <\frac{1}{2}|\left\langle \widehat{J}%
\,_{z}\right\rangle |\,\approx \,\frac{\pi }{16}N$, so the spin squeezing
test is satisfied for this entangled state of modes $\widehat{c}$ and $%
\widehat{d}$.

\subsubsection{Hillery Entanglement Test - Multi-Mode Case}

It turns out that the Hillery spin variance test can also be applied in
multi-mode situations, where the spin operators are defined as in SubSection %
\ref{SubSection - Spin Operators Multimode Case}. As explained in SubSection %
\ref{SubSection - SpinSqgEnt MultiMode} three cases occur in regard to
specifying the sub-systems. For \emph{Case 1}, where there are \emph{two
sub-systems} each consisting of all the modes $\widehat{a}_{i}$ or all the
modes $\widehat{b}_{i}$. the Hillery spin variance test as in (\ref%
{Eq.HillerySpinEntTest}) applies. The proof is set out in Appendix \ref%
{Appendix = Hillery Spin MultiMode}, and again does not require the
sub-system density operators to be local SSR\ compliant. Also, for \emph{%
Case 2 }where there are $2n$ subsystems consisting of \emph{all} modes $%
\widehat{a}_{i}$ and \emph{all} modes $\widehat{b}_{i}$ the Hillery spin
variance test as in (\ref{Eq.HillerySpinEntTest}) applies. The proof is set
out in Appendix \ref{Appendix = Hillery Spin MultiMode}, and again does not
require the sub-system density operators to be local SSR\ compliant.
However, for \emph{Case 3 }where there are $n$ sub-systems consisting of 
\emph{all} mode pairs $\widehat{a}_{i}$ and $\widehat{b}_{i}$ the Hillery
spin variance test does \emph{not} apply. The reason is explained in
Appendix \ref{Appendix = Hillery Spin MultiMode}. Basically, it is because
specific sub-system density operators $\widehat{\rho }_{R}^{ab(i)}$ (see (%
\ref{Eq.SepStatesMultiModeCase3})) could be \emph{entangled} states of the
modes $\widehat{a}_{i}$ and $\widehat{b}_{i}$ all of which \emph{do} satisfy
the Hillery test involving $\left\langle \widehat{N}_{i}\right\rangle _{R}$
for this $i$th sub-system. If we choose a special separable state of the
form (\ref{Eq.SepStatesMultiModeCase3}) with just \emph{one} term (no sum
over $R$), it is easy to see that the Hillery test will be satisfied for the
full system. However, the full system state involving these sub-systems is
still a \emph{separable} state, showing that satisfying the Hillery spin
variance test does not always require the state to be entangled.

\subsection{He et al 2012}

\label{SubSection - He 2012}

In two papers dealing with EPR entanglement He et al \cite{He11a}, \cite%
{He12a} a \emph{four mode} system associated with a double well potential is
considered. In the left well $1$ there are two localised modes with
annihilation operators $\widehat{a}_{1}$, $\widehat{b}_{1}$ and in the right
well $2$ there are two localised modes with annihilation operators $\widehat{%
a}_{2}$, $\widehat{b}_{2}$. The modes in each well are associated with two
different internal states $A$\ and $B$. Note that we use a different
notation to \cite{He11a}, \cite{He12a}.\textbf{\ }This four mode system
provides for the possibility of entanglement of \emph{two sub-systems} each
consisting of \emph{pairs of modes}. We can therefore still consider \emph{%
bipartite} entanglement however. With four modes there are three different
choices of such sub-systems but perhaps the most interesting from the point
of view of entanglement of spatially separated modes - and hence
implications for EPR entanglement - would be to have the two\emph{\ left well%
} modes $\widehat{a}_{1}$, $\widehat{b}_{1}$\ as sub-system $1$\ and the two 
\emph{right well} modes $\widehat{a}_{2}$, $\widehat{b}_{2}$\ as sub-system $%
2$. This is an example of the general Case 3 considered for multi-modes in
SubSection \ref{SubSection - SpinSqgEnt MultiMode}. Consistent with the
requirement that the sub-system density operators $\widehat{\rho }%
_{R}^{ab(1)}$, $\widehat{\rho }_{R}^{ab(2)}$ conform to the symmetrisation
principle and the super-selection rule, these density operators will not in
general represent separable states for their single mode sub-systems $%
\widehat{a}_{1}$, $\widehat{b}_{1}$ or $\widehat{a}_{2}$, $\widehat{b}_{2}$
- and may even be entangled states. As a result when considering \emph{%
non-entangled} states for the \textbf{pair of} sub-systems $1$ and $2$ we
now have 
\begin{equation}
\left\langle (\widehat{a}_{i}^{\dag }\widehat{b}_{i})^{n}\right\rangle
_{ab(i)}=Tr(\widehat{\rho }_{R}^{ab(i)}(\widehat{a}_{i}^{\dag }\widehat{b}%
_{i})^{n})\neq 0\qquad i=1,2  \label{Eq.MeanAnnihCreatOprsModePairs}
\end{equation}%
in general. In this case where the sub-systems are \emph{pairs} of modes the
spin squeezing entanglement tests as in Eqs.(\ref{Eq.SpinSqEntTestOrigModes}%
) - (\ref{Eq.SpinSqEntTestOrigModesC}) for sub-systems consisting of \emph{%
single} modes cannot be applied, as explained for Case 3 in SubSection \ref%
{SubSection - SpinSqgEnt MultiMode} unless there is only one boson in each
sub-system. Nevertheless, there are still tests of bipartite entanglement
involving spin operators. We next examine entanglement tests in Refs. \cite%
{He11a}, \cite{He12a} to see if any changes occur when we invoke the
definition of entanglement based on SSR compliance.

\subsubsection{Spin Operator Tests for Entanglement}

There are numerous choices for defining spin operators, but the most useful
would be the \emph{local spin operators} for each well \cite{He12a} defined
by 
\begin{eqnarray}
\widehat{S}_{x}^{1} &=&(\widehat{b}_{1}^{\dag }\widehat{a}_{1}+\widehat{a}%
_{1}^{\dag }\widehat{b}_{1})/2\qquad \widehat{S}_{y}^{1}=(\widehat{b}%
_{1}^{\dag }\widehat{a}_{1}-\widehat{a}_{1}^{\dag }\widehat{b}_{1})/2i\qquad 
\widehat{S}_{z}^{1}=(\widehat{b}_{1}^{\dag }\widehat{b}_{1}-\widehat{a}%
_{1}^{\dag }\widehat{a}_{1})/2  \nonumber \\
\widehat{S}_{x}^{2} &=&(\widehat{b}_{2}^{\dag }\widehat{a}_{2}+\widehat{a}%
_{2}^{\dag }\widehat{b}_{2})/2\qquad \widehat{S}_{y}^{2}=(\widehat{b}%
_{2}^{\dag }\widehat{a}_{2}-\widehat{a}_{2}^{\dag }\widehat{b}_{2})/2i\qquad 
\widehat{S}_{z}^{2}=(\widehat{b}_{2}^{\dag }\widehat{b}_{2}-\widehat{a}%
_{2}^{\dag }\widehat{a}_{2})/2  \nonumber \\
&&  \label{Eq.LocalSpinOprs}
\end{eqnarray}%
These satisfy the usual angular momentum commutation rules and those or the
different wells commute. The squares of the local vector spin operators are
related to the total number operators $\widehat{N}_{1}=\widehat{b}_{1}^{\dag
}\widehat{b}_{1}+\widehat{a}_{1}^{\dag }\widehat{a}_{1}$ and $\widehat{N}%
_{2}=\widehat{b}_{2}^{\dag }\widehat{b}_{2}+\widehat{a}_{2}^{\dag }\widehat{a%
}_{2}$ as $\tsum\limits_{\alpha }(\widehat{S}_{\alpha }^{1})^{2}=($ $%
\widehat{N}_{1}/2)(\widehat{N}_{1}/2+1)$ and $\tsum\limits_{\alpha }(%
\widehat{S}_{\alpha }^{2})^{2}=($ $\widehat{N}_{2}/2)(\widehat{N}_{2}/2+1)$.
The\emph{\ total spin operators} are 
\begin{equation}
\widehat{S}_{\alpha }=\widehat{S}_{\alpha }^{1}+\widehat{S}_{\alpha
}^{2}\qquad \alpha =x,y,z  \label{Eq.TotalSpinOprs}
\end{equation}%
and these satisfy the usual angular momentum commutation rules. Hence there
may be cases of spin squeezing, but these do not in general provide
entanglement tests.

For the local spin operators we have in general 
\begin{equation}
\left\langle \widehat{S}_{\alpha }^{1}\right\rangle _{ab(1)}=Tr(\widehat{%
\rho }_{R}^{ab(1)}\,\widehat{S}_{\alpha }^{1})\neq 0\qquad \left\langle 
\widehat{S}_{\alpha }^{2}\right\rangle _{ab(2)}=Tr(\widehat{\rho }%
_{R}^{ab(2)}\,\widehat{S}_{\alpha }^{2})\neq 0\qquad \alpha =x,y,z
\label{Eq.MeanSpinOprsModePairs}
\end{equation}%
based on (\ref{Eq.MeanAnnihCreatOprsModePairs}), and applying (\ref%
{Eq.MeanResult}) we see that \emph{in general} $\left\langle \widehat{S}%
_{\alpha }\right\rangle \neq 0$ for separable states. Thus the Bloch vector
test for entanglement does not apply.

Furthermore, there is no spin squeezing test either. Following a similar
approach as in Section \ref{Section - Relationship Spin Squeezing &
Entanglement} we can obtain the following inequalities for separable states
of the sub-systems $1$ and $2$%
\begin{eqnarray}
&&\left\langle \Delta \widehat{S}\,_{x}^{2}\right\rangle -\frac{1}{2}%
|\left\langle \widehat{S}\,_{z}\right\rangle |\,  \nonumber \\
&\geq &\,\dsum\limits_{R}P_{R}\,(\left\langle (\Delta \widehat{S}%
\,_{x}^{1})^{2}\right\rangle -\frac{1}{2}|\left\langle \widehat{S}%
\,_{z}^{1}\right\rangle |)+\dsum\limits_{R}P_{R}\,(\left\langle (\Delta 
\widehat{S}\,_{x}^{2})^{2}\right\rangle -\frac{1}{2}|\left\langle \widehat{S}%
\,_{z}^{2}\right\rangle |)  \nonumber \\
&& \\
&&\left\langle \Delta \widehat{S}\,_{y}^{2}\right\rangle -\frac{1}{2}%
|\left\langle \widehat{S}\,_{z}\right\rangle |\,  \nonumber \\
&\geq &\,\dsum\limits_{R}P_{R}\,(\left\langle (\Delta \widehat{S}%
\,_{y}^{1})^{2}\right\rangle -\frac{1}{2}|\left\langle \widehat{S}%
\,_{z}^{1}\right\rangle |)+\dsum\limits_{R}P_{R}\,(\left\langle (\Delta 
\widehat{S}\,_{y}^{2})^{2}\right\rangle -\frac{1}{2}|\left\langle \widehat{S}%
\,_{z}^{2}\right\rangle |)  \nonumber \\
&&
\end{eqnarray}%
Similar inequalities can be obtained for other pairs of spin operators. In
neither case can we state that the right sides are always non-negative, For
example, each $\widehat{\rho }_{R}^{ab(1)}$ may be a spin squeezed state for 
$\widehat{S}\,_{x}^{1}$ versus $\widehat{S}\,_{y}^{1}$ and each $\widehat{%
\rho }_{R}^{ab(2)}$ may be a spin squeezed state for $\widehat{S}\,_{x}^{2}$
versus $\widehat{S}\,_{y}^{2}$. In this case the right side of the first
inequality is a negative quantity, so we cannot conclude that the total $%
\widehat{S}\,_{x}$ is \emph{not} squeezed versus $\widehat{S}\,_{y}$ for 
\emph{all }separable states. As the $\widehat{\rho }_{R}^{ab(1)}$ and $%
\widehat{\rho }_{R}^{ab(2)}$ can be chosen independently we see that
separable states for the sub-systems $1$ and $2$ may \emph{be} spin
squeezed, so the presence of spin squeezing in a\emph{\ total} spin operator
is not a test for bipartite entanglement in this four mode system. This does
not of course preclude tests for bipartite entanglement involving spin
operators, as we will now see.

In SubSection \textbf{2.8 }of paper\textbf{\ 1} it was shown that $%
|\left\langle \widehat{\Omega }_{A}^{\dag }\widehat{\Omega }%
_{B}\right\rangle |^{2}\leq \left\langle \widehat{\Omega }_{A}^{\dag }\,%
\widehat{\Omega }_{A}\,\widehat{\Omega }_{B}^{\dag }\,\widehat{\Omega }%
_{B}\right\rangle $ for a non-entangled state of general sub-systems\textbf{%
\ }$A$ \emph{and}\textbf{\ }$B$, so with $\widehat{\Omega }_{A}\rightarrow 
\widehat{S}_{-}^{1}=\widehat{S}_{x}^{1}-i\widehat{S}_{y}^{1}$ and $\widehat{%
\Omega }_{B}\rightarrow \widehat{S}_{-}^{2}=\widehat{S}_{x}^{2}-i\widehat{S}%
_{y}^{2}=(\widehat{S}_{+}^{2})^{\dag }$ to give 
\begin{equation}
|\left\langle \widehat{S}_{+}^{1}\,\widehat{S}_{-}^{2}\right\rangle
|^{2}\leq \left\langle \widehat{S}_{+}^{1}\,\widehat{S}_{-}^{1}\,\widehat{S}%
_{+}^{2}\,\widehat{S}_{-}^{2}\,\right\rangle
\label{Eq.InequalitySpinOprsNonEntStatesModePairs}
\end{equation}%
for a non-entangled state of sub-systems $1$ and $2$. For the non-entangled
state of these two sub-systems we have 
\begin{equation}
\left\langle \widehat{S}_{+}^{1}\,\widehat{S}_{-}^{2}\right\rangle
=\dsum\limits_{R}P_{R}\,\left\langle \widehat{S}_{+}^{1}\right\rangle
_{ab(1)}^{R}\left\langle \widehat{S}_{-}^{2}\right\rangle _{ab(2)}^{R}
\end{equation}%
which in general is non-zero from Eq.(\ref{Eq.MeanSpinOprsModePairs}).

Hence a valid \emph{entanglement test} involving \emph{spin operators} for
sub-systems $1$ and $2$ - \emph{each} consisting of \emph{two modes}
localised in each well exists, \textbf{so if} 
\begin{equation}
|\left\langle \widehat{S}_{+}^{1}\,\widehat{S}_{-}^{2}\right\rangle
|^{2}>\left\langle \widehat{S}_{+}^{1}\,\widehat{S}_{-}^{1}\,\widehat{S}%
_{+}^{2}\,\widehat{S}_{-}^{2}\,\right\rangle
\label{Eq.SpinEntTestPairsModes}
\end{equation}%
then the two sub-systems are entangled. A similar conclusion is stated in 
\cite{He12a}, where the criterion was predicted to be satisfied for four
mode two well BEC systems. This test for entanglement involves the local
spin operators, though it is not then the same as spin squeezing criteria.
It is referred to there as \emph{spin entanglement}. Other similar tests may
be obtained via different choices of $\widehat{\Omega }_{A}$ and $\widehat{%
\Omega }_{B}$.

\subsection{Raymer et al 2003}

\label{SubSection - Raymer 2003}

In a paper also dealing with bipartite entanglement where the sub-systems
each consist of two modes, Raymer et al \cite{Raymer03a} derive entanglement
tests involving spin operators for the sub-systems defined in (\ref%
{Eq.LocalSpinOprs}). With Hermitian operators $\widehat{\Omega }_{A},%
\widehat{\Lambda }_{A}$ and $\widehat{\Omega }_{B},\widehat{\Lambda }_{B}$
for the two sub-systems we consider 
\begin{equation}
\widehat{U}=\alpha \widehat{\Omega }_{A}+\beta \widehat{\Omega }_{B}\qquad 
\widehat{V}=\alpha \widehat{\Lambda }_{A}-\beta \widehat{\Lambda }_{B}
\label{Eq.DefnUV}
\end{equation}%
where $\alpha ,\beta $ are real. Then with $\widehat{\rho }=\sum_{R}P_{R}\,%
\widehat{\rho }_{R}$ and $\widehat{\rho }_{R}=\widehat{\rho }_{R}^{A}\otimes 
\widehat{\rho }_{R}^{B}$ and using (\ref{Eq.VarianceResult}) it can first be
shown that%
\begin{equation}
\left\langle \Delta \widehat{U}^{2}\right\rangle \geq
\dsum\limits_{R}P_{R}\,\left\langle \Delta \widehat{U}_{R}^{2}\right\rangle
\qquad \left\langle \Delta \widehat{V}^{2}\right\rangle \geq
\dsum\limits_{R}P_{R}\,\left\langle \Delta \widehat{V}_{R}^{2}\right\rangle
\label{Eq.VariancesUV}
\end{equation}%
where $\Delta \widehat{U}_{R}=\widehat{U}-\left\langle \widehat{U}%
\right\rangle _{R}$, $\Delta \widehat{V}_{R}=\widehat{V}-\left\langle 
\widehat{V}\right\rangle _{R}$ with $\left\langle \widehat{U}\right\rangle
_{R}=Tr(\widehat{U}\,\widehat{\rho }_{R})$, $\left\langle \widehat{V}%
\right\rangle _{R}=Tr(\widehat{V}\,\widehat{\rho }_{R})$.

Substituting for $\widehat{U}$ and $\widehat{V}$ from (\ref{Eq.DefnUV}) and
using $\widehat{\rho }_{R}=\widehat{\rho }_{R}^{A}\otimes \widehat{\rho }%
_{R}^{B}$ we can then evaluate the various terms as follows.%
\begin{eqnarray}
\left\langle \widehat{U}^{2}\right\rangle _{R} &=&\alpha ^{2}\left\langle 
\widehat{\Omega }_{A}^{2}\right\rangle _{A}^{R}+\beta ^{2}\left\langle 
\widehat{\Omega }_{B}^{2}\right\rangle _{B}^{R}+2\alpha \beta \left\langle 
\widehat{\Omega }_{A}\right\rangle _{A}^{R}\left\langle \widehat{\Omega }%
_{B}\right\rangle _{B}^{R}  \nonumber \\
\left\langle \widehat{U}\right\rangle _{R} &=&\alpha \left\langle \widehat{%
\Omega }_{A}\right\rangle _{A}^{R}+\beta \left\langle \widehat{\Omega }%
_{B}\right\rangle _{B}^{R}  \nonumber \\
\left( \left\langle \widehat{U}\right\rangle _{R}\right) ^{2} &=&\alpha
^{2}\left( \left\langle \widehat{\Omega }_{A}\right\rangle _{A}^{R}\right)
^{2}+\beta ^{2}\left( \left\langle \widehat{\Omega }_{B}\right\rangle
_{B}^{R}\right) ^{2}+2\alpha \beta \left\langle \widehat{\Omega }%
_{A}\right\rangle _{A}^{R}\left\langle \widehat{\Omega }_{B}\right\rangle
_{B}^{R}  \nonumber \\
\left\langle \Delta \widehat{U}_{R}^{2}\right\rangle &=&\alpha ^{2}\left(
\left\langle \widehat{\Omega }_{A}^{2}\right\rangle _{A}^{R}-\left(
\left\langle \widehat{\Omega }_{A}\right\rangle _{A}^{R}\right) ^{2}\right)
+\beta ^{2}\left( \left\langle \widehat{\Omega }_{B}^{2}\right\rangle
_{B}^{R}-\left( \left\langle \widehat{\Omega }_{B}\right\rangle
_{B}^{R}\right) ^{2}\right)  \nonumber \\
&&  \label{Eq.VarianceUR}
\end{eqnarray}%
with a similar result for $\left\langle \Delta \widehat{V}%
_{R}^{2}\right\rangle $. Here for sub-system $A$ we define$\left\langle 
\widehat{\Omega }_{A}^{2}\right\rangle _{A}^{R}=Tr(\widehat{\Omega }%
_{A}^{2}\,\widehat{\rho }_{R}^{A})$, $\left\langle \widehat{\Omega }%
_{A}\right\rangle _{A}^{R}=Tr(\widehat{\Omega }_{A}\,\widehat{\rho }%
_{R}^{A}) $ and $\left\langle \widehat{\Lambda }_{A}^{2}\right\rangle
_{A}^{R}=Tr(\widehat{\Lambda }_{A}^{2}\,\widehat{\rho }_{R}^{A})$, $%
\left\langle \widehat{\Lambda }_{A}\right\rangle _{A}^{R}=Tr(\widehat{%
\Lambda }_{A}\,\widehat{\rho }_{R}^{A})$ with analogous expressions for
sub-system $B$.

We thus have 
\begin{eqnarray}
\left\langle \Delta \widehat{U}^{2}\right\rangle &\geq &\alpha
^{2}\dsum\limits_{R}P_{R}\,\left\langle \Delta \widehat{\Omega }%
_{AR}^{2}\right\rangle _{A}^{R}+\beta
^{2}\dsum\limits_{R}P_{R}\,\left\langle \Delta \widehat{\Omega }%
_{BR}^{2}\right\rangle _{B}^{R}  \nonumber \\
\left\langle \Delta \widehat{V}^{2}\right\rangle &\geq &\alpha
^{2}\dsum\limits_{R}P_{R}\,\left\langle \Delta \widehat{\Lambda }%
_{AR}^{2}\right\rangle _{A}^{R}+\beta
^{2}\dsum\limits_{R}P_{R}\,\left\langle \Delta \widehat{\Lambda }%
_{BR}^{2}\right\rangle _{B}^{R}
\end{eqnarray}%
where $\Delta \widehat{\Omega }_{AR}=\widehat{\Omega }_{A}-\left\langle 
\widehat{\Omega }_{A}\right\rangle _{A}^{R}$, $\Delta \widehat{\Omega }_{BR}=%
\widehat{\Omega }_{B}-\left\langle \widehat{\Omega }_{B}\right\rangle
_{B}^{R}$, $\Delta \widehat{\Lambda }_{AR}=\widehat{\Lambda }%
_{A}-\left\langle \widehat{\Lambda }_{A}\right\rangle _{A}^{R}$ and $\Delta 
\widehat{\Lambda }_{BR}=\widehat{\Lambda }_{B}-\left\langle \widehat{\Lambda 
}_{B}\right\rangle _{B}^{R}$.

Adding the two results gives 
\begin{eqnarray}
&&\left\langle \Delta \widehat{U}^{2}\right\rangle +\left\langle \Delta 
\widehat{V}^{2}\right\rangle  \nonumber \\
&\geq &\alpha ^{2}\dsum\limits_{R}P_{R}\,\left( \left\langle \Delta \widehat{%
\Omega }_{AR}^{2}\right\rangle _{A}^{R}+\left\langle \Delta \widehat{\Lambda 
}_{AR}^{2}\right\rangle _{A}^{R}\right)  \nonumber \\
&&+\beta ^{2}\dsum\limits_{R}P_{R}\,\left( \left\langle \Delta \widehat{%
\Omega }_{BR}^{2}\right\rangle _{B}^{R}+\left\langle \Delta \widehat{\Lambda 
}_{BR}^{2}\right\rangle _{B}^{R}\right)  \label{Eq.GeneralVarianceResult}
\end{eqnarray}%
a general variance inequality for separable states.

This last result can be developed further based on the \emph{commutation
rules} 
\begin{equation}
\lbrack \widehat{\Omega }_{A},\widehat{\Lambda }_{A}]=i\widehat{\Theta }%
_{A}\qquad \lbrack \widehat{\Omega }_{B},\widehat{\Lambda }_{B}]=i\widehat{%
\Theta }_{B}  \label{Eq.CommRules}
\end{equation}%
The \emph{Schwarz} inequalities - valid for all real $\lambda _{A}$ and $%
\lambda _{B}$ 
\begin{eqnarray}
\left\langle (\Delta \widehat{\Omega }_{AR}-i\lambda _{A}\Delta \widehat{%
\Lambda }_{AR})\,\widehat{\rho }_{R}^{A}\,(\Delta \widehat{\Omega }%
_{AR}+i\lambda _{A}\Delta \widehat{\Lambda }_{AR})\right\rangle _{A}^{R}
&\geq &0  \nonumber \\
\left\langle (\Delta \widehat{\Omega }_{BR}-i\lambda _{B}\Delta \widehat{%
\Lambda }_{BR})\,\widehat{\rho }_{R}^{B}\,(\Delta \widehat{\Omega }%
_{BR}+i\lambda _{B}\Delta \widehat{\Lambda }_{BR})\right\rangle _{B}^{R}
&\geq &0  \label{Eq.SchwarzB}
\end{eqnarray}%
lead to the following inequalities 
\begin{eqnarray}
\left\langle \Delta \widehat{\Omega }_{AR}^{2}\right\rangle _{A}^{R}+\lambda
_{A}\left\langle \widehat{\Theta }_{A}\right\rangle _{A}^{R}+\lambda
_{A}^{2}\left\langle \Delta \widehat{\Lambda }_{AR}^{2}\right\rangle
_{A}^{R} &\geq &0  \nonumber \\
\left\langle \Delta \widehat{\Omega }_{BR}^{2}\right\rangle _{B}^{R}+\lambda
_{B}\left\langle \widehat{\Theta }_{B}\right\rangle _{B}^{R}+\lambda
_{B}^{2}\left\langle \Delta \widehat{\Lambda }_{BR}^{2}\right\rangle
_{B}^{R} &\geq &0  \label{Eq.InequalitiesAB}
\end{eqnarray}%
so by taking $\lambda _{A,B}=1$ or $-1$ we have 
\begin{eqnarray}
\left\langle \Delta \widehat{\Omega }_{AR}^{2}\right\rangle
_{A}^{R}+\left\langle \Delta \widehat{\Lambda }_{AR}^{2}\right\rangle
_{A}^{R} &\geq &|\left\langle \widehat{\Theta }_{A}\right\rangle _{A}^{R}| 
\nonumber \\
\left\langle \Delta \widehat{\Omega }_{BR}^{2}\right\rangle
_{B}^{R}+\left\langle \Delta \widehat{\Lambda }_{BR}^{2}\right\rangle
_{B}^{R} &\geq &|\left\langle \widehat{\Theta }_{B}\right\rangle _{B}^{R}|
\label{Eq.KeyInequalities}
\end{eqnarray}%
The Heisenberg Uncertainty principle results $\left\langle \Delta \widehat{%
\Omega }_{AR}^{2}\right\rangle _{A}^{R}\left\langle \Delta \widehat{\Lambda }%
_{AR}^{2}\right\rangle _{A}^{R}\geq |\left\langle \widehat{\Theta }%
_{A}\right\rangle _{A}^{R}|^{2}/4$ etc also follow from (\ref%
{Eq.InequalitiesAB}).

Noting that $\dsum\limits_{R}P_{R}\,|\left\langle \widehat{\Theta }%
_{A}\right\rangle _{A}^{R}|\;\geq |\dsum\limits_{R}P_{R}\,\left\langle 
\widehat{\Theta }_{A}\right\rangle _{A}^{R}|\;=|\left\langle \widehat{\Theta 
}_{A}\right\rangle |$ and $\dsum\limits_{R}P_{R}\,|\left\langle \widehat{%
\Theta }_{B}\right\rangle _{B}^{R}|\;\geq
|\dsum\limits_{R}P_{R}\,\left\langle \widehat{\Theta }_{B}\right\rangle
_{A}^{R}|\;=|\left\langle \widehat{\Theta }_{B}\right\rangle |$ since the
modulus of a sum is never greater than the sum of the moduli, we finally
arrive at the final inequality for \emph{separable} states 
\begin{equation}
\left\langle \Delta (\alpha \widehat{\Omega }_{A}+\beta \widehat{\Omega }%
_{B})^{2}\right\rangle +\left\langle \Delta (\alpha \widehat{\Lambda }%
_{A}-\beta \widehat{\Lambda }_{B})^{2}\right\rangle \geq \alpha
^{2}|\left\langle \widehat{\Theta }_{A}\right\rangle |\;+\beta
^{2}|\left\langle \widehat{\Theta }_{B}\right\rangle |
\label{Eq.InequalityVarianceSumSepStates}
\end{equation}%
This leads to the following test for \emph{entanglement} 
\begin{equation}
\left\langle \Delta (\alpha \widehat{\Omega }_{A}+\beta \widehat{\Omega }%
_{B})^{2}\right\rangle +\left\langle \Delta (\alpha \widehat{\Lambda }%
_{A}-\beta \widehat{\Lambda }_{B})^{2}\right\rangle <\alpha
^{2}|\left\langle \widehat{\Theta }_{A}\right\rangle |\;+\beta
^{2}|\left\langle \widehat{\Theta }_{B}\right\rangle |
\label{Eq.GeneralVarianceEntanglementTest}
\end{equation}%
which is usually based on choices where $\alpha ^{2}=\beta ^{2}=1$.

We now choose $\widehat{\Omega }_{A}=\widehat{S}\,_{x}^{1}$, $\widehat{%
\Omega }_{B}=\widehat{S}\,_{x}^{2}$, $\widehat{\Lambda }_{A}=\widehat{S}%
\,_{y}^{1}$ and $\widehat{\Lambda }_{B}=\widehat{S}\,_{y}^{2}$ as in Eq. (%
\ref{Eq.LocalSpinOprs}) along with $\alpha =\beta =1$ \ Here $\widehat{%
\Theta }_{A}=\widehat{S}\,_{z}^{1}$ and $\widehat{\Theta }_{\backslash B}=%
\widehat{S}\,_{z}^{2}.$Here sub-systems $A=1$, $B=2$ consist of modes $%
\widehat{a}_{1}$, $\widehat{b}_{1}$ and $\widehat{a}_{2}$, $\widehat{b}_{2}$
respectively. Hence if we have%
\begin{equation}
\left\langle \Delta (\widehat{S}\,_{x}^{1}+\widehat{S}\,_{x}^{2})^{2}\right%
\rangle +\left\langle \Delta (\widehat{S}\,_{y}^{1}-\widehat{S}%
\,_{y}^{2})^{2}\right\rangle <|\left\langle \widehat{S}_{z}^{1}\right\rangle
|\;+|\left\langle \widehat{S}_{z}^{2}\right\rangle |  \label{Eq.RaymerTest}
\end{equation}%
then \emph{bipartite entanglement} is established. Note that this test did
not require local particle number SSR compliance, but still will apply if
this is invoked. Other tests involving a cyclic interchange of $x,y,z$ can
also be established, as can other tests where the signs within the left
terms are replaced by $(-,+),(+,+),(-,-)$ via appropriate choices of $\alpha 
$, $\beta $. . These tests involve mean values of powers of \emph{local}
spin operators. Similar to tests in SubSections \ref{SubSection - Hillery
2006}, \ref{SubSection - He 2012}, this test also does not require SSR
compliance.

\subsection{Sorensen et al 2001}

\label{SubSection - Sorensen 2001}

\subsubsection{Sorensen Spin Squeezing Entanglement Test}

In a paper entitled "Many-particle entanglement with Bose-Einstein
condensates" Sorensen et al \cite{Sorensen01a} consider the implications for
spin squeezing for non-entangled states of the form 
\begin{equation}
\widehat{\rho }=\sum_{R}P_{R}\,\widehat{\rho }_{R}^{1}\otimes \widehat{\rho }%
_{R}^{2}\otimes \widehat{\rho }_{R}^{3}\otimes ...
\label{Eq.NonEntStateIdenticalAtoms}
\end{equation}%
where $\widehat{\rho }_{R}^{i}$ is a density operator for particle $i$. As
discussed previously, a density operator of this general form is not
consistent with the symmetrisation principle - having separate density
operators $\widehat{\rho }_{R}^{i}$ for specific particles $i$ in an
identical particle system (such as for a BEC) is not compatible with the
indistinguishability of such particles. It is modes that are
distinguishable, not identical particles, so the basis for applying their
results to systems of identical bosons is flawed. However, they derive an
inequality for the spin variance $\left\langle \Delta \widehat{S}%
\,_{z}^{2}\right\rangle $%
\begin{equation}
\left\langle \Delta \widehat{S}\,_{z}^{2}\right\rangle \geq \frac{1}{N}%
\left( \left\langle \widehat{S}\,_{x}\right\rangle ^{2}+\left\langle 
\widehat{S}\,_{y}\right\rangle ^{2}\right)  \label{Eq.SorensenInequality}
\end{equation}%
that applies in the case of non-entangled states. Key steps in their
derivation are stated in the Appendix to \cite{Sorensen01a}, but as the
justification of these steps is not obvious for completeness the full
derivation is given in Appendix \ref{Appendix - Sorensen Results} of the
present paper. This inequality (\ref{Eq.SorensenInequality}) establishes
that if 
\begin{equation}
\xi ^{2}=\frac{\left\langle \Delta \widehat{S}\,_{z}^{2}\right\rangle }{%
\left( \left\langle \widehat{S}\,_{x}\right\rangle ^{2}+\left\langle 
\widehat{S}\,_{y}\right\rangle ^{2}\right) }<\frac{1}{N}
\label{Eq.SpinSqueezingMeasure2}
\end{equation}%
then the state is entangled, so that if the condition for spin squeezing
analogous to that in Eq. (\ref{Eq.NewCriterionSpinSqueezing}) is satisfied,
then entanglement is required if spin squeezing for $\widehat{S}\,_{z}$ to
occur. Spin squeezing is then a test for entanglement in terms of their
definition of an entangled state.

If the Bloch vector is close to the Bloch sphere, for example with $%
\left\langle \widehat{S}\,_{x}\right\rangle =0$ and $\left\langle \widehat{S}%
\,_{y}\right\rangle =N/2$ then the condition (\ref{Eq.SpinSqueezingMeasure2}%
) is equivalent to 
\begin{equation}
\left\langle \Delta \widehat{S}\,_{z}^{2}\right\rangle <\frac{1}{2}%
|\left\langle \widehat{S}\,_{y}\right\rangle |  \label{Eq.SorensenSqg}
\end{equation}%
which is the condition for squeezing in $\widehat{S}\,_{z}$ compared to $%
\widehat{S}\,_{x}$. Spin squeezing is then a test for entanglement in terms
of their definition of an entangled state. Note that the condition (\ref%
{Eq.SorensenSqg}) requires the Bloch vector to be in the $xy$ plane and
close to the Bloch sphere of radius $N/2$. By comparison with (\ref%
{Eq.NewCriterionSpinSqueezing}) we see that the \emph{Sorensen spin squeezing%
} test is that if there is squeezing in $\widehat{S}_{z}$ with respect to
any spin component in the $xy$ plane \emph{and} the Bloch vector is close to
the Bloch sphere, then the state is entangled.

As explained above, the proof of Sorensen et al really applies only when the
individual spins are \emph{distinguishable}. It is possible however to
modify the work of Sorensen et al \cite{Sorensen01a} to apply to a system of
identical bosons in accordance with the symmetrization and super-selection
rules if the index $i$ is \emph{re-interpreted} as specifying diffferent
modes, for example modes localised on \emph{optical lattice} sites $%
i=1,2,..,n$ or distinct free space \emph{momentum states} listed $i=1,2,..,n$%
. On each lattice site or for each momentum state there would be two modes $%
a,b$ - for example associated with two different \emph{internal states} - so
the sub-system density operator $\widehat{\rho }_{R}^{i}$ then applies to
the two modes on site $i$. However the proof of (\ref{Eq.SorensenInequality}%
) requires the $\widehat{\rho }_{R}^{i}$ to be restricted to the case where
there is exactly \emph{one} identical boson on each site or in a momentum
state. Such a localisation process in position or momentum space has the
effect of enabling the identical bosons to be treated \emph{as if} they are
distinguishable. Details are given in the next SubSections. A similar
modification has been carried out by Hyllus et al \cite{Hyllus12a}.

However, as we have seen in SubSection \ref{SubSection - Spin Squeezing
Original Modes} it does in fact turn out for two mode systems of identical
bosons that showing that $\widehat{S}\,_{z}$ is spin squeezed compared to $%
\widehat{S}\,_{x}$ or $\widehat{S}\,_{y}$ \emph{is} sufficient to prove that
the quantum state is entangled. There are \emph{no} restrictions either on
the mean number of bosons occupying each mode. The proof is based on
applying the requirement of local particle number SSR compliance to the
separable states in the present case of massive bosons and treating modes
(not particles) as sub-systems. In SubSection \ref{SubSection - SpinSqgEnt
MultiMode} we have also shown that the same result applies to multi-mode
situations in cases where the sub-systems are all single modes (Case2) or
where there are two sub-systems each containing all modes for a single
component (Case1). So the spin squeezing test is still \emph{valid }for many
particle BEC, though the justification is not as in the proof of Sorensen et
al \cite{Sorensen01a} - which was derived for systems of distinguishable
particles, with each sub-system being a single two state particle.

\subsubsection{Revising Sorensen Spin Squeezing Entanglement Test -
Localised Modes}

The work of Sorensen et al really applies only when the individual spins are
distinguishable. It is possible however to modify the work of Sorensen et al 
\cite{Sorensen01a} to apply to a system of identical bosons in accordance
with the symmetrisation and super-selection rules if the index $i$ is \emph{%
re-interpreted} as specifying diffferent modes, for example modes localised
on \emph{optical lattice} sites or in different \emph{momentum states} $%
i=1,2,..,n$. Another example would be single two state ions with each ion
being trapped in a different spatial region. The revised approach draws on
the results established for multi-mode cases in Appendices \ref{Appendix =
MultiMo Spin Sq Choice 1} and \ref{Appendix - Revised Sorensen}. With two
single particle states $a,b$ available on each site (these could be two
different internal atomic states or two distinct spatial modes localised on
the site) the modes would then be labelled $\left\vert \phi _{\alpha
i}\right\rangle $ with $\alpha =a,b$. The mode orthogonality and
completeness relations would then be 
\begin{eqnarray}
\left\langle \phi _{\alpha \,i}|\phi _{\beta \,j}\right\rangle &=&\delta
_{\alpha \beta }\delta _{ij}  \nonumber \\
\tsum\limits_{\alpha i}\left\vert \phi _{\alpha \,i}\right\rangle
\left\langle \phi _{\alpha \,i}\right\vert &=&\widehat{1}
\label{Eq.ModesRevisedSorensen}
\end{eqnarray}%
With the particles now labelled $K=1,2,3,...$one can define spin operators
in first quantization via 
\begin{eqnarray}
\widehat{S}_{x} &=&\sum_{K}\sum_{i}(\left\vert \phi _{b\,i}(K)\right\rangle
\left\langle \phi _{a\,i}(K)\right\vert +\left\vert \phi
_{a\,i}(K)\right\rangle \left\langle \phi _{b\,i}(K)\right\vert )/2 
\nonumber \\
\widehat{S}_{y} &=&\sum_{K}\sum_{i}(\left\vert \phi _{b\,i}(K)\right\rangle
\left\langle \phi _{a\,i}(K)\right\vert -\left\vert \phi
_{a\,i}(K)\right\rangle \left\langle \phi _{b\,i}(K)\right\vert )/2i 
\nonumber \\
\widehat{S}_{z} &=&\sum_{K}\sum_{i}(\left\vert \phi _{b\,i}(K)\right\rangle
\left\langle \phi _{b\,i}(K)\right\vert -\left\vert \phi
_{a\,i}(K)\right\rangle \left\langle \phi _{a\,i}(K)\right\vert )/2
\label{Eq.SpinOprsRevisedSorensen}
\end{eqnarray}%
In second quantization if the annihilation, creation operators for the modes 
$\left\vert \phi _{ai}\right\rangle $ ,$\left\vert \phi _{bi}\right\rangle $
are $\widehat{a}_{i}$, $\widehat{b}_{i}$ and $\widehat{a}_{i}^{\dag }$, $%
\widehat{b}_{i}^{\dag }$ respectively, then the Schwinger spin operators are
just%
\begin{eqnarray}
\widehat{S}_{x} &=&\sum_{i}(\widehat{b}_{i}^{\dag }\widehat{a}_{i}+\widehat{a%
}_{i}^{\dag }\widehat{b}_{i})/2=\sum_{i}\widehat{S}_{x}^{i}  \nonumber \\
\widehat{S}_{y} &=&\sum_{i}(\widehat{b}_{i}^{\dag }\widehat{a}_{i}-\widehat{a%
}_{i}^{\dag }\widehat{b}_{i})/2i=\sum_{i}\widehat{S}_{y}^{i}  \nonumber \\
\widehat{S}_{z} &=&\sum_{i}(\widehat{b}_{i}^{\dag }\widehat{b}_{i}-\widehat{a%
}_{i}^{\dag }\widehat{a}_{i})/2=\sum_{i}\widehat{S}_{z}^{i}
\label{Eq.SchwingerSpinOprsRevisedSorensen}
\end{eqnarray}%
It is easy to confirm that the overall spin operators $\widehat{S}_{\alpha }$
and the spin operators $\widehat{S}_{\alpha }^{i}$ for the separate \emph{%
pairs} of \emph{modes} $\left\vert \phi _{ai}\right\rangle $, $\left\vert
\phi _{bi}\right\rangle $ (or $\widehat{a}_{i}$, $\widehat{b}_{i}$ for
short) satisfy the same commutation rules as Sorensen et al \cite%
{Sorensen01a} have for the overall spin operators and those for the separate 
\emph{particles}. With this modification the non-entangled state in Eq. (\ref%
{Eq.NonEntStateIdenticalAtoms})\textbf{\ }could be interpreted as being a
non-entangled state where the subsystems are actually \emph{pairs }of\emph{\
modes} $\left\vert \phi _{ai}\right\rangle $ ,$\left\vert \phi
_{bi}\right\rangle $ and the density operators $\widehat{\rho }_{R}^{i}$
would then refer to a subsystem consisting of these pairs of modes. This
corresponds to Case 3 discussed in SubSection \ref{SubSection - SpinSqgEnt
MultiMode}. It is to be noted that entanglement of \emph{pairs} of modes is
different to entanglement of \emph{all separate} modes - Case 2 discussed in
SubSection \ref{SubSection - SpinSqgEnt MultiMode}. It is an example of a
special kind of \emph{multimode entanglement} - since the modes $\left\vert
\phi _{ai}\right\rangle $ ,$\left\vert \phi _{bi}\right\rangle $ may
themselves be entangled we may have "entanglement of entanglement". In terms
of the present paper the density operators $\widehat{\rho }_{R}^{i}$ would
be restricted by the super-selection rule to statistical mixtures of states
with specific total numbers $N_{i}$ of bosons in the pair of modes $%
\left\vert \phi _{ai}\right\rangle $ ,$\left\vert \phi _{bi}\right\rangle $.
In terms of Fock states $\left\vert n_{a\,i}\right\rangle ,\left\vert
n_{b\,i}\right\rangle $ for this pair of modes the allowed quantum states
for the sub-system will be%
\begin{equation}
\left\vert \Phi _{N_{i}}\right\rangle
=\tsum\limits_{k=0}^{N_{i}}A_{k}^{N_{i}}\,\left\vert k\right\rangle
_{a\,i}\left\vert N_{i}-k\right\rangle _{b\,i}
\label{Eq.TwoModeQuantumSuperposition}
\end{equation}%
so at this stage the general mixed physical state for the two mode system 
\emph{could} be 
\begin{equation}
\widehat{\rho }_{R}^{i}=\tsum\limits_{N_{i}=0}^{\infty }\tsum\limits_{\Phi
}P_{\Phi
N_{i}}\,\tsum\limits_{k=0}^{N_{i}}\tsum\limits_{l=0}^{N_{i}}A_{k}^{N_{i}}%
\,(A_{l}^{N})^{\ast }\left\vert k\right\rangle _{a\,i}\left\langle
l\right\vert _{a\,i}\otimes \left\vert N_{i}-k\right\rangle
_{b\,i}\left\langle N_{i}-l\right\vert _{b\,i}
\label{Eq.GeneralDensityOprModePair}
\end{equation}%
This state has no coherences between states of the two mode subsystem with
differing total boson number $N_{i}$ for the pair of modes. However this is
still an entangled states for the two modes $\left\vert \phi
_{ai}\right\rangle $ ,$\left\vert \phi _{bi}\right\rangle $, so the overall
state in Eq. (\ref{Eq.GeneralDensityOprModePair}) is not a separable state
if the subsystems were to consist of \emph{all} the distinct modes.

\subsubsection{Revising Sorensen Spin Squeezing Entanglement Test -
Separable State of Single Modes}

It is possible however to link spin squeezing and entanglement in the case
where the sub-systems consist of \emph{all} the distinct modes (Case2 in
SubSection \ref{SubSection - SpinSqgEnt MultiMode}). To obtain a \emph{fully
non-entangled state} of \emph{all} the modes $\left\vert \phi
_{ai}\right\rangle $ ,$\left\vert \phi _{bi}\right\rangle $ the density
operator $\widehat{\rho }_{R}^{i}$ must then be a product of density
operators for modes $\left\vert \phi _{ai}\right\rangle $ and $\left\vert
\phi _{bi}\right\rangle $%
\begin{equation}
\widehat{\rho }_{R}^{i}=\widehat{\rho }_{R}^{a\,i}\otimes \widehat{\rho }%
_{R}^{b\,i}  \label{Eq.DensityOprModesAiBi}
\end{equation}%
giving the full density operator as 
\begin{equation}
\widehat{\rho }=\sum_{R}P_{R}\,\left( \widehat{\rho }_{R}^{a\,1}\otimes 
\widehat{\rho }_{R}^{b\,1}\right) \otimes \left( \widehat{\rho }%
_{R}^{a\,2}\otimes \widehat{\rho }_{R}^{b\,2}\right) \otimes \left( \widehat{%
\rho }_{R}^{a\,3}\otimes \widehat{\rho }_{R}^{b\,3}\right) \otimes .
\label{Eq.RevisedSorensenDensityOprNonEnt}
\end{equation}%
as is required for a general non-entangled state all $2N$ modes.
Furthermore, as previously the density operators for the individual modes
must represent possible physical states for such modes, so the
super-selection rule for atom number will apply and we have 
\begin{eqnarray}
\left\langle (\widehat{a}_{i})^{n}\right\rangle _{a\,i} &=&Tr(\widehat{\rho }%
_{R}^{a\,i}(\widehat{a}_{i})^{n})=0\qquad \left\langle (\widehat{a}%
_{i}^{\dag })^{n}\right\rangle _{a\,i}=Tr(\widehat{\rho }_{R}^{a\,i}(%
\widehat{a}_{i}^{\dag })^{n})=0  \nonumber \\
\left\langle (\widehat{b}_{i})^{m}\right\rangle _{b\,i} &=&Tr(\widehat{\rho }%
_{R}^{b\,i}(\widehat{b}_{i})^{m})=0\qquad \left\langle (\widehat{b}%
_{i}^{\dag })^{m}\right\rangle _{b\,i}=Tr(\widehat{\rho }_{R}^{b\,i}(%
\widehat{b}_{i}^{\dag })^{m})=0  \nonumber \\
&&  \label{Eq.RevisedSorensenAverages}
\end{eqnarray}

The question is whether this reformulation will lead to a useful inequality
for the spin variances such as $\left\langle \Delta \widehat{S}%
\,_{x}^{2}\right\rangle $. This issue is dealt with in Appendix \ref%
{Appendix - Revised Sorensen} and it is found that we can indeed show for
the general \emph{fully non-entangled} state (\ref%
{Eq.RevisedSorensenDensityOprNonEnt}) that 
\begin{equation}
\left\langle \Delta \widehat{S}\,_{x}^{2}\right\rangle \geq \frac{1}{2}%
|\left\langle \widehat{S}\,_{z}\right\rangle |\quad and\quad \left\langle
\Delta \widehat{S}\,_{y}^{2}\right\rangle \geq \frac{1}{2}|\left\langle 
\widehat{S}\,_{z}\right\rangle |  \label{Eq.VarianceInequalRevisedSorensen}
\end{equation}%
This shows that if there is spin squeezing in \emph{either} $\widehat{S}%
\,_{x}$ \emph{or} $\widehat{S}\,_{y}$ then the state must be entangled. Note
that this result depends on the general non-entangled state being
non-entangled for \emph{all} modes and that the density operator for each
mode $\widehat{a}_{i}$ or $\widehat{b}_{i}$ being a physical state with no
coherences between mode Fock states with differing atom numbers. In terms of
the revised interpretation of the density operator to refer to a multi-mode
system with modes $\left\vert \phi _{ai}\right\rangle $ ,$\left\vert \phi
_{bi}\right\rangle $ the statement that spin squeezing for systems of
identical massive bosons requires all the modes to be entangled is correct.
However superposition states of the form (\ref%
{Eq.TwoModeQuantumSuperposition}) that are consistent with the
super-selection rule applying to pure states of a two mode system are
precluded, and such states ought to be allowed if entanglement of \emph{pairs%
} of modes rather than of \emph{separate} modes is to be considered.

In addition, we can show that if either $\left\langle \widehat{S}%
\,_{x}\right\rangle $ or $\left\langle \widehat{S}\,_{y}\right\rangle $ is
non-zero, then the state must be entangled - the \emph{Bloch vector} test.
Finally, if it is found that if there is spin squeezing in $\widehat{S}%
\,_{z} $ then the state must be entangled. Thus spin squeezing in \emph{any}
spin component confirms entanglement of the $2n$ individual modes.

\subsubsection{Revising Sorensen Spin Squeezing Entanglement Test -
Separable State of Pairs of Modes with One Boson Occupancy}

It is also possible however to link spin squeezing and entanglement in the
case where the subsystems consist of \emph{pairs} of modes (Case3 in
SubSection \ref{SubSection - SpinSqgEnt MultiMode}), but only if \emph{%
further restrictions} are applied. The general \emph{non-entangled} state of
the \emph{pairs} of \emph{modes} would actually be of the form (see (\ref%
{Eq.SepStatesMultiModeCase3}), here the $ab$ dropped for simplicity) 
\begin{equation}
\widehat{\rho }=\sum_{R}P_{R}\,\widehat{\rho }_{R}^{1}\otimes \widehat{\rho }%
_{R}^{2}\otimes \widehat{\rho }_{R}^{3}\otimes ...
\label{Eq.NonEntStateModePairs}
\end{equation}%
where the $\widehat{\rho }_{R}^{i}$ are now of the form given in Eq. (\ref%
{Eq.GeneralDensityOprModePair}) and no longer are density operators for the $%
i$th identical particle. Unlike in (\ref{Eq.RevisedSorensenAverages}) we now
have expectation values $\left\langle (\widehat{a}_{i})^{n}\right\rangle
_{\,i}=Tr(\widehat{\rho }_{R}^{\,i}(\widehat{a}_{i})^{n})$ etc that are
non-zero, so considerations of the link between spin squeezing and
entanglement - now entanglement of pairs of modes, willl be different.

If the density operators $\widehat{\rho }_{R}^{i}$ associated with the \emph{%
pair} of modes $\widehat{a}_{i}$, $\widehat{b}_{i}$ are all \emph{restricted}
to be associated with \emph{one boson states} then this density operator is
of the form 
\begin{eqnarray}
\widehat{\rho }_{R}^{i} &=&\rho _{aa}^{i}(\left\vert 1\right\rangle
_{ia}\left\langle 1\right\vert _{ia}\otimes \left\vert 0\right\rangle
_{ib}\left\langle 0\right\vert _{ib})+\rho _{ab}^{i}(\left\vert
1\right\rangle _{ia}\left\langle 0\right\vert _{ia}\otimes \left\vert
0\right\rangle _{ib}\left\langle 1\right\vert _{ib})  \nonumber \\
&&+\rho _{ba}^{i}(\left\vert 0\right\rangle _{ia}\left\langle 1\right\vert
_{ia}\otimes \left\vert 1\right\rangle _{ib}\left\langle 0\right\vert
_{ib})+\rho _{bb}^{i}(\left\vert 0\right\rangle _{ia}\left\langle
0\right\vert _{ia}\otimes \left\vert 1\right\rangle _{ib}\left\langle
1\right\vert _{ib})  \nonumber \\
&&  \label{Eq.DensityOprModePair}
\end{eqnarray}%
where the $\rho _{ef}^{i}$ are density matrix elements. With this
restriction the pair of modes $\widehat{a}_{i}$, $\widehat{b}_{i}$ behave
like \emph{distinguishable} two state particles, essentially the case that
Sorensen et al \cite{Sorensen01a} implicitly considered. The expectation
values for the spin operators $\widehat{S}_{x}^{i}$, $\widehat{S}_{y}^{i}$
and $\widehat{S}_{z}^{i}$ associated with the $i$th pair of modes are then 
\begin{eqnarray}
\left\langle \widehat{S}_{x}^{i}\right\rangle _{R} &=&\frac{1}{2}\left( \rho
_{ab}^{i}+\rho _{ba}^{i}\right) \qquad \left\langle \widehat{S}%
_{y}^{i}\right\rangle _{R}=\frac{1}{2i}\left( \rho _{ab}^{i}-\rho
_{ba}^{i}\right)  \nonumber \\
\left\langle \widehat{S}_{z}^{i}\right\rangle _{R} &=&\frac{1}{2}\left( \rho
_{bb}^{i}-\rho _{aa}^{i}\right)  \label{Eq.ExpnValues}
\end{eqnarray}

If in addition Hermitiancy, positivity, unit trace $Tr(\widehat{\rho }%
_{R}^{i})=1$ and $Tr(\widehat{\rho }_{R}^{i})^{2}\leq 1$ are used (see
Appendix \ref{Appendix - Sorensen Results}) then we can show that $\rho
_{bb}^{i}$ and $\rho _{aa}^{i}$ are real and positive, $\rho _{ab}^{i}=(\rho
_{ba}^{i})^{\ast }$ and $\rho _{aa}^{i}\rho _{bb}^{i}-|\rho
_{ab}^{i}|^{2}\geq 0$. The condition $Tr(\widehat{\rho }_{R}^{i})=1$ leads
to $\rho _{aa}^{i}+\rho _{bb}^{i}=1$, from which $Tr(\widehat{\rho }%
_{R}^{i})^{2}\leq 1$ follows using the previous positivity results. These
results enable the matrix elements in (\ref{Eq.DensityOprModePair}) to be
parameterised in the form%
\begin{eqnarray}
\rho _{aa}^{i} &=&\sin ^{2}\alpha _{i}\qquad \rho _{bb}^{i}=\cos ^{2}\alpha
_{i}  \nonumber \\
\rho _{ab}^{i} &=&\sqrt{\sin ^{2}\alpha _{i}\,\cos ^{2}\alpha _{i}}\,\sin
^{2}\beta _{i}\,\exp (+i\phi _{i})\qquad \rho _{ba}^{i}=\sqrt{\sin
^{2}\alpha _{i}\,\cos ^{2}\alpha _{i}}\,\sin ^{2}\beta _{i}\,\exp (-i\phi
_{i})  \nonumber \\
&&  \label{Eq.ParamTwoModeDensityME}
\end{eqnarray}%
where $\alpha _{i}$, $\beta _{i}$ and $\phi _{i}$ are real. In terms of
these quantities we then have 
\begin{eqnarray}
\left\langle \widehat{S}_{x}^{i}\right\rangle _{R} &=&\frac{1}{2}\sin
2\alpha _{i}\,\sin ^{2}\beta _{i}\,\cos \phi _{i}\qquad \left\langle 
\widehat{S}_{y}^{i}\right\rangle _{R}=\frac{1}{2}\sin 2\alpha _{i}\,\sin
^{2}\beta _{i}\,\sin \phi _{i}  \nonumber \\
\left\langle \widehat{S}_{z}^{i}\right\rangle _{R} &=&\frac{1}{2}\cos
2\alpha _{i}\,  \label{Eq.ExpnValuesB}
\end{eqnarray}%
and then a key inequality 
\begin{equation}
\left\langle \widehat{S}_{x}^{i}\right\rangle _{R}^{2}+\left\langle \widehat{%
S}_{y}^{i}\right\rangle _{R}^{2}+\left\langle \widehat{S}_{z}^{i}\right%
\rangle _{R}^{2}=\frac{1}{4}-\frac{1}{4}\sin ^{2}2\alpha _{i}\,(1-\sin
^{4}\beta _{i}\,)\leq \frac{1}{4}  \label{Eq.SorensenInequalityB}
\end{equation}%
follows. This result depends on the density operators $\widehat{\rho }%
_{R}^{i}$ being for one boson states, as in (\ref{Eq.DensityOprModePair}).
The same steps as in Sorensen et al \cite{Sorensen01a} (see Appendix \ref%
{Appendix - Sorensen Results}) leads to the result%
\begin{equation}
\left\langle \Delta \widehat{S}\,_{z}^{2}\right\rangle \geq \frac{1}{N}%
\left( \left\langle \widehat{S}\,_{x}\right\rangle ^{2}+\left\langle 
\widehat{S}\,_{y}\right\rangle ^{2}\right)
\label{Eq.SpinSqgCondModifiedSorensen}
\end{equation}%
for non-entangled \emph{pair} of modes $\widehat{a}_{i}$, $\widehat{b}_{i}$.
Thus when the interpretation is changed so that are the separate sub-systems
are these pairs of modes \emph{and} the sub-systems are in one boson states,
it follows that spin squeezing requires entanglement of all the mode pairs.

A similar proof extending the test of Sorensen et al \cite{Sorensen01a} to
appply to systems of identical bosons is given by Hyllus et al \cite%
{Hyllus12a} based on a particle entanglement approach. In their approach
bosons in differing external modes (analogous to differing $i$ here) are
treated as distinguishable, and the symmetrization principle is ignored for
such bosons.

\subsection{Benatti et al 2011}

\label{SubSection - Benatti et al 2011}

In earlier work Toth and Gunhe \cite{Toth09a} derived several spin operator
based inequalities for separable states for two mode particle systems based
on the assumption that the particles were \emph{distinguishable}. As in Eq.(%
\ref{Eq.NonEntStateIdenticalAtoms}), the density operator was not required
to satisfy the symmetrisation principle. Tests for entanglement involving
the mean values and variances for two mode spin operators resulted.
Subsequently, Benatti et al \cite{Benatti11a} \ considered whether these
tests would still apply if the particles were \emph{indistinguishable}.
Their work involves considering states with $N$ bosons.

For \emph{separable} states they found (see Eq.(10)) that for three
orthogonal spin operators $\widehat{J}_{n1}$, $\widehat{J}_{n2}$ and $%
\widehat{J}_{n3}$ 
\begin{equation}
\left\langle \widehat{J}_{n1}^{2}\right\rangle +\left\langle \widehat{J}%
_{n2}^{2}\right\rangle +\left\langle \widehat{J}_{n3}^{2}\right\rangle \leq 
\frac{N(N+2)}{4}
\end{equation}%
from which it might be concluded that if the left side exceeded $N(N+2)/4$
then the state must be entangled. However, since $\widehat{J}_{n1}^{2}+%
\widehat{J}_{n1}^{2}+\widehat{J}_{n1}^{2}=\widehat{S}_{x}^{2}+\widehat{S}%
_{y}^{2}+\widehat{S}_{z}^{2}=\widehat{N}(\widehat{N}+2)/4$ the left side is
always equal to $N(N+2)/4$ for all states with $N$ bosons, so no
entanglement test results. This outcome is for similar reasons as for the
failed entanglement test $\left\langle \Delta \widehat{S}\,_{x}^{2}\right%
\rangle +\left\langle \Delta \widehat{S}\,_{y}^{2}\right\rangle
<|\left\langle \widehat{S}\,_{z}\right\rangle |$ discussed in SubSection \ref%
{SubSection - Hillery 2006}.

For \emph{separable} states they also found (see Eq.(11)) that for three
orthogonal spin operators $\widehat{J}_{n1}$, $\widehat{J}_{n2}$ and $%
\widehat{J}_{n3}$%
\begin{equation}
\left\langle \Delta \widehat{J}_{n1}^{2}\right\rangle +\left\langle \Delta 
\widehat{J}_{n2}^{2}\right\rangle +\left\langle \Delta \widehat{J}%
_{n3}^{2}\right\rangle \geq \frac{N}{2}
\end{equation}%
so that if 
\begin{equation}
\left\langle \Delta \widehat{J}_{n1}^{2}\right\rangle +\left\langle \Delta 
\widehat{J}_{n2}^{2}\right\rangle +\left\langle \Delta \widehat{J}%
_{n3}^{2}\right\rangle <\frac{N}{2}  \label{Eq.BennattiTest1}
\end{equation}%
then the state must be entangled. This test is an extended form of the
Hillery spin variance test (\ref{Eq.HillerySpinEntTest}). To prove this
result we note that $\left\langle \Delta \widehat{J}_{n1}^{2}\right\rangle
+\left\langle \Delta \widehat{J}_{n2}^{2}\right\rangle +\left\langle \Delta 
\widehat{J}_{n3}^{2}\right\rangle =\left\langle \Delta \widehat{S}%
_{x}^{2}\right\rangle +\left\langle \Delta \widehat{S}_{y}^{2}\right\rangle
+\left\langle \Delta \widehat{S}_{z}^{2}\right\rangle =$.$\left\langle 
\widehat{S}_{x}^{2}\right\rangle +\left\langle \widehat{S}%
_{y}^{2}\right\rangle +\left\langle \widehat{S}_{z}^{2}\right\rangle
-\left\langle \widehat{S}_{x}\right\rangle ^{2}-\left\langle \widehat{S}%
_{y}\right\rangle ^{2}-\left\langle \widehat{S}_{z}\right\rangle ^{2}=$ $%
N(N+2)/4-\left\langle \widehat{S}_{x}\right\rangle ^{2}-\left\langle 
\widehat{S}_{y}\right\rangle ^{2}-\left\langle \widehat{S}_{z}\right\rangle
^{2}$ for all states with $N$ bosons. For separable states we have $%
\left\langle \widehat{S}_{x}\right\rangle =\left\langle \widehat{S}%
_{y}\right\rangle =0$ so that $\left\langle \Delta \widehat{J}%
_{n1}^{2}\right\rangle +\left\langle \Delta \widehat{J}_{n2}^{2}\right%
\rangle +\left\langle \Delta \widehat{J}_{n3}^{2}\right\rangle
=N(N+2)/4-\left\langle \widehat{S}_{z}\right\rangle ^{2}$. As the
eigenvalues for $\widehat{S}_{z}$ lie between $-N/2$ and $+N/2$ we have $%
\left\langle \widehat{S}_{z}\right\rangle ^{2}\leq N^{2}/4$. Thus $%
\left\langle \Delta \widehat{J}_{n1}^{2}\right\rangle +\left\langle \Delta 
\widehat{J}_{n2}^{2}\right\rangle +\left\langle \Delta \widehat{J}%
_{n3}^{2}\right\rangle \geq \frac{N}{2}$ as required. \ The test in (\ref%
{Eq.BennattiTest1}) is quite useful in that it applies to any three
orthogonal spin operators, though it would be harder to satisfy compared to
the Hillery spin variance test because of the additional .$\left\langle
\Delta \widehat{S}_{z}^{2}\right\rangle $ term.

For \emph{separable} states they also found (see Eq.(13)) that for three
orthogonal spin operators $\widehat{J}_{n1}$, $\widehat{J}_{n2}$ and $%
\widehat{J}_{n3}$%
\begin{equation}
(N-1)\left( \left\langle \Delta \widehat{J}_{n1}^{2}\right\rangle
+\left\langle \Delta \widehat{J}_{n2}^{2}\right\rangle \right) -\left\langle 
\widehat{J}_{n3}^{2}\right\rangle \geq \frac{N(N-2)}{4}
\end{equation}%
so that if 
\begin{equation}
(N-1)\left( \left\langle \Delta \widehat{J}_{n1}^{2}\right\rangle
+\left\langle \Delta \widehat{J}_{n2}^{2}\right\rangle \right) -\left\langle 
\widehat{J}_{n3}^{2}\right\rangle <\frac{N(N-2)}{4}  \label{Eq.BennattiTest2}
\end{equation}%
then the state must be entangled. To prove this result for $n1=%
\overrightarrow{x}$, $n2=\overrightarrow{y}$ and $n3=\overrightarrow{z}$, we
use the result (\ref{Eq.InequalityXYVariancesNonEntState}) for separable
states that $\left\langle \Delta \widehat{S}_{x}^{2}\right\rangle
+\left\langle \Delta \widehat{S}_{y}^{2}\right\rangle \geq \sum_{R}P_{R}\,%
\frac{1}{2}(\left\langle \widehat{b}^{\dag }\widehat{b}\right\rangle
_{R}+\left\langle \widehat{a}^{\dag }\widehat{a}\right\rangle
_{R})+\sum_{R}P_{R}\,(\left\langle \widehat{a}^{\dag }\widehat{a}%
\right\rangle _{R}\left\langle \widehat{b}^{\dag }\widehat{b}\right\rangle
_{R})=N/2+\sum_{R}P_{R}\,(\left\langle \widehat{a}^{\dag }\widehat{a}%
\right\rangle _{R}\left\langle \widehat{b}^{\dag }\widehat{b}\right\rangle
_{R})$. It is straightforward to show that $\widehat{S}_{z}^{2}=(\widehat{b}%
^{\dag }\widehat{b}+\widehat{a}^{\dag }\widehat{a})^{2}/4-\widehat{b}^{\dag }%
\widehat{b}.\widehat{a}^{\dag }\widehat{a}$, so that $\left\langle \widehat{S%
}_{z}^{2}\right\rangle =N^{2}/4-\sum_{R}P_{R}\,(\left\langle \widehat{a}%
^{\dag }\widehat{a}\right\rangle _{R}\left\langle \widehat{b}^{\dag }%
\widehat{b}\right\rangle _{R})$. Hence for separable states $%
(N-1)(\left\langle \Delta \widehat{S}_{x}^{2}\right\rangle +\left\langle
\Delta \widehat{S}_{y}^{2}\right\rangle )-\left\langle \widehat{S}%
_{z}^{2}\right\rangle \geq (N-1)N/2+(N-1)\sum_{R}P_{R}\,(\left\langle 
\widehat{a}^{\dag }\widehat{a}\right\rangle _{R}\left\langle \widehat{b}%
^{\dag }\widehat{b}\right\rangle _{R})-N^{2}/4+\sum_{R}P_{R}\,(\left\langle 
\widehat{a}^{\dag }\widehat{a}\right\rangle _{R}\left\langle \widehat{b}%
^{\dag }\widehat{b}\right\rangle _{R})$. Thus $(N-1)(\left\langle \Delta 
\widehat{S}_{x}^{2}\right\rangle +\left\langle \Delta \widehat{S}%
_{y}^{2}\right\rangle )-\left\langle \widehat{S}_{z}^{2}\right\rangle \geq 
\frac{N(N-2)}{4}+N\sum_{R}P_{R}\,(\left\langle \widehat{a}^{\dag }\widehat{a}%
\right\rangle _{R}\left\langle \widehat{b}^{\dag }\widehat{b}\right\rangle
_{R})$. As the second term on the right side is always positive the required
inequality follows.

Finally, they considered another inequality (see Eq. (12)) found to apply
for \emph{separable} states involving \emph{distinguishable} particles in
Ref. \cite{Toth09a}. 
\begin{equation}
\left( \left\langle \widehat{J}_{n1}^{2}\right\rangle +\left\langle \widehat{%
J}_{n2}^{2}\right\rangle \right) -\frac{N}{2}-(N-1)\left\langle \Delta 
\widehat{J}_{n3}^{2}\right\rangle \leq 0
\end{equation}%
so the question is whether an entanglement test $\left( \left\langle 
\widehat{J}_{n1}^{2}\right\rangle +\left\langle \widehat{J}%
_{n2}^{2}\right\rangle \right) -N/2-(N-1)\left\langle \Delta \widehat{J}%
_{n3}^{2}\right\rangle >0$ applies for the case of indistinguishable
particles. For the case where $n1=\overrightarrow{x}$, $n2=\overrightarrow{y}
$ and $n3=\overrightarrow{z}$, $\left( \left\langle \widehat{S}%
_{x}^{2}\right\rangle +\left\langle \widehat{S}_{y}^{2}\right\rangle \right)
-N/2-(N-1)\left\langle \Delta \widehat{S}_{z}^{2}\right\rangle =\left(
\left\langle \widehat{S}_{x}^{2}\right\rangle +\left\langle \widehat{S}%
_{y}^{2}\right\rangle +\left\langle \widehat{S}_{z}^{2}\right\rangle \right)
-N/2-(N)\left\langle \widehat{S}_{z}^{2}\right\rangle +(N-1)\left\langle 
\widehat{S}_{z}\right\rangle ^{2}=N(N+2)/4-N/2-N\left(
N^{2}/4-\sum_{R}P_{R}\,(\left\langle \widehat{a}^{\dag }\widehat{a}%
\right\rangle _{R}\left\langle \widehat{b}^{\dag }\widehat{b}\right\rangle
_{R})\right) +(N-1)\left\langle \widehat{S}_{z}\right\rangle ^{2}$. As $%
\left\langle \widehat{S}_{z}\right\rangle ^{2}\leq N^{2}/4$ we see that $%
\left( \left\langle \widehat{S}_{x}^{2}\right\rangle +\left\langle \widehat{S%
}_{y}^{2}\right\rangle \right) -N/2-(N-1)\left\langle \Delta \widehat{S}%
_{z}^{2}\right\rangle \leq N\sum_{R}P_{R}\,(\left\langle \widehat{a}^{\dag }%
\widehat{a}\right\rangle _{R}\left\langle \widehat{b}^{\dag }\widehat{b}%
\right\rangle _{R})$, which is certainly $\geq 0$ and not $\leq 0$ as
required. However, perhaps an entanglement test such that if is could be
shown that 
\begin{equation}
\left( \left\langle \widehat{S}_{x}^{2}\right\rangle +\left\langle \widehat{S%
}_{y}^{2}\right\rangle \right) -N/2-(N-1)\left\langle \Delta \widehat{S}%
_{z}^{2}\right\rangle >N\sum_{R}P_{R}\,(\left\langle \widehat{a}^{\dag }%
\widehat{a}\right\rangle _{R}\left\langle \widehat{b}^{\dag }\widehat{b}%
\right\rangle _{R})  \label{Eq.NonTest}
\end{equation}%
always applies then it could be included that the state is entangled.
Unfortunately the right side could be too large for the left side to always
exceed the right side for some separable states. Noting that $\left\langle 
\widehat{a}^{\dag }\widehat{a}\right\rangle _{R}+\left\langle \widehat{b}%
^{\dag }\widehat{b}\right\rangle _{R}=N$ for the $N$ bosons states being
considered we find that the right side is maximised when $\left\langle 
\widehat{a}^{\dag }\widehat{a}\right\rangle _{R}=\left\langle \widehat{b}%
^{\dag }\widehat{b}\right\rangle _{R}=N/2$ for all $P_{R}$, giving a maximum
for the right side of $N^{3}/2$ - and this can occur for some separable
states. To show that the state is entangled the left side must exceed this
value, otherwise the state might be one of the separable states. However,
the left side is at most of order $N^{2}$ from the firsr two terms and the
negative terms only make the left side smaller. Hence there is no
entanglement test of the form (\ref{Eq.NonTest}).$\left\langle \widehat{a}%
^{\dag }\widehat{a}\right\rangle _{R}=\left\langle \widehat{b}^{\dag }%
\widehat{b}\right\rangle _{R}$

Hence Benatti et al \cite{Benatti11a} have demonstrated two further
entanglement tests (\ref{Eq.BennattiTest1}) and (\ref{Eq.BennattiTest2}) for
two mode systems of identical particle that involve spin operators. Again,
these tests do not involve invoking the local particle number SSR for
separable states.

\subsection{Sorensen and Molmer 2001}

\label{SubSection - Sorensen and Molmer 2001}

In a paper entitled "Entanglement and Extreme Spin Squeezing" Sorensen and
Molmer \cite{Sorensen01b} first consider the limits imposed by the
Heisenberg uncertainty principle on the variance $\left\langle \Delta 
\widehat{S}\,_{x}^{2}\right\rangle $ considered as a function of $%
|\left\langle \widehat{S}\,_{z}\right\rangle |$ for states with $N$ two mode
bosons where the spin operators are chosen such that $\left\langle \widehat{S%
}\,_{x}\right\rangle =\left\langle \widehat{S}\,_{y}\right\rangle =0$. Note
that such spin operators can always be chosen so that the Bloch vector does
lie along the $z$ axis, even if the spin operators are not principal spin
operators. Their treatment is based on combining the result from the Schwarz
inequality 
\begin{equation}
\left\langle \widehat{S}\,_{x}^{2}\right\rangle +\left\langle \widehat{S}%
\,_{y}^{2}\right\rangle +\left\langle \widehat{S}\,_{z}\right\rangle
^{2}\leq J(J+1)  \label{Eq.SchwarzResult}
\end{equation}%
where $J=N/2$, and the Heisenberg uncertainty principle 
\begin{equation}
\left\langle \Delta \widehat{S}\,_{x}^{2}\right\rangle \left\langle \Delta 
\widehat{S}\,_{y}^{2}\right\rangle =\xi \frac{1}{4}|\left\langle \widehat{S}%
\,_{z}\right\rangle |^{2}  \label{Eq.HUP}
\end{equation}%
where $\xi \geq 1$. In fact two inequalities can be obtained 
\begin{eqnarray}
\left\langle \Delta \widehat{S}\,_{x}^{2}\right\rangle &\geq &\frac{1}{2}%
\left\{ \left( J(J+1)-\left\langle \widehat{S}\,_{z}\right\rangle
^{2}\right) -\sqrt{\left( J(J+1)-\left\langle \widehat{S}\,_{z}\right\rangle
^{2}\right) ^{2}-\xi \left\langle \widehat{S}\,_{z}\right\rangle ^{2}}%
\right\}  \nonumber \\
&&  \label{Eq.HUPRestriction1} \\
\left\langle \Delta \widehat{S}\,_{x}^{2}\right\rangle &\leq &\frac{1}{2}%
\left\{ \left( J(J+1)-\left\langle \widehat{S}\,_{z}\right\rangle
^{2}\right) +\sqrt{\left( J(J+1)-\left\langle \widehat{S}\,_{z}\right\rangle
^{2}\right) ^{2}-\xi \left\langle \widehat{S}\,_{z}\right\rangle ^{2}}%
\right\}  \nonumber \\
&&  \label{Eq.HUPRestriction2}
\end{eqnarray}%
which restricts the region in a $\left\langle \Delta \widehat{S}%
\,_{x}^{2}\right\rangle $ versus $|\left\langle \widehat{S}%
\,_{z}\right\rangle |$ plane that applies for states that are consistent
with the Heisenberg uncertainty principle. Note that in the first inequality
the minimum value for $\left\langle \Delta \widehat{S}\,_{x}^{2}\right%
\rangle $ occurs for $\xi =1$, and in the second inequality the maximum
value for $\left\langle \Delta \widehat{S}\,_{x}^{2}\right\rangle $ also
occurs for $\xi =1$ - the minimum HUP case. \ The first of these two
inequalities is given as Eq. (3) in \cite{Sorensen01b}. For states in which $%
\widehat{S}\,_{x}$ is squeezed relative to $\widehat{S}\,_{y}$ the points in
the$\left\langle \Delta \widehat{S}\,_{x}^{2}\right\rangle $ versus $%
|\left\langle \widehat{S}\,_{z}\right\rangle |$ plane must also satisfy 
\begin{equation}
\left\langle \Delta \widehat{S}\,_{x}^{2}\right\rangle \leq \frac{1}{2}%
|\left\langle \widehat{S}\,_{z}\right\rangle |  \label{Eq.SpinSqRestriction}
\end{equation}%
Note that as $\widehat{J}\,_{z}$ is a spin angular momentum component we
always have $|\left\langle \widehat{S}\,_{z}\right\rangle |\,\leq J$, which
places an overall restriction on $|\left\langle \widehat{S}%
\,_{z}\right\rangle |$. However, for $\xi >1$ there are values of $%
|\left\langle \widehat{S}\,_{z}\right\rangle |$ which are excluded via the
Heisenberg uncertainty principle, since the quantity $\left(
J(J+1)-\left\langle \widehat{S}\,_{z}\right\rangle ^{2}\right) ^{2}-\xi
\left\langle \widehat{S}\,_{z}\right\rangle ^{2}$ then becomes negative.
This effect is seen in Figure 4.\pagebreak

The question is: Is it possible to find values for $\left\langle \Delta 
\widehat{S}\,_{x}^{2}\right\rangle $ and $|\left\langle \widehat{S}%
\,_{z}\right\rangle |$ in which all three inequalities are satisfied? The
answer is yes. Results showing the regions in the $\left\langle \Delta 
\widehat{S}\,_{x}^{2}\right\rangle $ versus $|\left\langle \widehat{S}%
\,_{z}\right\rangle |$ plane corresponding to the three inequalities are
shown in Figures 4 and 5 for the cases where $J=1000$ and with $\xi =1.0$
and $\xi =10.0$ respectively. The quantities for which the regions are shown
are the scaled variance and mean $\left\langle \Delta \widehat{S}%
\,_{x}^{2}\right\rangle /J$ and $|\left\langle \widehat{S}%
\,_{z}\right\rangle |/J$, with $\left\langle \Delta \widehat{S}%
\,_{x}^{2}\right\rangle $ given as a function of $|\left\langle \widehat{S}%
\,_{z}\right\rangle |$ via (\ref{Eq.HUPRestriction1}), (\ref%
{Eq.HUPRestriction2}) and (\ref{Eq.SpinSqRestriction}). The spin squeezing
region is always consistent with the second Heisenberg inequality (\ref%
{Eq.HUPRestriction2}) and for large $J=1000$ there is a large region of
overlap with the first inequality (\ref{Eq.HUPRestriction1}). For small $J$
and large $\xi $ the region of overlap becomes much smaller, as the result
in Figure 6 for $J=1$ and with $\xi =10.0$ shows. As the derivation of the
Heisenberg principle inequalities is not obvious, this is set out in
Appendix.\ref{Appendix - Heisenberg Uncertainty Principle Results}.

\bigskip

\begin{center}
\includegraphics[height=4.4745in]{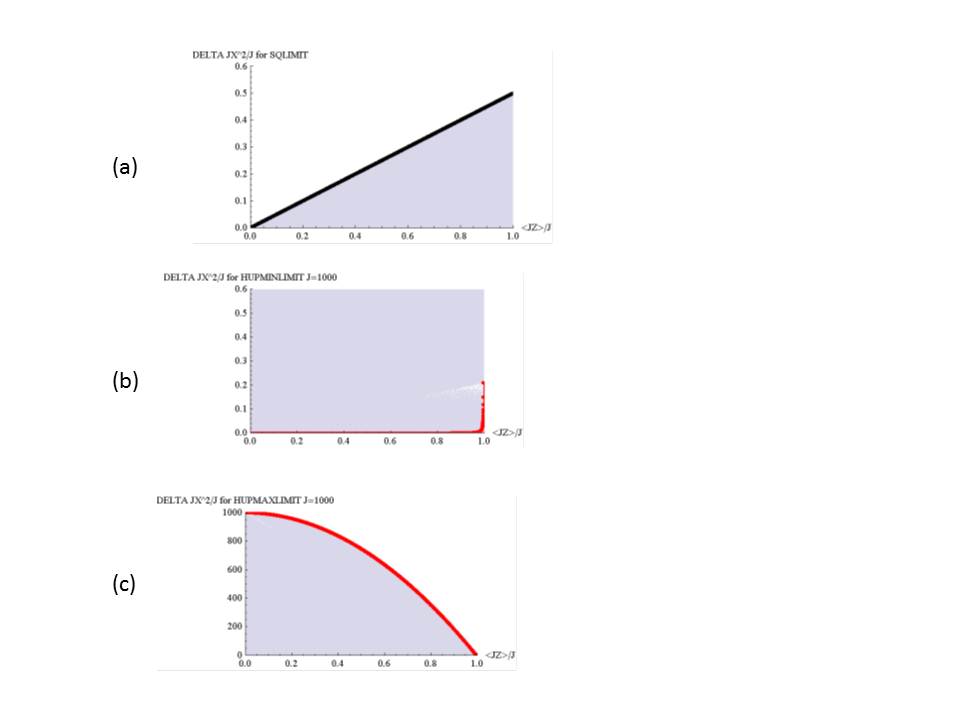}
%
%
%
%
\medskip \medskip

Figure 4. Regions in the $<\Delta \widehat{S}_{x}^{2}>$ versus $|<\widehat{S}%
_{z}>|$ plane (shown shaded) for states that satisfy (a) the spin squeezing
inequality Eq. (\ref{Eq.SpinSqRestriction}) (b) the smaller Heisenberg
uncertainty principle inequality Eq. (\ref{Eq.HUPRestriction1}) and (c) the
larger HUP inequality Eq. (\ref{Eq.HUPRestriction2}). The case shown is for $%
J=1000$ and HUP factor $\xi =1$. Both $<\Delta \widehat{S}_{x}^{2}>$ and $|<%
\widehat{S}_{z}>|$ are in units of $J$. The spin operators are chosen so
that $<\widehat{S}_{x}>=<\widehat{S}_{y}>=0$.
\end{center}

\bigskip \pagebreak

\includegraphics[height=3.9643in]{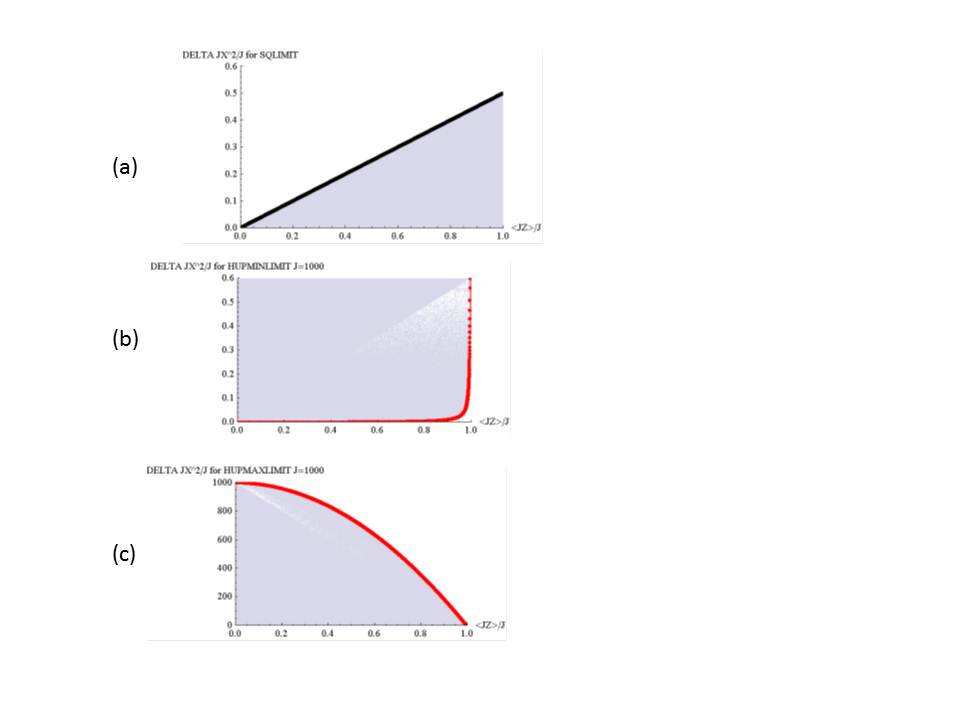}
%
\bigskip

\begin{center}
Figure 5. As in Figure 4, but with $J=1000$ and HUP factor $\xi =10.0$.
\end{center}

\bigskip \pagebreak

\includegraphics[height=4.0689in]{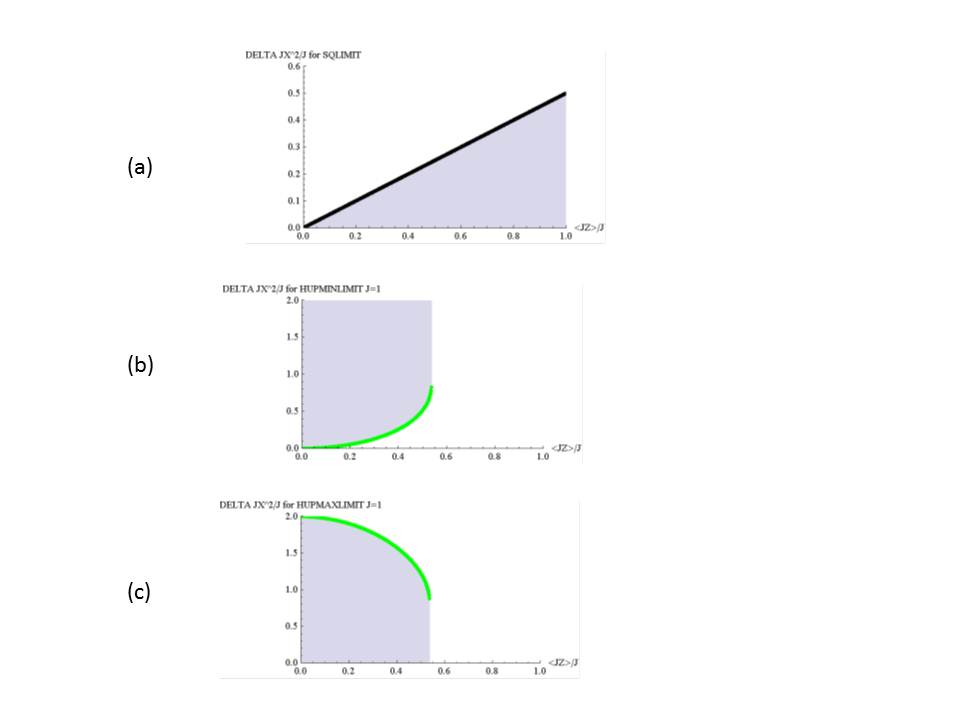}


\bigskip

\begin{center}
Figure 6. As in Figure 4, but with $J=1$ and HUP factor $\xi =10.0$.\medskip
\end{center}

Sorensen and Molmer \cite{Sorensen01b} also determine the minimum for $%
\left\langle \Delta \widehat{S}\,_{x}^{2}\right\rangle =\left\langle 
\widehat{S}\,_{x}^{2}\right\rangle $ as a function of $|\left\langle 
\widehat{S}\,_{z}\right\rangle |$ for various choices of $J$, subject to the
constraints $\left\langle \widehat{S}\,_{x}\right\rangle =\left\langle 
\widehat{S}\,_{y}\right\rangle =0$. The results show again that there is a
region in the $\left\langle \Delta \widehat{S}\,_{x}^{2}\right\rangle $
versus $|\left\langle \widehat{S}\,_{z}\right\rangle |$ plane which is
compatible with spin squeezing.

So although these considerations show that the Heisenberg uncertainty
principle does not rule out extreme spin squeezing, nothing is yet directly
determined about whether the spin squeezed states are entangled states for 
\emph{modes} $\widehat{a}$, $\widehat{b}$, where the $\widehat{S}_{\alpha }$
are given as in Eq. (\ref{Eq.OldSpinOprs}). The discussion in \cite%
{Sorensen01b} regarding entanglement is also based on using a density
operator for non-entangled states as in Eq. (\ref%
{Eq.NonEntStateIdenticalAtoms}) which only applies to distinguishable
particles (see SubSection \ref{SubSection - Sorensen 2001}). Sorensen \cite%
{Sorensen01b} also showed that for higher $J$ the amount of squeezing
attainable could be greater. This fact enables a conclusion to be drawn from
the measured spin variance about the minimum number of particles that
participate in the non-separable component of an entangled state \cite%
{Dalton16a}\textbf{. }

\section{Correlation Tests for Entanglement}

\label{Section - Correlation Tests for Entanglement}

\ In SubSection \textbf{2.4 }of the accompanying paper I it was shown that
for separable states the inequality $|\left\langle \widehat{\Omega }%
_{A}\otimes \widehat{\Omega }_{B}^{\dag }\right\rangle |^{2}\leq
\left\langle \widehat{\Omega }_{A}^{\dag }\widehat{\Omega }_{A}\otimes 
\widehat{\Omega }_{B}^{\dag }\widehat{\Omega }_{B}\right\rangle $ applies,
so that if 
\begin{equation}
|\left\langle \widehat{\Omega }_{A}\otimes \widehat{\Omega }_{B}^{\dag
}\right\rangle |^{2}>\left\langle \widehat{\Omega }_{A}^{\dag }\widehat{%
\Omega }_{A}\otimes \widehat{\Omega }_{B}^{\dag }\widehat{\Omega }%
_{B}\right\rangle  \label{Eq.GenCorrTest}
\end{equation}%
then the state is entangled. This is a general \emph{correlation} test.

As will be seen the correlation tests can be re-expressed in terms of spin
operators when dealing with SSR compliant states.\textbf{\ }

\subsection{Dalton et al 2014}

\label{SubSection - Dalton 2014}

\subsubsection{Weak Correlation Test for Local SSR Compliant Non-Entangled
States}

For a non-entangled state based on \emph{SSR compliant} $\widehat{\rho }%
_{R}^{A},\widehat{\rho }_{R}^{B}$ for modes $\widehat{a}$ and $\widehat{b}$
where the SSR is satisfied we have with $\widehat{\Omega }_{A}=(\widehat{a}%
)^{m}$ and $\widehat{\Omega }_{B}=(\widehat{b})^{n}$ 
\begin{equation}
\left\langle (\widehat{a})^{m}\,(\widehat{b}^{\dag })^{n}\right\rangle
=\tsum\limits_{R}P_{R}\,\left\langle (\widehat{a})^{m}\,(\widehat{b}^{\dag
})^{n}\right\rangle _{R}=\tsum\limits_{R}P_{R}\,\left\langle (\widehat{a}%
)^{m}\right\rangle _{R}\left\langle (\widehat{b}^{\dag })^{n}\right\rangle
_{R}=0  \label{Eq.ResultTrueNonEntState}
\end{equation}%
since from Eqs. analogous to (\ref{Eq.CondNonEntStateCD}) $\left\langle (%
\widehat{a})^{m}\right\rangle _{R}=\left\langle (\widehat{b}^{\dag
})^{n}\right\rangle _{R}=0$. Hence for a SSR compliant non-entangled state
as defined in the present paper the inequality becomes 
\begin{equation}
0\leq \left\langle (\widehat{a}^{\dag })^{m}(\widehat{a})^{m}\,(\widehat{b}%
^{\dag })^{n}(\widehat{b})^{n}\right\rangle
\label{Eq.InequalNonEntPhysState}
\end{equation}%
which is trivially true and applies for \emph{any} state, entangled or not.

Since $\left\langle (\widehat{a})^{m}\,(\widehat{b}^{\dag
})^{n}\right\rangle $ is zero for non-entangled states it follows that it is
merely necessary to show that this quantity is non-zero to establish that
the state is entangled. Hence an \emph{entanglement test} \cite{Dalton14a}
in the case of sub-systems consisting of single modes $\widehat{a}$ and $%
\widehat{b}$ becomes 
\begin{equation}
|\left\langle (\widehat{a})^{m}\,(\widehat{b}^{\dag })^{n}\right\rangle
|^{2}>0  \label{Eq.EntangTest}
\end{equation}%
for a non-entangled state based on \emph{SSR compliant} $\widehat{\rho }%
_{R}^{A},\widehat{\rho }_{R}^{B}$. Note that for globally compliant states $%
\left\langle (\widehat{a})^{m}\,(\widehat{b}^{\dag })^{n}\right\rangle =0$
unless $n=m$, so only that case is of interest. This is a useful \emph{weak
correlation} test for entanglement in terms of the definition of
entanglement in the present paper. A related but different test is that of
Hillery et al \cite{Hillery06a} - discussed in the next SubSection.

For the case where $n=m=1$ the weak correlation test is 
\begin{equation}
|\left\langle (\widehat{a}\,\widehat{b}^{\dag }\right\rangle |^{2}>0
\label{Eq.WeakCorrelTest}
\end{equation}%
which is equivalent to $\left\langle \widehat{S}_{x}\right\rangle \neq 0$
and/or $\left\langle \widehat{S}_{y}\right\rangle \neq 0$, the Bloch vector
test.

\subsection{Hillery et al 2006, 2009}

\label{SubSection - Hillery 2009}

\subsubsection{Hillery Strong Correlation Entanglement Test}

In a later paper entitled "Detecting entanglement with non-Hermitian
operators" Hillery et al \cite{Hillery09a} apply other inequalities for
determining entanglement derived in the earlier paper \cite{Hillery06a} but
now also to systems of massive identical bosons, while still retaining
.density operators $\widehat{\rho }_{R}^{A}$, $\widehat{\rho }_{R}^{B}$ that
contain coherences between states with differing boson numbers. In
particular, for a non-entangled state the following family of inequalities -
originally derived in \cite{Hillery06a}, is invoked. 
\begin{equation}
|\left\langle (\widehat{a})^{m}\,(\widehat{b}^{\dag })^{n}\right\rangle
|^{2}\leq \left\langle (\widehat{a}^{\dag })^{m}(\widehat{a})^{m}\,(\widehat{%
b}^{\dag })^{n}(\widehat{b})^{n}\right\rangle
\label{Eq.GeneralHilleryNonEntState}
\end{equation}%
This is just a special case of (\ref{Eq.GenCorrTest}) with $\widehat{\Omega }%
_{A}=(\widehat{a})^{m}$ and $\widehat{\Omega }_{B}=(\widehat{b})^{n}$. Thus
if 
\begin{equation}
|\left\langle (\widehat{a})^{m}\,(\widehat{b}^{\dag })^{n}\right\rangle
|^{2}>\left\langle (\widehat{a}^{\dag })^{m}(\widehat{a})^{m}\,(\widehat{b}%
^{\dag })^{n}(\widehat{b})^{n}\right\rangle  \label{Eq.HilleryEntangTest}
\end{equation}%
then the state is entangled.The \emph{Hillery} et al \cite{Hillery06a}
entanglement test (\ref{Eq.HilleryEntangTest}) is a valid test for
entanglement and is actually a \emph{more stringent test} than merely
showing that $|\left\langle (\widehat{a})^{m}\,(\widehat{b}^{\dag
})^{n}\right\rangle |^{2}>0$, since the quantity $|\left\langle (\widehat{a}%
)^{m}\,(\widehat{b}^{\dag })^{n}\right\rangle |^{2}$ is now required to be 
\emph{larger}. In a paper by He et al \cite{He12a} (see SubSection \ref%
{subSection - He et al 2012 Correlation Tests}) the Hillery et al \cite%
{Hillery06a} entanglement test $|\left\langle (\widehat{a})^{m}\,(\widehat{b}%
^{\dag })^{n}\right\rangle |^{2}>\left\langle (\widehat{a}^{\dag })^{m}(%
\widehat{a})^{m}\,(\widehat{b}^{\dag })^{n}(\widehat{b})^{n}\right\rangle $
is applied for the case where $A$ and $B$ \emph{each} consist of \emph{one
mode} localised in each well of a double well potential. This test whilst
applicable could be replaced by the more easily satisfied test $%
|\left\langle (\widehat{a})^{m}\,(\widehat{b}^{\dag })^{n}\right\rangle
|^{2}>0$ (see (\ref{Eq.EntangTest})). However, as will be seen below in
SubSection \ref{subSection - He et al 2012 Correlation Tests}, the Hillery
et al \cite{Hillery06a} entanglement criterion is needed if the sub-systems
each consist of \emph{pairs of modes}, as treated in \cite{He11a}, \cite%
{He12a}.

Note that if $n\neq m$\ the left side is zero for states that are globally
SSR compliant. In this case we can always substitute for two mode systems

\begin{eqnarray}
(\widehat{a}\,\widehat{b}^{\dag })^{n} &=&(\widehat{S}_{x}-i\widehat{S}%
_{y})^{n}  \nonumber \\
(\widehat{a}^{\dag })^{n}(\widehat{a})^{n} &=&(\widehat{a}\widehat{a}^{\dag
})^{n}=\left( 1+\frac{\widehat{N}}{2}-\widehat{S}_{z}\right) ^{n}  \nonumber
\\
(\widehat{b}^{\dag })^{n}(\widehat{b})^{n} &=&(\widehat{b}\widehat{b}^{\dag
})^{n}=\left( 1+\frac{\widehat{N}}{2}+\widehat{S}_{z}\right) ^{n}
\label{Eq.SpinOprSubstn}
\end{eqnarray}%
to write both the Hillery and the weak correlation test in terms of spin
operators.

A particular case for $n=m=1$ is the test $|\left\langle \widehat{a}\,%
\widehat{b}^{\dag }\right\rangle |^{2}>\left\langle \widehat{n}_{a}\,%
\widehat{n}_{b}\right\rangle $ for an entangled state. To put this result in
context, for a general quantum state and any operator $\widehat{\Omega }$ we
have $\left\langle \widehat{\Omega }^{\dag }\right\rangle =\left\langle 
\widehat{\Omega }\right\rangle ^{\ast }$ and $\left\langle \left( \widehat{%
\Omega }^{\dag }-\left\langle \widehat{\Omega }^{\dag }\right\rangle \right)
\left( \widehat{\Omega }-\left\langle \widehat{\Omega }\right\rangle \right)
\right\rangle \geq 0$, hence leading to the Schwarz inequality $%
|\left\langle \widehat{\Omega }\right\rangle |^{2}=|\left\langle \widehat{%
\Omega }^{\dag }\right\rangle |^{2}\leq \left\langle \widehat{\Omega }^{\dag
}\,\widehat{\Omega }\right\rangle $. Taking $\widehat{\Omega }=\widehat{a}\,%
\widehat{b}^{\dag }$ leads to the inequality $|\left\langle \widehat{a}\,%
\widehat{b}^{\dag }\right\rangle |^{2}\leq \left\langle \widehat{n}_{a}\,(%
\widehat{n}_{b}+1)\right\rangle $, whilst choosing $\widehat{\Omega }=%
\widehat{b}\,\widehat{a}^{\dag }$ leads to the inequality $|\left\langle 
\widehat{a}\,\widehat{b}^{\dag }\right\rangle |^{2}\leq \left\langle (%
\widehat{n}_{a}+1)\,\widehat{n}_{b}\right\rangle $ for \emph{all} quantum
states. In both cases the right side of the inequality is greater than $%
\left\langle \widehat{n}_{a}\,\widehat{n}_{b}\right\rangle $, so \emph{if }%
it was found that $|\left\langle \widehat{a}\,\widehat{b}^{\dag
}\right\rangle |^{2}>\left\langle \widehat{n}_{a}\,\widehat{n}%
_{b}\right\rangle $ (though of course still $\leq \left\langle \widehat{n}%
_{a}\,(\widehat{n}_{b}+1)\right\rangle $ and $\leq \left\langle (\widehat{n}%
_{a}+1)\,\widehat{n}_{b}\right\rangle $) then it could be concluded that the
state was entangled. However, as we will see the left side $|\left\langle 
\widehat{a}\,\widehat{b}^{\dag }\right\rangle |^{2}$ actually works out to
be zero if physical states for $\widehat{\rho }_{R}^{A}$, $\widehat{\rho }%
_{R}^{B}$ are involved in defining non-entangled states, so that for a
non-entangled state defined as in the present paper the true inequality
replacing $|\left\langle \widehat{a}\,\widehat{b}^{\dag }\right\rangle
|^{2}\leq \left\langle \widehat{n}_{a}\,\widehat{n}_{b}\right\rangle $ is
just $0\leq \left\langle \widehat{n}_{a}\,\widehat{n}_{b}\right\rangle $,
which is trivially true for any quantum state.

For the case where $n=m=1$\ we can write the test (\ref{Eq.HilleryEntangTest}%
)\ in terms of spin operators using $\widehat{a}\,\widehat{b}^{\dag }=%
\widehat{S}_{x}-i\widehat{S}_{y}$ as 
\begin{equation}
\left\langle \widehat{S}_{x}\right\rangle _{\rho }^{2}+\left\langle \widehat{%
S}_{y}\right\rangle _{\rho }^{2}>\frac{1}{4}\left\langle \widehat{N}%
^{2}\right\rangle _{\rho }-\left\langle \widehat{S}_{z}^{2}\right\rangle
_{\rho }
\end{equation}%
which when combined with the general result $\widehat{S}_{x}{}^{2}+\widehat{S%
}_{y}{}^{2}+\widehat{S}_{z}{}^{2}=(\widehat{N}/2)(\widehat{N}/2+1)$\ leads
to the test 
\begin{equation}
\left\langle \Delta \widehat{S}_{x}{}^{2}\right\rangle _{\rho }+\left\langle
\Delta \widehat{S}_{y}{}^{2}\right\rangle _{\rho }<\frac{1}{2}\left\langle 
\widehat{N}\right\rangle _{\rho }  \label{Eq.HillSimpleCorrTest}
\end{equation}%
This is the same as the Hillery spin variance test (\ref%
{Eq.HillerySpinEntTest}) , so the Hillery first order correlation test does
not add a further test for demonstrating non-SSR compliant entanglement. The
Hillery correlation test for $n=2$\ leads to complex conditions involving
higher powers of spin operators.

\subsubsection{Applications of Correlation Tests for Entanglement}

As an example of applying these tests consider the \emph{mixed two mode
coherent states} described in Appendix \ref{Appendix - Separable but Non
Local States}, whose density operator for the two mode $\widehat{a}$, $%
\widehat{b}$ system is given in Eq. (\ref{Eq.TwoModeCoherentStateMixture}).
We can now examine the Hillery et al \cite{Hillery09a} entanglement test in
Eq.(\ref{Eq.HilleryEntangTest}) and the entanglement test in Eq.(\ref%
{Eq.EntangTest}) for the case where $m=n=1$. It is straight-forward to show
that 
\begin{eqnarray}
|\left\langle \widehat{a}\,\widehat{b}^{\dag }\right\rangle |^{2} &=&|\alpha
|^{4}  \nonumber \\
\left\langle (\widehat{a}^{\dag }\widehat{a})\,(\widehat{b}^{\dag }\widehat{b%
})\right\rangle &=&|\alpha |^{4}  \label{Eq.TestResults}
\end{eqnarray}%
so that $|\left\langle \widehat{a}\,\widehat{b}^{\dag }\right\rangle
|^{2}=\left\langle (\widehat{a}^{\dag }\widehat{a})\,(\widehat{b}^{\dag }%
\widehat{b})\right\rangle $. A non-entangled state defined in terms of the
SSR requirement for the separate modes satisfies $|\left\langle \widehat{a}\,%
\widehat{b}^{\dag }\right\rangle |^{2}=0$, whilst for a non-entangled state
in which the SSR requirement for separate modes is not specifically required
merely satisfies $|\left\langle \widehat{a}\,\widehat{b}^{\dag
}\right\rangle |^{2}\leq \left\langle (\widehat{a}^{\dag }\widehat{a})\,(%
\widehat{b}^{\dag }\widehat{b})\right\rangle $. Hence the test for
entanglement of modes $A$, $B$ in the present paper $|\left\langle \widehat{a%
}\,\widehat{b}^{\dag }\right\rangle |^{2}>0$ is satisfied, whilst the
Hillery et al \cite{Hillery09a} test $|\left\langle \widehat{a}\,\widehat{b}%
^{\dag }\right\rangle |^{2}>\left\langle (\widehat{a}^{\dag }\widehat{a})\,(%
\widehat{b}^{\dag }\widehat{b})\right\rangle $ is not.

In terms of the definition of non-entangled states in the present paper, the
mixture of two mode coherent states given in Eq.(\ref%
{Eq.TwoModeCoherentStateMixture}) is \emph{not} a \emph{separable }state,
not a separable state. As discussed in Paper 1 (see \textbf{Section 3.4.3})
this is because a coherent state gives rise to a non-zero coherence ($%
\left\langle \widehat{a}\right\rangle \neq 0$) and thus cannot represent a
physical state for the SSR compliant states involving identical massive
bosons (as in BECs). However, in terms of the definition of non-entangled
states in other papers such as those of Hillery et al \cite{Hillery06a}, 
\cite{Hillery09a} the mixture of two mode coherent states would be a \emph{%
non-entangled} state. It is thus a useful state for providing an example of
the different outcomes of definitions where the local SSR is applied or not.

A further example of applying correlation tests is provided by the \emph{%
NOON state }defined in (\ref{Eq.GeneralNOONState}) where here we consider
modes $A$, $B$. All matrix elements of the form $\left\langle (\widehat{a}%
^{\dag })^{m}(\widehat{a})^{m}\,(\widehat{b}^{\dag })^{n}(\widehat{b}%
)^{n}\right\rangle $ are zero for all $m,n$ because both terms contain one
mode with zero bosons. Matrix elements of the form $\left\langle (\widehat{a}%
)^{m}\,(\widehat{b}^{\dag })^{n}\right\rangle $ are all zero unless $m=n=N$
and in this case 
\begin{eqnarray}
\left\langle (\widehat{a})^{N}\,(\widehat{b}^{\dag })^{N}\right\rangle
&=&\left\langle (\widehat{S}_{x}-i\widehat{S}_{y})^{N}\right\rangle 
\nonumber \\
&=&\cos \theta \,\sin \theta \,\left\langle N,0\right\vert (\widehat{S}%
_{-})^{N}\left\vert 0,N\right\rangle  \nonumber \\
&=&\cos \theta \,\sin \theta \,\sqrt{N}\sqrt{2N-2}\sqrt{3N-6}\sqrt{4N-12}...%
\sqrt{N}  \label{Eq.NOONTestResult}
\end{eqnarray}%
which is non-zero in general. Hence $|\left\langle (\widehat{a})^{N}\,(%
\widehat{b}^{\dag })^{N}\right\rangle |^{2}>0$ as required for both the weak
and strong correlation tests, confirming that the NOON state is \emph{%
entangled}. Carrying out this entanglement test experimentally for large $N$
would involve measuring expectation values of high powers of the spin
operators $\widehat{S}_{x}$ and $\widehat{S}_{y}$, which is difficult at
present.

\subsection{He et al 2012}

\label{subSection - He et al 2012 Correlation Tests}

For the \emph{four mode} system associated with a double well \ described in
SubSection \ref{SubSection - He 2012} (see \cite{He12a}), the inequalities
derived by Hillery et al \cite{Hillery09a} (see SubSection \ref{SubSection -
Hillery 2009}) 
\begin{equation}
|\left\langle (\widehat{a}_{i})^{m}\,(\widehat{b}_{j}^{\dag
})^{n}\right\rangle |^{2}\leq \left\langle (\widehat{a}_{i}^{\dag })^{m}(%
\widehat{a}_{i})^{m}\,(\widehat{b}_{j}^{\dag })^{n}(\widehat{b}%
_{j})^{n}\right\rangle  \label{Eq.HilleryIneqNonEntModePairs}
\end{equation}%
that apply for two non-entangled sub-systems $A$ and $B$ can now be usefully
applied, since in this case the quantities $\left\langle (\widehat{a}%
_{i})^{m}\,(\widehat{b}_{j}^{\dag })^{n}\right\rangle $ are in general no
longer zero for separable states. Thus there is an \emph{entanglement test }%
for two sub-systems consisting of \emph{pairs of modes}. If 
\begin{eqnarray}
|\left\langle (\widehat{a}_{i})^{m}\,(\widehat{b}_{j}^{\dag
})^{n}\right\rangle |^{2} &>&\left\langle (\widehat{a}_{i}^{\dag })^{m}(%
\widehat{a}_{i})^{m}\,(\widehat{b}_{j}^{\dag })^{n}(\widehat{b}%
_{j})^{n}\right\rangle \qquad  \nonumber \\
for\;any\;of\;i,j &=&1,2  \label{Eq.HilleryEntTestPairsModes}
\end{eqnarray}%
then the quantum state for two sub-systems $A$ and $B$ - \emph{each}
consisting of \emph{two modes} localised in each well - is entangled. Again
only the case where $m=n$ is relevant for states that are global SSR
compliant.

\section{Quadrature Tests for Entanglement}

\label{Section - Quadrature Tesrs for Entanglement}

In this Section we discuss tests for two mode entanglement involving so
called \emph{quadrature} operators - \emph{position} and \emph{momentum}
being particular examples of such operators. These tests are distinct from
those involving \emph{spin} operators or \emph{correlation} tests - the
latter have been shown to be closely related to spin operator tests. The
issue of \emph{measurement} of the quadrature variances involved in these
tests for the case of two mode systems involving identical massive bosons
will be briefly discussed in Section \ref{section - Interferometry in Two
Mode BEC}. Again we have a situation where tests derived in which local
particle number SSR compliance for separable states is ignored are still
valid when it is taken into account. However, when local particle number SSR
compliance for separable states is actually included new entanglement tests
arise. The \emph{two mode quadrature squeezing} test in (\ref%
{Eq.TwoModeQuadSqgTest}) is an example, though this test is not very useful
as it could be replaced by the \emph{Bloch vector} test. The quadrature
correlation test in (\ref{Eq,CorrelTest}) also applies and is equivalent to
the \emph{Bloch vector} test. However the non-existent \emph{quadrature
variance} test in (\ref{Eq.ApplnDuanTest}) is an example where there is no
generalisation of the previous entanglement test (see (\ref%
{Eq.QuadratureVarianceEntTest})) that applied when the SSR\ were irrelevant.

\subsection{Duan et al 2000}

\label{Duan et al 2000}

\subsubsection{Two Distinguishable Particles}

A further inequality aimed at providing a signature for entanglement is set
out in the papers by Duan et al \cite{Duan00a}, Toth et al \cite{Toth03a}.
Duan et al \cite{Duan00a} considered a general situation where the system
consisted of two disinguishable sub-systems $A$ and $B$, for which \emph{%
position} and \emph{momentum} Hermitian operators $\widehat{x}_{A},\widehat{p%
}_{A}$ and $\widehat{x}_{B},\widehat{p}_{B}$ were involved that satisfied
the standard commutation rules $[\widehat{x}_{A},\widehat{p}_{A}]=[\widehat{x%
}_{B},\widehat{p}_{B}]=i$ in units where $\hbar =1$. These sub-systems were
quite general and could be two \emph{distinguishable} quantum particles $A$
and $B$, but other situations can also be treated. An inequality was
obtained for a two sub-system non-entangled state involving the variances
for the commuting observables $\widehat{x}_{A}+\widehat{x}_{B}$ and $%
\widehat{p}_{A}-\widehat{p}_{B}$%
\begin{equation}
\left\langle \Delta (\widehat{x}_{A}+\widehat{x}_{B})^{2}\right\rangle
+\left\langle \Delta (\widehat{p}_{A}-\widehat{p}_{B})^{2}\right\rangle \geq
2  \label{Eq.DuanInequalityNonEntState}
\end{equation}%
which could be used to establish a \emph{variance test} for entangled states
of the $A$ and $B$ sub-systems, so that if%
\begin{equation}
\left\langle \Delta (\widehat{x}_{A}+\widehat{x}_{B})^{2}\right\rangle
+\left\langle \Delta (\widehat{p}_{A}-\widehat{p}_{B})^{2}\right\rangle <2
\label{Eq.QuadratureVarianceEntTest}
\end{equation}%
then the sub-systems are entangled. For the case of distinguishable
particles such states are possible - consider for example any simultaneous
eigenstate of the commuting observables $\widehat{x}_{A}+\widehat{x}_{B}$
and $\widehat{p}_{A}-\widehat{p}_{B}$. For such a state $\left\langle \Delta
(\widehat{x}_{A}+\widehat{x}_{B})^{2}\right\rangle $ and $\left\langle
\Delta (\widehat{p}_{A}-\widehat{p}_{B})^{2}\right\rangle $ are both zero,
so the simultaneous eigenstates are entangled states of \emph{particles} $A$%
, $B$. For simplicity we only set out the case for which $a=1$ in \cite%
{Duan00a}. The proof given in \cite{Duan00a} considered separable states of
the general form as in Eq.(\ref{Eq.NonEntStateModesCD}) for two sub-systems
but where $\widehat{\rho }_{R}^{A}$ and $\widehat{\rho }_{R}^{B}$ are
possible states for sub-systems $A$, $B$. Consequently, a first quantization
case involving \emph{one particle states} could be involved, where
super-selection rules were\emph{\ not} relevant. As explained in the
Introduction, the two distinguishable quantum particles are each equivalent
to a whole set of single particle states (momentum eigenstates, harmonic
oscillator states, ..) that each quantum particle can occupy, and because
both $\widehat{\rho }_{R}^{A}$ and $\widehat{\rho }_{R}^{B}$ represent
states for one particle we have $[\widehat{n}_{A},\widehat{\rho }_{R}^{A}]=[%
\widehat{n}_{B},\widehat{\rho }_{R}^{B}]=0$. Because $\widehat{\rho }$
represent a state for the two particles $[\widehat{n}_{A}+\widehat{n}_{B},%
\widehat{\rho }]=0$, the SSR are still true, though irrelevant in the case
of distinguishable quantum particles $A$ and $B$.

Another inequality that can be established is 
\begin{equation}
\left\langle \Delta (\widehat{x}_{A}-\widehat{x}_{B})^{2}\right\rangle
+\left\langle \Delta (\widehat{p}_{A}+\widehat{p}_{B})^{2}\right\rangle \geq
2  \label{Eq.DuanIneq}
\end{equation}%
which could also be used to establish a variance test for entangled states
of the $A$\ and $B$\ sub-systems, so that if%
\begin{equation}
\left\langle \Delta (\widehat{x}_{A}-\widehat{x}_{B})^{2}\right\rangle
+\left\langle \Delta (\widehat{p}_{A}+\widehat{p}_{B})^{2}\right\rangle <2
\label{Eq.QuadratureVarianceEntTestB}
\end{equation}%
then the sub-systems are entangled.

\subsubsection{Two Mode Systems of Identical Bosons}

\label{SubSubSection - Duan}

However, we can also consider cases of systems of \emph{identical }bosons
with two \emph{modes} $A$, $B$ rather than two \emph{distinguishable}
quantum particles $A$ and $B$. In this case both the sub-systems may involve
arbitrary numbers of particles, so it is of interest to see what
implications follow from the physical sub-system states $\widehat{\rho }%
_{R}^{A}$ and $\widehat{\rho }_{R}^{B}$ now being required to satisfy the
local particle number SSR, and all quantum states $\widehat{\rho }$
satisfying the global particle number SSR. It is well-known that in two mode
boson systems \emph{quadrature} \emph{operators} can be defined via 
\begin{eqnarray}
\widehat{x}_{A} &=&\frac{1}{\sqrt{2}}(\widehat{a}+\widehat{a}^{\dag })\qquad 
\widehat{p}_{A}=\frac{1}{\sqrt{2}i}(\widehat{a}-\widehat{a}^{\dag }) 
\nonumber \\
\widehat{x}_{B} &=&\frac{1}{\sqrt{2}}(\widehat{b}+\widehat{b}^{\dag })\qquad 
\widehat{p}_{B}=\frac{1}{\sqrt{2}i}(\widehat{b}-\widehat{b}^{\dag })
\label{Eq.PositionMtmOprs}
\end{eqnarray}%
which have the same commutation rules as the position and momentum operators
for distinguishable particles. Thus $[\widehat{x}_{A},\widehat{p}_{A}]=[%
\widehat{x}_{B},\widehat{p}_{B}]=i$ as for cases where $A$, $B$ were
distinguishable particles.

Since the proof of Eq. (\ref{Eq.DuanInequalityNonEntState}) in \cite{Duan00a}
did not involve invoking the SSR, then\textbf{\ }\emph{if} the inequality in
Eq.(\ref{Eq.QuadratureVarianceEntTest}) is satisfied, then the state \emph{%
would be} an entangled state for \emph{modes} $A$, $B$. as well as for
distinguishable particles $A$, $B$. The situation would \emph{then} be
similar to that for the Hillery et al \cite{Hillery06a}, \cite{Hillery09a}
tests - the SSR compliant sub-system states are just a particular case of
the set of all sub-system states. However, in regard to spin squeezing and
correlation tests for entanglement, new tests were found when the SSR were
explicitly considered and it is \emph{possible} that this could occur here.
This turns out not to be the case.

As we will see, the inequality (\ref{Eq.DuanInequalityNonEntState}) is
replaced by an \emph{equation} that is satisfied by\emph{\ all }quantum
states for two mode systems of identical bosons where the global particle
number SSR applies. This equation is the same irrespective of whether the
state is separable or entangled. To see this we evaluate $\left\langle
\Delta (\widehat{x}_{A}+\widehat{x}_{B})^{2}\right\rangle +\left\langle
\Delta (\widehat{p}_{A}-\widehat{p}_{B})^{2}\right\rangle $ for states that
are global SSR compliant.

Firstly, 
\begin{equation}
\left\langle (\widehat{x}_{A}+\widehat{x}_{B})\right\rangle =\left\langle (%
\widehat{p}_{A}-\widehat{p}_{B})\right\rangle =0
\label{Eq.MeansGlobSSRState}
\end{equation}%
since $\left\langle \widehat{a}\right\rangle =\left\langle \widehat{b}%
\right\rangle =\left\langle \widehat{a}^{\dag }\right\rangle =\left\langle 
\widehat{b}^{\dag }\right\rangle =0$ for SSR compliant states.

Secondly,\textbf{\ }%
\[
\left\langle (\widehat{x}_{A}+\widehat{x}_{B})^{2}\right\rangle =\frac{1}{2}%
\left( 
\begin{array}{c}
\left\langle \widehat{a}\,\widehat{a}^{\dag }\right\rangle +\left\langle 
\widehat{a}\,\widehat{b}^{\dag }\right\rangle +\left\langle \widehat{b}\,%
\widehat{a}^{\dag }\right\rangle +\left\langle \widehat{b}\,\widehat{b}%
^{\dag }\right\rangle \\ 
+\left\langle \widehat{a}^{\dag }\,\widehat{a}\right\rangle +\left\langle 
\widehat{a}^{\dag }\,\widehat{b}\right\rangle +\left\langle \widehat{b}%
^{\dag }\,\widehat{a}\right\rangle +\left\langle \widehat{b}^{\dag }\,%
\widehat{b}\right\rangle%
\end{array}%
\right) 
\]%
using $\left\langle \widehat{a}^{2}\right\rangle =\left\langle (\widehat{a}%
^{\dag })^{2}\right\rangle =\left\langle \widehat{b}^{2}\right\rangle
=\left\langle (\widehat{b}^{\dag })^{2}\right\rangle =\left\langle \widehat{a%
}\,\widehat{b}\right\rangle =\left\langle \widehat{a}^{\dag }\,\widehat{b}%
^{\dag }\right\rangle =0$ for global SSR compliant states. Hence using the
commutation rules, introducing the number operator $\widehat{N}$ and the
spin operator $\widehat{S}_{x}$ and using (\ref{Eq.MeansGlobSSRState}) we
find that 
\begin{eqnarray}
\left\langle \Delta (\widehat{x}_{A}+\widehat{x}_{B})^{2}\right\rangle
&=&\left\langle (\widehat{x}_{A}+\widehat{x}_{B})^{2}\right\rangle  \nonumber
\\
&=&1+\left\langle \widehat{a}^{\dag }\,\widehat{a}\right\rangle
+\left\langle \widehat{b}^{\dag }\,\widehat{b}\right\rangle +\left\langle 
\widehat{b}^{\dag }\,\widehat{a}\right\rangle +\left\langle \widehat{a}%
^{\dag }\,\widehat{b}\right\rangle  \nonumber \\
&=&1+\left\langle \widehat{N}\right\rangle +2\left\langle \widehat{S}%
_{x}\right\rangle  \label{Eq.VarSumXGlobSSRState}
\end{eqnarray}%
Similarly 
\begin{equation}
\left\langle \Delta (\widehat{p}_{A}-\widehat{p}_{B})^{2}\right\rangle
=1+\left\langle \widehat{N}\right\rangle -2\left\langle \widehat{S}%
_{x}\right\rangle  \label{Eq.VarDiffPGlobnSSRState}
\end{equation}

Thus we have for all globally SSR compliant states%
\begin{equation}
\left\langle \Delta (\widehat{x}_{A}+\widehat{x}_{B})^{2}\right\rangle
+\left\langle \Delta (\widehat{p}_{A}-\widehat{p}_{B})^{2}\right\rangle
=2+2\left\langle \widehat{N}\right\rangle
\label{Eq.QuadVarEqualitysSSRSepState}
\end{equation}%
Since $\left\langle \widehat{N}\right\rangle \geq 0$ for all quantum states
we see that the Duan et al inequality (\ref{Eq.DuanInequalityNonEntState})
for separable states is still satisfied, but because (\ref%
{Eq.QuadVarEqualitysSSRSepState}) applies for all states irrespective of
whether or not they are separable, we see that there is\textbf{\ }\emph{no}%
\textbf{\ }quadrature variance entanglement test of the form 
\begin{equation}
\left\langle \Delta (\widehat{x}_{A}+\widehat{x}_{B})^{2}\right\rangle
+\left\langle \Delta (\widehat{p}_{A}-\widehat{p}_{B})^{2}\right\rangle
<2+2\left\langle \widehat{N}\right\rangle  \label{Eq.ApplnDuanTest}
\end{equation}%
for the case of two mode systems of identical massive bosons. The situation
is similar to the non-existent test $\left\langle \Delta \widehat{S}%
\,_{x}^{2}\right\rangle +\left\langle \Delta \widehat{S}\,_{y}^{2}\right%
\rangle <|\left\langle \widehat{S}\,_{z}\right\rangle |\,$\ in Section \ref%
{SubSubSection - Non Applicable Test}. The situation contrasts that in
Section \ref{SubSection - Raymer 2003}, where a test $\left\langle \Delta (%
\widehat{S}\,_{x}^{1}+\widehat{S}\,_{x}^{2})^{2}\right\rangle +\left\langle
\Delta (\widehat{S}\,_{y}^{1}-\widehat{S}\,_{y}^{2})^{2}\right\rangle
<|\left\langle \widehat{S}_{z}\right\rangle |$ establishes entanglement
between two sub-systems ($1$ and $2$) - but in this case each consisting of
two modes.

We can also show for all globally SSR compliant states that%
\begin{eqnarray}
\left\langle \Delta (\widehat{x}_{A}-\widehat{x}_{B})^{2}\right\rangle
&=&1+\left\langle \widehat{N}\right\rangle -2\left\langle \widehat{S}%
_{x}\right\rangle  \label{Eq.VarDiffXGlobSSRState} \\
\left\langle \Delta (\widehat{p}_{A}+\widehat{p}_{B})^{2}\right\rangle
&=&1+\left\langle \widehat{N}\right\rangle +2\left\langle \widehat{S}%
_{x}\right\rangle  \label{Eq.VarSumPGlobnSSRState}
\end{eqnarray}%
and hence 
\begin{equation}
\left\langle \Delta (\widehat{x}_{A}-\widehat{x}_{B})^{2}\right\rangle
+\left\langle \Delta (\widehat{p}_{A}+\widehat{p}_{B})^{2}\right\rangle
=2+2\left\langle \widehat{N}\right\rangle
\label{Eq.QuadVarEqualitysSSRSepStateB}
\end{equation}%
but again no entanglement test results.

The universal result (\ref{Eq.QuadVarEqualitysSSRSepState}) for the
quadrature variance sum may seem paradoxical in view of the operators $(%
\widehat{x}_{A}+\widehat{x}_{B})$ and $(\widehat{p}_{A}-\widehat{p}_{B})$
commuting. Mathematically, this would imply that they would then have a
complete set of simultaneous eigenvectors $\left\vert
X_{A,B},\,P_{A,B}\right\rangle $\ such that $(\widehat{x}_{A}+\widehat{x}%
_{B})\,\left\vert X_{A,B},\,P_{A,B}\right\rangle =X_{A,B}\,\left\vert
X_{A,B},\,P_{A,B}\right\rangle $ and $(\widehat{p}_{A}-\widehat{p}%
_{B})\,\left\vert X_{A,B},\,P_{A,B}\right\rangle =P_{A,B}\,\left\vert
X_{A,B},\,P_{A,B}\right\rangle $. For these eigenstates $\left\langle \Delta
(\widehat{x}_{A}+\widehat{x}_{B})^{2}\right\rangle =\left\langle \Delta (%
\widehat{p}_{A}-\widehat{p}_{B})^{2}\right\rangle =0$\ which contradicts (%
\ref{Eq.QuadVarEqualitysSSRSepState}) for such states. However, no such
eigenstates exist that are globally SSR compliant. For SSR compliant states $%
\left\vert X_{A,B},\,P_{A,B}\right\rangle $ must be an eigenstate of $%
\widehat{N}$\ \ and for eigenvalue $N$ we see that $(\widehat{x}_{A}+%
\widehat{x}_{B})$\ $\left\vert X_{A,B},\,P_{A,B}\right\rangle _{N}$ is a
linear combination of eigenstates of $\widehat{N}$\ \ with eigenvalues $N\pm
1$. Hence $(\widehat{x}_{A}+\widehat{x}_{B})$\ $\left\vert
X_{A,B},\,P_{A,B}\right\rangle _{N}\neq X_{A,B}\,\left\vert
X_{A,B},\,P_{A,B}\right\rangle _{N}$ so simultaneous eigenstates that are
SSR compliant do not exist and there is no paradox. As pointed out above,
this issue does not arise for the case of two distinguishable particles
where the operators $\widehat{x}_{A},\widehat{x}_{B},\widehat{p}_{A}$ and $%
\widehat{p}_{B}$ are\textbf{\ }\emph{not}\textbf{\ }related to mode
annihilation and creation operators - as in the present case.

We can also derive \emph{inequalities} for \emph{separable} states involving 
$\widehat{x}_{A},\widehat{p}_{A}$ and $\widehat{x}_{B},\widehat{p}_{B}$
based on the approach in Section \ref{SubSection - Raymer 2003}. Starting
with Eq. (\ref{Eq.InequalityVarianceSumSepStates}) we choose $\widehat{%
\Omega }_{A}=\widehat{x}\,_{A}$, $\widehat{\Omega }_{B}=\widehat{x}\,_{B}$, $%
\widehat{\Lambda }_{A}=\widehat{p}\,_{A}$ and $\widehat{\Lambda }_{B}=%
\widehat{p}\,_{B}$. Here $\widehat{\Theta }_{A}=\widehat{1}\,_{A}$ and $%
\widehat{\Theta }_{\backslash B}=\widehat{1}\,_{B}$ For \emph{separable}
states we have from (\ref{Eq.InequalityVarianceSumSepStates})%
\begin{equation}
\left\langle \Delta (\alpha \widehat{x}_{A}+\beta \widehat{x}%
_{B})^{2}\right\rangle +\left\langle \Delta (\alpha \widehat{p}_{A}-\beta 
\widehat{p}_{B})^{2}\right\rangle \geq \alpha ^{2}+\beta ^{2}
\end{equation}%
With the choice of $\alpha ^{2}=\beta ^{2}=1$ we then find the following
inequalities for separable states%
\begin{eqnarray}
\left\langle \Delta (\widehat{x}_{A}+\widehat{x}_{B})^{2}\right\rangle
+\left\langle \Delta (\widehat{p}_{A}-\widehat{p}_{B})^{2}\right\rangle
&\geq &2  \nonumber \\
\left\langle \Delta (\widehat{x}_{A}-\widehat{x}_{B})^{2}\right\rangle
+\left\langle \Delta (\widehat{p}_{A}+\widehat{p}_{B})^{2}\right\rangle
&\geq &2  \nonumber \\
\left\langle \Delta (\widehat{x}_{A}+\widehat{x}_{B})^{2}\right\rangle
+\left\langle \Delta (\widehat{p}_{A}+\widehat{p}_{B})^{2}\right\rangle
&\geq &2  \nonumber \\
\left\langle \Delta (\widehat{x}_{A}-\widehat{x}_{B})^{2}\right\rangle
+\left\langle \Delta (\widehat{p}_{A}-\widehat{p}_{B})^{2}\right\rangle
&\geq &2  \label{Eq.DuanIneqC}
\end{eqnarray}%
depending on the choice of $\alpha $ and $\beta $. With $\alpha =\beta =1$
the first result is obtained and is the same as in (\ref%
{Eq.DuanInequalityNonEntState}). This result is consistent with (\ref%
{Eq.QuadVarEqualitysSSRSepState}). However using (\ref%
{Eq.VarSumXGlobSSRState}), (\ref{Eq.VarSumPGlobnSSRState}), (\ref%
{Eq.VarDiffXGlobSSRState}) and (\ref{Eq.VarDiffPGlobnSSRState}) we have for 
\emph{global SSR compliant} states - separable \emph{and} non-separable%
\begin{eqnarray}
\left\langle \Delta (\widehat{x}_{A}+\widehat{x}_{B})^{2}\right\rangle
+\left\langle \Delta (\widehat{p}_{A}-\widehat{p}_{B})^{2}\right\rangle
&=&2+2\left\langle \widehat{N}\right\rangle  \nonumber \\
\left\langle \Delta (\widehat{x}_{A}-\widehat{x}_{B})^{2}\right\rangle
+\left\langle \Delta (\widehat{p}_{A}+\widehat{p}_{B})^{2}\right\rangle
&=&2+2\left\langle \widehat{N}\right\rangle  \nonumber \\
\left\langle \Delta (\widehat{x}_{A}+\widehat{x}_{B})^{2}\right\rangle
+\left\langle \Delta (\widehat{p}_{A}+\widehat{p}_{B})^{2}\right\rangle
&=&2+2\left\langle \widehat{N}\right\rangle +4\left\langle \widehat{S}%
_{x}\right\rangle  \nonumber \\
\left\langle \Delta (\widehat{x}_{A}-\widehat{x}_{B})^{2}\right\rangle
+\left\langle \Delta (\widehat{p}_{A}-\widehat{p}_{B})^{2}\right\rangle
&=&2+2\left\langle \widehat{N}\right\rangle -4\left\langle \widehat{S}%
_{x}\right\rangle
\end{eqnarray}%
The implications for the first two equalities have been discussed above. In
the case of the $(+,+)$ and $(-,-)$ cases, we note that for states with
eigenvalue $N$ for $\widehat{N}$ the eigenvalues for $\widehat{S}_{x}$ lie
in the range $-N/2$ to $+N/2$ and hence $\left\langle \widehat{N}%
\right\rangle \pm 2\left\langle \widehat{S}_{x}\right\rangle $ is always $%
\geq 0$. Thus (\ref{Eq.DuanIneqC}) will apply for both separable \emph{and}
entangled states. Hence for global SSR compliant states none of \ (\ref%
{Eq.DuanIneqC}) lead to an entanglement test.

\subsubsection{Non SSR Compliant States}

On the other hand if \emph{neither }the sub-system \emph{nor }the overall
system states are \emph{required} to be SSR compliant - though they may be -%
\textbf{\ }we find that for\textbf{\ }\emph{separable}\textbf{\ }states 
\begin{equation}
\left\langle \Delta (\widehat{x}_{A}\pm \widehat{x}_{B})^{2}\right\rangle
_{\rho }+\left\langle \Delta (\widehat{p}_{A}\mp \widehat{p}%
_{B})^{2}\right\rangle _{\rho }\geq 2+2\left\langle \widehat{N}\right\rangle
_{\rho }+2(\left\langle \widehat{a}\,\widehat{b}\right\rangle +\left\langle 
\widehat{a}^{\dag }\,\widehat{b}^{\dag }\right\rangle )-2|\left\langle 
\widehat{a}\right\rangle _{\rho }+\left\langle \widehat{b}^{\dag
}\right\rangle _{\rho }|^{2}
\end{equation}%
so entanglement based on ignoring local particle number SSR in the separable
states is now shown if 
\begin{equation}
\left\langle \Delta (\widehat{x}_{A}\pm \widehat{x}_{B})^{2}\right\rangle
_{\rho }+\left\langle \Delta (\widehat{p}_{A}\mp \widehat{p}%
_{B})^{2}\right\rangle _{\rho }<2+2\left\langle \widehat{N}\right\rangle
_{\rho }+2(\left\langle \widehat{a}\,\widehat{b}\right\rangle _{\rho
}+\left\langle \widehat{a}^{\dag }\,\widehat{b}^{\dag }\right\rangle _{\rho
})-2|\left\langle \widehat{a}\right\rangle _{\rho }+\left\langle \widehat{b}%
^{\dag }\right\rangle _{\rho }|^{2}  \label{Eq.QuadVarTestModesNonSSR}
\end{equation}%
However, even if local particle number SSR compliance is ignored for the
sub-system states (as in Ref \cite{Verstraete03a}), global particle number
SSR compliance is still required for the overall quantum state. This applies
to both the separable states and to states that are being tested for
entanglement. In this case the quantities $\left\langle \widehat{a}\,%
\widehat{b}\right\rangle _{\rho }$, $\left\langle \widehat{a}^{\dag }\,%
\widehat{b}^{\dag }\right\rangle _{\rho }$, $\left\langle \widehat{a}%
\right\rangle _{\rho }$, $\left\langle \widehat{a}^{\dag }\right\rangle
_{\rho }$, $\left\langle \widehat{b}\right\rangle _{\rho }$ and $%
\left\langle \widehat{b}^{\dag }\right\rangle _{\rho }$\ are all zero, so
the entanglement test in (\ref{Eq.QuadVarTestModesNonSSR}) would become the
same as the\textbf{\ }\emph{hypothetical}\textbf{\ }entanglement test (\ref%
{Eq.ApplnDuanTest}).

For the sceptic (see Section \ref{SubSubSection - Significance of Spin
Squeezing Test}) who wishes to completely disregard the SSR (both locally
and globally) and proposes to use tests based on quadrature variances such
as (\ref{Eq.QuadVarTestModesNonSSR}) to establish entanglement, the
challenge will be to find a way of measuring the allegedly non-zero
quantities $\left\langle \widehat{a}\,\widehat{b}\right\rangle _{\rho }$.. $%
\left\langle \widehat{b}^{\dag }\right\rangle _{\rho }$. This would require
some sort of system with a well-defined\textbf{\ }\emph{phase}\textbf{\ }%
reference. Such a measurement is not possible with the beam splitter
interferometer discussed in this paper, and the lack of such a detector
system would preclude establishing SSR neglected entanglement for systems of
identical bosons. Essentially the same problem arises in testing whether
states that are non-SSR compliant\ exist in single mode systems of massive
bosons.

As mentioned previously, the result in Eq. (\ref%
{Eq.DuanInequalityNonEntState}) was established in Ref. \cite{Duan00a} \emph{%
without} requiring the sub-system states $\widehat{\rho }_{R}^{A}$, $%
\widehat{\rho }_{R}^{B}$ to be compliant with the local particle number SSR
or the density operator $\widehat{\rho }$ for the state being tested to
comply with the global particle number SSR, as would be the case for
physical sub-system and system states of identical bosons. However, in Ref. 
\cite{Duan00a} it was pointed out that \emph{two mode squeezed vacuum}
states of the form $\left\vert \Phi \right\rangle =\exp (-r(\widehat{a}%
^{\dag }\widehat{b}^{\dag }-\widehat{a}\widehat{b}))\,\left\vert
0\right\rangle $ satisfy the entanglement test. However, such stand alone
two-mode states are\emph{\ not} allowed quantum states for massive identical
boson systems. as they are not compliant with the global particle number
SSR. To create states with correlated pairs of bosons in modes $a$ and $b$
processes such as the \emph{dissociation }of a bosonic \emph{molecular BEC}
in a mode $M$ into \emph{pair} of bosonic atoms in modes $a$ and $b$ can
indeed occur, but would involve interaction Hamiltonians such as $\widehat{V}%
=\kappa (\widehat{a}^{\dag }\widehat{b}^{\dag }\widehat{M}+\widehat{M}^{\dag
}\widehat{a}\widehat{b})$. The state produced would be an entangled state of
the atoms plus molecules which would be compliant with the global total
boson number SSR - taking into account the boson particle content of the
molecule via $\widehat{N}=2\widehat{n}_{M}+\widehat{n}_{a}+\widehat{n}_{b}$.
It would not be a state of the form $\left\vert \Phi \right\rangle =\exp (-r(%
\widehat{a}^{\dag }\widehat{b}^{\dag }-\widehat{a}\widehat{b}))\,\left\vert
0\right\rangle $.

\subsection{Reid 1989}

Another test involves the \emph{general quadrature operators} defined as in 
\cite{Reid89a}, for which those in (\ref{Eq.PositionMtmOprs}) are special
cases 
\begin{eqnarray}
\widehat{X}_{a}^{\theta } &=&\frac{1}{\sqrt{2}}(\widehat{a}\exp (-i\theta )+%
\widehat{a}^{\dag }\exp (+i\theta ))  \nonumber \\
\widehat{X}_{b}^{\phi } &=&\frac{1}{\sqrt{2}}(\widehat{b}\exp (-i\phi )+%
\widehat{b}^{\dag }\exp (+i\phi ))  \label{Eq.QuadOprs}
\end{eqnarray}%
These operators are Hermitian. The conjugate operators are 
\begin{eqnarray}
\widehat{P}_{a}^{\theta } &=&\frac{1}{\sqrt{2}i}(\widehat{a}\exp (-i\theta )-%
\widehat{a}^{\dag }\exp (+i\theta ))=\widehat{X}_{a}^{\theta +\pi /2} 
\nonumber \\
\widehat{P}_{b}^{\phi } &=&\frac{1}{\sqrt{2}i}(\widehat{b}\exp (-i\phi )-%
\widehat{b}^{\dag }\exp (+i\phi ))=\widehat{X}_{b}^{\phi +\pi /2}
\label{Eq.ConjQuadOprs}
\end{eqnarray}%
where $[\widehat{X}_{a}^{\theta },\widehat{P}_{a}^{\theta }]=[\widehat{X}%
_{b}^{\phi },\widehat{P}_{b}^{\phi }]=i$.

Noting that for any state we have $\left\langle (\widehat{X}_{a}^{\theta
}-\lambda \widehat{X}_{b}^{\phi })^{2}\right\rangle \geq 0$ for all real $%
\lambda $ establishes the Cauchy inequality for all quantum states%
\begin{equation}
C_{ab}^{\theta \phi }=\frac{|\left\langle \widehat{X}_{a}^{\theta }\widehat{X%
}_{b}^{\phi }\right\rangle |^{2}}{\left\langle (\widehat{X}_{a}^{\theta
})^{2}\right\rangle \left\langle (\widehat{X}_{b}^{\phi })^{2}\right\rangle }%
\leq 1  \label{Eq.CorrelnCoeft}
\end{equation}%
The quantity $C_{ab}^{\theta \phi }$ is a \emph{correlation coefficient}.
For SSR compliant separable states $\left\langle \widehat{X}_{a}^{\theta }%
\widehat{X}_{b}^{\phi }\right\rangle =\dsum\limits_{R}P_{R}\left\langle 
\widehat{X}_{a}^{\theta }\right\rangle _{R}\left\langle \widehat{X}%
_{b}^{\phi }\right\rangle _{R}=0$, whilst for all globally SSR compliant
states $\left\langle (\widehat{X}_{a}^{\theta })^{2}\right\rangle
=\left\langle \widehat{n}_{a}\right\rangle +\frac{1}{2}>\frac{1}{2}$ and $%
\left\langle (\widehat{X}_{b}^{\phi })^{2}\right\rangle =\left\langle 
\widehat{n}_{b}\right\rangle +\frac{1}{2}>\frac{1}{2}$.\ Hence for SSR
compliant separable states the correlation coefficient is zero. A \emph{%
quadrature correlation} test\textbf{\ }for entanglement based on locally SSR
compliant sub-system states is then 
\begin{equation}
C_{ab}^{\theta \phi }\neq 0  \label{Eq,CorrelTest}
\end{equation}%
However, it is not difficult to show that for states that are globally SSR
compliant 
\begin{equation}
\left\langle \widehat{X}_{a}^{\theta }\widehat{X}_{b}^{\phi }\right\rangle
=\left\langle \widehat{S}_{x}\right\rangle \cos (\theta -\phi )+\left\langle 
\widehat{S}_{y}\right\rangle \sin (\theta -\phi )  \label{Eq.QuadCorrelation}
\end{equation}%
so that the entanglement test based on locally SSR compliant sub-system
states is equivalent to finding one of $\left\langle \widehat{S}%
_{x}\right\rangle $ or $\left\langle \widehat{S}_{y}\right\rangle $ to be
non-zero. This is the \emph{same} as the previous \emph{Bloch vector} test
in Eq.(\ref{Eq.BlochVectorEntTest}) or the weak correlation test in Eq.(\ref%
{Eq.WeakCorrelTest}). \pagebreak

\subsection{Two Mode Quadrature Squeezing}

From Eq. (\ref{Eq.QuadOprs}) we can define \emph{two mode quadrature
operators} as%
\begin{eqnarray}
\widehat{X}_{\theta }(+) &=&\frac{1}{\sqrt{2}}(\widehat{X}_{a}^{\theta }+%
\widehat{X}_{b}^{\theta })=\frac{1}{2}(\widehat{a}\exp (-i\theta )+\widehat{b%
}^{\dag }\exp (+i\theta )+\widehat{a}^{\dag }\exp (+i\theta )+\widehat{b}%
\exp (-i\theta ))  \nonumber \\
\widehat{P}_{\theta }(+) &=&\frac{1}{\sqrt{2}}(\widehat{P}_{a}^{\theta }+%
\widehat{P}_{b}^{\theta })=\frac{1}{2i}(\widehat{a}\exp (-i\theta )-\widehat{%
b}^{\dag }\exp (+i\theta )-\widehat{a}^{\dag }\exp (+i\theta )+\widehat{b}%
\exp (-i\theta ))  \nonumber \\
&=&\widehat{X}_{\theta +\pi /2}(+)  \nonumber \\
&&  \label{Eq.TwoModeQuadOprs}
\end{eqnarray}%
where we have $[$ $\widehat{X}_{\theta }(+),\widehat{P}_{\theta }(+)]=i$.
Note that $\widehat{X}_{0}(+)=(\widehat{x}_{A}+\widehat{x}_{B})/\sqrt{2}$
and $\widehat{P}_{0}(+)=(\widehat{p}_{A}+\widehat{p}_{B})/\sqrt{2}$ unlike
the operators considered in Section \ref{SubSubSection - Duan}. As we have
seen there is no entanglement test for systems of identical bosons of the
form\textbf{\ }$\left\langle \Delta (\widehat{x}_{A}+\widehat{x}%
_{B})^{2}\right\rangle +\left\langle \Delta (\widehat{p}_{A}-\widehat{p}%
_{B})^{2}\right\rangle <2+2\left\langle \widehat{N}\right\rangle $\textbf{. }%
The Heisenberg uncertainty principle gives $\left\langle \Delta \widehat{X}%
_{\theta }^{2}(+)\right\rangle \left\langle \Delta \widehat{P}_{\theta
}^{2}(+)\right\rangle \geq 1/4$, so a state is squeezed in $\widehat{X}%
_{\theta }(+)$ if $\left\langle \Delta \widehat{X}_{\theta
}^{2}(+)\right\rangle <1/2$, and similarly for squeezing in $\widehat{P}%
_{\theta }(+)$.

We can show that for separable states both $\left\langle \Delta \widehat{X}%
_{\theta }^{2}(+)\right\rangle \geq 1/2$ and $\left\langle \Delta \widehat{P}%
_{\theta }^{2}(+)\right\rangle \geq 1/2$, so two mode quadrature squeezing
in either $\widehat{X}_{\theta }(+)$ or $\widehat{P}_{\theta }(+)$ is a test
for two mode entanglement. Firstly, for SSR compliant sub-system states 
\begin{equation}
\left\langle \widehat{X}_{\theta }(+)\right\rangle =\frac{1}{\sqrt{2}}%
\sum_{R}P_{R}(\left\langle \widehat{X}_{a}^{\theta }\right\rangle
_{R}+\left\langle \widehat{X}_{b}^{\theta }\right\rangle _{R})=0
\label{Eq.MeanTwoModeQuadOpr}
\end{equation}%
since$\left\langle \widehat{a}\right\rangle _{R}=\left\langle \widehat{b}%
\right\rangle _{R}=0$. Secondly,%
\begin{eqnarray}
&&\left\langle \widehat{X}_{\theta }^{2}(+)\right\rangle  \nonumber \\
&=&\frac{1}{4}\sum_{R}P_{R}(\left\langle \widehat{a}\widehat{a}^{\dag
}\right\rangle _{R}+\left\langle \widehat{a}^{\dag }\widehat{a}\right\rangle
_{R}+\left\langle \widehat{b}\widehat{b}^{\dag }\right\rangle
_{R}+\left\langle \widehat{b}^{\dag }\widehat{b}\right\rangle _{R}) 
\nonumber \\
&=&\sum_{R}P_{R}(\frac{1}{2}+\frac{1}{2}(\left\langle \widehat{n}%
_{a}\right\rangle _{R}+\left\langle \widehat{n}_{b}\right\rangle _{R}) 
\nonumber \\
&=&\frac{1}{2}+\frac{1}{2}\left\langle \widehat{N}\right\rangle  \nonumber \\
&\geq &\frac{1}{2}
\end{eqnarray}%
where for local SSR compliant states other terms involving $\left\langle 
\widehat{a}^{2}\right\rangle _{R},\left\langle \widehat{b}^{2}\right\rangle
_{R},\left\langle \widehat{a}\widehat{b}\right\rangle _{R}=\left\langle 
\widehat{a}\right\rangle _{R}\left\langle \widehat{b}\right\rangle
_{R},\left\langle \widehat{a}\widehat{b}^{\dag }\right\rangle
_{R}=\left\langle \widehat{a}\right\rangle _{R}\left\langle \widehat{b}%
^{\dag }\right\rangle _{R}$ etc. are all zero. Hence%
\begin{equation}
\left\langle \Delta \widehat{X}_{\theta }^{2}(+)\right\rangle =\left\langle 
\widehat{X}_{\theta }^{2}(+)\right\rangle -\left\langle \widehat{X}_{\theta
}(+)\right\rangle ^{2}\geq \frac{1}{2}  \label{Eq.VarTwoModeQuadOpr}
\end{equation}%
which establishes the result. Since $\widehat{P}_{\theta }(+)=\widehat{X}%
_{\theta +\pi /2}(+)$ we also have $\left\langle \Delta \widehat{P}_{\theta
}^{2}(+)\right\rangle =\frac{1}{2}+\frac{1}{2}\left\langle \widehat{N}%
\right\rangle \geq \frac{1}{2}$ for a separable state. Hence the\textbf{\ }%
\emph{two mode quadrature squeezing}\textbf{\ }test. If%
\begin{equation}
\left\langle \Delta \widehat{X}_{\theta }^{2}(+)\right\rangle <\frac{1}{2}%
\qquad or\qquad \left\langle \Delta \widehat{P}_{\theta
}^{2}(+)\right\rangle <\frac{1}{2}  \label{Eq.TwoModeQuadSqgTest}
\end{equation}%
then the state is entangled. Obviously $\widehat{X}_{\theta }(+)$ and $%
\widehat{P}_{\theta }(+)$ cannot both be squeezed for the same state.

We can also define additional two mode quadrature operators as%
\begin{eqnarray}
\widehat{X}_{\theta }(-) &=&\frac{1}{\sqrt{2}}(\widehat{X}_{a}^{\theta }-%
\widehat{X}_{b}^{\theta })=\frac{1}{2}(\widehat{a}\exp (-i\theta )-\widehat{b%
}^{\dag }\exp (+i\theta )+\widehat{a}^{\dag }\exp (+i\theta )-\widehat{b}%
\exp (-i\theta ))  \nonumber \\
\widehat{P}_{\theta }(-) &=&\frac{1}{\sqrt{2}}(\widehat{P}_{a}^{\theta }-%
\widehat{P}_{b}^{\theta })=\frac{1}{2i}(\widehat{a}\exp (-i\theta )+\widehat{%
b}^{\dag }\exp (+i\theta )-\widehat{a}^{\dag }\exp (+i\theta )-\widehat{b}%
\exp (-i\theta ))  \nonumber \\
&=&\widehat{X}_{\theta +\pi /2}(-)  \nonumber \\
&&  \label{Eq.TwoModeQuadOprsB}
\end{eqnarray}%
where we also have $[$\ $\widehat{X}_{\theta }(-),\widehat{P}_{\theta
}(-)]=i $. Again $\left\langle \Delta \widehat{X}_{\theta
}^{2}(-)\right\rangle \left\langle \Delta \widehat{P}_{\theta
}^{2}(-)\right\rangle \geq 1/4$, so a state is squeezed in $\widehat{X}%
_{\theta }(-)$\ if $\left\langle \Delta \widehat{X}_{\theta
}^{2}(-)\right\rangle <1/2$, and similarly for squeezing in $\widehat{P}%
_{\theta }(-)$.

A similar proof shows that for separable states both $\left\langle \Delta 
\widehat{X}_{\theta }^{2}(-)\right\rangle =\frac{1}{2}+\frac{1}{2}%
\left\langle \widehat{N}\right\rangle \geq 1/2$\ and $\left\langle \Delta 
\widehat{P}_{\theta }^{2}(-)\right\rangle =\frac{1}{2}+\frac{1}{2}%
\left\langle \widehat{N}\right\rangle \geq 1/2$, so if%
\begin{equation}
\left\langle \Delta \widehat{X}_{\theta }^{2}(-)\right\rangle <\frac{1}{2}%
\qquad or\qquad \left\langle \Delta \widehat{P}_{\theta
}^{2}(-)\right\rangle <\frac{1}{2}  \label{Eq.TwoModeQuadSqgTestB}
\end{equation}%
then the state is entangled. Hence\textbf{\ }\emph{any}\textbf{\ }one of $%
\widehat{X}_{\theta }(+),\widehat{P}_{\theta }(+),\widehat{X}_{\theta }(-),%
\widehat{P}_{\theta }(-)$\ being squeezed will demonstate two mode
entanglement.

The question then arises - Can two of the four two mode quadrature operators
be squeezed? For simplicity we only discuss $\theta =0$ cases in detail.
Obviously pairs such as $\widehat{X}_{0}(+),\widehat{P}_{0}(+)$ or $\widehat{%
X}_{0}(-),\widehat{P}_{0}(-)$ cannot. Next, we consider $\widehat{X}_{0}(+)\ 
$and $\widehat{P}_{0}(-)$. We note that for all global SSR compliant states $%
\left\langle \Delta \widehat{X}_{0}^{2}(+)\right\rangle +\left\langle \Delta 
\widehat{P}_{0}^{2}(-)\right\rangle =\frac{1}{2}\left( \left\langle \Delta (%
\widehat{x}_{A}+\widehat{x}_{B})^{2}\right\rangle +\left\langle \Delta (%
\widehat{p}_{A}-\widehat{p}_{B})^{2}\right\rangle \right) =1+\left\langle 
\widehat{N}\right\rangle $ using (\ref{Eq.QuadVarEqualitysSSRSepState}), so
that if $\widehat{X}_{0}(+)$ is squeezed $\left\langle \Delta \widehat{X}%
_{0}^{2}(+)\right\rangle <\frac{1}{2}$ then $\left\langle \Delta \widehat{P}%
_{0}^{2}(-)\right\rangle >\frac{1}{2}+\left\langle \widehat{N}\right\rangle $%
, showing that both $\widehat{X}_{0}(+)\ $and $\widehat{P}_{0}(-)\ $cannot
both be squeezed - in spite of the operators commuting. A similar conclusion
applies to $\widehat{X}_{0}(-)\ $and $\widehat{P}_{0}(+)$. For the pair $%
\widehat{X}_{0}(+)$\ and $\widehat{X}_{0}(-)$\ we have $\left\langle \Delta 
\widehat{X}_{0}^{2}(+)\right\rangle +\left\langle \Delta \widehat{X}%
_{0}^{2}(-)\right\rangle =\frac{1}{2}\left( \left\langle \Delta (\widehat{x}%
_{A}+\widehat{x}_{B})^{2}\right\rangle +\left\langle \Delta (\widehat{x}_{A}-%
\widehat{x}_{B})^{2}\right\rangle \right) =1+\left\langle \widehat{N}%
\right\rangle $ using (\ref{Eq.VarSumXGlobSSRState}) and (\ref%
{Eq.VarDiffXGlobSSRState}), so the same situation as for $\widehat{X}%
_{0}(+)\ $and $\widehat{P}_{0}(-)$ applies, and thus $\widehat{X}_{0}(+)$\
and $\widehat{X}_{0}(-)$\ cannot both be squeezed. A similar conclusion
applies to $\widehat{P}_{0}(-)\ $and $\widehat{P}_{0}(+)$. In general, only%
\textbf{\ }\emph{one}\textbf{\ }of $\widehat{X}_{\theta }(+),\widehat{P}%
_{\theta }(+),\widehat{X}_{\theta }(-),\widehat{P}_{\theta }(-)$ can be
squeezed.

Further questions are: What quantities need to be measured in order to test
whether two mode quadrature squeezing occurs and how useful would it be to
detect entanglement? It is straight-forward to show from (\ref%
{Eq.TwoModeQuadOprs}) and (\ref{Eq.TwoModeQuadOprsB}) that for states that
are global SSR compliant%
\begin{eqnarray}
\left\langle \widehat{X}_{\theta }(+)\right\rangle  &=&0\qquad \left\langle 
\widehat{X}_{\theta }(-)\right\rangle =0  \label{Eq.MeanTwoModeQuadSSRState}
\\
\left\langle \Delta \widehat{X}_{\theta }^{2}(+)\right\rangle 
&=&\left\langle \widehat{X}_{\theta }^{2}(+)\right\rangle =\frac{1}{2}\left(
\left\langle \widehat{N}\right\rangle +2\left\langle \widehat{S}%
_{x}\right\rangle \right)   \nonumber \\
\left\langle \Delta \widehat{X}_{\theta }^{2}(-)\right\rangle 
&=&\left\langle \widehat{X}_{\theta }^{2}(-)\right\rangle =\frac{1}{2}\left(
\left\langle \widehat{N}\right\rangle -2\left\langle \widehat{S}%
_{x}\right\rangle \right)   \label{Eq.VarTwoModeQuadSSRState}
\end{eqnarray}%
since terms such as $\left\langle \widehat{a}^{2}\right\rangle $, $%
\left\langle \widehat{a}\widehat{b}\right\rangle $ etc are all zero for SSR
compliant states. As explained in Section \ref{SubSection - Var Sz Two Mode
SubSystems}, both $\left\langle \widehat{N}\right\rangle +2\left\langle 
\widehat{S}_{x}\right\rangle $ and $\left\langle \widehat{N}\right\rangle
-2\left\langle \widehat{S}_{x}\right\rangle $ are always non-negative, but
the entanglement test would require 
\begin{eqnarray}
\left\langle \widehat{N}\right\rangle +2\left\langle \widehat{S}%
_{x}\right\rangle  &<&1\qquad for\;squeezing\;in\;\widehat{X}_{\theta }(+) 
\nonumber \\
\left\langle \widehat{N}\right\rangle -2\left\langle \widehat{S}%
_{x}\right\rangle  &<&1\qquad for\;squeezing\;in\;\widehat{X}_{\theta }(-)
\label{Eq.TwoModeQuadSqgTests}
\end{eqnarray}%
This shows that the two mode quadrature squeezing test involves measuring $%
\left\langle \widehat{N}\right\rangle $ and $\left\langle \widehat{S}%
_{x}\right\rangle $, so that once again measurements of boson number and the
mean value of a spin operator are involved. Similar conclusions apply for $%
\widehat{P}_{\theta }(+)$ and $\widehat{P}_{\theta }(-)$. However, since the
test requires $\left\langle \widehat{S}_{x}\right\rangle $ to be non-zero it
would simpler to use the \emph{Bloch vector} test (see (\ref%
{Eq.BlochVectorEntTest})) which merely requires showing that one of $%
\left\langle \widehat{S}_{x}\right\rangle $ or $\left\langle \widehat{S}%
_{y}\right\rangle $ to be non-zero. 

In most cases the inequalities in (\ref{Eq.TwoModeQuadSqgTests}) will not be
satisfied, since both $\left\langle \widehat{N}\right\rangle $ and $%
\left\langle \widehat{S}_{x}\right\rangle $ are $O(N)$. However, for the
binomial state in (\ref{Eq.BinomialState}) with $\theta =3\pi /4$ and $\chi
=0$ we have for $\left\vert \Phi \right\rangle =\left( \frac{-\widehat{a}%
^{\dag }+\widehat{b}^{\dag }}{\sqrt{2}}\right) ^{N}\left\vert 0\right\rangle
/\sqrt{N!}$\ the results $\left\langle \widehat{N}\right\rangle =N$ and $%
\left\langle \widehat{S}_{x}\right\rangle =-N/2$\ (see (163) in Ref. \cite%
{Dalton12a}). Hence spin squeezing in $\widehat{X}_{\theta }(+)$ occurs,
confirming that this particular binomial state is entangled. Note that the
test does not confirm entanglement for almost all other binomial states
(those where $\left\langle \widehat{S}_{x}\right\rangle $\ is different from 
$\pm N/2$), though these are actually entangled. \textbf{\pagebreak }

\section{Interferometry in Bosonic Systems}

\label{section - Interferometry in Two Mode BEC}

In this section we discuss how interferometers in two mode bosonic systems
operate. This topic has of course been discussed many times before, but for
completeness we present it here. Our approach is essentially the same as in
earlier papers, for example that of Yurke et al \cite{Yurke86a}. Before
discussing interferometry in two mode bosonic systems, we first need to set
out the general Hamiltonian for the two mode systems that could be of
interest. The two modes may be associated with two distinct single boson 
\emph{spatial} states, such as in a double well potential in which case the
coupling between the two modes is associated with \emph{quantum tunneling}.
Or they may be associated with two different atomic internal\emph{\ hyperfine%
} states in a single well, which may be coupled via \emph{classical fields}
in the form of very short pulses, for which the time dependent amplitude is$%
\mathcal{A(}t)$, the centre frequency is $\omega _{0}$ and the \emph{phase}
is $\phi $. Since this coupling process is much easier to control than
quantum tunneling, we will mainly focus on the case of two modes associated
with different hyperfine states, though the approach might also be applied
to the case of two spatial modes. The free atoms occupying the two modes are
associated with energies $\hbar \omega _{a}$, $\hbar \omega _{b}$, the \emph{%
transition frequency} $\omega _{ba}=\omega _{b}-\omega _{a}$ being close to 
\emph{resonance} with $\omega _{0}$. It is envisaged that a large \emph{%
number} $N$ of \emph{bosonic atoms} occupy the two modes. The bosonic atoms
may also interact with each other via \emph{spin conserving}, \emph{zero
range} interatomic potentials. We will show that measurements on the mean
and variance for the \emph{population difference} determine the \emph{mean
values} and \emph{covariance matrix} for the spin operators involved in 
\emph{entanglement} tests.

For interferometry involving \emph{multi-mode} systems, a straightforward
generalisation of the two mode case is possible, based on the reasonable
assumption the interferometer process couples the modes in a pairwise
manner. This is based on the operation of \emph{selection rules}, as will be
explained below.

However, although in the present section we show that two mode
interferometers can be used to measure the\emph{\ mean values}\textbf{\ }and%
\textbf{\ }\emph{covariance matrix}\textbf{\ }for the\textbf{\ }\emph{spin}%
\textbf{\ }operators involved in\textbf{\ }\emph{entanglement}\textbf{\ }%
tests for systems of \emph{massive}\textbf{\ }bosons, the issue of how to
measure\textbf{\ }\emph{mean values}\textbf{\ }and\textbf{\ }\emph{variances}%
\textbf{\ }for the\textbf{\ \emph{quadrature} }operators involved in other
entanglement tests for massive bosons is still to be established. Such a
measurement is not possible with the beam splitter interferometer discussed
in this paper. An interferometer involving some sort of\emph{\ phase
reference}\textbf{\ }would seem to be needed. Proposals based on\textbf{\ }%
\emph{homodyne}\textbf{\ }measurements have been made by Olsen et al \cite%
{Ferris08a}, \cite{Chiaca15a}, but these are based on hypothetical reference
systems with large numbers of bosons in\textbf{\ }\emph{Glauber coherent}%
\textbf{\ }states, and such states are prohibited via the global particle
number SSR.

\subsection{Simple Two Mode Interferometer}

\label{SubSection - Simple Model}

A simple description of the two mode system is provided by the\emph{\
Josephson model}, where the overall Hamiltonian is of the form \cite%
{Dalton12a}%
\begin{equation}
\widehat{H}_{Joseph}=\widehat{H}_{0}+\widehat{V}+\widehat{V}_{\func{col}}
\label{Eq.JosephHam}
\end{equation}%
with 
\begin{eqnarray}
\widehat{H}_{0} &=&\hbar \omega _{a}\,\widehat{a}^{\dag }\widehat{a}+\hbar
\omega _{b}\,\widehat{b}^{\dag }\widehat{b}  \nonumber \\
\widehat{V} &=&\mathcal{A(}t)\exp (-i\omega _{0}t)\,\exp (i\phi )\,\widehat{b%
}^{\dag }\widehat{a}+\mathcal{A(}t)\exp (+i\omega _{0}t)\,\exp (-i\phi )\,%
\widehat{a}^{\dag }\widehat{b}  \nonumber \\
\widehat{V}_{\func{col}} &=&\chi (\widehat{b}^{\dag }\widehat{b}-\widehat{a}%
^{\dag }\widehat{a})^{2}  \label{Eq.JosephHamTerms}
\end{eqnarray}%
where $\widehat{H}_{0}$ is the free boson Hamiltonian, $\widehat{V}$ gives
the interaction with the classical field and $\widehat{V}_{\func{col}}$ is
the collisional interaction term. For the case of quantum tunneling between
two distinct \emph{spatial} modes, the interaction term $\widehat{V}$ can
also be described in the Josephson model (see \cite{Dalton12a} for details),
in which case the factors multiplying $\widehat{b}^{\dag }\widehat{a}$ or $%
\widehat{a}^{\dag }\widehat{b}$ involve the \emph{trapping potential} and
the two spatial mode functions. A time dependent amplitude and phase might
be obtained via adding a suitable time dependent field to the trapping
potential - this would be experimentally difficult. The Hamiltonian can also
be written in terms of spin operators as 
\begin{eqnarray}
\widehat{H}_{0} &=&1/2(\hbar \omega _{a}+\hbar \omega _{b})\widehat{N}-\hbar
\omega _{ab}\widehat{S}_{z}  \nonumber \\
\widehat{V} &=&\mathcal{A(}t)\exp (-i\omega _{0}t)\,\exp (i\phi )\,(\widehat{%
S}_{x}+i\widehat{S}_{y})+H.C  \nonumber \\
\widehat{V}_{\func{col}} &=&4\chi \widehat{S}_{z}^{2}  \label{Eq.JHSpinOprs}
\end{eqnarray}

The coupling effect in a \emph{simple two mode interferometer} can be
described via the classical interaction term $\widehat{V}$, where now the
amplitude $\mathcal{A(}t)$ is only non-zero over a short time interval. The
pulsed classical field is associated with an\emph{\ area variable} $s$,
defined by 
\begin{equation}
s=\dint_{t_{0}}^{t}dt\,_{1}\mathcal{A(}t_{1})/\hbar  \label{Eq.PulseArea}
\end{equation}%
the integral eventually being over the pulse's duration. The \emph{%
interferometer frequency} $\omega _{0}$ is assumed for simplicity to be in 
\emph{resonance} with the \emph{transition frequency} $\omega _{ba}=\omega
_{b}-\omega _{a}$. The classical field is also associated with a phase $\phi 
$, so the simple two mode interferometer is described by two \emph{%
interferometric variables} $2s=\theta $ giving the pulse area and $\phi $
specifying the phase. Changing these variables leads to a range of differing
applications of the interferometer. When acting as a \emph{beam splitter}
(BS)\ a $2s=\pi /2$ pulse is involved and $\phi $ is variable, but for a 
\emph{phase changer} a $2s=\pi $ pulse is involved ($\phi $ is arbitrary).
For \emph{state tomography} in the $yz$ plane we choose $2s=\theta $
(variable) and $\phi =0$ or $\pi $. The beam splitter enables state
tomography in the $xy$ plane to be carried out. Generally speaking the
effect of \emph{collisions} can be \emph{ignored} during the short classical
pulse and we will do so here.

\subsection{General Two Mode Interferometers}

More complex two mode bosonic interferometers applied to a specific input
quantum state will involve specific \emph{arrangements} of simple two mode
interferometers such as beam splitters, phase changers and free evolution
intervals, followed by final measurement of the mean population diference
between the modes and its variance. \emph{Ramsey interferometry} involves%
\emph{\ two} beam splitters separated by a controllable free evolution \emph{%
time interval }$T$. During such an interval in which free evolution occurs,
the interaction of the classical beam splitter field with the two mode
system can be ignored, but the effect of collisions and coupling to external
systems may be important if collision parameters are to be measured using
the interferometer. The overall behaviour of such multi-element
interferometers will also depend on the initial two mode quantum\emph{\
input }state that acts as the\emph{\ input state} for the interferometer, as
well as important variables such as the phase $\phi $, the centre frequency $%
\omega _{0}$, the area variable $s$ for the classical pulses used, and also
the the free evolution intervals (if any). The behaviour also will depend on
the characteristic parameters such as the transition frequency $\omega _{ab}$%
, collision parameter $\chi $ and total boson number $N$ for the two mode
system used in the interferometry. The variables that describe the
interaction with other systems whose properties are to be measured using the
interferometer must also affect its behaviour if the interferometer is to be
useful. Finally, a choice must be made for the interferometer physical
quantity whose \emph{mean value} and \emph{quantum fluctuation} is to be
measured - referred to as the\emph{\ measurable}. The outcome of such
measurements can be studied as a function of one or more of the variables on
which the interferometer behaviour depends - referred to as the \emph{%
interferometric variable}. There are obviously a wide range of possible two
mode\emph{\ interferometers types} that could be studied, depending on the
application envisaged. Interferometers also have a wide range of \emph{uses}%
, including determining the properties of the input two mode state - such as
squeezing or entanglement. For a suitable known input state they can be used
to measure interferometric variables - such as the classical phase $\phi $
of the pulsed field associated with a beam splitter or a parameter
associated with an external system coupled to the interferometer. On the
other hand, in a Ramsey interferometer the interferometric could be the
collision parameter $\chi $, obtainable if the free evolution period $T$ is
known. No attempt to be comprehensive will be made here.

The \emph{Ramsey interferometer} is illustrated in Figure 7.

\includegraphics[width=\textwidth]{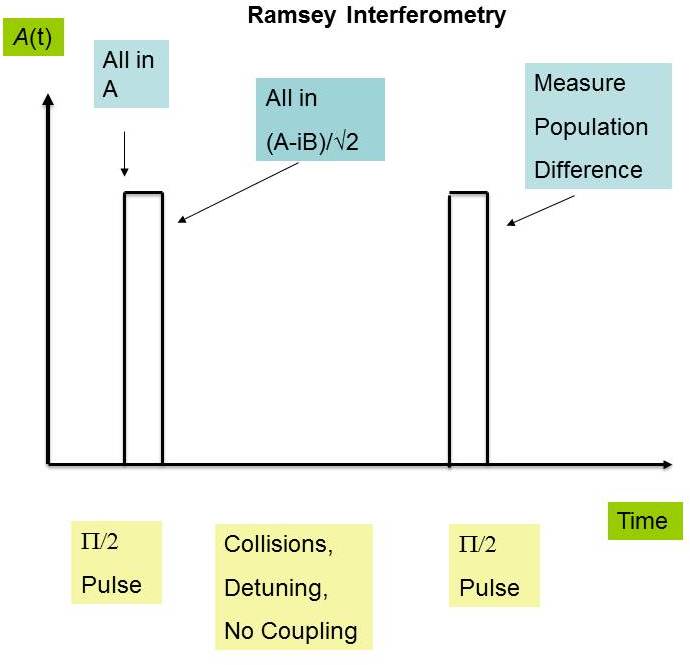}


\medskip

\begin{center}
Figure 7. Ramsey Interferometry. Two $\pi /2$ beam splitters separated by a
free evolution region.

\medskip
\end{center}

\pagebreak

For the purpose of considering entanglement tests a \emph{simple two mode
interferometer} operating under conditions of exact \emph{resonance} $\omega
_{0}=$ $\omega _{ab}$ will be treated, and its behaviour for $N$ large when
the phase $\phi $ is changed and for different choices of the input state $%
\widehat{\rho }$ will be examined. Measurements appropriate to detecting%
\emph{\ }entanglement via \emph{spin squeezing} and \emph{correlation} will
be discussed. The measurable chosen will initially be half the population
difference $(\widehat{b}^{\dag }\widehat{b}-\widehat{a}^{\dag }\widehat{a}%
)/2 $ - which equals $\widehat{S}_{z}$ - generally measured after the two
mode system has interacted with the simple interferometer, but also without
this interaction. The phase $\phi $ will act as the interferometric
variable, as will the pulse area $2s=\theta $. As we will see, different
choices of input state ranging from \emph{separable} to \emph{entangled}
states lead to markedly different behaviours. In particular, the behaviour
of \emph{relative phase eigenstates} as input states will be examined. Later
we will also consider measurements involving the square of $\widehat{S}_{z}$.

\subsection{Measurements in Simple Two Mode Interferometer}

\label{SubSection - Simple Two Mode Interferometer}

As discussed in the previous paragraph, the initial choice of \emph{%
measurable} is 
\begin{equation}
\widehat{M}=\frac{1}{2}(\widehat{b}^{\dag }\widehat{b}-\widehat{a}^{\dag }%
\widehat{a})=\widehat{S}_{z}  \label{Eq.MeasurableSz}
\end{equation}%
and we will determine its mean and variance for the state $\widehat{\rho }%
^{\#}$ given by 
\begin{equation}
\widehat{\rho }^{\#}=\widehat{U}\,\widehat{\rho }\,\widehat{U}^{-1}
\label{Eq.EvolvedState}
\end{equation}%
where the \emph{output state} $\widehat{\rho }^{\#}$ has evolved from the
initial \emph{input state} $\widehat{\rho }$ due to the effect of the \emph{%
simple two mode interferometer}. $\widehat{U}$ is the unitary \emph{%
evolution operator} describing evolution during the time the short classical
pulse is applied. Collision terms and interactions with other systems will
be ignored during the short time interval involved.

We note that for an $N$ boson state the eigenvalues of $\widehat{M}$ range
from $-N/2$ to $+N/2$ in integer steps. For more general states the possible
values for $\widehat{M}$ are any integer or half integer. When $\widehat{M}$
is measured the result will be one of these eigenvalues, but the average of
repeated measurements will be $\left\langle \widehat{M}\right\rangle $ which
must also lie in the range $-N/2$ to $+N/2$. The variance of the results for
the repeated measurements of $\widehat{M}$ is also experimentally
measureable and will not exceed $(N/2)^{2}$, and apart from $NOON$ states
will be much less than this. The experimentally determinable results for
both $\left\langle \widehat{M}\right\rangle $ and $\left\langle \Delta 
\widehat{M}^{2}\right\rangle $ will depend on $\widehat{\rho }\,$\ and on
the interferometer variables such as the phase $\phi $ and the pulse area $%
2s=\theta $.

The Hamiltonian governing the evolution in the simple two mode
interferometer\ will be $\widehat{H}_{0}+\widehat{V}$. For the \emph{output
state} the mean value and variance are 
\begin{eqnarray}
\left\langle \widehat{M}\right\rangle &=&Tr(\widehat{M}\,\widehat{\rho }%
^{\#})  \nonumber \\
\left\langle \Delta \widehat{M}^{2}\right\rangle &=&Tr(\left\{ \widehat{M}%
-\left\langle \widehat{M}\right\rangle \right\} ^{2}\,\widehat{\rho }^{\#})
\label{Eq.OutputStateMeanVariance}
\end{eqnarray}%
These will be evaluated at the end of the pulse. If the input state is
measured\emph{\ directly} without applying the interferometer, then the mean
value and variance are as in the last equations but with $\widehat{\rho }%
^{\#}$ replaced by $\widehat{\rho }$.

The derivation of the results is set out in Appendix \ref{Appendix -
Derivatio n of Interferometer Result} and are given by the same form as (\ref%
{Eq.OutputStateMeanVariance}), but with $\widehat{\rho }^{\#}$ replaced by $%
\widehat{\rho }$ and with $\widehat{M}$ replaced by the interaction picture
Heisenberg operator $\widehat{M}_{H}(2s,\phi )$ at the end of the pulse,
which is given by 
\begin{eqnarray}
\widehat{M}_{H}(2s,\phi ) &=&\frac{1}{2}(\widehat{b}_{H}^{\dag }(s,\phi )%
\widehat{b}_{H}(s,\phi )-\widehat{a}_{H}^{\dag }(s,\phi )\widehat{a}%
_{H}(s,\phi ))  \nonumber \\
&=&\sin 2s\,(\sin \phi \,\widehat{S}_{x}+\cos \phi \,\widehat{S}_{y})+\cos
2s\,\widehat{S}_{z}  \label{Eq.GeneralResultMeasurableHeisenberg}
\end{eqnarray}%
with 
\begin{equation}
\widehat{b}_{H}(s,\phi )=\cos s\,\widehat{b}-i\exp (i\phi )\,\sin s\,%
\widehat{a}\qquad \widehat{a}_{H}(s,\phi )=-i\exp (-i\phi )\,\sin s\,%
\widehat{b}+\,\cos s\,\widehat{a}  \label{Eq.GeneralResultModeOprsHeisenberg}
\end{equation}%
The versatility of the measurement follows from the range of possible
choices of the pulse area $2s=\theta $ and the phase $\phi $. These results
are valid for both \emph{bosonic} and \emph{fermionic} modes.

We then find that the \emph{general result} for the \emph{mean} value is 
\begin{equation}
\left\langle \widehat{M}\right\rangle =\sin \theta \,\sin \phi
\,\left\langle \widehat{S}_{x}\right\rangle _{\rho }+\sin \theta \,\cos \phi
\,\left\langle \widehat{S}_{y}\right\rangle _{\rho }+\cos \theta
\,\left\langle \widehat{S}_{z}\right\rangle _{\rho }
\label{Eq.GeneralResultMean}
\end{equation}%
and for the\emph{\ variance} is 
\begin{eqnarray}
&&\left\langle \Delta \widehat{M}^{2}\right\rangle  \nonumber \\
&=&\frac{(1-\cos 2\theta )}{2}\frac{(1-\cos 2\phi )}{2}C(\widehat{S}_{x},%
\widehat{S}_{x})+\frac{(1-\cos 2\theta )}{2}\frac{(1+\cos 2\phi )}{2}C(%
\widehat{S}_{y},\widehat{S}_{y})  \nonumber \\
&&+\frac{(1+\cos 2\theta )}{2}C(\widehat{S}_{z},\widehat{S}_{z})  \nonumber
\\
&&+\frac{(1-\cos 2\theta )}{2}\sin 2\phi \,C(\widehat{S}_{x},\widehat{S}%
_{y})+\sin 2\theta \,\cos \phi \,C(\widehat{S}_{y},\widehat{S}_{z})+\sin
2\theta \,\sin \phi \,C(\widehat{S}_{z},\widehat{S}_{x})  \nonumber \\
&&  \label{Eq.GeneralResultVariance}
\end{eqnarray}%
where the \emph{mean value} of the spin operators are $\left\langle \widehat{%
S}_{\alpha }\right\rangle _{\rho }=Tr(\widehat{S}_{\alpha }\,\widehat{\rho }%
) $ and the \emph{covariance matrix} elements are given by $C(\widehat{S}%
_{\alpha },\widehat{S}_{\beta })=1/2\left\langle (\widehat{S}_{\alpha }%
\widehat{S}_{\beta }+\widehat{S}_{\beta }\widehat{S}_{\alpha })\right\rangle
_{\rho }-\left\langle \widehat{S}_{\alpha }\right\rangle _{\rho
}\left\langle \widehat{S}_{\beta }\right\rangle _{\rho }$. The diagonal
elements $C(\widehat{S}_{\alpha },\widehat{S}_{\alpha })=\left\langle 
\widehat{S}_{\alpha }{}^{2}\right\rangle _{\rho }-\left\langle \widehat{S}%
_{\alpha }\right\rangle _{\rho }^{2}=\left\langle \Delta \widehat{S}_{\alpha
}^{2}\right\rangle $ is the variance. By making appropriate choises of the
interferometer variables $\theta $ (half the the pulse area) and $\phi $
(the phase) the mean values of all the spin operators and all elements of
the covariance matrrix can be measured. \emph{Tomography} for the spin
operators in any selected plane can be carried out.

\subsubsection{Tomography in xy Plane - Beam Splitter}

In the \emph{beam splitter case} (for state tomography in the $xy$ plane) we
choose $2s=\pi /2$ and $\phi $ (variable) giving 
\begin{equation}
\widehat{M}_{H}(\frac{\pi }{2},\phi )=\sin \phi \,\widehat{S}_{x}+\cos \phi
\,\widehat{S}_{y}  \label{Eq.BeamSplitterHeis}
\end{equation}%
and we find that for the output state of the \emph{BS\ interferometer} the
mean value and variance of $\widehat{M}$ are given by 
\begin{eqnarray}
\left\langle \widehat{M}\right\rangle &=&\sin \phi \,\left\langle \widehat{S}%
_{x}\right\rangle _{\rho }+\cos \phi \,\left\langle \widehat{S}%
_{y}\right\rangle _{\rho }  \label{Eq.FormOutputMean} \\
\left\langle \Delta \widehat{M}^{2}\right\rangle &=&\frac{1}{2}(1-\cos 2\phi
)\,C(\widehat{S}_{x},\widehat{S}_{x})+\frac{1}{2}(1+\cos 2\phi )\,C(\widehat{%
S}_{y},\widehat{S}_{y})+\sin 2\phi \,\,C(\widehat{S}_{x},\widehat{S}_{y}) 
\nonumber \\
&&  \label{Eq.FormOutputVariance}
\end{eqnarray}%
showing the mean value for the measureable $\widehat{M}$ depends
sinusoidally on the phase $\phi $ and the \emph{mean values} of the spin
operators $\widehat{S}_{x},\widehat{S}_{y}$. The variance for the measurable
depends sinusoidally on $2\phi $ and on the \emph{covariance matrix} of the
same spin operators. Both the means and covariances of the spin operators $%
\widehat{S}_{x},\widehat{S}_{y}$ now depend on the input state $\widehat{%
\rho }$ for the interferometer rather than the output state $\widehat{\rho }%
^{\#}$.

\subsubsection{Tomography in yz Plane}

For state tomography in the $yz$ plane we obtain the means and covariances
of the spin operators $\widehat{S}_{y},\widehat{S}_{z}$. To do this we
choose $2s=\theta $ (variable) and $\phi =0$ so that 
\begin{equation}
\widehat{M}_{H}(\theta ,0)=\sin \theta \,\widehat{S}_{y}+\cos \theta \,%
\widehat{S}_{z}  \label{Eq.TomogYZHeis}
\end{equation}%
and find that for the output state the mean value and variance of $\widehat{M%
}$ are given by%
\begin{eqnarray}
\left\langle \widehat{M}\right\rangle &=&\sin \theta \,\left\langle \widehat{%
S}_{y}\right\rangle _{\rho }+\cos \theta \,\left\langle \widehat{S}%
_{z}\right\rangle _{\rho }  \label{Eq.FormOutputMeanYZTomog} \\
\left\langle \Delta \widehat{M}^{2}\right\rangle &=&\frac{1}{2}(1-\cos
2\theta )\,C(\widehat{S}_{y},\widehat{S}_{y})+\frac{1}{2}(1+\cos 2\theta
)\,C(\widehat{S}_{z},\widehat{S}_{z})+\sin 2\theta \,\,C(\widehat{S}_{y},%
\widehat{S}_{z})  \nonumber \\
&&  \label{Eq.formOutputVarianceYZTomog}
\end{eqnarray}

A \emph{single} measurement does not of course determine the mean value $%
\left\langle \widehat{M}\right\rangle $. An average over a large number of 
\emph{independent repetitions} of the measurement is needed to accurately
determine $\left\langle \widehat{M}\right\rangle $ which can then be
compared to theoretical predictions. This is a well-known practical issue
for the experimenter that we need not dwell on here. A brief account of the
issues involved is included in Appendix \ref{Appendix - Limits on
Interferometry Tests}.

\subsubsection{Phase Changer}

In this case we choose $2s=\theta =\pi $ and $\phi $ (arbitrary) giving%
\begin{equation}
\widehat{M}_{H}(\pi ,\phi )=-\,\widehat{S}_{z}  \label{Eq.PhaseChanger}
\end{equation}%
and for the output state the mean value and variance of $\widehat{M}$ are
given by%
\begin{eqnarray}
\left\langle \widehat{M}\right\rangle &=&-\,\left\langle \widehat{S}%
_{z}\right\rangle _{\rho }  \label{Eq.FormOutputMeanPhaseChanger} \\
\left\langle \Delta \widehat{M}^{2}\right\rangle &=&\left\langle \Delta 
\widehat{S}_{z}^{2}\right\rangle  \label{Eq.FormOutputVariancePhaseChanger}
\end{eqnarray}%
so the phase changer measures the negative of the population difference.
Effectively the phase changer interchanges the modes $\widehat{a}\rightarrow 
\widehat{b}$ and $\widehat{b}\rightarrow \widehat{a}$ and this is its role
rather than being directly involved in a measurement. Phase changers are
often included in complex interferometers at the midpoint of free evolution
regions to cancel out unwanted causes of phase change.

\subsubsection{Other Measurements in Simple Two Mode Interferometer}

Another useful choice of measureable is the \emph{square} of the population
difference 
\begin{equation}
\widehat{M}_{2}=\left( \frac{1}{2}(\widehat{b}^{\dag }\widehat{b}-\widehat{a}%
^{\dag }\widehat{a})\right) ^{2}=\widehat{S}_{z}^{2}  \label{Eq.Measurable2}
\end{equation}%
For the beam splitter case with $2s=\pi /2$ and $\phi $ (variable), we can
easily show (see Appendix \ref{Appendix - Derivatio n of Interferometer
Result}) that the mean value of $\widehat{M}_{2}$ for the output state is
given by 
\begin{equation}
\left\langle \widehat{M}_{2}\right\rangle =\sin ^{2}\phi \,\left\langle (%
\widehat{S}_{x})^{2}\right\rangle +\cos ^{2}\phi \,\left\langle (\widehat{S}%
_{y})^{2}\right\rangle +\sin \phi \,\cos \phi \,\left\langle (\widehat{S}_{x}%
\widehat{S}_{y}+\widehat{S}_{y}\widehat{S}_{x})\right\rangle
\label{Eq.FormOutputMeanM2}
\end{equation}%
showing that the mean for the new observable $\widehat{M}_{2}$ is a
sinusoidal function of the BS interferometer variable $\phi $ with
coefficients that depend on the means of $\widehat{S}_{x}^{2}$, $\widehat{S}%
_{y}^{2}$ and $\widehat{S}_{x}\widehat{S}_{y}+\widehat{S}_{y}\widehat{S}_{x}$%
.

Choosing special cases for the interferometer variable yields the following
useful results%
\begin{eqnarray}
\left\langle \widehat{M}_{2}\right\rangle _{\phi =0} &=&\left\langle (%
\widehat{S}_{y})^{2}\right\rangle _{\rho }\qquad \left\langle \widehat{M}%
_{2}\right\rangle _{\phi =\pi /2}=\left\langle (\widehat{S}%
_{x})^{2}\right\rangle _{\rho }  \nonumber \\
\left\langle \widehat{M}_{2}\right\rangle _{\phi =\pi /4}-\left\langle 
\widehat{M}_{2}\right\rangle _{\phi =-\pi /4} &=&\left\langle (\widehat{S}%
_{x}\widehat{S}_{y}+\widehat{S}_{y}\widehat{S}_{x})\right\rangle _{\rho }
\label{Eq.OutputMeanM2Results}
\end{eqnarray}%
Hence all three quantities $\left\langle (\widehat{S}_{x})^{2}\right\rangle $%
, $\left\langle (\widehat{S}_{y})^{2}\right\rangle $ and $\left\langle (%
\widehat{S}_{x}\widehat{S}_{y}+\widehat{S}_{y}\widehat{S}_{x})\right\rangle $
can be measured. We note that just measuring $\left\langle \widehat{M}%
_{2}\right\rangle $ does not add to the results obtained by measuring the
mean and variance of the original measureable $\widehat{M}$, since $%
\left\langle \widehat{M}_{2}\right\rangle =\left\langle \Delta \widehat{M}%
^{2}\right\rangle +\left\langle \widehat{M}\right\rangle ^{2}$. The variance 
$\left\langle \Delta \widehat{M}_{2}^{2}\right\rangle $ does of course
depend on higher moments, for example with $\phi =0$ $\left\langle \Delta 
\widehat{M}_{2}^{2}\right\rangle =\left\langle \Delta \left( \widehat{S}%
_{y}^{2}\right) ^{2}\right\rangle $ and $\phi =\pi /2$ $\left\langle \Delta 
\widehat{M}_{2}^{2}\right\rangle =\left\langle \Delta \left( \widehat{S}%
_{x}^{2}\right) ^{2}\right\rangle $, so these also could be measured.

\subsection{Multi-Mode Interferometers}

\label{SubSection - MultiMode Interferometers}

For the multi-mode case we consider two sets of modes $\widehat{a}_{i}$ and $%
\widehat{b}_{i}$ as descibed in SubSection \ref{SubSection - Spin Operators
Multimode Case}. These may be different modes associated with two different
hyperfine states or they may be modes associated with two separated
potential wells. The Hamiltonian analogous to that in (\ref%
{Eq.JosephHamTerms}) for the two mode case is 
\begin{eqnarray}
\widehat{H}_{0} &=&\sum_{i}\hbar (\omega _{a}+\omega _{i})\,\widehat{a}%
_{i}^{\dag }\widehat{a}_{i}+\sum_{i}\hbar (\omega _{b}+\omega _{i})\,%
\widehat{b}_{i}^{\dag }\widehat{b}_{i}  \nonumber \\
\widehat{V} &=&\mathcal{A(}t)\exp (-i\omega _{0}t)\,\exp (i\phi )\,\sum_{i}%
\widehat{b}_{i}^{\dag }\widehat{a}_{i}+\mathcal{A(}t)\exp (+i\omega
_{0}t)\,\exp (-i\phi )\,\sum_{i}\widehat{a}_{i}^{\dag }\widehat{b}_{i} 
\nonumber \\
&&  \label{Eq.JosephHamMultiMode}
\end{eqnarray}%
where the collision terms are ignored since we are only considering the
effect of the short interferometer coupling pulse. Here we have assumed that
the energy for the mode $\widehat{a}_{i}$ is $\hbar (\omega _{a}+\omega
_{i}) $, which is the sum of a basic energy for all $a$ modes - $\hbar
\omega _{a}$, and an energy term $\hbar \omega _{i}$ that distinguishes
differing modes $\widehat{a}_{i}$ (and similarly for the mode $\widehat{b}%
_{i}$). In addition, we assume selection rules lead to pairwise coupling $%
\widehat{a}_{i}\leftrightarrow \widehat{b}_{i}$. In the case where coupling
is due to pulsed external fields (microwave and RF) we can assume that the
momenta ($\sim \sqrt{m\hbar \omega _{trap}}$) associated with trapped modes $%
\widehat{a}_{i}$ and $\widehat{b}_{i}$ are the same, since the momenta
associated with the low frequency photons ($\sim \hbar \omega _{RF}/c$)
involved can be ignored. The spin operators for the multi-mode system are
set out in (\ref{Eq.SpinOprs2}) in terms of the mode operators.

As in SubSection \ref{SubSection - Simple Model} the \emph{interferometer
frequency} $\omega _{0}$ is assumed for simplicity to be in \emph{resonance}
with the \emph{transition frequency} $\omega _{ba}=\omega _{b}-\omega _{a}$.
Following the treatment in SubSection \ref{SubSection - Simple Two Mode
Interferometer}, the choice of \emph{measurable} is the half the total
population difference between the two sets of modes 
\begin{equation}
\widehat{M}=\frac{1}{2}\sum_{i}(\widehat{b}_{i}^{\dag }\widehat{b}_{i}-%
\widehat{a_{i}}^{\dag }\widehat{a}_{i})=\widehat{S}_{z}
\label{Eq.MeasurableMultiMode}
\end{equation}%
and we will determine its mean and variance for the state $\widehat{\rho }%
^{\#}$ given by 
\begin{equation}
\widehat{\rho }^{\#}=\widehat{U}\,\widehat{\rho }\,\widehat{U}^{-1}
\label{Eq.OuputStateMultiMode}
\end{equation}%
where the \emph{output state} $\widehat{\rho }^{\#}$ has evolved from the
initial \emph{input state} $\widehat{\rho }$ due to the effect of the \emph{%
multi- mode interferometer}. $\widehat{U}$ is the unitary \emph{evolution
operator} describing evolution during the time the short classical pulse is
applied. Collision terms and interactions with other systems will be ignored
during the short time interval involved.

As in the two mode interferometer case, the results for the mean and
variance of $\widehat{M}$ depend on the\emph{\ pulse area} $2s=\theta $ and
the \emph{phase} $\phi $.of the interferometer coupling pulse. They have the
same dependence on the \emph{mean values} and \emph{covariance matrix} for
the \emph{multi-mode spin operators} $\widehat{S}_{x}$, $\,\widehat{S}_{y}$
and $\widehat{S}_{z}$ for the input state $\widehat{\rho }$ $\ $as in the
two mode interferometer. Thus we then find that the \emph{general result}
for the \emph{mean} value is 
\begin{equation}
\left\langle \widehat{M}\right\rangle =\sin \theta \,\sin \phi
\,\left\langle \widehat{S}_{x}\right\rangle _{\rho }+\sin \theta \,\cos \phi
\,\left\langle \widehat{S}_{y}\right\rangle _{\rho }+\cos \theta
\,\left\langle \widehat{S}_{z}\right\rangle _{\rho }
\label{Eq.MeanMultiMode}
\end{equation}%
and for the\emph{\ variance} is 
\begin{eqnarray}
&&\left\langle \Delta \widehat{M}^{2}\right\rangle  \nonumber \\
&=&\frac{(1-\cos 2\theta )}{2}\frac{(1-\cos 2\phi )}{2}C(\widehat{S}_{x},%
\widehat{S}_{x})+\frac{(1-\cos 2\theta )}{2}\frac{(1+\cos 2\phi )}{2}C(%
\widehat{S}_{y},\widehat{S}_{y})  \nonumber \\
&&+\frac{(1+\cos 2\theta )}{2}C(\widehat{S}_{z},\widehat{S}_{z})  \nonumber
\\
&&+\frac{(1-\cos 2\theta )}{2}\sin 2\phi \,C(\widehat{S}_{x},\widehat{S}%
_{y})+\sin 2\theta \,\cos \phi \,C(\widehat{S}_{y},\widehat{S}_{z})+\sin
2\theta \,\sin \phi \,C(\widehat{S}_{z},\widehat{S}_{x})  \nonumber \\
&&  \label{Eq.VarianceMultiMode}
\end{eqnarray}%
Details of the derivation are set out in Appendix \ref{Appendix - Derivatio
n of Interferometer Result}.

\subsection{Application to Spin Squeezing Tests for Entanglement}

\label{SubSection - Appn to Spin Sq Tests}

Unless stated otherwise, we now focus on spin squeezing tests for \emph{SSR
compliant entanglement} based on the beam splitter measurements (the simple
two mode interferometer with $2s=\theta =\pi /2$). By choosing the phase $%
\phi =0$ we see that $\left\langle \widehat{M}\right\rangle =\left\langle 
\widehat{S}_{y}\right\rangle _{\rho }$ and $\left\langle \Delta \widehat{M}%
^{2}\right\rangle =C(\widehat{S}_{y},\widehat{S}_{y})=\left\langle \{%
\widehat{S}_{y}-\left\langle \widehat{S}_{y}\right\rangle _{\rho
}\}^{2}\right\rangle _{\rho }$ giving the mean and variance for the spin
operator $\widehat{S}_{y}$. By choosing the phase $\phi =\pi /2$ we see that 
$\left\langle \widehat{M}\right\rangle =\left\langle \widehat{S}%
_{x}\right\rangle _{\rho }$ and $\left\langle \Delta \widehat{M}%
^{2}\right\rangle =C(\widehat{S}_{x},\widehat{S}_{x})=\left\langle \{%
\widehat{S}_{x}-\left\langle \widehat{S}_{x}\right\rangle _{\rho
}\}^{2}\right\rangle _{\rho }$ giving the mean and variance for the spin
operator $\widehat{S}_{x}$. If the measurement of $\left\langle \widehat{M}%
\right\rangle $ were carried out without the beam splitter being present
then $\left\langle \widehat{M}\right\rangle =\left\langle \widehat{S}%
_{z}\right\rangle _{\rho }$. Combining all these results then enables us to
see whether or not $\widehat{S}_{x}$ is squeezed with respect to $\widehat{S}%
_{y}\,$or vice versa. This illustrates the \emph{use }of the interferometer
in seeing if a state $\widehat{\rho }$ is \emph{squeezed}. Squeezing in $%
\widehat{S}_{z}$ with respect to $\widehat{S}_{y}$ (or $\widehat{S}_{x}$) or
vice versa also demonstrates entanglement and again the simple two mode
interferometer with appropriate choices of $\theta $ and $\phi $ can be used
to measure the means and variances of the relevant spin operators.

As the presence of spin squeezing shows that the state must be entangled 
\cite{Dalton14a} the use of the interferometer for squeezing tests is
important. Note that we still need to consider whether fluctuations due to a
finite measurement sample could mask the test. However, as spin squeezing
has been demonstrated in two mode systems of bosonic atoms this approach to
demonstrating entanglement is clearly useful.

\subsubsection{Spin Squeezing in $\widehat{S}_{x}$, $\widehat{S}_{y}$}

To demonstrate spin squeezing in $\widehat{S}_{x}$ with respect to $\widehat{%
S}_{y}$ we need to measure the variances $\left\langle \Delta \widehat{S}%
_{x}{}^{2}\right\rangle _{\rho }$ and $\left\langle \Delta \widehat{S}%
_{y}{}^{2}\right\rangle _{\rho }$ and the mean $\left\langle \widehat{S}%
_{z}\right\rangle _{\rho }$ and show that 
\begin{equation}
\left\langle \Delta \widehat{S}_{x}{}^{2}\right\rangle _{\rho }<\frac{1}{2}%
|\left\langle \widehat{S}_{z}\right\rangle _{\rho }|\qquad \left\langle
\Delta \widehat{S}_{y}{}^{2}\right\rangle _{\rho }>\frac{1}{2}|\left\langle 
\widehat{S}_{z}\right\rangle _{\rho }|  \label{Eq.SpinSqgTest}
\end{equation}%
As we have seen, the variances in $\widehat{S}_{y}$, $\widehat{S}_{x}$ are
obtained by measuring the \emph{fluctuation} in the measureable $\widehat{M}$
\emph{after} applying the interferometer to the state $\widehat{\rho }$,
with the interferometer phase set to $\phi =0$ or $\phi =\pi /2$ for the two
cases respectively. The mean $\left\langle \widehat{S}_{z}\right\rangle
_{\rho }$ is obtained by a direct measurement of the measureable $\widehat{M}
$ \emph{without} applying the interferometer to the state $\widehat{\rho }$.

\subsubsection{Spin Squeezing in $xy$ Plane}

As shown in Section \ref{Section - Relationship Spin Squeezing &
Entanglement} (see \cite{Dalton14a}) squeezing in $\widehat{S}_{x}$ compared
to $\widehat{S}_{y}$ or vice versa is sufficient to show that the state is
entangled. However, as the interferometer measures the variance for the
state $\widehat{\rho }$ in the quantity 
\begin{equation}
\widehat{M}_{H}(\pi /2,\phi )=\sin \phi \,\widehat{S}_{x}+\cos \phi \,%
\widehat{S}_{y}=\widehat{S}_{x}^{\#}(\frac{3\pi }{2}+\phi )
\label{Eq.OutputOprSx}
\end{equation}%
corresponding to the $x$ component of the spin vector operator $%
\underrightarrow{(\widehat{S})}$ after it has been rotated about the $z$
axis through an angle $\frac{3\pi }{2}+\phi $, it is desirable to\emph{\
extend} the entanglement test to consider the squeezing of $\widehat{S}%
_{x}^{\#}(\frac{3\pi }{2}+\phi )$ with respect to the corresponding $y$
component $\widehat{S}_{y}^{\#}(\frac{3\pi }{2}+\phi )$ - and vice versa,
where%
\begin{equation}
\widehat{S}_{y}^{\#}(\frac{3\pi }{2}+\phi )=-\cos \phi \,\widehat{S}%
_{x}+\sin \phi \,\widehat{S}_{y}  \label{Eq.OutputOprSy}
\end{equation}%
The variance in $\widehat{S}_{y}^{\#}(\frac{3\pi }{2}+\phi )$ can be
obtained by changing the interferometer phase to $\phi +\frac{\pi }{2}$.
Clearly $[\widehat{S}_{x}^{\#}(\frac{3\pi }{2}+\phi ),\widehat{S}_{y}^{\#}(%
\frac{3\pi }{2}+\phi )]=i\,\widehat{S}_{z}$, as before.

The question is - does squeezing in either $\widehat{S}_{x}^{\#}(\frac{3\pi 
}{2}+\phi )$ or $\widehat{S}_{y}^{\#}(\frac{3\pi }{2}+\phi )$ demonstrate
entanglement of the modes $\widehat{a}$ and $\widehat{b}$ ? The answer is
that it does.

For a \emph{separable} state we have 
\begin{equation}
\left\langle \widehat{S}_{x}^{\#}(\frac{3\pi }{2}+\phi )\right\rangle _{\rho
}=\left\langle \widehat{S}_{y}^{\#}(\frac{3\pi }{2}+\phi )\right\rangle
_{\rho }=0  \label{Eq.NewMeanSpins}
\end{equation}%
as before, since $\left\langle \widehat{S}_{x,y}^{\#}(\frac{3\pi }{2}+\phi
)\right\rangle _{\rho }$ are just linear combinations of the zero $%
\left\langle \widehat{S}_{x,y}\right\rangle _{\rho }$.

Since $[$ $\widehat{S}_{x}^{\#}(\frac{3\pi }{2}+\phi ),$ $\widehat{S}%
_{y}^{\#}(\frac{3\pi }{2}+\phi )]=i\widehat{S}_{z}$ the Heisenberg
uncertainty principle shows that $\left\langle \Delta \widehat{S}_{x}^{\#}(%
\frac{3\pi }{2}+\phi )^{2}\right\rangle _{\rho }\left\langle \Delta \widehat{%
S}_{y}^{\#}(\frac{3\pi }{2}+\phi )^{2}\right\rangle _{\rho }\geq \frac{1}{4}%
|\left\langle \widehat{S}_{z}\right\rangle _{\rho }|^{2}$ so spin squeezing
in $\widehat{S}_{x}^{\#}(\frac{3\pi }{2}+\phi )$ with respect to $\widehat{S}%
_{y}^{\#}(\frac{3\pi }{2}+\phi )$ or vice versa requires us to show that 
\begin{equation}
\left\langle \Delta \widehat{S}_{x}^{\#}(\frac{3\pi }{2}+\phi
)^{2}\right\rangle _{\rho }<\frac{1}{2}|\left\langle \widehat{S}%
_{z}\right\rangle _{\rho }|\qquad or\qquad \left\langle \Delta \widehat{S}%
_{y}^{\#}(\frac{3\pi }{2}+\phi )^{2}\right\rangle _{\rho }<\frac{1}{2}%
|\left\langle \widehat{S}_{z}\right\rangle _{\rho }|
\label{Eq.SpinSqgTestNewSx}
\end{equation}%
Since we measure $\left\langle \Delta \widehat{S}_{x}^{\#}(\frac{3\pi }{2}%
+\phi )^{2}\right\rangle _{\rho }$ the spin squeezing test is$\left\langle
\Delta \widehat{S}_{x}^{\#}(\frac{3\pi }{2}+\phi )^{2}\right\rangle _{\rho }<%
\frac{1}{2}|\left\langle \widehat{S}_{z}\right\rangle _{\rho }|$.

Now for $\widehat{S}_{x}^{\#}(\frac{3\pi }{2}+\phi )$ we have 
\begin{eqnarray}
\left\langle \Delta \widehat{S}_{x}^{\#}(\frac{3\pi }{2}+\phi
)^{2}\right\rangle _{\rho } &=&\left\langle \widehat{S}_{x}^{\#}(\frac{3\pi 
}{2}+\phi )^{2}\right\rangle _{\rho }  \label{Eq.NewVarianceSx} \\
&=&\sin ^{2}\phi \,\left\langle \widehat{S}_{x}^{2}\right\rangle +\cos
^{2}\phi \,\left\langle \widehat{S}_{y}^{2}\right\rangle +\sin \phi \,\cos
\phi \,\left\langle \widehat{S}_{x}\widehat{S}_{y}+\widehat{S}_{y}\widehat{S}%
_{x}\right\rangle _{\rho }  \nonumber
\end{eqnarray}%
and for a separable state we have from SubSection \ref{SubSection - Spin
Squeezing New Spin Oprs} 
\begin{eqnarray}
\left\langle \widehat{S}_{x}^{2}\right\rangle &=&\left\langle \Delta 
\widehat{S}_{x}^{2}\right\rangle \geq \frac{1}{2}|\left\langle \widehat{S}%
_{z}\right\rangle _{\rho }|  \nonumber \\
\left\langle \widehat{S}_{y}^{2}\right\rangle &=&\left\langle \Delta 
\widehat{S}_{y}^{2}\right\rangle \geq \frac{1}{2}|\left\langle \widehat{S}%
_{z}\right\rangle _{\rho }|
\end{eqnarray}%
whilst for the remaining term 
\begin{eqnarray}
\left\langle \widehat{S}_{x}\widehat{S}_{y}+\widehat{S}_{y}\widehat{S}%
_{x}\right\rangle _{\rho } &=&\frac{1}{2i}\left\langle \{(\widehat{b}^{\dag
})^{2}(\widehat{a})^{2}-(\widehat{a}^{\dag })^{2}(\widehat{b}%
)^{2}\}\right\rangle  \nonumber \\
&=&\frac{1}{2i}\dsum\limits_{R}P_{R}\{\left\langle (\widehat{b}^{\dag
})^{2}\right\rangle _{\rho _{R}^{B}}\left\langle (\widehat{a}%
)^{2}\right\rangle _{\rho _{R}^{A}}-\left\langle (\widehat{a}^{\dag
})^{2}\right\rangle _{\rho _{R}^{A}}\left\langle (\widehat{b}%
)^{2}\right\rangle _{\rho _{R}^{B}}\}  \nonumber \\
&=&0
\end{eqnarray}%
using the \emph{local particle number} SSR.

Thus as $\sin ^{2}\phi +\cos ^{2}\phi =1$ and applying similar
considerations to $\left\langle \Delta \widehat{S}_{y}^{\#}(\frac{3\pi }{2}%
+\phi )^{2}\right\rangle _{\rho }$ 
\begin{eqnarray}
\left\langle \Delta \widehat{S}_{x}^{\#}(\frac{3\pi }{2}+\phi
)^{2}\right\rangle _{\rho } &\geq &\frac{1}{2}|\left\langle \widehat{S}%
_{z}\right\rangle _{\rho }|  \nonumber \\
\left\langle \Delta \widehat{S}_{y}^{\#}(\frac{3\pi }{2}+\phi
)^{2}\right\rangle _{\rho } &\geq &\frac{1}{2}|\left\langle \widehat{S}%
_{z}\right\rangle _{\rho }|  \label{Eq.NewSpinSqgTest}
\end{eqnarray}%
showing that for a separable state there is no squeezing for $\widehat{S}%
_{x}^{\#}(\frac{3\pi }{2}+\phi )$ compared to $\widehat{S}_{y}^{\#}(\frac{%
3\pi }{2}+\phi )$ or vice versa. Hence squeezing in either $\widehat{S}%
_{x}^{\#}(\frac{3\pi }{2}+\phi )$ or $\widehat{S}_{y}^{\#}(\frac{3\pi }{2}%
+\phi )$ demonstrates entanglement of the modes $\widehat{a}$ and $\widehat{b%
}$.

\subsubsection{Measurement of $\left\langle \widehat{S}_{z}\right\rangle _{%
\protect\rho }$}

The question remaining is whether the mean value $\left\langle \widehat{S}%
_{z}\right\rangle _{\rho }$ can be measured accurately enough to apply the
test for entanglement. With an infinite number of repeated measurements this
is always possible, since then both the variances $\left\langle \Delta 
\widehat{S}_{x}^{\#}(\frac{3\pi }{2}+\phi )^{2}\right\rangle _{\rho }$ and
the mean $\left\langle \widehat{S}_{z}\right\rangle _{\rho }$. would become
error free. For a finite number of measurements $R$ the measurement of $%
\left\langle \widehat{S}_{z}\right\rangle _{\rho }$ requires a consideration
of the variance in $\widehat{S}_{z}$. For \emph{general} entangled states
general considerations indicate that this mean will be of order $N$ and the
variance will be at worst of order $N^{2}$, Hence the variance $\left\langle
\Delta \left\langle \widehat{S_{z}}\right\rangle ^{2}\right\rangle _{sample}$
in determining the mean $\left\langle \widehat{S}_{z}\right\rangle $ for $R$
repetitions of the measurement would be $\sim N^{2}/R$, giving a fluctuation
of $\sim N/\sqrt{R}$. For this to be small compared to $\sim N$ we merely
require $R\gg 1$, which is not unexpected. This result indicates that the
application of the spin squeezing test via interferometric measurement of
both the variances $\left\langle \Delta \widehat{S}_{x}^{\#}(\frac{3\pi }{2}%
+\phi )^{2}\right\rangle _{\rho }$ and the mean $\left\langle \widehat{S}%
_{z}\right\rangle _{\rho }$ looks feasible.

\subsubsection{Spin Squeezing in $\widehat{S}_{z}$, $\widehat{S}_{y}$}

To demonstrate spin squeezing in $\widehat{S}_{z}$ with respect to $\widehat{%
S}_{y}$ we need to measure the variances $\left\langle \Delta \widehat{S}%
_{z}{}^{2}\right\rangle _{\rho }$ and $\left\langle \Delta \widehat{S}%
_{y}{}^{2}\right\rangle _{\rho }$ and the mean $\left\langle \widehat{S}%
_{x}\right\rangle _{\rho }$ and show that 
\begin{equation}
\left\langle \Delta \widehat{S}_{z}{}^{2}\right\rangle _{\rho }<\frac{1}{2}%
|\left\langle \widehat{S}_{x}\right\rangle _{\rho }|\qquad \left\langle
\Delta \widehat{S}_{y}{}^{2}\right\rangle _{\rho }>\frac{1}{2}|\left\langle 
\widehat{S}_{x}\right\rangle _{\rho }|  \label{Eq.SzSpinSqueezTest}
\end{equation}%
As we have seen, the variances in $\widehat{S}_{z}$, $\widehat{S}_{y}$ are
obtained by measuring the \emph{fluctuation} in the measureable $\widehat{M}$
\emph{after} applying the interferometer to the state $\widehat{\rho }$,
with the interferometer phase set to $\phi =0$ and the pulse area $2s=\theta 
$ made variable. From Eq.(\ref{Eq.formOutputVarianceYZTomog}) we see that
choosing $\theta =0$ gives $\left\langle \Delta \widehat{S}%
_{z}{}^{2}\right\rangle _{\rho }$ and choosing $\theta =\frac{\pi }{2}$
gives $\left\langle \Delta \widehat{S}_{y}{}^{2}\right\rangle _{\rho }$.
From Eq.(\ref{Eq.FormOutputMean}) the mean $\left\langle \widehat{S}%
_{x}\right\rangle _{\rho }$ is obtained by a measurement of the \emph{mean}
in the measureable $\widehat{M}$ \emph{after} applying the interferometer to
the state $\widehat{\rho }$, with the interferometer phase set to $\phi =\pi
/2$ and the pulse area $2s=\pi /2$.

\subsection{Application to Correlation Tests for Entanglement}

\label{Subsection - Appn to Correln Tests}

\subsubsection{First Order Correlation Test}

Unless stated otherwise, we again focus on correlation tests for \emph{SSR
compliant entanglement} \ For the beam splitter case and by choosing the
phase $\phi =0$ we see that $\left\langle \widehat{M}\right\rangle
=\left\langle \widehat{S}_{y}\right\rangle _{\rho }$ and by choosing the
phase $\phi =\pi /2$ we see that $\left\langle \widehat{M}\right\rangle
=\left\langle \widehat{S}_{x}\right\rangle _{\rho }$. The simplest form of
the correlation test with $n=m=1$ requires 
\begin{equation}
\left\langle \widehat{S}_{x}\right\rangle _{\rho }\neq 0\qquad \left\langle 
\widehat{S}_{y}\right\rangle _{\rho }\neq 0  \label{Eq.SimpleCorrTestResult}
\end{equation}%
for establishing that the state is entangled. For the separable state with $%
\widehat{M}_{H}=\sin \phi \,\widehat{S}_{x}+\cos \phi \,\widehat{S}_{y}=%
\widehat{S}_{x}^{\#}(\frac{3\pi }{2}+\phi )$%
\begin{equation}
\left\langle \widehat{M}\right\rangle =\left\langle \widehat{M}%
_{H}\right\rangle _{\rho }=0
\end{equation}%
so that the mean value of the measureable is zero and independent of the
beam splitter phase $\phi $.for all $\phi $. Finding any non-zero value for $%
\left\langle \widehat{M}_{H}\right\rangle _{\rho }$ would then show that the
state $\widehat{\rho }$ is entangled. More importantly from the general
result, $\left\langle \widehat{M}_{H}\right\rangle _{\rho }$ would show a 
\emph{sinusoidal dependence} on the interferometer phase $\phi $, so the
appearance of such a dependence would be an indication that the state was
entangled.

The question remaining is whether the mean values $\left\langle \widehat{S}%
_{x,y}\right\rangle _{\rho }$ can be measured accurately enough to apply the
test for entanglement. With an infinite number of repeated measurements this
is always possible, since then both the variances $\left\langle \Delta 
\widehat{S}_{x,y}^{2}\right\rangle _{\rho }$ and the means $\left\langle 
\widehat{S}_{x,y}\right\rangle _{\rho }$. would become error free. For a
finite number of measurements the measurement of $\left\langle \widehat{S}%
_{x,y}\right\rangle _{\rho }$ requires a consideration of the variances in $%
\widehat{S}_{x,y}$ (or $\widehat{M}_{H}$ to cover both cases). For a general
entangled state we can assume that $\left\langle \widehat{M}%
_{H}\right\rangle _{\rho }\sim N/2$ and the variance will be at worst of
order $N^{2}$, Hence the variance $\left\langle \Delta \left\langle \widehat{%
M}_{H}\right\rangle ^{2}\right\rangle _{sample}$ in determining the mean $%
\left\langle \widehat{S}_{x,y}\right\rangle $ for $R$ repetitions of the
measurement would be $\sim N^{2}/R$, giving a fluctuation of $\sim N/\sqrt{R}
$. For this to be small compared to $\sim N$ we merely require $R\gg 1$,
which is not unexpected. This result indicates that the application of the
simple correlation test via interferometric measurement of $\left\langle 
\widehat{S}_{x}\right\rangle _{\rho }$ and $\left\langle \widehat{S}%
_{y}\right\rangle _{\rho }$ looks feasible.

\subsubsection{Second Order Correlation Test}

For the second order form of the correlation test with $n=m=2$ requires 
\begin{eqnarray}
\left\langle \Delta \widehat{S}_{x}{}^{2}\right\rangle _{\rho }+\left\langle 
\widehat{S}_{x}\right\rangle _{\rho }^{2}-\left\langle \Delta \widehat{S}%
_{y}{}^{2}\right\rangle _{\rho }-\left\langle \widehat{S}_{y}\right\rangle
_{\rho }^{2} &\neq &0  \nonumber \\
(C(\widehat{S}_{x},\widehat{S}_{y})+\left\langle \widehat{S}%
_{x}\right\rangle _{\rho }\left\langle \widehat{S}_{y}\right\rangle _{\rho
}) &\neq &0
\end{eqnarray}%
We have already shown using Eqs (\ref{Eq.FormOutputMean}) and (\ref%
{Eq.FormOutputVariance}) that the variances $\left\langle \Delta \widehat{S}%
_{y}{}^{2}\right\rangle _{\rho }$ and $\left\langle \Delta \widehat{S}%
_{x}{}^{2}\right\rangle _{\rho }$ and the means $\left\langle \widehat{S}%
_{y}\right\rangle _{\rho }$ and $\left\langle \widehat{S}_{x}\right\rangle
_{\rho }$ can be obtained via the BS\ interferometer.from the mean $%
\left\langle \widehat{M}\right\rangle $ and the variance $\left\langle
\Delta \widehat{M}^{2}\right\rangle $ for the choices of $\phi =0$ and $\phi
=\pi /2$. To obtain the covariance matrix element $C(\widehat{S}_{x},%
\widehat{S}_{y})$ we see that if we choose $\phi =\pi /4$ then $\left\langle
\Delta \widehat{M}^{2}\right\rangle =\frac{1}{2}\,\left\langle \Delta 
\widehat{S}_{x}{}^{2}\right\rangle _{\rho }+\frac{1}{2}\,\left\langle \Delta 
\widehat{S}_{y}{}^{2}\right\rangle _{\rho }+\,C(\widehat{S}_{x},\widehat{S}%
_{y})$, from which the covariance can be measured. Thus the second order
correlation test can be applied.

Alternately, if the measurement quantity for the BS interferometer is the
square of the population difference then we see from (\ref%
{Eq.OutputMeanM2Results}) that the mean value of $\widehat{M}_{2}$ for
certain choices of the BS variable $\phi $ measures $\left\langle \Delta 
\widehat{S}_{x}{}^{2}\right\rangle _{\rho }+\left\langle \widehat{S}%
_{x}\right\rangle _{\rho }^{2}=\left\langle \widehat{S}_{x}{}^{2}\right%
\rangle _{\rho }$, $\left\langle \Delta \widehat{S}_{y}{}^{2}\right\rangle
_{\rho }+\left\langle \widehat{S}_{y}\right\rangle _{\rho }^{2}=\left\langle 
\widehat{S}_{y}{}^{2}\right\rangle _{\rho }$ and $(C(\widehat{S}_{x},%
\widehat{S}_{y})+\left\langle \widehat{S}_{x}\right\rangle _{\rho
}\left\langle \widehat{S}_{y}\right\rangle _{\rho })=\left\langle (\widehat{S%
}_{x}\widehat{S}_{y}+\widehat{S}_{y}\widehat{S}_{x})\right\rangle _{\rho }$.
The second order correlation test is that if 
\begin{eqnarray}
\left\langle \widehat{S}_{x}{}^{2}\right\rangle _{\rho } &\neq &\left\langle 
\widehat{S}_{y}{}^{2}\right\rangle _{\rho }  \nonumber \\
\left\langle (\widehat{S}_{x}\widehat{S}_{y}+\widehat{S}_{y}\widehat{S}%
_{x})\right\rangle _{\rho } &\neq &0  \label{Eq.SecondOrderCorrTestEquiv}
\end{eqnarray}%
then the state is entangled.

\subsection{Application to Quadrature Tests for Entanglement}

\label{SubSection - Appn to Quad Tests}

As we saw previously, no useful quadrature test for SSR\ compliant
entanglement in two mode systems of identical bosons of the form $%
\left\langle \Delta (\widehat{x}_{A}\pm \widehat{x}_{B})^{2}\right\rangle
_{\rho }+\left\langle \Delta (\widehat{p}_{A}\mp \widehat{p}%
_{B})^{2}\right\rangle _{\rho }<2+2\left\langle \widehat{N}\right\rangle
_{\rho }$\ results if the overall system state is globally SSR compliant.
However, if the\emph{\ separable}\textbf{\ }states are non-compliant then
showing that 
\begin{equation}
\left\langle \Delta (\widehat{x}_{A}\pm \widehat{x}_{B})^{2}\right\rangle
_{\rho }+\left\langle \Delta (\widehat{p}_{A}\mp \widehat{p}%
_{B})^{2}\right\rangle _{\rho }<2+2\left\langle \widehat{N}\right\rangle
_{\rho }+2(\left\langle \widehat{a}\,\widehat{b}\right\rangle _{\rho
}+\left\langle \widehat{a}^{\dag }\,\widehat{b}^{\dag }\right\rangle _{\rho
})-2|\left\langle \widehat{a}\right\rangle _{\rho }+\left\langle \widehat{b}%
^{\dag }\right\rangle _{\rho }|^{2}.
\end{equation}%
would demonstrate entanglement. This test requires measuring $\left\langle 
\widehat{N}\right\rangle _{\rho }$\ together with $\left\langle \widehat{a}\,%
\widehat{b}\right\rangle _{\rho }$, $\left\langle \widehat{a}\right\rangle
_{\rho }$\ and $\left\langle \widehat{b}^{\dag }\right\rangle _{\rho }$.
Although the first can be measured using the BS interferometer the
quantities $\left\langle \widehat{a}\,\widehat{b}\right\rangle _{\rho }$, $%
\left\langle \widehat{a}\right\rangle _{\rho }$\ and $\left\langle \widehat{b%
}^{\dag }\right\rangle _{\rho }$\ cannot. Another technique involving a
measuring system where there is a well-defined\textbf{\ }\emph{phase
reference} is therefore required if quadrature tests for SSR neglected
entanglement are to be undertaken. Furthermore, the\textbf{\ }\emph{overall}%
\textbf{\ }state must still be globally SSR compliant, and hence $%
\left\langle \widehat{a}\,\widehat{b}\right\rangle _{\rho }$, $\left\langle 
\widehat{a}\right\rangle _{\rho }$\ and $\left\langle \widehat{b}^{\dag
}\right\rangle _{\rho }$\ are all zero, even for entangled states, so the
test reduces to $\left\langle \Delta (\widehat{x}_{A}\pm \widehat{x}%
_{B})^{2}\right\rangle _{\rho }+\left\langle \Delta (\widehat{p}_{A}\mp 
\widehat{p}_{B})^{2}\right\rangle _{\rho }<2+2\left\langle \widehat{N}%
\right\rangle _{\rho }$. Since for\textbf{\ }\emph{all}\textbf{\ }states $%
\left\langle \Delta (\widehat{x}_{A}\pm \widehat{x}_{B})^{2}\right\rangle
_{\rho }+\left\langle \Delta (\widehat{p}_{A}\mp \widehat{p}%
_{B})^{2}\right\rangle _{\rho }=2+2\left\langle \widehat{N}\right\rangle
_{\rho }$ this test must fail anyway. We have also seen that finding the 
\emph{correlation coefficient} - defined in terms of generalised quadrature
operators (\ref{Eq.QuadOprs}) in Eq. (\ref{Eq.CorrelnCoeft}) - to be
non-zero does not lead to a new test for SSR\ compliant entanglement. The
tests involving\textbf{\ }\emph{two mode quadrature squeezing}\textbf{\ }%
look more promising, assuming the relevant quadrature variances can be
measured. \medskip \pagebreak

\section{Experiments on Spin Squeezing}

\label{Section - Experiments}

We now examine a number of recent experimental papers involving squeezing
and entanglement in BEC with \emph{large} numbers of \emph{identical}
bosons. Their notation will be modified to be the same as here. There are
really two questions to consider. One is whether squeezing has been created
(and of which type). The second is whether or not this demonstrates
entanglement of the modes involved. Here we define entanglement for
identical bosons as set out in Section \textbf{3 of paper I}. Many of these
experiments involve Ramsey interferometers and the aim was to demonstrate
spin squeezing created via the collisional interaction between the bosons.
Obviously, in demonstrating \emph{spin squeezing} they would hope to have
created an entangled state, though in most cases an entangled state had
already been created via the interaction with the first beam splitter.
Although the criterion for entanglement used in most cases was based on an
experimental\emph{\ proposal} \cite{Sorensen01a}, \cite{Sorensen01b} which
regarded identical particles as \emph{distinguishable }sub-systems, the spin
squeezing test based on $\widehat{S}_{z}$ does turn out to be a.valid test
for two mode entanglement, as explained in Section \ref{Section -
Relationship Spin Squeezing & Entanglement}. However, it should be noted
that all the papers discussed have a different viewpoint regarding what
exactly is entangled - generally referring to entanglement of \emph{atoms}
or \emph{particles} rather than modes. All the experiments discussed below
establish entanglement, though often this was already created in a first $%
\pi /2$ coupling pulse. Most are based on the spin squeezing test involving $%
\widehat{S}_{z}$, that of Gross et al \cite{Gross11a} involved \emph{%
population difference} \emph{squeezing }rather than spin squeezing (see
SubSection \ref{SubSection - Test for Number Difference}). The other
experiment of Gross et al \cite{Gross10a} shows (see Fig 2b in \cite%
{Gross10a}) that the mean value of one of the two spin operators $\widehat{S}%
_{x}$, $\widehat{S}_{y}$ is non-zero, as measurement results such as in (\ref%
{Eq.FormOutputMeanYZTomog}) for the simple two mode interferometer with $%
2s=\theta $, $\phi =0$ can determine. This is sufficient to demonstrate two
mode entanglement, as (\ref{Eq.BlochVectorEntTest}) shows.

A key result of the present paper (and \cite{Dalton14a}) is that the
conclusion that experiments which have demonstrated spin squeezing in $%
\widehat{S}_{z}$ have thereby demonstrated two mode entanglement,\ no longer
has to be justified on the basis of a proof that clearly does not apply to a
system of identical bosons.

\subsection{Esteve et al. (2008) \protect\cite{Esteve08a}}

$\bullet $ Stated emphasis - generation of spin squeezed states suitable for
atom interferometry, demonstration of\emph{\ particle} entanglement.

$\bullet $ System - Rb$^{87}$ in two hyperfine states.

$\bullet $ BEC\ of Rb$^{87}$ trapped in optical lattice superposed on
harmonic trap.

$\bullet $ Occupation number per site $100$ to $1100$ atoms - macroscopic.

$\bullet $ Situation where atoms trapped in just two sites treated - two
mode entanglement?

$\bullet $ Claimed observed (see Fig 1 in \cite{Esteve08a}) spin squeezing
based on $N\left\langle \Delta \widehat{S}_{z}^{2}\right\rangle
/(\left\langle \widehat{S}_{x}^{2}\right\rangle +\left\langle \widehat{S}%
_{y}^{2}\right\rangle )<1$ (see (\ref{Eq.SpinSqueezingMeasure2})).

$\bullet $ Claimed entanglement of identical atoms.

$\bullet $ Spin squeezing test is based on assumption that Bloch vector is
on Bloch sphere, a result not established.

$\bullet $ Comment - spin squeezing in $\widehat{S}_{z}$ (almost)
demonstrated (see (\ref{Eq.SpinSqueezingMeasure2})), so entanglement is
established.

\subsection{Riedel et al. (2010) \protect\cite{Riedel10a}}

$\bullet $ Stated emphasis - generation of spin squeezed states suitable for
atom interferometry, demonstration of \emph{multi-particle} entanglement.

$\bullet $ System - Rb$^{87}$ in two hyperfine states.

$\bullet $ BEC\ of Rb$^{87}$ trapped in harmonic trap with non-zero magnetic
field - Zeeman splitting of levels.

$\bullet $ Number of atoms $1200$ - macroscopic.

$\bullet $ Process involves Ramsey interferometry - starts with all atoms in
one state, $\pi /2$ pulse (duration $\pi /2\Omega $ ?) generates coherent
spin state $(\widehat{a}^{\dag }+\widehat{b}^{\dag })^{N}\left\vert
0\right\rangle $ (entangled), free evolution with collisions (causes
squeezing), second pulse with area $2s=\theta $ and phase $\pi $ (or $0$)
followed by detection of population difference - associated with operator $%
\widehat{S}_{z}.$

$\bullet $ Evolution described using Josephson Hamiltonian $\widehat{H}%
=\delta \widehat{S}_{z}+\Omega \widehat{S}_{\phi }+\chi \widehat{S}_{z}^{2}$
where $\widehat{S}_{\phi }=\cos \phi \,\widehat{S}_{x}-\sin \phi \,\widehat{S%
}_{y}$, $\Omega $ is Rabi frequency, $\phi $ is phase of RF-microwave field, 
$\delta $ is detuning, $\chi $ describes collisions. Interaction picture and
on resonance ?

$\bullet $ During free evolution including collisions spatial modes for
internal states pushed apart so that $\chi $ becomes much bigger in order to
give larger squeezing.

$\bullet $ Final pulse enables state tomography in the $yz$ plane to be
carried out - measures spin squeezing for spin operator $\widehat{S}_{\theta
}=\cos \theta \,\widehat{S}_{z}-\sin \theta \,\widehat{S}_{y}$ in this plane
(see (\ref{Eq.VarianceMultiMode}) herein with $\phi =\pi $).

$\bullet $ Claimed observed spin squeezing based on $\left\langle \Delta 
\widehat{S}_{\theta }^{2}\right\rangle $ being less than standard quantum
limit $N/4$. (see Fig 2 in \cite{Riedel10a}).

$\bullet $ No measurement made to show that $|\left\langle \widehat{S}%
_{x}\right\rangle |\approx N/2$ as required to justify spin squeezing test.
Spin squeezing test is based on assumption that Bloch vector is on Bloch
sphere, a result not established.

$\bullet $ Claim that state of atoms at end of free evolution is
four-partite entangled based on spin squeezing test is not substantiated,
also an entangled state was already created by first $\pi /2$ pulse.

$\bullet $ Comment - spin squeezing in $\widehat{S}_{z}$ (almost)
demonstrated (see (\ref{Eq.SpinSqueezingMeasure2})), so entanglement is
established. An entangled state was of course already created by first $\pi
/2$ pulse, and then modified via the collisional effects.

\subsection{Gross et al. (2010) \protect\cite{Gross10a}}

$\bullet $ Stated emphasis - generation of non-classical spin squeezed
states for non-linear atom inteferometry, demonstration of entanglement
between\emph{\ atoms}.

$\bullet $ System - Rb$^{87}$ in two hyperfine states.

$\bullet $ Six independent BECs\ of Rb$^{87}$ trapped in six separate wells
in a optical lattice.

$\bullet $ Number of atoms $2300$ - macroscopic, down to ca $170$ in each
well.

$\bullet $ Evolution described using Josephson Hamiltonian $\widehat{H}%
=\Delta \omega _{0}\widehat{S}_{z}+\Omega \widehat{S}_{\gamma }+\chi 
\widehat{S}_{z}^{2}$ where (in the present notation) $\widehat{S}_{\gamma
}=\cos \gamma \,\widehat{S}_{x}+\sin \gamma \,\widehat{S}_{y}$, $\Omega $ is
Rabi frequency, $\gamma $ is phase of RF-microwave field, $\Delta \omega
_{0} $ is detuning, $\chi $ describes collisions. Interaction picture and on
resonance ?

$\bullet $ During free evolution plus collision evolution Feshach resonance
used so that $\chi $ becomes much bigger in order to give larger squeezing.

$\bullet $ One process involves Ramsey interferometry - starts with all
atoms in one state, $\pi /2$ pulse (duration $\pi /2\Omega $ ?) generates
coherent spin state $(\widehat{a}^{\dag }+\widehat{b}^{\dag })^{N}\left\vert
0\right\rangle $ (entangled) with $\left\langle \widehat{S}_{z}\right\rangle
=0$, free evolution with collisions (causes squeezing) and with spin echo
pulse applied, second $\pi /2$ pulse with phase $\phi $ followed by
detection of population difference - associated with operator $\widehat{S}%
_{z}.$

$\bullet $ Population difference measured after last $\pi /2$ pulse shows a
sinusoidal dependence on phase $\phi $ (see Fig 2b in \cite{Gross10a}). This
shows that $\left\langle \widehat{S}_{x}\right\rangle $ and $\left\langle 
\widehat{S}_{y}\right\rangle $ are non-zero, thereby showing that the state
created just prior to last pulse is entangled (see Bloch vector test (\ref%
{Eq.BlochVectorEntTest})). This does not of course show that the state is
spin squeezed.

$\bullet $ Another process involves generation of coherent spin state $(%
\widehat{a}^{\dag }+\widehat{b}^{\dag })^{N}\left\vert 0\right\rangle $
(entangled) with $\left\langle \widehat{S}_{z}\right\rangle =0$, then free
evolution with collisions (causes squeezing), followed by coupling pulse to
rotate Bloch vector through angle $\alpha $ thereby crossing $xy$ plane. The
variance in $\widehat{S}_{z}$ is then measured as $\alpha $ changes.

$\bullet $ Claimed observed spin squeezing based on $N\left\langle \Delta 
\widehat{S}_{z}^{2}\right\rangle /(\left\langle \widehat{S}_{x}\right\rangle
^{2}+\left\langle \widehat{S}_{y}\right\rangle ^{2})$ being less than $1$
(see Fig 3 in \cite{Gross10a}).

$\bullet $ Spin squeezing test is based on assumption that Bloch vector is
on Bloch sphere, a result not established since $\left\langle \widehat{S}%
_{x}\right\rangle $ and $\left\langle \widehat{S}_{y}\right\rangle $ not
measured.

$\bullet $ Claimed entanglement of ca $170$ atoms.

$\bullet $ Comment - spin squeezing in $\widehat{S}_{z}$ (almost)
demonstrated (see (\ref{Eq.SpinSqueezingMeasure2})), so entanglement is
established. An entangled state was of course already created by first $\pi
/2$ pulse, and then modified via the collisional effects.

\subsection{Gross et al. (2011) \protect\cite{Gross11a}}

$\bullet $ Stated emphasis - continuous variable entangled \emph{twin-atom
states}.

$\bullet $ System - Rb$^{87}$ in several hyperfine states.

$\bullet $ Independent BECs\ of Rb$^{87}$ trapped in separate wells in a
optical lattice.

$\bullet $ Number of atoms macroscopic, ca few $100$ in each well.

$\bullet $ Spin dynamics in Zeeman hyperfine states $(2,0)$, $(1,\pm 1)$.

$\bullet $ Initially have BEC in $(2,0)$ hyperfine state - acts as pump mode.

$\bullet $ Spin conserving collisional coupling to $(1,\pm 1)$ hyperfine
states - which act as the two mode system.

$\bullet $ One boson created in each of $(1,\pm 1)$ hyperfine states with
two bosons lost from $(2,0)$ hyperfine state due spin conserving collisions.

$\bullet $ OPA type situation associated with spin changing collisions with $%
(1,\pm 1)$ hyperfine states acting as idler, signal modes .

$\bullet $ Mean and variance of population difference between $(1,+1)$ and $%
(1,-1)$ hyperfine states measured. Total population also measured.

$\bullet $ Entanglement test is that if the variance in population
difference is small, but that in the total boson number is large then the
state is entangled (see (\ref{Eq.SzIneqality}), (\ref{Eq.NumberInequal})).

$\bullet $ Measurements (see Fig 1c in \cite{Gross11a}) show noise in
population difference is small, but that in the total boson number is large.

$\bullet $ Further entanglement test is that if there is two mode quadrature
squeezing then the state is entangled.

$\bullet $ Comment - Number squeezing and two mode quadrature squeezing
demonstrated and entanglement confirmed.\medskip

\pagebreak

\section{Discussion and Summary of Key Results}

\label{Section - Discussion & Summary of Key Results}

The two accompanying papers are concerned with mode entanglement for systems
of identical massive bosons and the relationship to spin squeezing and other
quantum correlation effects. These bosons may be atoms or molecules as in
cold quantum gases. The previous paper I\textbf{\ }dealt with the general
features of quantum entanglement and its specific definition in the case of
systems of identical bosons. In defining entanglement for systems of
identical massive particles, it was concluded that the single particle
states or modes are the most appropriate choice for sub-systems that are
distinguishable, that the general quantum states must comply both with the
symmetrisation principle and the super-selection rules (SSR) that forbid
quantum superpositions of states with differing total particle number
(global SSR compliance), and that in the separable states quantum
superpositions of sub-system states with differing sub-system particle
number (local SSR compliance) also do not occur \cite{Dalton14a}. The
present paper II has examined possible tests for two mode entanglement based
on the treatment of entanglement set out in paper I.

The present paper first defines \emph{spin squeezing} in \emph{two mode}
systems for the original spin operators $\widehat{S}_{x},\widehat{S}_{y},%
\widehat{S}_{z}$, which are defined in terms of the original mode
annihilation and creation operators $\widehat{a},\widehat{b}$ and $\widehat{a%
}^{\dag },\widehat{b}^{\dag }$. Spin squeezing for the \emph{principal spin
operators} $\widehat{J}_{x},\widehat{J}_{y},\widehat{J}_{z}$ for which the 
\emph{covariance matrix} is diagonal, rather than via the original spin
operators is then discussed. It is seen that the two sets of spin operators
are related via a rotation operator and the principal spin operators are
given in terms of \emph{new mode operators} $\widehat{c},\widehat{d}$ and $%
\widehat{c}^{\dag },\widehat{d}^{\dag }$, with $\widehat{c},\widehat{d}$
obtained as linear combinations of the original mode operators $\widehat{a},%
\widehat{b}$ and hence defining two new modes. Finally, we consider spin
squeezing in the context of \emph{multi-mode} systems.

The consequence for the case of two mode systems of identical bosons of the
present approach to defining entangled states is that spin squeezing in 
\emph{any} of the spin operators $\widehat{S}_{x}$, $\widehat{S}_{y}$ or $%
\widehat{S}_{z}$ \emph{requires} entanglement of the original modes $%
\widehat{a},\widehat{b}$. Similarly, spin squeezing in \emph{any} of the new
spin operators $\widehat{J}_{x}$, $\widehat{J}_{y}$ or $\widehat{J}_{z}$
requires entanglement of the new modes $\widehat{c},\widehat{d}$. The full
proof of these results has been presented here. A typical and \emph{simple} 
\emph{spin squeezing} test for entanglement is $\left\langle \Delta \widehat{%
S}\,_{x}^{2}\right\rangle <|\left\langle \widehat{S}_{z}\right\rangle |/2$
or $\left\langle \Delta \widehat{S}\,_{y}^{2}\right\rangle <|\left\langle 
\widehat{S}_{z}\right\rangle |/2$. We also found a simple \emph{Bloch vector}
test $\left\langle \widehat{S}_{x}\right\rangle \neq 0$ or $\left\langle 
\widehat{S}_{y}\right\rangle \neq 0$. It was noted that though spin
squeezing requires entanglement, the opposite is not the case and the $\emph{%
NOON}$ state provided an example of an entangled physical state that is not
spin squeezed. Also, the \emph{binomial state} provided an example of a
state that is entangled and spin squeezed for one choice of mode sub-systems
but is non-entangled and not spin squeezed for another choice. The \emph{%
relative phase state} provided an example that is entangled for new modes $%
\widehat{c},\widehat{d}$ and is highly spin squeezed in $\widehat{J}_{y}$
and very unsqueezed in $\widehat{J}_{x}$. We then showed that in certain 
\emph{multi-mode} cases, spin squeezing in any spin component confirmed
entanglement. In the multi-mode case this test applied in the bipartite case
(\emph{Case 1}) where the \emph{two sub-systems} each consisted of all the
modes $\widehat{a}_{i}$ \emph{or} all the modes $\widehat{b}_{i}$ or in the 
\emph{single modes} case (\emph{Case 2})where there were $2n$ sub-systems
consisting of all the modes $\widehat{a}_{i}$ \emph{and} all the modes $%
\widehat{b}_{i}$. For the \emph{mode pairs} case (\emph{Case 3}) where there
were $n$ sub-systems consisting of all the pairs of modes $\widehat{a}_{i}$
and $\widehat{b}_{i}$, a spin squeezing entanglement test was found in the
situation where for separable states each mode pair involved a \emph{single
boson}. The connection between spin squeezing and entanglement was regarded
as well-known, but up to now the only existing proofs were based on
non-entangled states that disregarded either the symmetrization principle or
the sub-system super-selection rules, placing the connection between spin
squeezing and entanglement on a somewhat shaky basis. On the other hand, the
proof given here is based on a definition of non-entangled (and hence
entangled) states that is compatible with both these requirements.

There are several papers that have obtained \emph{different tests} for
whether a state is entangled from those involving \emph{spin operators}, the
proofs often being based on a definition of non-entangled states that
ignores the sub-system SSR. Hillery et al \cite{Hillery06a} obtained
criteria of this type, such as the \emph{spin variance} entanglement test $%
\left\langle \Delta \widehat{S}\,_{x}^{2}\right\rangle +\left\langle \Delta 
\widehat{S}\,_{y}^{2}\right\rangle $ $<\frac{1}{2}\left\langle \widehat{N}%
\right\rangle $. The proof of this test has also been set out here, and the
test is also seen to be valid if the non-entangled state definition is
consistent with the SSR. The test $\left\langle \Delta \widehat{S}%
\,_{x}^{2}\right\rangle +\left\langle \Delta \widehat{S}\,_{y}^{2}\right%
\rangle $ $<|\left\langle \widehat{S}_{z}\right\rangle |$ suggested by the
requirement that $\left\langle \Delta \widehat{S}\,_{x}^{2}\right\rangle
+\left\langle \Delta \widehat{S}\,_{y}^{2}\right\rangle \geq |\left\langle 
\widehat{S}_{z}\right\rangle |$ for non-entangled states - since both $%
\left\langle \Delta \widehat{S}\,_{x}^{2}\right\rangle \geq |\left\langle 
\widehat{S}_{z}\right\rangle |/2$ and $\left\langle \Delta \widehat{S}%
\,_{y}^{2}\right\rangle \geq |\left\langle \widehat{S}_{z}\right\rangle |/2$
is of no use, since $\left\langle \Delta \widehat{S}\,_{x}^{2}\right\rangle
+\left\langle \Delta \widehat{S}\,_{y}^{2}\right\rangle $ $\geq
|\left\langle \widehat{S}_{z}\right\rangle |$ for all states. However as
previously noted, showing that either $\left\langle \Delta \widehat{S}%
\,_{x}^{2}\right\rangle <|\left\langle \widehat{S}_{z}\right\rangle |/2$ or $%
\left\langle \Delta \widehat{S}\,_{y}^{2}\right\rangle <|\left\langle 
\widehat{S}_{z}\right\rangle |/2$ - or the analogous tests for other pairs
of spin operators - already provides a test for the entanglement of the
original modes $\widehat{a},\widehat{b}$. This test is a different test for
entanglement than that of Hillery et al \cite{Hillery06a}. In fact the case
of the \emph{relative phase eigenstate} is an example of an entangled state
in which the simple spin squeezing test for entanglement \emph{succeeds}
whereas that of Hillery et al \cite{Hillery06a} \emph{fails}. The
consequences of applying both the simple spin squeezing and the Hillery spin
operator test were examined with the aim of seeing whether the results could
determine whether or not the local particle number SSR applied to separable
states. The conclusion was negative as all outcomes were consistent with
both possibilities. In addition, the Hillery spin variance test was also
shown to apply to the multi-mode situation, in the Cases 1 and 2 described
above, but did not apply in Case 3. Other entanglement tests of Benatti et
al \cite{Benatti11a} involving variances of two mode spin operators were
also found to apply for identical bosons.

The present paper also considered \emph{correlation tests} for entanglement.
Inequalities found by Hillery et al \cite{Hillery09a} for non-entangled
states which also do not depend on whether non-entangled states satisfy the
super-selection rule include $|\left\langle (\widehat{a})^{m}\,(\widehat{b}%
^{\dag })^{n}\right\rangle |^{2}\leq \left\langle (\widehat{a}^{\dag })^{m}(%
\widehat{a})^{m}\,(\widehat{b}^{\dag })^{n}(\widehat{b})^{n}\right\rangle $,
giving a valid \emph{strong correlation} test $|\left\langle (\widehat{a}%
)^{m}\,(\widehat{b}^{\dag })^{n}\right\rangle |^{2}>\left\langle (\widehat{a}%
^{\dag })^{m}(\widehat{a})^{m}\,(\widehat{b}^{\dag })^{n}(\widehat{b}%
)^{n}\right\rangle $ for an entangled state. However, with entanglement
defined as in the present paper we have $|\left\langle (\widehat{a})^{m}\,(%
\widehat{b}^{\dag })^{n}\right\rangle |^{2}=0$ for a non-entangled state, so
we have also proved a \emph{weak correlation} test for entanglement in the
form $|\left\langle (\widehat{a})^{m}\,(\widehat{b}^{\dag
})^{n}\right\rangle |^{2}>0$. This test is less stringent than the strong
correlation test of Hillery et al \cite{Hillery09a}., as $|\left\langle (%
\widehat{a})^{m}\,(\widehat{b}^{\dag })^{n}\right\rangle |^{2}$ is then
required to be larger. In all these cases For $n\neq m$ none of these cases
are of interest since for global SSR compliant states $\left\langle (%
\widehat{a})^{m}\,(\widehat{b}^{\dag })^{n}\right\rangle $ would be zero. In
the cases where $n=m$ we show that \emph{all} the correlation tests can be
expressed in terms of \emph{spin operators}, so they reduce to tests
involving powers of spin operators. For the case $n=m=1$ the weak
correlation test is the same as the Bloch vector test.

Work by other authors on bipartite entanglement tests has also been examined
here. He et al \cite{He11a}, \cite{He12a} considered a \emph{four mode}
system, with two modes localised in each well of a double well potential. If
the two sub-systems $A$ and $B$ each consist of two modes - with $\widehat{a}%
_{1}$, $\widehat{a}_{2}$ as sub-system $A$ and $\widehat{b}_{1}$, $\widehat{b%
}_{2}$ as sub-system $B$, then tests of bipartite entanglement of the two
sub-systems of the Hillery \cite{Hillery09a} type $|\left\langle (\widehat{a}%
_{i})^{m}\,(\widehat{b}_{j}^{\dag })^{n}\right\rangle |^{2}>\left\langle (%
\widehat{a}_{i}^{\dag })^{m}(\widehat{a}_{i})^{m}\,(\widehat{b}_{j}^{\dag
})^{n}(\widehat{b}_{j})^{n}\right\rangle $ for any $i$, $j=1$, $2$ or
involving local spin operators $|\left\langle \widehat{S}_{+}^{A}\,\widehat{S%
}_{-}^{B}\right\rangle |^{2}>\left\langle \widehat{S}_{+}^{A}\,\widehat{S}%
_{-}^{A}\,\widehat{S}_{+}^{B}\,\widehat{S}_{-}^{B}\,\right\rangle $ apply.
Raymer et al \cite{Raymer03a} have also considered such a four mode system
and derived bipartite entanglement tests such as\textbf{\ }$\left\langle
\Delta (\widehat{S}\,_{x}^{A}\pm \widehat{S}\,_{x}^{B})^{2}\right\rangle
+\left\langle \Delta (\widehat{S}\,_{y}^{A}\mp \widehat{S}%
\,_{y}^{B})^{2}\right\rangle <|\left\langle \widehat{S}_{z}\right\rangle |$%
\textbf{\ }that involve local spin operators for the two sub-systems.

We also considered the work of Sorensen et al \cite{Sorensen01a}, who showed
that spin squeezing is a test for a state being entangled, but defined
non-entangled states for identical particle systems (such as BECs) in a form
that is \emph{inconsistent} with the symmetrisation principle - the
sub-systems being regarded as individual identical particles. However, the
treatment of Sorensen et al \cite{Sorensen01a} can be modified to apply to a
system of identical bosons if the particle index $i$ is \emph{re-interpreted}
as specifying diffferent modes, for example modes localised on optical
lattice sites $i=1,2,..,n$ or localised in momentum space. With two single
particle states $\left\vert \phi _{ai}\right\rangle $, $\left\vert \phi
_{bi}\right\rangle $ with annihilation operators $a_{i},b_{i}$ available on
each site, there would then be $2n$ modes involved, but spin operators can
still be defined. This is just a particular case of the multi-mode situation
described above. If the definitions of non--entangled and entangled states
in the present paper are applied, it can be shown that spin squeezing in
either of the spin operators $\widehat{S}_{x}$ or $\widehat{S}_{y}$ requires
entanglement of \emph{all} the original modes $\widehat{a}_{i},\widehat{b}%
_{i}$ (Case 1, above). Alternatively, if the sub-systems are \emph{pairs} of
modes $\widehat{a}_{i},\widehat{b}_{i}$ \emph{and} the sub-system density
operators $\widehat{\rho }_{R}^{i}$ were restricted to states with exactly 
\emph{one boson}, then it can be shown that spin squeezing in $\widehat{S}%
_{z}$ requires entanglement of all the pairs of modes (Case 3, above). With
this restriction the pair of modes $\widehat{a}_{i}$, $\widehat{b}_{i}$
behave like \emph{distinguishable} two state particles, which was
essentially the case that Sorensen et al \cite{Sorensen01a} implicitly
considered. This type of entanglement is a multi-mode entanglement of a
special type - since the modes $\widehat{a}_{i}$, $\widehat{b}_{i}$ may
themselves be entangled there is an "entanglement of entanglement". So with
either of these revisions, the work of Sorensen et al \cite{Sorensen01a}
could be said to show that spin squeezing requires entanglement. However,
neither of these revisions really deals with the case of entanglement in%
\emph{\ two mode} systems, and here the proof given in this paper showing
that spin squeezing in $\widehat{S}_{z}$ requires entanglement of the two
modes provides the justification of this result \emph{without} treating
identical particles as distinguishable sub-systems. Sorensen and Molmer \cite%
{Sorensen01b} have also deduced an inequality involving $\left\langle \Delta 
\widehat{S}\,_{x}^{2}\right\rangle $ and $|\left\langle \widehat{S}%
\,_{z}\right\rangle |$ for states where $\left\langle \widehat{S}%
\,_{x}\right\rangle =\left\langle \widehat{S}\,_{y}\right\rangle =0$ based
on just the Heisenberg uncertainty principle. This is useful in terms of
confirming that states do exist that are spin squeezed still conform to this
principle.

Entanglement tests involving quadrature variables have also been published,
so we have also examined these. Duan et al \cite{Duan00a}, Toth et al \cite%
{Toth03a} devised a test for entanglement based on the sum of the \emph{%
quadrature variances} $\left\langle \Delta (\widehat{x}_{A}\pm \widehat{x}%
_{B})^{2}\right\rangle +\left\langle \Delta (\widehat{p}_{A}\mp \widehat{p}%
_{B})^{2}\right\rangle \geq 2$ for separable states, which involve
quadrature components $\widehat{x}_{A},\widehat{p}_{A},\widehat{x}_{B},%
\widehat{p}_{B}$ constructed from the original mode annihilation, creation
operators for modes $A$, $B$. Their conclusion that if the\textbf{\ }%
quadrature variances sum is less than $2$ then the state is entangled is
valid both for the present definition of entanglement and for that in which
the application of the super-selection rule is ignored. However, for quantum
states for systems of identical bosons that are global SSR compliant $%
\left\langle \Delta (\widehat{x}_{A}\pm \widehat{x}_{B})^{2}\right\rangle
+\left\langle \Delta (\widehat{p}_{A}\mp \widehat{p}_{B})^{2}\right\rangle
=2+2\left\langle \widehat{N}\right\rangle $\ for\textbf{\ }\emph{all}\textbf{%
\ }such states - both separable and entangled, and although this is
consistent with \cite{Duan00a}, \cite{Toth03a} we have concluded that the
quadrature variance test can \emph{never} confirm entanglement. A more
general test \cite{Reid89a} involving quadrature operators $\widehat{X}%
_{A}^{\theta },\widehat{X}_{B}^{\theta }$ required showing that $%
\left\langle \widehat{X}_{A}^{\theta }\,\widehat{X}_{B}^{\phi }\right\rangle
\neq 0$. This was shown to be equivalent to showing that $\left\langle 
\widehat{S}_{x}\right\rangle \neq 0$ or $\left\langle \widehat{S}%
_{y}\right\rangle \neq 0$, the \emph{Bloch vector} or \emph{weak correlation}
test. A \emph{two mode quadrature squeezing} test was also obtained, but
found to be less useful than the \emph{Bloch} \emph{vector} test. 

Overall then, all of the \emph{entanglement tests} (spin squeezing and
other) in the other papers discussed here are \emph{still valid} when
reconsidered in accord with the definition of entanglement based on the
symmetrisation and super-selection rules, though in one case Sorensen et al 
\cite{Sorensen01a} a re-definition of the sub-systems is required to satisfy
the symmetrization principle. However, \emph{further} tests for entanglement
are obtained in the present paper based on non-entangled states that are
consistent with the symmetrization and super-selection rules. In some cases
they are less stringent - the correlation test in Eq.(\ref{Eq.EntangTest})
being easier to satisfy than that of Hillery et al \cite{Hillery09a} in Eq. (%
\ref{Eq.HilleryEntangTest}). The tests introduced here are certainly \emph{%
different} to others previously discovered.

The theory for a simple\emph{\ two mode interferometer} was then presented
and it was shown that such an interferometer can be used to measure the mean
values and covariance matrix for the spin operators involved in entanglement
tests for the two mode bosonic system. The treatmebnt was also generalised
to \emph{multi-mode} interferometry. The interferometer involved a \emph{%
pulsed classical field} characterised by a \emph{phase} variable $\phi $ and
an \emph{area} variable $2s=\theta $ defined by the time integral of the
field amplitude, and leads to a coupling between the two modes. For
simplicity the centre frequency was chosen to be \emph{resonant} with the
mode transition frequency. Measuring the mean and variance of the\emph{\
population} \emph{difference} between the two modes for the \emph{output}
state of the interferometer for various choices of $\phi $ and $\theta $
enabled the mean values and covariance matrix for the spin operators for the 
\emph{input} quantum state of the two mode system to be determined. More
complex interferometers were seen to involve combinations of simple
interferometers separated by time intervals during which \emph{free evolution%
} of the two mode system can occur, including the effect of \emph{collisions}%
.

Experiments have been carried out demonstrating that spin squeezing occurs,
which according to theory requires entanglement. An analysis of these
experiments has been presented here. However, since no results for
entanglement \emph{measures} are presented or other\emph{\ independent}
tests for entanglement carried out, the entanglement presumably created in
the experiments has not been independently confirmed. \pagebreak

\section{Acknowledgements}

The authors thank S. M. Barnett, F. Binder, Th. Busch, J. F. Corney, P. D.
Drummond, M. Hall, L. Heaney, J. Jeffers, U. Marzolini,\textbf{\ }K. Molmer,
D. Oi, M. Plenio, K. Rzazewski, T. Rudolph, J. A. Vaccaro, V. Vedral and H.
W. Wiseman for helpful discussions. BJD thanks the Science Foundation of
Ireland for funding this research via an E\ T S Walton Visiting Fellowship
and E Hinds for the hospitality of the Centre for Cold Matter, Imperial
College, London during this work. MDR acknowledges support from the
Australian Research Council via a Discovery Project Grant. \pagebreak

\section{Appendix A - Spin Squeezing Test for Bipartite Multi-Mode Case}

\label{Appendix = MultiMo Spin Sq Choice 1}

We now consider spin squeezing for the multi-mode spin operators given in
Eqs. (\ref{Eq.SpinFieldOprs}) and (\ref{Eq.SpinOprs2}) in SubSection \ref%
{SubSection - Spin Operators Multimode Case}. We consider separable states
for \emph{Case 1}, the density operator being given in Eq. (\ref%
{Eq.SepStatesMultiModeCase1}). In this \emph{bipartite case} the two
subsystems consist of \emph{all} modes $\widehat{a}_{i}$ and \emph{all}
modes $\widehat{b}_{i}$. The development involves expressions such as $%
\left\langle \widehat{\Psi }_{c}(\mathbf{r})\right\rangle _{R}^{C}=Tr_{C}(%
\widehat{\Psi }_{c}(\mathbf{r})\widehat{\rho }_{R}^{C})$, $\left\langle 
\widehat{\Psi }_{c}^{\dag }(\mathbf{r})\right\rangle _{R}^{C}=Tr_{C}(%
\widehat{\Psi }_{c}^{\dag }(\mathbf{r})\widehat{\rho }_{R}^{C})$ and $%
\left\langle \widehat{\Psi }_{c}^{\dag }(\mathbf{r})\widehat{\Psi }%
_{c}^{\dag }(\mathbf{r}^{\prime })\right\rangle _{R}^{C}=Tr_{C}(\widehat{%
\Psi }_{c}^{\dag }(\mathbf{r})\widehat{\Psi }_{c}^{\dag }(\mathbf{r}^{\prime
})\widehat{\rho }_{R}^{C})$, $\left\langle \widehat{\Psi }_{c}(\mathbf{r})%
\widehat{\Psi }_{c}(\mathbf{r}^{\prime })\right\rangle _{R}^{C}=Tr_{C}(%
\widehat{\Psi }_{c}(\mathbf{r})\widehat{\Psi }_{c}(\mathbf{r}^{\prime })%
\widehat{\rho }_{R}^{C})$, $\left\langle \widehat{\Psi }_{c}^{\dag }(\mathbf{%
r})\widehat{\Psi }_{c}(\mathbf{r}^{\prime })\right\rangle _{R}^{C}=Tr_{C}(%
\widehat{\Psi }_{c}^{\dag }(\mathbf{r})\widehat{\Psi }_{c}(\mathbf{r}%
^{\prime })\widehat{\rho }_{R}^{C})$, where $C=A,B$.

Firstly, we have 
\begin{equation}
\left\langle \widehat{S}\,_{x}\right\rangle _{R}=\frac{1}{2}\tint d\mathbf{%
r\,}\left( \left\langle \widehat{\Psi }_{b}^{\dag }(\mathbf{r})\right\rangle
_{R}^{B}\left\langle \widehat{\Psi }_{a}(\mathbf{r})\right\rangle
_{R}^{A}+\left\langle \widehat{\Psi }_{a}^{\dag }(\mathbf{r})\right\rangle
_{R}^{A}\left\langle \widehat{\Psi }_{b}(\mathbf{r})\right\rangle
_{R}^{B}\right) =0
\end{equation}%
since from the local particle number SSR for sub-systems $A$ and $B$ we have 
$\left\langle \widehat{\Psi }_{b}^{\dag }(\mathbf{r})\right\rangle
_{R}^{B}=\left\langle \widehat{\Psi }_{a}(\mathbf{r})\right\rangle
_{R}^{A}=0 $. A similar result applies to $\left\langle \widehat{S}%
\,_{y}\right\rangle _{R}$ so it then follows that 
\begin{equation}
\left\langle \widehat{S}\,_{x}\right\rangle =\left\langle \widehat{S}%
\,_{y}\right\rangle =0  \label{Eq.BlochVectorCase1}
\end{equation}%
This immediately yields the \emph{Bloch vector} entanglement test. It also
leads to the \emph{spin squeezing} in $\widehat{S}\,_{z}$ entanglement test,
namely if $\widehat{S}\,_{z}$ is squeezed with respect to $\widehat{S}\,_{x}$
or $\widehat{S}\,_{y}$ (or vice versa), then the state must be entangled.
The question then is: Does spin squeezing in $\widehat{S}\,_{x}$ with
respect to $\widehat{S}\,_{y}$ (or vice versa) require the state to be
entangled for the two $n$ mode sub-systems $A$ and $B$?

To obtain an inequality for the variance in $\widehat{S}\,_{x}$, we see that 
\begin{eqnarray}
\left\langle \widehat{S}\,_{x}^{2}\right\rangle _{R} &=&\frac{1}{4}\tiint d%
\mathbf{r\,}d\mathbf{r}^{\prime }\mathbf{\,}\times \{\left\langle \widehat{%
\Psi }_{b}^{\dag }(\mathbf{r})\widehat{\Psi }_{b}^{\dag }(\mathbf{r}^{\prime
})\right\rangle _{R}^{B}\left\langle \widehat{\Psi }_{a}(\mathbf{r})\widehat{%
\Psi }_{a}(\mathbf{r}^{\prime })\right\rangle _{R}^{A}  \nonumber \\
&&+\left\langle \widehat{\Psi }_{b}^{\dag }(\mathbf{r})\widehat{\Psi }_{b}(%
\mathbf{r}^{\prime })\right\rangle _{R}^{B}\left\langle \widehat{\Psi }_{a}(%
\mathbf{r})\widehat{\Psi }_{a}^{\dag }(\mathbf{r}^{\prime })\right\rangle
_{R}^{A}+\left\langle \widehat{\Psi }_{b}(\mathbf{r})\widehat{\Psi }%
_{b}^{\dag }(\mathbf{r}^{\prime })\right\rangle _{R}^{B}\left\langle 
\widehat{\Psi }_{a}^{\dag }(\mathbf{r})\widehat{\Psi }_{a}(\mathbf{r}%
^{\prime })\right\rangle _{R}^{A}  \nonumber \\
&&+\left\langle \widehat{\Psi }_{b}(\mathbf{r})\widehat{\Psi }_{b}(\mathbf{r}%
^{\prime })\right\rangle _{R}^{B}\left\langle \widehat{\Psi }_{a}^{\dag }(%
\mathbf{r})\widehat{\Psi }_{a}^{\dag }(\mathbf{r}^{\prime })\right\rangle
_{R}^{A}\}
\end{eqnarray}%
From the local particle number SSR for sub-systems $A$ and $B$ we have $%
\left\langle \widehat{\Psi }_{b}^{\dag }(\mathbf{r})\widehat{\Psi }%
_{b}^{\dag }(\mathbf{r}^{\prime })\right\rangle _{R}^{B}=\left\langle 
\widehat{\Psi }_{a}(\mathbf{r})\widehat{\Psi }_{a}(\mathbf{r}^{\prime
})\right\rangle _{R}^{A}=0$, so the first and fourth terms are zero. Using
the field operator commutation rules we then obtain 
\begin{eqnarray}
\left\langle \widehat{S}\,_{x}^{2}\right\rangle _{R} &=&\frac{1}{2}\tiint d%
\mathbf{r\,}d\mathbf{r}^{\prime }\mathbf{\,}\left\langle \widehat{\Psi }%
_{b}^{\dag }(\mathbf{r})\widehat{\Psi }_{b}(\mathbf{r}^{\prime
})\right\rangle _{R}^{B}\left\langle \widehat{\Psi }_{a}^{\dag }(\mathbf{r}%
^{\prime })\widehat{\Psi }_{a}(\mathbf{r})\right\rangle _{R}^{A}  \nonumber
\\
&&+\frac{1}{4}\tint d\mathbf{r\,\,\{}\left\langle \widehat{\Psi }_{b}^{\dag
}(\mathbf{r})\widehat{\Psi }_{b}(\mathbf{r})\right\rangle
_{R}^{B}+\left\langle \widehat{\Psi }_{a}^{\dag }(\mathbf{r})\widehat{\Psi }%
_{a}(\mathbf{r})\right\rangle _{R}^{A}\}
\end{eqnarray}%
so that 
\begin{eqnarray}
\left\langle \Delta \widehat{S}\,_{x}^{2}\right\rangle _{R} &=&\frac{1}{2}%
\tiint d\mathbf{r\,}d\mathbf{r}^{\prime }\mathbf{\,}\left\langle \widehat{%
\Psi }_{b}^{\dag }(\mathbf{r})\widehat{\Psi }_{b}(\mathbf{r}^{\prime
})\right\rangle _{R}^{B}\left\langle \widehat{\Psi }_{a}^{\dag }(\mathbf{r}%
^{\prime })\widehat{\Psi }_{a}(\mathbf{r})\right\rangle _{R}^{A}  \nonumber
\\
&&+\frac{1}{4}\tint d\mathbf{r\,\,\{}\left\langle \widehat{\Psi }_{b}^{\dag
}(\mathbf{r})\widehat{\Psi }_{b}(\mathbf{r})\right\rangle
_{R}^{B}+\left\langle \widehat{\Psi }_{a}^{\dag }(\mathbf{r})\widehat{\Psi }%
_{a}(\mathbf{r})\right\rangle _{R}^{A}\}
\end{eqnarray}

Hence from (\ref{Eq.VarianceResult}) 
\begin{eqnarray}
&&\left\langle \Delta \widehat{S}\,_{x}^{2}\right\rangle  \nonumber \\
&\geq &\tsum\limits_{R}P_{R}\{\frac{1}{2}\tiint d\mathbf{r\,}d\mathbf{r}%
^{\prime }\mathbf{\,}\left\langle \widehat{\Psi }_{b}^{\dag }(\mathbf{r})%
\widehat{\Psi }_{b}(\mathbf{r}^{\prime })\right\rangle _{R}^{B}\left\langle 
\widehat{\Psi }_{a}^{\dag }(\mathbf{r}^{\prime })\widehat{\Psi }_{a}(\mathbf{%
r})\right\rangle _{R}^{A}  \nonumber \\
&&+\frac{1}{4}\tint d\mathbf{r\,\,\{}\left\langle \widehat{\Psi }_{b}^{\dag
}(\mathbf{r})\widehat{\Psi }_{b}(\mathbf{r})\right\rangle
_{R}^{B}+\left\langle \widehat{\Psi }_{a}^{\dag }(\mathbf{r})\widehat{\Psi }%
_{a}(\mathbf{r})\right\rangle _{R}^{A}\}  \label{Eq.IneqVarSxMultiMode}
\end{eqnarray}%
The same result applies to $\left\langle \Delta \widehat{S}%
\,_{y}^{2}\right\rangle $.

Now we can easily show that 
\begin{equation}
\left\langle \widehat{S}\,_{z}\right\rangle =\tsum\limits_{R}P_{R}\frac{1}{2}%
\tint d\mathbf{r\,\{}\left\langle \widehat{\Psi }_{b}^{\dag }(\mathbf{r})%
\widehat{\Psi }_{b}(\mathbf{r})\right\rangle _{R}^{B}-\left\langle \widehat{%
\Psi }_{a}^{\dag }(\mathbf{r})\widehat{\Psi }_{a}(\mathbf{r})\right\rangle
_{R}^{A}\}
\end{equation}%
so that 
\begin{equation}
\frac{1}{2}|\left\langle \widehat{S}\,_{z}\right\rangle |\leq
\tsum\limits_{R}P_{R}\frac{1}{4}\tint d\mathbf{r\,\{}\left\langle \widehat{%
\Psi }_{b}^{\dag }(\mathbf{r})\widehat{\Psi }_{b}(\mathbf{r})\right\rangle
_{R}^{B}+\left\langle \widehat{\Psi }_{a}^{\dag }(\mathbf{r})\widehat{\Psi }%
_{a}(\mathbf{r})\right\rangle _{R}^{A}\}
\end{equation}%
as $\left\langle \widehat{\Psi }_{b}^{\dag }(\mathbf{r})\widehat{\Psi }_{b}(%
\mathbf{r})\right\rangle _{R}^{B}$ and $\left\langle \widehat{\Psi }%
_{a}^{\dag }(\mathbf{r})\widehat{\Psi }_{a}(\mathbf{r})\right\rangle
_{R}^{A} $ are real and positive.

Hence we find that 
\begin{eqnarray}
&&\left\langle \Delta \widehat{S}\,_{x}^{2}\right\rangle -\frac{1}{2}%
|\left\langle \widehat{S}\,_{z}\right\rangle |  \nonumber \\
&\geq &\tsum\limits_{R}P_{R}\frac{1}{2}\tiint d\mathbf{r\,}d\mathbf{r}%
^{\prime }\mathbf{\,}\left\langle \widehat{\Psi }_{b}^{\dag }(\mathbf{r})%
\widehat{\Psi }_{b}(\mathbf{r}^{\prime })\right\rangle _{R}^{B}\left\langle 
\widehat{\Psi }_{a}^{\dag }(\mathbf{r}^{\prime })\widehat{\Psi }_{a}(\mathbf{%
r})\right\rangle _{R}^{A}  \label{Eq.IneqForm1} \\
&=&\tsum\limits_{R}P_{R}\frac{1}{2}\tiint d\mathbf{r\,}d\mathbf{r}^{\prime }%
\mathbf{\,}Tr_{B}\{\widehat{\Psi }_{b}(\mathbf{r}^{\prime })\,\widehat{\rho }%
_{R}^{B}\;\widehat{\Psi }_{b}^{\dag }(\mathbf{r})\}\,Tr_{A}\{\widehat{\Psi }%
_{a}(\mathbf{r})\,\widehat{\rho }_{R}^{A}\;\widehat{\Psi }_{a}^{\dag }(%
\mathbf{r}^{\prime })\}  \label{Eq.IneqForm2} \\
&=&\frac{1}{2}\tiint d\mathbf{r\,}d\mathbf{r}^{\prime }\mathbf{\,}Tr\left\{ 
\widehat{\Psi }_{a}(\mathbf{r})\,\widehat{\Psi }_{b}(\mathbf{r}^{\prime })\,%
\widehat{\rho }_{sep}\,\widehat{\Psi }_{a}^{\dag }(\mathbf{r}^{\prime })\,%
\widehat{\Psi }_{b}^{\dag }(\mathbf{r})\right\}  \label{Eq.IneqForm3}
\end{eqnarray}%
giving three forms that the inequality for $\left\langle \Delta \widehat{S}%
\,_{x}^{2}\right\rangle -\frac{1}{2}|\left\langle \widehat{S}%
\,_{z}\right\rangle |$ has to satisfy in the case of a separable state. The
last form involves a double space integral of a quantum correlation
function. Note the order of $\mathbf{r}$ and $\mathbf{r}^{\prime }$. It is
straightforward to show that the right side of the inequality is real, but
to achieve an entanglement test involving spin squeezing for $\widehat{S}%
\,_{x}$ we need to show that it is non-negative. Identical inequalities can
be found for $\left\langle \Delta \widehat{S}\,_{y}^{2}\right\rangle -\frac{1%
}{2}|\left\langle \widehat{S}\,_{z}\right\rangle |$.

\subsubsection{Mode Expansions}

If we use Eq.(\ref{Eq.FieldOprs}) to expand the field operators then using
Eq.(\ref{Eq.IneqForm2}) we have%
\begin{eqnarray}
&&\left\langle \Delta \widehat{S}\,_{x}^{2}\right\rangle -\frac{1}{2}%
|\left\langle \widehat{S}\,_{z}\right\rangle |  \nonumber \\
&\geq &\tsum\limits_{R}P_{R}\frac{1}{2}\tsum\limits_{ij}\tsum\limits_{kl}%
\tiint d\mathbf{r\,}d\mathbf{r}^{\prime }\mathbf{\,}  \nonumber \\
&&\times \left\{ \phi _{i}(\mathbf{r})\phi _{j}^{\ast }(\mathbf{r}^{\prime
})\phi _{k}(\mathbf{r}^{\prime })\phi _{l}^{\ast }(\mathbf{r})\right\} 
\nonumber \\
&&\times \left\{ Tr_{A}\{\widehat{a}_{i}\,\widehat{\rho }_{R}^{A}\;\widehat{a%
}_{j}^{\dag }\}\;Tr_{B}\{\widehat{b}_{k}\,\widehat{\rho }_{R}^{B}\;\widehat{b%
}_{l}^{\dag }\}\right\}  \nonumber \\
&=&\tsum\limits_{R}P_{R}\frac{1}{2}\tsum\limits_{ij}\left\{ Tr_{A}\{\widehat{%
a}_{i}\,\widehat{\rho }_{R}^{A}\;\widehat{a}_{j}^{\dag }\}\;Tr_{B}\{\widehat{%
b}_{j}\,\widehat{\rho }_{R}^{B}\;\widehat{b}_{i}^{\dag }\}\right\}  \nonumber
\\
&=&\tsum\limits_{R}P_{R}\frac{1}{4}\tsum\limits_{ij}(A_{ij}^{R}%
\;B_{ji}^{R}+B_{ij}^{R}\;A_{ji}^{R})  \label{Eq.InequalityModeExpn1} \\
&=&\tsum\limits_{R}P_{R}\frac{1}{4}Tr\{A^{R}B^{R}+B^{R}A^{R}\}
\label{Eq.InequalityModeMatrixForm}
\end{eqnarray}%
where mode orthogonality has been used and we have introduced \emph{matrices}
$A^{R}$ and $B^{R}$ whose elements are 
\begin{equation}
A_{ij}^{R}=Tr_{A}\{\widehat{a}_{i}\,\widehat{\rho }_{R}^{A}\;\widehat{a}%
_{j}^{\dag }\}\qquad B_{ji}^{R}=Tr_{B}\{\widehat{b}_{j}\,\widehat{\rho }%
_{R}^{B}\;\widehat{b}_{i}^{\dag }\}  \label{Eq.ModeMatrices}
\end{equation}%
It is easy to show that $A_{ij}^{R}=(A_{ji}^{R})^{\ast \text{ }}$and $%
B_{ij}^{R}=(B_{ji}^{R})^{\ast \text{ }}$showing that the matrices $A^{R}$
and $B^{R}$ are \emph{Hermitian}, as is $A^{R}B^{R}+B^{R}A^{R}$. The
quantity $\tsum\limits_{ij}(A_{ij}^{R}\;B_{ji}^{R}+B_{ij}^{R}\;A_{ji}^{R})$
is \emph{real}. The question is: Is it also \emph{positive} ?

For the simple case where there is only \emph{one} spatial mode for each
component the right side of the inequality is just equal to $%
\tsum\limits_{R}P_{R}\frac{1}{2}\left\{ Tr_{A}\{\widehat{a}\,\widehat{\rho }%
_{R}^{A}\;\widehat{a}^{\dag }\}\;Tr_{B}\{\widehat{b}\,\widehat{\rho }%
_{R}^{B}\;\widehat{b}^{\dag }\}\right\} =\tsum\limits_{R}P_{R}\frac{1}{2}%
N_{R}^{A}\,N_{R}^{B}$ , where $N_{R}^{A}$ and $N_{R}^{B}$ give the mean
numbers of bosons in sub-systems $A$ and $B$ for the states $\widehat{\rho }%
_{R}^{A}$ and $\widehat{\rho }_{R}^{B}$. The right side of the inequality is
positive, showing that the separable state is not spin squeezed for $%
\widehat{S}\,_{x}$ with respect to $\widehat{S}\,_{y}$ (or vice versa),
leading as before to the test that such spin squeezing requires entanglement.

\subsubsection{Positive Definiteness}

For the multi-mode case we now take into account that the sub-system density
operators $\widehat{\rho }_{R}^{A}$ and $\widehat{\rho }_{R}^{B}$ are \emph{%
positive-definite}. Their eigenvalues $\pi _{\lambda }^{AR}$ and $\pi _{\mu
}^{BR}$ are real and non-negative as well as summing to unity, and we can
write the density operators in terms of their orthonormal eigenvectors $%
\left\vert AR,\lambda \right\rangle $ and $\left\vert BR,\mu \right\rangle $
as 
\begin{equation}
\widehat{\rho }_{R}^{A}=\tsum\limits_{\lambda }\pi _{\lambda
}^{AR}\,\left\vert AR,\lambda \right\rangle \left\langle AR,\lambda
\right\vert \qquad \widehat{\rho }_{R}^{B}=\tsum\limits_{\mu }\pi _{\mu
}^{BR}\,\left\vert BR,\mu \right\rangle \left\langle BR,\mu \right\vert
\label{Eq.SubSystDensityOprs}
\end{equation}

Then from (\ref{Eq.ModeMatrices}) 
\begin{equation}
A_{ij}^{R}=\sum_{\lambda }\pi _{\lambda }^{AR}\left\langle AR,\lambda
\right\vert \;\widehat{a}_{j}^{\dag }\widehat{a}_{i}\,\left\vert AR,\lambda
\right\rangle \qquad B_{ji}^{R}=\sum_{\mu }\pi _{\mu }^{BR}\left\langle
BR,\mu \right\vert \;\widehat{b}_{i}^{\dag }\widehat{b}_{j}\,\left\vert
BR,\mu \right\rangle  \label{Eq.MatricesAandB}
\end{equation}%
Consider a $1\times n$ row matrix $\xi ^{\dag }=\{\xi _{1}^{\ast },\xi
_{2}^{\ast },..,\xi _{n}^{\ast }$ $\}$ 
\begin{eqnarray}
\xi ^{\dag }A^{R}\xi &=&\sum_{ij}\xi _{i}^{\ast }\,A_{ij}^{R}\,\xi _{j} 
\nonumber \\
&=&\sum_{\lambda }\pi _{\lambda }^{AR}\sum_{ij}\xi _{i}^{\ast
}\,\left\langle AR,\lambda \right\vert \;\widehat{a}_{j}^{\dag }\widehat{a}%
_{i}\,\left\vert AR,\lambda \right\rangle \,\xi _{j}  \nonumber \\
&=&\sum_{\lambda }\pi _{\lambda }^{AR}\;\left\langle AR,\lambda \right\vert
\;\widehat{\Omega }_{A}^{\dag }\,\widehat{\Omega }_{A}\,\left\vert
AR,\lambda \right\rangle \,
\end{eqnarray}%
where we have introduced the operator $\widehat{\Omega }_{A}=\sum_{i}\xi
_{i}^{\ast }\,\widehat{a}_{i}$. Since $\xi ^{\dag }A^{R}\xi $ is always
non-negative for all $\xi $, this shows that $A^{R\text{ }}$is a \emph{%
positive definite} matrix. Similarly, considering a $1\times n$ row matrix $%
\eta ^{\dag }=\{\eta _{1}^{\ast },\eta _{2}^{\ast },..,\eta _{n}^{\ast }$ $%
\} $ and introducing the operator $\widehat{\Omega }_{B}=\sum_{i}\eta _{i}\,%
\widehat{b}_{i}$ we find that 
\begin{equation}
\eta ^{\dag }B^{R}\eta =\sum_{\mu }\pi _{\mu }^{BR}\;\left\langle BR,\mu
\right\vert \;\widehat{\Omega }_{B}^{\dag }\,\widehat{\Omega }%
_{B}\,\left\vert BR,\mu \right\rangle \,
\end{equation}%
which is also always non-negative, showing that $B^{R\text{ }}$is also a 
\emph{positive definite} matrix.

We can then express the positive definite Hermitian matrices $A^{R\text{ }}$%
and $B^{R\text{ }}$ in terms of their normalised column eigenvectors $\theta
_{\alpha }^{A}$ and $\zeta _{\beta \text{ }}^{B}$ respectively, where the
corresponding real, positive eigenvalues are $\nu _{\alpha }$ and $\sigma
_{\beta }$ Thus we have (for ease of notation $R$ will be left understood) 
\begin{eqnarray}
A^{R\text{ }}\theta _{\alpha }^{A} &=&\nu _{\alpha }\theta _{\alpha
}^{A}\,\qquad (\theta _{\alpha }^{A})^{\dag }\theta _{\gamma }^{A}=\delta
_{\alpha \gamma }\qquad A^{R\text{ }}=\sum_{\alpha }\nu _{\alpha }\theta
_{\alpha }^{A}\,(\theta _{\alpha }^{A})^{\dag }  \nonumber \\
B^{R\text{ }}\zeta _{\beta }^{B} &=&\sigma _{\beta }\zeta _{\beta
}^{B}\,\qquad (\zeta _{\beta }^{B})^{\dag }\zeta _{\epsilon }^{B}=\delta
_{\beta \epsilon }\qquad B^{R\text{ }}=\sum_{\beta }\sigma _{\beta }\zeta
_{\beta }^{B}\,(\zeta _{\beta }^{B})^{\dag }  \label{Eq.MatrixEigevectorAanB}
\end{eqnarray}%
Then 
\begin{eqnarray}
&&Tr\{A^{R}B^{R}+B^{R}A^{R}\}  \nonumber \\
&=&Tr\{\sum_{\alpha }\sum_{\beta }\nu _{\alpha }\sigma _{\beta }\,\theta
_{\alpha }^{A}\,(\theta _{\alpha }^{A})^{\dag }\zeta _{\beta }^{B}\,(\zeta
_{\beta }^{B})^{\dag }\}  \nonumber \\
&&+Tr\{\sum_{\alpha }\sum_{\beta }\nu _{\alpha }\sigma _{\beta }\,\zeta
_{\beta }^{B}\,(\zeta _{\beta }^{B})^{\dag }\,\theta _{\alpha }^{A}\,(\theta
_{\alpha }^{A})^{\dag }\}  \nonumber \\
&=&\sum_{\alpha }\sum_{\beta }\nu _{\alpha }\sigma _{\beta }\,[(\theta
_{\alpha }^{A})^{\dag }\zeta _{\beta }^{B}]\;[(\zeta _{\beta }^{B})^{\dag
}\theta _{\alpha }^{A}]  \nonumber \\
&&+\sum_{\alpha }\sum_{\beta }\nu _{\alpha }\sigma _{\beta }\,[(\zeta
_{\beta }^{B})^{\dag }\theta _{\alpha }^{A}]\;[(\theta _{\alpha }^{A})^{\dag
}\zeta _{\beta }^{B}]\;  \nonumber \\
&=&2\sum_{\alpha }\sum_{\beta }\nu _{\alpha }\sigma _{\beta }\,|\,[(\theta
_{\alpha }^{A})^{\dag }\zeta _{\beta }^{B}]\,|^{2}  \label{Eq.KeyResult}
\end{eqnarray}

Hence we have using (\ref{Eq.InequalityModeMatrixForm})%
\begin{eqnarray}
&&\left\langle \Delta \widehat{S}\,_{x}^{2}\right\rangle -\frac{1}{2}%
|\left\langle \widehat{S}\,_{z}\right\rangle |  \nonumber \\
&\geq &\tsum\limits_{R}P_{R}\frac{1}{2}\sum_{\alpha }\sum_{\beta }\nu
_{\alpha }\sigma _{\beta }\,|\,[(\theta _{\alpha }^{A})^{\dag }\zeta _{\beta
}^{B}]\,|^{2}  \label{Eq.FinalInequality}
\end{eqnarray}%
where the right side of the inequality is non-negative. The same result
applies to $\left\langle \Delta \widehat{S}\,_{y}^{2}\right\rangle -\frac{1}{%
2}|\left\langle \widehat{S}\,_{z}\right\rangle |$. Thus separable states are 
\emph{not} spin squeezed in $\widehat{S}\,_{x}$ or in $\widehat{S}\,_{y}$.

Thus we have established the \emph{spin squeezing} test for the multi--mode
Case 1 - states that are spin squeezed in $\widehat{S}\,_{x}$ compared to $%
\widehat{S}\,_{y}$.(or vice versa) must be entangled states for the two
subsystems consisting of all modes $\widehat{a}_{i}$ and all modes $\widehat{%
b}_{i}$.

For the other spin components, the Bloch vector result in (\ref%
{Eq.BlochVectorCase1}) that $\left\langle \widehat{S}\,_{x}\right\rangle
=\left\langle \widehat{S}\,_{y}\right\rangle =0$ for \emph{separable} states
enables us to show that if $\widehat{S}\,_{z}$ is squeezed compared to $%
\widehat{S}\,_{x}$.(or vice versa) or if $\widehat{S}\,_{z}$ is squeezed
compared to $\widehat{S}\,_{y}$.(or vice versa) then the state must be
entangled. Thus spin squeezing in \emph{any} spin component requires the
state to be entangled, justt as for the two mode case. \pagebreak

\section{Appendix B - Spin Squeezing Tests for Other Multi-Mode Cases}

\label{Appendix - Revised Sorensen}

\subsection{Single Mode Sub-Systems}

\label{SubSection - Var Sz One Mode Subsystems}

We now consider spin squeezing for the multi-mode spin operators given in
Eqs. (\ref{Eq.SpinFieldOprs}) and (\ref{Eq.SpinOprs2}) in SubSection \ref%
{SubSection - Spin Operators Multimode Case}. We consider separable states
for \emph{Case 2}, the density operator being given in Eq. (\ref%
{Eq.SepStatesMultiModeCase2}). In this \emph{single mode sub-system case}
there are $2n$ subsystems consist of \emph{all} modes $\widehat{a}_{i}$ and 
\emph{all} modes $\widehat{b}_{i}$.

This case is that involved in the modified approach to Sorensen et al and we
will see that it leads to a useful inequality for $\left\langle \Delta 
\widehat{S}\,_{x}^{2}\right\rangle $ or.$\left\langle \Delta \widehat{S}%
\,_{y}^{2}\right\rangle $ that applies when non-entangled states are those
when \emph{all} the separate modes $\widehat{a}_{i}$ and $\widehat{b}_{i}$
are the sub-systems . We will follow the approach used for the simple two
mode case in Section \ref{Section - Relationship Spin Squeezing &
Entanglement}.

Firstly, the \emph{variance} for a Hermitian operator $\widehat{\Omega }$ in
a mixed state 
\begin{equation}
\widehat{\rho }=\sum_{R}P_{R}\,\widehat{\rho }_{R}
\end{equation}%
is always greater than or equal to the the average of the variances for the
separate components 
\begin{equation}
\left\langle \Delta \widehat{\Omega }\,^{2}\right\rangle \geq
\sum_{R}P_{R}\,\left\langle \Delta \widehat{\Omega }{}^{2}\right\rangle _{R}
\label{Eq.VarianceResultB}
\end{equation}%
where $\left\langle \Delta \widehat{\Omega }\,^{2}\right\rangle =Tr(\widehat{%
\rho }\,\Delta \widehat{\Omega }\,^{2})$ with $\Delta \widehat{\Omega }=%
\widehat{\Omega }-\left\langle \widehat{\Omega }\right\rangle $ and $%
\left\langle \Delta \widehat{\Omega }\,^{2}\right\rangle _{R}=Tr(\widehat{%
\rho }_{R}\,\Delta \widehat{\Omega }_{R}\,^{2})$ with $\Delta \widehat{%
\Omega }_{R}=\widehat{\Omega }-\left\langle \widehat{\Omega }\right\rangle
_{R}$ . The proof is straight-forward and given in Ref. \cite{Hoffmann03a}.

Next we calculate $\left\langle \Delta \widehat{S}\,_{x}^{2}\right\rangle
_{R}$, $\left\langle \Delta \widehat{S}\,_{y}^{2}\right\rangle _{R}$ and $%
\left\langle \widehat{S}_{x}\right\rangle _{R}$, $\left\langle \widehat{S}%
_{y}\right\rangle _{R}$, $\left\langle \widehat{S}_{z}\right\rangle _{R}$
for the case where 
\begin{equation}
\widehat{\rho }=\sum_{R}P_{R}\,\left( \widehat{\rho }_{R}^{a\,1}\otimes 
\widehat{\rho }_{R}^{b\,1}\right) \otimes \left( \widehat{\rho }%
_{R}^{a\,2}\otimes \widehat{\rho }_{R}^{b\,2}\right) \otimes \left( \widehat{%
\rho }_{R}^{a\,3}\otimes \widehat{\rho }_{R}^{b\,3}\right) \otimes ....
\label{Eq.RevisedSorensenDensityOprNonEntB}
\end{equation}%
as is required for a \emph{general non-entangled} state \emph{all} $2n$
modes. This situation is that of Choice 2 for the sub-systems, as described
in SubSection \ref{SubSection - SpinSqgEnt MultiMode}. As the density
operators for the individual modes must represent possible physical states
for such modes, so super-selection rule for atom number applies and we have 
\begin{eqnarray}
\left\langle (\widehat{a}_{i})^{p}\right\rangle _{a\,i} &=&Tr(\widehat{\rho }%
_{R}^{a\,i}(\widehat{a}_{i})^{p})=0\qquad \left\langle (\widehat{a}%
_{i}^{\dag })^{p}\right\rangle _{a\,i}=Tr(\widehat{\rho }_{R}^{a\,i}(%
\widehat{a}_{i}^{\dag })^{p})=0  \nonumber \\
\left\langle (\widehat{b}_{i})^{m}\right\rangle _{b\,i} &=&Tr(\widehat{\rho }%
_{R}^{b\,i}(\widehat{b}_{i})^{m})=0\qquad \left\langle (\widehat{b}%
_{i}^{\dag })^{m}\right\rangle _{b\,i}=Tr(\widehat{\rho }_{R}^{b\,i}(%
\widehat{b}_{i}^{\dag })^{m})=0  \nonumber \\
&&  \label{Eq.RevisedSorensenAveragesB}
\end{eqnarray}

The Schwinger spin operators are%
\begin{eqnarray}
\widehat{S}_{x} &=&\sum_{i}(\widehat{b}_{i}^{\dag }\widehat{a}_{i}+\widehat{a%
}_{i}^{\dag }\widehat{b}_{i})/2=\sum_{i}\widehat{S}_{x}^{i}  \nonumber \\
\widehat{S}_{y} &=&\sum_{i}(\widehat{b}_{i}^{\dag }\widehat{a}_{i}-\widehat{a%
}_{i}^{\dag }\widehat{b}_{i})/2i=\sum_{i}\widehat{S}_{y}^{i}  \nonumber \\
\widehat{S}_{z} &=&\sum_{i}(\widehat{b}_{i}^{\dag }\widehat{b}_{i}-\widehat{a%
}_{i}^{\dag }\widehat{a}_{i})/2=\sum_{i}\widehat{S}_{z}^{i}
\label{Eq.NewSpinOprsB}
\end{eqnarray}%
where $\widehat{a}_{i}$, $\widehat{b}_{i}$ and $\widehat{a}_{i}^{\dag }$, $%
\widehat{b}_{i}^{\dag }$ respectively are mode annihilation, creation
operators. Note that this expression for the spin operators is the same as (%
\ref{Eq.SpinOprs2}) for the multi-mode case treated in SubSection \ref%
{SubSection - Spin Operators Multimode Case}. From Eqs. (\ref%
{Eq.NewSpinOprsB}) we find that

\begin{equation}
\widehat{S}\,_{x}^{2}=\tsum\limits_{i}(\widehat{S}_{x}^{i})^{2}+\tsum%
\limits_{i\neq j}\widehat{S}_{x}^{i}\widehat{S}_{x}^{j}
\end{equation}%
so that on taking the trace with $\widehat{\rho }_{R}$ and using Eqs. (\ref%
{Eq.RevisedSorensenDensityOprNonEntB}) we get after applying the commutation
rules $[\widehat{e},\widehat{e}^{\dag }]=\widehat{1}$ ($\widehat{e}=\widehat{%
a}$ or $\widehat{b}$)%
\begin{equation}
\left\langle \widehat{S}\,_{x}^{2}\right\rangle
_{R}=\tsum\limits_{i}\left\langle (\widehat{S}_{x}^{i})^{2}\right\rangle
_{R}+\tsum\limits_{i\neq j}\left\langle \widehat{S}_{x}^{i}\right\rangle
_{R}\left\langle \widehat{S}_{x}^{j}\right\rangle _{R}
\end{equation}

As we also have%
\begin{equation}
\left\langle \widehat{S}\,_{x}\right\rangle
_{R}=\tsum\limits_{i}\left\langle \widehat{S}_{x}^{i}\right\rangle
_{R}\qquad \left\langle \widehat{S}\,_{x}\right\rangle
_{R}^{2}=\tsum\limits_{i}\left\langle \widehat{S}_{x}^{i}\right\rangle
_{R}^{2}+\tsum\limits_{i\neq j}\left\langle \widehat{S}_{x}^{i}\right\rangle
_{R}\left\langle \widehat{S}_{x}^{j}\right\rangle _{R}
\end{equation}%
using Eqs. (\ref{Eq.RevisedSorensenDensityOprNonEntB}) and we see finally
that the variance $\left\langle \Delta \widehat{S}\,_{x}^{2}\right\rangle
_{R}$ is%
\begin{equation}
\left\langle \Delta \widehat{S}\,_{x}^{2}\right\rangle
_{R}=\tsum\limits_{i}\left\langle (\widehat{S}_{x}^{i})^{2}\right\rangle
_{R}-\tsum\limits_{i}\left\langle \widehat{S}_{x}^{i}\right\rangle _{R}^{2}
\label{Eq.VarianceSxPairsRTerm}
\end{equation}%
all the terms with $i\neq j$ cancelling out. and therefore from Eq. (\ref%
{Eq.VarianceResultB})%
\begin{equation}
\left\langle \Delta \widehat{S}\,_{x}^{2}\right\rangle \geq
\sum_{R}P_{R}\,\tsum\limits_{i}\left( \left\langle (\widehat{S}%
_{x}^{i})^{2}\right\rangle _{R}-\left\langle \widehat{S}_{x}^{i}\right%
\rangle _{R}^{2}\right)  \label{Eq.VarianceSxInequalityPairs}
\end{equation}%
An analogous result applies for $\left\langle \Delta \widehat{S}%
\,_{y}^{2}\right\rangle $.

But using (\ref{Eq.RevisedSorensenAveragesB}) 
\begin{eqnarray}
(\widehat{S}_{x}^{i})^{2} &=&\frac{1}{4}(\widehat{b}_{i}^{\dag }\widehat{a}%
_{i}\widehat{b}_{i}^{\dag }\widehat{a}_{i}+\widehat{b}_{i}^{\dag }\widehat{a}%
_{i}\widehat{a}_{i}^{\dag }\widehat{b}_{i}+\widehat{a}_{i}^{\dag }\widehat{b}%
_{i}\widehat{b}_{i}^{\dag }\widehat{a}_{i}+\widehat{a}_{i}^{\dag }\widehat{b}%
_{i}\widehat{a}_{i}^{\dag }\widehat{b}_{i})  \nonumber \\
\left\langle (\widehat{S}_{x}^{i})^{2}\right\rangle _{R} &=&\frac{1}{4}%
(\left\langle (\widehat{b}^{\dag }\widehat{b})_{i}\right\rangle
_{R}+\left\langle (\widehat{a}^{\dag }\widehat{a})_{i}\right\rangle _{R})+%
\frac{1}{2}(\left\langle (\widehat{a}^{\dag }\widehat{a})_{i}\right\rangle
_{R}\left\langle (\widehat{b}^{\dag }\widehat{b})_{i}\right\rangle _{R})
\label{Eq.MeanSxiSquaredSingleModes}
\end{eqnarray}%
and 
\begin{equation}
\left\langle \widehat{S}_{x}^{i}\right\rangle _{R}=0
\label{Eq.MeanSxiSingleModes}
\end{equation}

It then follows that%
\begin{equation}
\left\langle \widehat{S}\,_{x}\right\rangle =\sum_{R}P_{R}\left\langle 
\widehat{S}\,_{x}\right\rangle _{R}=0\qquad \left\langle \widehat{S}%
\,_{y}\right\rangle =\sum_{R}P_{R}\left\langle \widehat{S}%
\,_{y}\right\rangle _{R}=0  \label{Eq.MeanNewSpinXYB}
\end{equation}%
so that 
\begin{equation}
\left\langle \Delta \widehat{S}\,_{x}^{2}\right\rangle \geq
\sum_{R}P_{R}\,\tsum\limits_{i}\left( \frac{1}{4}(\left\langle (\widehat{b}%
^{\dag }\widehat{b})_{i}\right\rangle _{R}+\left\langle (\widehat{a}^{\dag }%
\widehat{a})_{i}\right\rangle _{R})+\frac{1}{2}(\left\langle (\widehat{a}%
^{\dag }\widehat{a})_{i}\right\rangle _{R}\left\langle (\widehat{b}^{\dag }%
\widehat{b})_{i}\right\rangle _{R})\right)
\label{Eq.InequalVarSxSingleMulti}
\end{equation}%
The same result applies for $\left\langle \Delta \widehat{S}%
\,_{y}^{2}\right\rangle $.

Now using (\ref{Eq.RevisedSorensenAveragesB})%
\begin{equation}
\left\langle \widehat{S}_{z}^{i}\right\rangle _{R}=\frac{1}{2}(\left\langle (%
\widehat{b}^{\dag }\widehat{b})_{i}\right\rangle _{R}-\left\langle (\widehat{%
a}^{\dag }\widehat{a})_{i}\right\rangle _{R}))
\end{equation}%
\begin{eqnarray}
\left\langle \widehat{S}\,_{z}\right\rangle
&=&\sum_{R}P_{R}\,\tsum\limits_{i}\left\langle \widehat{S}%
_{z}^{i}\right\rangle _{R}  \nonumber \\
\frac{1}{2}|\left\langle \widehat{S}\,_{z}\right\rangle | &=&\frac{1}{2}%
\sum_{R}P_{R}\,|\tsum\limits_{i}\frac{1}{2}(\left\langle (\widehat{b}^{\dag }%
\widehat{b})_{i}\right\rangle _{R}-\left\langle (\widehat{a}^{\dag }\widehat{%
a})_{i}\right\rangle _{R}))|  \nonumber \\
&\leq &\sum_{R}P_{R}\,\frac{1}{4}\tsum\limits_{i}|(\left\langle (\widehat{b}%
^{\dag }\widehat{b})_{i}\right\rangle _{R}-\left\langle (\widehat{a}^{\dag }%
\widehat{a})_{i}\right\rangle _{R}))|  \nonumber \\
&\leq &\sum_{R}P_{R}\,\frac{1}{4}\tsum\limits_{i}(\left\langle (\widehat{b}%
^{\dag }\widehat{b})_{i}\right\rangle _{R}+\left\langle (\widehat{a}^{\dag }%
\widehat{a})_{i}\right\rangle _{R}))
\end{eqnarray}%
and thus%
\begin{eqnarray}
&&\left\langle \Delta \widehat{S}\,_{x}^{2}\right\rangle -\frac{1}{2}%
|\left\langle \widehat{S}\,_{z}\right\rangle |\;  \nonumber \\
&\geq &\sum_{R}P_{R}\,\tsum\limits_{i}\left( \frac{1}{4}(\left\langle (%
\widehat{b}^{\dag }\widehat{b})_{i}\right\rangle _{R}+\left\langle (\widehat{%
a}^{\dag }\widehat{a})_{i}\right\rangle _{R})+\frac{1}{2}(\left\langle (%
\widehat{a}^{\dag }\widehat{a})_{i}\right\rangle _{R}\left\langle (\widehat{b%
}^{\dag }\widehat{b})_{i}\right\rangle _{R})\right)  \nonumber \\
&&-\sum_{R}P_{R}\,\frac{1}{4}\tsum\limits_{i}(\left\langle (\widehat{b}%
^{\dag }\widehat{b})_{i}\right\rangle _{R}+\left\langle (\widehat{a}^{\dag }%
\widehat{a})_{i}\right\rangle _{R}))  \nonumber \\
&=&\sum_{R}P_{R}\,\frac{1}{2}\tsum\limits_{i}(\left\langle (\widehat{a}%
^{\dag }\widehat{a})_{i}\right\rangle _{R}\left\langle (\widehat{b}^{\dag }%
\widehat{b})_{i}\right\rangle _{R})  \nonumber \\
&\geq &0
\end{eqnarray}%
A similar proof shows that $\left\langle \Delta \widehat{S}%
\,_{y}^{2}\right\rangle -\frac{1}{2}|\left\langle \widehat{S}%
\,_{z}\right\rangle |\geq 0$ for the non-entangled state of all $2n$ modes.

This shows that for the general non-entangled state with all modes $\widehat{%
a}_{i}$ and $\widehat{b}_{i}$ as the sub-systems, the variances for two of
the spin fluctuations $\left\langle \Delta \widehat{S}\,_{x}^{2}\right%
\rangle $ and $\left\langle \Delta \widehat{S}\,_{y}^{2}\right\rangle $ are
both greater than $\frac{1}{2}|\left\langle \widehat{S}\,_{z}\right\rangle |$%
, and hence there is no spin squeezing for $\widehat{S}_{x}$ or $\widehat{S}%
_{y}$. Note that as $|\left\langle \widehat{S}\,_{y}\right\rangle |=0$, the
quantity $\sqrt{\left( |\left\langle \widehat{S}_{\perp \,1}\right\rangle
|^{2}+|\left\langle \widehat{S}_{\perp \,2}\right\rangle |^{2}\right) }$ is
the same as $|\left\langle \widehat{S}\,_{z}\right\rangle |$, so the
alternative criterion in Eq. (\ref{Eq.NewCriterionSpinSqueezing}) is the
same as that in Eq. (\ref{Eq.SpinSqueezingJXJY}) which is used here.

Hence we have shown that for a \emph{non-entangled} physical state for all
the $2n$ modes $\widehat{a}_{i}$ and $\widehat{b}_{i}$%
\begin{equation}
\left\langle \Delta \widehat{S}\,_{x}^{2}\right\rangle \geq \frac{1}{2}%
|\left\langle \widehat{S}\,_{z}\right\rangle |\quad and\quad \left\langle
\Delta \widehat{S}\,_{y}^{2}\right\rangle \geq \frac{1}{2}|\left\langle 
\widehat{S}\,_{z}\right\rangle |  \label{Eq.NonEntStateSpinSqCondnB}
\end{equation}%
so that spin squeezing in either $\widehat{S}_{x}$ or $\widehat{S}_{y}$
requires entanglement.

From (\ref{Eq.MeanNewSpinXYB}) we see that $\left\langle \widehat{S}%
\,_{x}\right\rangle =\left\langle \widehat{S}\,_{y}\right\rangle =0$ for the
general separable state, showing there is a \emph{Bloch vector test} for
entanglement such that if either $\left\langle \widehat{S}%
\,_{x}\right\rangle $ or $\left\langle \widehat{S}\,_{y}\right\rangle $ is
non-zero, then the state must be entangled.

Finally, if there is spin squeezing in $\widehat{S}_{z}$ with respect to $%
\widehat{S}_{x}$ or vice versa, or spin squeezing in $\widehat{S}_{z}$ with
respect to $\widehat{S}_{y}$ or vice versa, it follows that one of $%
\left\langle \widehat{S}\,_{x}\right\rangle $ or $\left\langle \widehat{S}%
\,_{y}\right\rangle $ is non-zero. But as both these quantities are zero for
a non-entangled state, if follows that spin squeezing in $\widehat{S}_{z}$
also requires entanglement.

Thus, spin squeezing in \emph{any} spin operator $\widehat{S}_{x},\widehat{S}%
_{y}$ or $\widehat{S}_{z}$ is a sufficiency test for entanglement of all the
separate mode sub-systems.

\subsection{Two Mode Sub-Systems}

\label{SubSection - Var Sz Two Mode SubSystems}

We now consider spin squeezing for the multi-mode spin operators given in
Eqs. (\ref{Eq.SpinFieldOprs}) and (\ref{Eq.SpinOprs2}) in SubSection \ref%
{SubSection - Spin Operators Multimode Case}. We consider separable states
for \emph{Case 3}, the density operator being given in Eq. (\ref%
{Eq.SepStatesMultiModeCase3}). In this \emph{mode pair sub-system case}
there are $n$ subsystems consist of \emph{all} \emph{pairs} of modes $%
\widehat{a}_{i}$ and $\widehat{b}_{i}$.

This case is also involved in a modified approach to Sorensen et al and we
show a useful inequality for $\left\langle \Delta \widehat{S}%
\,_{z}^{2}\right\rangle $ applies when non-entangled states are those when
the \emph{pairs} of modes $\widehat{a}_{i}$ and $\widehat{b}_{i}$ are the
separate sub-systems, but only in restricted situations. The pairs of modes
corresponding to localised modes on different lattice sites or pairs of
modes with the same momenta doe represent the closest way of simulating the
approach used by Sorensen et al where identical particles $i$ were regarded
as the sub-systems.

Now the general non-entangled state will be 
\begin{equation}
\widehat{\rho }=\sum_{R}P_{R}\,\widehat{\rho }_{R}^{1}\otimes \widehat{\rho }%
_{R}^{2}\otimes \widehat{\rho }_{R}^{3}\otimes ...
\end{equation}%
where the $\widehat{\rho }_{R}^{i}$ are now the density operators for
sub-system $i$ consisting of the pair of modes $\widehat{a}_{i}$ and $%
\widehat{b}_{i}$ (which are of the form given in Eq. (\ref%
{Eq.GeneralDensityOprModePair})) and the conditions in Eq. (\ref%
{Eq.RevisedSorensenAveragesB}) no longer apply. The Fock states are of the
form $\left\vert N_{ia}\right\rangle \otimes \left\vert N_{ib}\right\rangle $
for the pair of modes $\widehat{a}_{i}$ and $\widehat{b}_{i}$, and for this
Fock state the total occupancy of the pair of modes is $N_{i}=N_{ia}+N_{ib}$%
. From the super-selection rule the density operator $\widehat{\rho }%
_{R}^{i} $ for the $i$th pair of modes $\widehat{a}_{i}$ and $\widehat{b}%
_{i} $ is diagonal in the total occupancy. For $N_{i}$ $=0$ there is one non
zero matrix element $(\left\langle 0\right\vert _{ia}\otimes \left\langle
0\right\vert _{ib})\,\widehat{\rho }_{R}^{i}\,(\left\vert 0\right\rangle
_{ia}\otimes \left\vert 0\right\rangle _{ib})$. For $N_{i}$ $=1$ there are
four non zero matrix elements, which may be written%
\begin{eqnarray}
(\left\langle 1\right\vert _{ia}\otimes \left\langle 0\right\vert _{ib})\,%
\widehat{\rho }_{R}^{i}\,(\left\vert 1\right\rangle _{ia}\otimes \left\vert
0\right\rangle _{ib}) &=&\rho _{aa}^{i}  \nonumber \\
(\left\langle 1\right\vert _{ia}\otimes \left\langle 0\right\vert _{ib})\,%
\widehat{\rho }_{R}^{i}\,(\left\vert 0\right\rangle _{ia}\otimes \left\vert
1\right\rangle _{ib}) &=&\rho _{ab}^{i}  \nonumber \\
(\left\langle 0\right\vert _{ia}\otimes \left\langle 1\right\vert _{ib})\,%
\widehat{\rho }_{R}^{i}\,(\left\vert 1\right\rangle _{ia}\otimes \left\vert
0\right\rangle _{ib}) &=&\rho _{ba}^{i}  \nonumber \\
(\left\langle 0\right\vert _{ia}\otimes \left\langle 1\right\vert _{ib})\,%
\widehat{\rho }_{R}^{i}\,(\left\vert 0\right\rangle _{ia}\otimes \left\vert
1\right\rangle _{ib}) &=&\rho _{bb}^{i}  \label{Eq.MatrixElemPairModes}
\end{eqnarray}%
For $N_{i}$ $=2$ there are nine non zero matrix element $(\left\langle
2\right\vert _{ia}\otimes \left\langle 0\right\vert _{ib})\,\widehat{\rho }%
_{R}^{i}\,(\left\vert 2\right\rangle _{ia}\otimes \left\vert 0\right\rangle
_{ib})$, $...$, $(\left\langle 0\right\vert _{ia}\otimes \left\langle
2\right\vert _{ib})\,\widehat{\rho }_{R}^{i}\,(\left\vert 0\right\rangle
_{ia}\otimes \left\vert 2\right\rangle _{ib})$ and the number increases with 
$N_{i}$.

If we restrict ourselves to general entangled states for \emph{one particle}%
, where $N_{i}$ $=1$ for all pairs of modes, then the density operator $%
\widehat{\rho }_{R}^{i}$ is of then form 
\begin{eqnarray}
\widehat{\rho }_{R}^{i} &=&\rho _{aa}^{i}(\left\vert 1\right\rangle
_{ia}\left\langle 1\right\vert _{ia}\otimes \left\vert 0\right\rangle
_{ib}\left\langle 0\right\vert _{ib})+\rho _{ab}^{i}(\left\vert
1\right\rangle _{ia}\left\langle 0\right\vert _{ia}\otimes \left\vert
0\right\rangle _{ib}\left\langle 1\right\vert _{ib})  \nonumber \\
&&+\rho _{ba}^{i}(\left\vert 0\right\rangle _{ia}\left\langle 1\right\vert
_{ia}\otimes \left\vert 1\right\rangle _{ib}\left\langle 0\right\vert
_{ib})+\rho _{bb}^{i}(\left\vert 0\right\rangle _{ia}\left\langle
0\right\vert _{ia}\otimes \left\vert 1\right\rangle _{ib}\left\langle
1\right\vert _{ib})  \label{Eq.PairModesdensityOpr}
\end{eqnarray}%
In addition Hermitiancy, positivity, unit trace $Tr(\widehat{\rho }%
_{R}^{i})=1$ and $Tr(\widehat{\rho }_{R}^{i})^{2}\leq 1$ can be used as in
Eq (\ref{Eq.ParamMatrixElements}) to parameterise the matrix elements in (%
\ref{Eq.MatrixElemPairModes}).%
\begin{eqnarray}
\rho _{aa}^{i} &=&\sin ^{2}\alpha _{i}\qquad \rho _{bb}^{i}=\cos ^{2}\alpha
_{i}  \nonumber \\
\rho _{ab}^{i} &=&\sqrt{\sin ^{2}\alpha _{i}\,\cos ^{2}\alpha _{i}}\,\sin
^{2}\beta _{i}\,\exp (+i\phi _{i})\qquad \rho _{ba}^{i}=\sqrt{\sin
^{2}\alpha _{i}\,\cos ^{2}\alpha _{i}}\,\sin ^{2}\beta _{i}\,\exp (-i\phi
_{i})  \nonumber \\
&&
\end{eqnarray}

The expectation values for the spin operators $\widehat{S}_{x}^{i}$, $%
\widehat{S}_{y}^{i}$ and $\widehat{S}_{z}^{i}$ associated with the $i$th
pair of modes are then 
\begin{eqnarray}
\left\langle \widehat{S}_{x}^{i}\right\rangle _{R} &=&Tr(\widehat{\rho }%
_{R}^{i}\,\frac{1}{2}(\widehat{b}_{i}^{\dag }\widehat{a}_{i}+\widehat{a}%
_{i}^{\dag }\widehat{b}_{i})  \nonumber \\
&=&\frac{1}{2}\left( \rho _{ab}^{i}+\rho _{ba}^{i}\right)  \nonumber \\
\left\langle \widehat{S}_{y}^{i}\right\rangle _{R} &=&\frac{1}{2i}\left(
\rho _{ab}^{i}-\rho _{ba}^{i}\right)  \nonumber \\
\left\langle \widehat{S}_{z}^{i}\right\rangle _{R} &=&\frac{1}{2}\left( \rho
_{bb}^{i}-\rho _{aa}^{i}\right)
\end{eqnarray}%
which are of exactly the same form as in Eq. (\ref%
{Eq.SpinExpnValuesMatrixElements}) as in the Appendix \ref{Appendix -
Sorensen Results} derivation of the original Sorensen et al \cite%
{Sorensen01a} results based on treating identical particles as the
sub-systems. The proof however is now different and rests on restricting the
states $\widehat{\rho }_{R}^{i}$ to each containing exactly one boson.

The remainder of the proof is exactly the same as in Appendix \ref{Appendix
- Sorensen Results} and we find that 
\begin{equation}
\left\langle \Delta \widehat{S}\,_{z}^{2}\right\rangle \geq \frac{1}{N}%
\left( \left\langle \widehat{S}\,_{x}\right\rangle ^{2}+\left\langle 
\widehat{S}\,_{y}\right\rangle ^{2}\right)
\end{equation}%
for non-entangled \emph{pairs} of modes $\widehat{a}_{i}$ and $\widehat{b}%
_{i}$. Thus when the interpretation is changed so that are the separate
sub-systems are these pairs of modes, it follows that spin squeezing in $%
\widehat{S}\,_{z}$ with respect to $\widehat{S}\,_{x}$ or $\widehat{S}\,_{y}$
requires entanglement of all the mode pairs, but only if there is one
particle in each mode pair.

In general, spin squeezing in either $\widehat{S}\,_{x}$ or $\widehat{S}%
\,_{y}$ is not linked to entanglement for Case 3 sub-systems, as has been
pointed out in SubSection \ref{SubSection - SpinSqgEnt MultiMode} by a
counter-example involving the relative phase state. Also there is no Bloch
vector entanglement test. For we have in general 
\begin{eqnarray}
\left\langle \widehat{S}_{x}^{i}\right\rangle _{R} &=&Tr(\widehat{\rho }%
_{R}^{i}\,\frac{1}{2}(\widehat{b}_{i}^{\dag }\widehat{a}_{i}+\widehat{a}%
_{i}^{\dag }\widehat{b}_{i})  \nonumber \\
\left\langle \widehat{S}_{y}^{i}\right\rangle _{R} &=&Tr(\widehat{\rho }%
_{R}^{i}\,\frac{1}{2i}(\widehat{b}_{i}^{\dag }\widehat{a}_{i}-\widehat{a}%
_{i}^{\dag }\widehat{b}_{i})
\end{eqnarray}%
and the local particle number SSR does not require these quantities to be
zero for sub-systems consisting of pairs of modes $\widehat{a}_{i}$ and $%
\widehat{b}_{i}$. Thus in general $\left\langle \widehat{S}_{x}\right\rangle 
$ and $\left\langle \widehat{S}_{y}\right\rangle $ can be non-zero for a
separable state, so the Bloch vector entanglement test does not
apply.\pagebreak

\section{Appendix C - Hillery Spin Variance - Multi-Mode}

\label{Appendix = Hillery Spin MultiMode}

\subsection{Bipartite Case}

We first consider \emph{Case 1} where there are \emph{two sub-systems} each
consisting of all the modes $\widehat{a}_{i}$ or all the modes $\widehat{b}%
_{i}$. We use the results from (\ref{Eq.IneqVarSxMultiMode}) to find that
for a separable state 
\begin{eqnarray}
&&\left\langle \Delta \widehat{S}\,_{x}^{2}\right\rangle +\left\langle
\Delta \widehat{S}\,_{y}^{2}\right\rangle  \nonumber \\
&\geq &\tsum\limits_{R}P_{R}\{\tiint d\mathbf{r\,}d\mathbf{r}^{\prime }%
\mathbf{\,}\left\langle \widehat{\Psi }_{b}^{\dag }(\mathbf{r})\widehat{\Psi 
}_{b}(\mathbf{r}^{\prime })\right\rangle _{R}^{B}\left\langle \widehat{\Psi }%
_{a}^{\dag }(\mathbf{r}^{\prime })\widehat{\Psi }_{a}(\mathbf{r}%
)\right\rangle _{R}^{A}  \nonumber \\
&&+\frac{1}{2}\tint d\mathbf{r\,\,\{}\left\langle \widehat{\Psi }_{b}^{\dag
}(\mathbf{r})\widehat{\Psi }_{b}(\mathbf{r})\right\rangle
_{R}^{B}+\left\langle \widehat{\Psi }_{a}^{\dag }(\mathbf{r})\widehat{\Psi }%
_{a}(\mathbf{r})\right\rangle _{R}^{A}\}
\end{eqnarray}%
The same result would have occured if the local sub-system SSR had been
disregarded, the terms such as $\frac{1}{4}\tiint d\mathbf{r\,}d\mathbf{r}%
^{\prime }\mathbf{\,}\times \{\left\langle \widehat{\Psi }_{b}^{\dag }(%
\mathbf{r})\widehat{\Psi }_{b}^{\dag }(\mathbf{r}^{\prime })\right\rangle
_{R}^{B}\left\langle \widehat{\Psi }_{a}(\mathbf{r})\widehat{\Psi }_{a}(%
\mathbf{r}^{\prime })\right\rangle _{R}^{A}$ cancelling out.

The mean number of bosons is obtained from (\ref{Eq.NumberOprsFields}) and
hence%
\begin{equation}
\frac{1}{2}\left\langle \widehat{N}\right\rangle =\frac{1}{2}%
\sum_{R}P_{R}\tint d\mathbf{r\,}\left( \left\langle \widehat{\Psi }%
_{b}^{\dag }(\mathbf{r})\widehat{\Psi }_{b}(\mathbf{r})\right\rangle
_{R}^{B}+\left\langle \widehat{\Psi }_{a}^{\dag }(\mathbf{r})\widehat{\Psi }%
_{a}(\mathbf{r})\right\rangle _{R}^{A}\right)
\end{equation}

Thus we have 
\begin{eqnarray}
&&\left\langle \Delta \widehat{S}\,_{x}^{2}\right\rangle +\left\langle
\Delta \widehat{S}\,_{y}^{2}\right\rangle -\frac{1}{2}\left\langle \widehat{N%
}\right\rangle  \nonumber \\
&\geq &\tsum\limits_{R}P_{R}\tiint d\mathbf{r\,}d\mathbf{r}^{\prime }\mathbf{%
\,}\left\langle \widehat{\Psi }_{b}^{\dag }(\mathbf{r})\widehat{\Psi }_{b}(%
\mathbf{r}^{\prime })\right\rangle _{R}^{B}\left\langle \widehat{\Psi }%
_{a}^{\dag }(\mathbf{r}^{\prime })\widehat{\Psi }_{a}(\mathbf{r}%
)\right\rangle _{R}^{A}
\end{eqnarray}

Using the mode expansion (\ref{Eq.FieldOprs}) we then get%
\begin{eqnarray}
&&\left\langle \Delta \widehat{S}\,_{x}^{2}\right\rangle +\left\langle
\Delta \widehat{S}\,_{y}^{2}\right\rangle -\frac{1}{2}\left\langle \widehat{N%
}\right\rangle  \nonumber \\
&\geq &\tsum\limits_{R}P_{R}\sum_{ij}\sum_{kl}\tiint d\mathbf{r\,}d\mathbf{r}%
^{\prime }\mathbf{\,}\phi _{i}^{\ast }(\mathbf{r})\phi _{j}(\mathbf{r}%
^{\prime })\phi _{k}^{\ast }(\mathbf{r}^{\prime })\phi _{l}(\mathbf{r}%
)\left\langle \widehat{b}_{i}^{\dag }\widehat{b}_{j}\right\rangle
_{R}^{B}\left\langle \widehat{a}_{k}^{\dag }\widehat{a}_{l}\right\rangle
_{R}^{A}  \nonumber \\
&=&\tsum\limits_{R}P_{R}\sum_{ij}Tr_{A}\{\widehat{a}_{i}\,\widehat{\rho }%
_{R}^{A}\;\widehat{a}_{j}^{\dag }\}Tr_{B}\{\widehat{b}_{j}\,\widehat{\rho }%
_{R}^{B}\;\widehat{b}_{i}^{\dag }\} \\
&=&\tsum\limits_{R}P_{R}\frac{1}{2}Tr(A^{R}B^{R}+B^{R}A^{R})
\label{Eq.InequalSumVarMultiMode}
\end{eqnarray}%
after orthogonality is used and the matrix elements $A_{ij}^{R}$ and $%
B_{ji}^{R}$ are introduced from (\ref{Eq.ModeMatrices}).

Since we have shown in Appendix \ref{Appendix = MultiMo Spin Sq Choice 1}
that the right side of the last inequality is always non-negative, the \emph{%
Hillery spin variance} entanglement test follows that if 
\begin{equation}
\left\langle \Delta \widehat{S}\,_{x}^{2}\right\rangle +\left\langle \Delta 
\widehat{S}\,_{y}^{2}\right\rangle <\frac{1}{2}\left\langle \widehat{N}%
\right\rangle  \label{Eq.HillerySpinVarTestMulti}
\end{equation}%
then the quantum state must be an entangled state for the case of two
sub-systems each consisting of all the modes $\widehat{a}_{i}$ or all the
modes $\widehat{b}_{i}$.

\subsection{Single Modes Case}

We now consider separable states for \emph{Case 2}, the density operator
being given in Eq. (\ref{Eq.SepStatesMultiModeCase2}). In this \emph{single
mode sub-system case} there are $2n$ subsystems consisting of \emph{all}
modes $\widehat{a}_{i}$ and \emph{all} modes $\widehat{b}_{i}$. We use the
results from (\ref{Eq.InequalVarSxSingleMulti}) to find that for a separable
state%
\begin{eqnarray}
&&\left\langle \Delta \widehat{S}\,_{x}^{2}\right\rangle +\left\langle
\Delta \widehat{S}\,_{y}^{2}\right\rangle  \nonumber \\
&\geq &\sum_{R}P_{R}\,\tsum\limits_{i}\left( \frac{1}{2}(\left\langle (%
\widehat{b}^{\dag }\widehat{b})_{i}\right\rangle _{R}+\left\langle (\widehat{%
a}^{\dag }\widehat{a})_{i}\right\rangle _{R})+(\left\langle (\widehat{a}%
^{\dag }\widehat{a})_{i}\right\rangle _{R}\left\langle (\widehat{b}^{\dag }%
\widehat{b})_{i}\right\rangle _{R})\right)
\end{eqnarray}%
The same result would have occured if the local sub-system SSR had been
disregarded, the terms such as $\frac{1}{4}\left\langle \widehat{b}%
_{i}^{\dag }\widehat{b}_{i}^{\dag }\right\rangle _{Ri}\left\langle \widehat{a%
}_{i}\widehat{a}_{i}\right\rangle _{R}$ cancelling out.

The mean number of bosons is obtained from (\ref{Eq.NumberOprsFields}) 
\begin{equation}
\frac{1}{2}\left\langle \widehat{N}\right\rangle =\frac{1}{2}%
\sum_{R}P_{R}\tsum\limits_{i}(\left\langle (\widehat{b}^{\dag }\widehat{b}%
)_{i}\right\rangle _{R}+\left\langle (\widehat{a}^{\dag }\widehat{a}%
)_{i}\right\rangle _{R})
\end{equation}%
Thus we have%
\begin{eqnarray}
&&\left\langle \Delta \widehat{S}\,_{x}^{2}\right\rangle +\left\langle
\Delta \widehat{S}\,_{y}^{2}\right\rangle -\frac{1}{2}\left\langle \widehat{N%
}\right\rangle  \nonumber \\
&\geq &\sum_{R}P_{R}\,\tsum\limits_{i}\left( \left\langle (\widehat{a}^{\dag
}\widehat{a})_{i}\right\rangle _{R}\left\langle (\widehat{b}^{\dag }\widehat{%
b})_{i}\right\rangle _{R}\right)  \label{Eq.InequalSumVarSingleMulti}
\end{eqnarray}%
which is always non-negative.

The \emph{Hillery spin variance} entanglement test follows that if the
inequality in (\ref{Eq.HillerySpinVarTestMulti}) occurs then the quantum
state must be an entangled state for the case of $2n$ sub-systems consisting
of all the modes $\widehat{a}_{i}$ and all the modes $\widehat{b}_{i}$.

\subsection{Two Modes Case}

We now consider separable states for \emph{Case 3}, the density operators
being given in Eq. (\ref{Eq.SepStatesMultiModeCase3}). In this \emph{two
mode sub-system case} there are $n$ subsystems consisting of \emph{all} mode
pairs $\widehat{a}_{i}$ and $\widehat{b}_{i}$. We consider a \emph{special}
separable state with just one term where 
\begin{equation}
\widehat{\rho }_{sep}=\widehat{\rho }^{ab(1)}\otimes \widehat{\rho }%
^{ab(2)}\otimes ..\otimes \widehat{\rho }^{ab(i)}..\otimes \widehat{\rho }%
^{ab(n)}  \label{Eq.SpecialSepStateCase3}
\end{equation}%
We use the results from (\ref{Eq.VarianceSxPairsRTerm}) to find that

\begin{eqnarray}
&&\left\langle \Delta \widehat{S}\,_{x}^{2}\right\rangle +\left\langle
\Delta \widehat{S}\,_{y}^{2}\right\rangle  \nonumber \\
&=&\tsum\limits_{i}\left( \left\langle (\Delta \widehat{S}%
_{x}^{i})^{2}\right\rangle +\left\langle (\Delta \widehat{S}%
_{y}^{i})^{2}\right\rangle \right)  \label{Eq.IneqSpinVarCase3}
\end{eqnarray}%
where $\Delta \widehat{S}_{\alpha }^{i}=\widehat{S}_{\alpha
}^{i}-\left\langle \widehat{S}_{\alpha }^{i}\right\rangle _{R}$ for $\alpha
=x,y$. This result did not depend on applying the local SSR.

Now suppose each of the two mode states $\widehat{\rho }^{ab(i)}$ is an
entangled state of the modes $\widehat{a}_{i}$ and $\widehat{b}_{i}$ in
which the Hillery spin variance test is satisfied. Then 
\[
\left\langle (\Delta \widehat{S}_{x}^{i})^{2}\right\rangle +\left\langle
(\Delta \widehat{S}_{y}^{i})^{2}\right\rangle <\frac{1}{2}\left\langle 
\widehat{n}_{i}\right\rangle 
\]%
Hence 
\begin{eqnarray}
&&\left\langle \Delta \widehat{S}\,_{x}^{2}\right\rangle +\left\langle
\Delta \widehat{S}\,_{y}^{2}\right\rangle  \nonumber \\
&<&\tsum\limits_{i}\frac{1}{2}\left\langle \widehat{N}_{i}\right\rangle 
\nonumber \\
&=&\frac{1}{2}\left\langle \widehat{N}\right\rangle  \label{Eq.HilleryIneqal}
\end{eqnarray}%
where $\widehat{N}=\sum_{i}\widehat{N}_{i}$ is the total number operator and 
$\widehat{N}_{i}=\widehat{b}_{i}^{\dag }\widehat{b}_{i}+\widehat{a}%
_{i}^{\dag }\widehat{a}_{i}$

Thus the Hillery spin variance test is satisfied even though the state (\ref%
{Eq.SpecialSepStateCase3}) is separable, shoeing that the test cannot be
applied for multi-mode Case 3. \pagebreak

\section{Appendix D - Derivation of Sorensen et al Results}

\label{Appendix - Sorensen Results}

Sorensen et al \cite{Sorensen01a} derive a number of ineqalities from which
they deduce a further inequality for the spin squeezing parameter in the
case of a non-entangled state. From this result they conclude that spin
squeezing implies entanglement. The final inequality they obtain for a
non-entangled state is 
\begin{equation}
\left\langle \Delta \widehat{S}\,_{z}^{2}\right\rangle \geq \frac{1}{N}%
\left( \left\langle \widehat{S}\,_{x}\right\rangle ^{2}+\left\langle 
\widehat{S}\,_{y}\right\rangle ^{2}\right)
\end{equation}

Their approach is based on writing the density operator for a non-entangled
state of $N$ identical particles as in Eq. (\ref%
{Eq.NonEntStateIdenticalAtoms})

\begin{equation}
\widehat{\rho }=\sum_{R}P_{R}\,\widehat{\rho }_{R}^{1}\otimes \widehat{\rho }%
_{R}^{2}\otimes \widehat{\rho }_{R}^{3}\otimes ...=\sum_{R}P_{R}\,\widehat{%
\rho }_{R}
\end{equation}%
The spin operators are defined as 
\begin{eqnarray}
\widehat{S}_{x} &=&\tsum\limits_{i}\widehat{S}_{x}^{i}=\sum_{i}(\left\vert
\phi _{b}(i)\right\rangle \left\langle \phi _{a}(i)\right\vert +\left\vert
\phi _{a}(i)\right\rangle \left\langle \phi _{b}(i)\right\vert )/2  \nonumber
\\
\widehat{S}_{y} &=&\tsum\limits_{i}\widehat{S}_{y}^{i}=\sum_{i}(\left\vert
\phi _{b}(i)\right\rangle \left\langle \phi _{a}(i)\right\vert -\left\vert
\phi _{a}(i)\right\rangle \left\langle \phi _{b}(i)\right\vert )/2i 
\nonumber \\
\widehat{S}_{z} &=&\tsum\limits_{i}\widehat{S}_{z}^{i}=\sum_{i}(\left\vert
\phi _{b}(i)\right\rangle \left\langle \phi _{b}(i)\right\vert -\left\vert
\phi _{a}(i)\right\rangle \left\langle \phi _{a}(i)\right\vert )/2
\end{eqnarray}
where the sum $i$ is over the identical atoms and each atom is associated
with two states $\left\vert \phi _{a}\right\rangle $ and $\left\vert \phi
_{b}\right\rangle $. Clearly, the spin operators satisfy the standard
commutation rules for agular momentum operators.

Sorensen et al \ \cite{Sorensen01a} state that the variance for $\widehat{S}%
_{z}$ satisfies the result%
\begin{equation}
\left\langle \Delta \widehat{S}\,_{z}^{2}\right\rangle =\frac{N}{4}%
-\tsum\limits_{R}P_{R}\tsum\limits_{i}\left\langle \widehat{S}%
_{z}^{i}\right\rangle _{R}^{2}+\tsum\limits_{R}P_{R}\left\langle \widehat{S}%
_{z}\right\rangle _{R}^{2}-\left\langle \widehat{S}_{z}\right\rangle ^{2}
\label{Eq.VarianceSZ}
\end{equation}%
To prove this we have 
\begin{eqnarray}
\left\langle \widehat{S}\,_{z}^{2}\right\rangle &=&\sum_{R}P_{R}\,Tr(%
\widehat{\rho }_{R}\tsum\limits_{i}\tsum\limits_{j}\widehat{S}_{z}^{i}%
\widehat{S}_{z}^{j})  \nonumber \\
&=&\sum_{R}P_{R}\,\left( \tsum\limits_{i}\left\langle \left( \widehat{S}%
_{z}^{i}\right) ^{2}\right\rangle _{R}+\tsum\limits_{i\neq j}\left\langle 
\widehat{S}_{z}^{i}\right\rangle _{R}\left\langle \widehat{S}%
_{z}^{j}\right\rangle _{R}\right)  \nonumber \\
&=&\frac{N}{4}+\sum_{R}P_{R}\,\left( \tsum\limits_{i\neq j}\left\langle 
\widehat{S}_{z}^{i}\right\rangle _{R}\left\langle \widehat{S}%
_{z}^{j}\right\rangle _{R}\right)
\end{eqnarray}%
where we have used 
\begin{eqnarray}
\left( \widehat{S}_{z}^{i}\right) ^{2} &=&\frac{1}{4}(\left\vert \phi
_{b}(i)\right\rangle \left\langle \phi _{b}(i)\right\vert -\left\vert \phi
_{a}(i)\right\rangle \left\langle \phi _{a}(i)\right\vert )^{2}  \nonumber \\
&=&\frac{1}{4}(\left\vert \phi _{b}(i)\right\rangle \left\langle \phi
_{b}(i)|\phi _{b}(i)\right\rangle \left\langle \phi _{b}(i)\right\vert
-(\left\vert \phi _{b}(i)\right\rangle \left\langle \phi _{b}(i)|\phi
_{a}(i)\right\rangle \left\langle \phi _{a}(i)\right\vert )  \nonumber \\
&&+\frac{1}{4}(-(\left\vert \phi _{a}(i)\right\rangle \left\langle \phi
_{a}(i)|\phi _{b}(i)\right\rangle \left\langle \phi _{b}(i)\right\vert
+(\left\vert \phi _{a}(i)\right\rangle \left\langle \phi _{a}(i)|\phi
_{a}(i)\right\rangle \left\langle \phi _{a}(i)\right\vert )  \nonumber \\
&=&\frac{1}{4}((\left\vert \phi _{b}(i)\right\rangle \left\langle \phi
_{b}(i)\right\vert +(\left\vert \phi _{a}(i)\right\rangle \left\langle \phi
_{a}(i)\right\vert )  \nonumber \\
&=&\frac{1}{4}\widehat{1}_{i}
\end{eqnarray}%
a result based on the orthogonality, normalisation and completeness of the
states $\left\vert \phi _{a}(i)\right\rangle ,\left\vert \phi
_{b}(i)\right\rangle $. Also 
\begin{eqnarray}
\left\langle \widehat{S}\,_{z}\right\rangle _{R} &=&Tr(\widehat{\rho }%
_{R}\tsum\limits_{i}\widehat{S}_{z}^{i})  \nonumber \\
&=&\tsum\limits_{i}\left\langle \widehat{S}_{z}^{i}\right\rangle _{R} 
\nonumber \\
\tsum\limits_{R}P_{R}\left\langle \widehat{S}_{z}\right\rangle _{R}^{2}
&=&\tsum\limits_{R}P_{R}\left( \tsum\limits_{i}\left\langle \widehat{S}%
_{z}^{i}\right\rangle _{R}^{2}+\tsum\limits_{i\neq j}\left\langle \widehat{S}%
_{z}^{i}\right\rangle _{R}\left\langle \widehat{S}_{z}^{j}\right\rangle
_{R}\right)
\end{eqnarray}%
so eliminating the term $\sum_{R}P_{R}\,\left( \tsum\limits_{i\neq
j}\left\langle \widehat{S}_{z}^{i}\right\rangle _{R}\left\langle \widehat{S}%
_{z}^{j}\right\rangle _{R}\right) $ gives the required expression for $%
\left\langle \Delta \widehat{S}\,_{z}^{2}\right\rangle =\left\langle 
\widehat{S}\,_{z}^{2}\right\rangle -\left\langle \widehat{S}%
_{z}\right\rangle ^{2}$.

Next, Sorensen et al \cite{Sorensen01a} state that 
\begin{equation}
\left\langle \widehat{S}_{x}\right\rangle ^{2}\leq
N\tsum\limits_{R}P_{R}\tsum\limits_{i}\left\langle \widehat{S}%
_{x}^{i}\right\rangle _{R}^{2}\qquad \left\langle \widehat{S}%
\,_{y}\right\rangle ^{2}\leq N\sum_{R}P_{R}\,\tsum\limits_{i}|\left\langle 
\widehat{S}_{y}^{i}\right\rangle _{R}|^{2}  \label{Eq.InequalSX2}
\end{equation}%
To prove this we have 
\begin{eqnarray}
\left\langle \widehat{S}\,_{x}\right\rangle &=&\sum_{R}P_{R}\,Tr(\widehat{%
\rho }_{R}\tsum\limits_{i}\widehat{S}_{x}^{i})  \nonumber \\
&=&\sum_{R}P_{R}\,\tsum\limits_{i}\left\langle \widehat{S}%
_{x}^{i}\right\rangle _{R}  \nonumber \\
|\left\langle \widehat{S}\,_{x}\right\rangle | &\leq
&\sum_{R}P_{R}\,\tsum\limits_{i}|\left\langle \widehat{S}_{x}^{i}\right%
\rangle _{R}|
\end{eqnarray}%
since the modulus of a sum is less than or equal to the sum of the moduli.
Now

\begin{eqnarray}
\left\langle \widehat{S}\,_{x}\right\rangle ^{2} &=&|\left\langle \widehat{S}%
\,_{x}\right\rangle |^{2}\leq \left(
\sum_{R}P_{R}\,\tsum\limits_{i}|\left\langle \widehat{S}_{x}^{i}\right%
\rangle _{R}|\right) ^{2}  \nonumber \\
&\leq &\sum_{R}P_{R}\,\left( \tsum\limits_{i}|\left\langle \widehat{S}%
_{x}^{i}\right\rangle _{R}|\right) ^{2}
\end{eqnarray}%
using the general result that $\left( \tsum\limits_{R}P_{R}\,\sqrt{C_{R}}%
\right) ^{2}\leq \tsum\limits_{R}P_{R}\,C_{R}$, where $\tsum%
\limits_{R}P_{R}=1$ with here $\sqrt{C_{R}}=\tsum\limits_{i}|\left\langle 
\widehat{S}_{x}^{i}\right\rangle _{R}|$. Next consider 
\begin{eqnarray}
y &=&N\tsum\limits_{i}|\left\langle \widehat{S}_{x}^{i}\right\rangle
_{R}|^{2}  \nonumber \\
z &=&\left( \tsum\limits_{i}|\left\langle \widehat{S}_{x}^{i}\right\rangle
_{R}|\right) ^{2}=\left( \tsum\limits_{i}|\left\langle \widehat{S}%
_{x}^{i}\right\rangle _{R}|\right) ^{2}  \nonumber \\
y-z &=&\tsum\limits_{i<j}(|\left\langle \widehat{S}_{x}^{i}\right\rangle
_{R}|-|\left\langle \widehat{S}_{x}^{j}\right\rangle _{R}|)^{2}\geq 0
\end{eqnarray}%
so that 
\begin{equation}
\left\langle \widehat{S}\,_{x}\right\rangle ^{2}\leq
N\sum_{R}P_{R}\,\tsum\limits_{i}|\left\langle \widehat{S}_{x}^{i}\right%
\rangle _{R}|^{2}\qquad \left\langle \widehat{S}\,_{y}\right\rangle ^{2}\leq
N\sum_{R}P_{R}\,\tsum\limits_{i}|\left\langle \widehat{S}_{y}^{i}\right%
\rangle _{R}|^{2}
\end{equation}%
which is the required result. The inequality for $\left\langle \widehat{S}%
\,_{y}\right\rangle ^{2}$ is proved similarly.

Another inequality is stated \cite{Sorensen01a} for $\left\langle \widehat{S}%
\,_{z}\right\rangle ^{2}$. This is 
\begin{equation}
\left\langle \widehat{S}\,_{z}\right\rangle ^{2}\leq
\sum_{R}P_{R}\,\left\langle \widehat{S}_{z}\right\rangle _{R}^{2}
\label{Eq.IneqalSZ2}
\end{equation}%
To show this we have 
\begin{eqnarray}
\left\langle \widehat{S}\,_{z}\right\rangle &=&\sum_{R}P_{R}\,Tr(\widehat{%
\rho }_{R}\tsum\limits_{i}\widehat{S}_{z}^{i})  \nonumber \\
&=&\sum_{R}P_{R}\,\tsum\limits_{i}\left\langle \widehat{S}%
_{z}^{i}\right\rangle _{R}  \nonumber \\
&=&\sum_{R}P_{R}\,\left\langle \widehat{S}_{z}\right\rangle _{R}  \nonumber
\\
|\left\langle \widehat{S}\,_{z}\right\rangle | &\leq
&\sum_{R}P_{R}\,|\left\langle \widehat{S}_{z}\right\rangle _{R}|
\end{eqnarray}%
so that 
\begin{eqnarray}
\left\langle \widehat{S}\,_{z}\right\rangle ^{2} &=&|\left\langle \widehat{S}%
\,_{z}\right\rangle |^{2}\leq \left( \sum_{R}P_{R}\,|\left\langle \widehat{S}%
_{z}\right\rangle _{R}|\right) ^{2}  \nonumber \\
&\leq &\sum_{R}P_{R}\,|\left\langle \widehat{S}_{z}\right\rangle _{R}|^{2} 
\nonumber \\
&=&\sum_{R}P_{R}\,\left\langle \widehat{S}_{z}\right\rangle _{R}^{2}
\end{eqnarray}%
using the general result that $\left( \tsum\limits_{R}P_{R}\,\sqrt{C_{R}}%
\right) ^{2}\leq \tsum\limits_{R}P_{R}\,C_{R}$, where $\tsum%
\limits_{R}P_{R}=1$ with here $\sqrt{C_{R}}=|\left\langle \widehat{S}%
_{z}\right\rangle _{R}|$.

Finally, we find that 
\begin{eqnarray}
\sum_{R}P_{R}\,\tsum\limits_{i}\left( \left\langle \widehat{S}%
_{x}^{i}\right\rangle _{R}^{2}+\left\langle \widehat{S}_{y}^{i}\right\rangle
_{R}^{2}+\left\langle \widehat{S}_{z}^{i}\right\rangle _{R}^{2}\right) &\leq
&\frac{1}{4}N  \nonumber \\
-\sum_{R}P_{R}\,\tsum\limits_{i}\left( \left\langle \widehat{S}%
_{z}^{i}\right\rangle _{R}^{2}\right) &\geq &-\frac{1}{4}N+\sum_{R}P_{R}\,%
\tsum\limits_{i}\left( \left\langle \widehat{S}_{x}^{i}\right\rangle
_{R}^{2}+\left\langle \widehat{S}_{y}^{i}\right\rangle _{R}^{2}\right) 
\nonumber \\
&&  \label{Eq.InequalBloch}
\end{eqnarray}%
To show this we use the properties of the density operator $\widehat{\rho }%
_{R}^{i}$ for the $i$th particle of Hermitiancy, positiveness, unit trace $%
Tr(\widehat{\rho }_{R}^{i})=1$ and $Tr(\widehat{\rho }_{R}^{i})^{2}\leq 1$.
In terms of matrix elements of the density operator $\widehat{\rho }_{R}^{i}$
between the two states $\left\vert \phi _{a}(i)\right\rangle $, $\left\vert
\phi _{b}(i)\right\rangle $ the quantities $\left\langle \widehat{S}%
_{x}^{i}\right\rangle _{R}$, $\left\langle \widehat{S}_{y}^{i}\right\rangle
_{R}$ and $\left\langle \widehat{S}_{z}^{i}\right\rangle _{R}$ are%
\begin{eqnarray}
\left\langle \widehat{S}_{x}^{i}\right\rangle _{R} &=&Tr(\widehat{\rho }%
_{R}^{i}\,\frac{1}{2}(\left\vert \phi _{b}(i)\right\rangle \left\langle \phi
_{a}(i)\right\vert +\left\vert \phi _{a}(i)\right\rangle \left\langle \phi
_{b}(i)\right\vert ))  \nonumber \\
&=&\frac{1}{2}\left( \rho _{ab}^{i}+\rho _{ba}^{i}\right)  \nonumber \\
\left\langle \widehat{S}_{y}^{i}\right\rangle _{R} &=&\frac{1}{2i}\left(
\rho _{ab}^{i}-\rho _{ba}^{i}\right)  \nonumber \\
\left\langle \widehat{S}_{z}^{i}\right\rangle _{R} &=&\frac{1}{2}\left( \rho
_{bb}^{i}-\rho _{aa}^{i}\right)  \label{Eq.SpinExpnValuesMatrixElements}
\end{eqnarray}%
where $\rho _{cd}^{i}=\left\langle \phi _{c}(i)\right\vert \widehat{\rho }%
_{R}^{i}$ $\left\vert \phi _{d}(i)\right\rangle $. The Hermitiancy and
positiveness of $\widehat{\rho }_{R}^{i}$ show that $\rho _{bb}^{i}$ and $%
\rho _{aa}^{i}$ are real and positive, $\rho _{ab}^{i}=(\rho
_{ba}^{i})^{\ast }$ and $\rho _{aa}^{i}\rho _{bb}^{i}-|\rho
_{ab}^{i}|^{2}\geq 0$. The condition $Tr(\widehat{\rho }_{R}^{i})=1$ leads
to $\rho _{aa}^{i}+\rho _{bb}^{i}=1$, from which $Tr(\widehat{\rho }%
_{R}^{i})^{2}\leq 1$ follows using the previous positivity results. Taken
together these conditions lead to the following useful parametrisation of
the density matrix elements 
\begin{eqnarray}
\rho _{aa}^{i} &=&\sin ^{2}\alpha _{i}\qquad \rho _{bb}^{i}=\cos ^{2}\alpha
_{i}  \nonumber \\
\rho _{ab}^{i} &=&\sqrt{\sin ^{2}\alpha _{i}\,\cos ^{2}\alpha _{i}}\,\sin
^{2}\beta _{i}\,\exp (+i\phi _{i})\qquad \rho _{ba}^{i}=\sqrt{\sin
^{2}\alpha _{i}\,\cos ^{2}\alpha _{i}}\,\sin ^{2}\beta _{i}\,\exp (-i\phi
_{i})  \nonumber \\
&&  \label{Eq.ParamMatrixElements}
\end{eqnarray}%
where $\alpha _{i}$, $\beta _{i}$ and $\phi _{i}$ are real. In terms of
these quantities we then have 
\begin{eqnarray}
\left\langle \widehat{S}_{x}^{i}\right\rangle _{R} &=&\frac{1}{2}\sin
2\alpha _{i}\,\sin ^{2}\beta _{i}\,\cos \phi _{i}  \nonumber \\
\left\langle \widehat{S}_{y}^{i}\right\rangle _{R} &=&\frac{1}{2}\sin
2\alpha _{i}\,\sin ^{2}\beta _{i}\,\sin \phi _{i}  \nonumber \\
\left\langle \widehat{S}_{z}^{i}\right\rangle _{R} &=&\frac{1}{2}\cos
2\alpha _{i}\,  \label{Eq.SpinExpnValuesParam}
\end{eqnarray}%
It is then easy to show that 
\begin{eqnarray}
\left\langle \widehat{S}_{x}^{i}\right\rangle _{R}^{2}+\left\langle \widehat{%
S}_{y}^{i}\right\rangle _{R}^{2}+\left\langle \widehat{S}_{z}^{i}\right%
\rangle _{R}^{2} &=&\frac{1}{4}-\frac{1}{4}\sin ^{2}2\alpha _{i}\,(1-\sin
^{4}\beta _{i}\,)  \nonumber \\
&\leq &\frac{1}{4}  \label{Eq.InequalSquaresSpinExpnVals}
\end{eqnarray}%
and the final inequality (\ref{Eq.InequalBloch}) then follows by taking the
sum over particles $i$ and then using $\sum_{R}P_{R}=1$. If only the Schwarz
inequality is used instead of the more detailed consequences of Hermtiancy,
positiveness etc it can be shown that $\left\langle \widehat{S}%
_{x}^{i}\right\rangle _{R}^{2}+\left\langle \widehat{S}_{y}^{i}\right\rangle
_{R}^{2}+\left\langle \widehat{S}_{z}^{i}\right\rangle _{R}^{2}$ $\leq \frac{%
3}{4}$, which though correct is not useful.

Combining the inequalities in Eqs. (\ref{Eq.InequalSX2}), (\ref{Eq.IneqalSZ2}%
) and (\ref{Eq.InequalBloch}) into Eq. (\ref{Eq.VarianceSZ}) shows that 
\begin{eqnarray}
\left\langle \Delta \widehat{S}\,_{z}^{2}\right\rangle &=&\frac{N}{4}%
-\tsum\limits_{R}P_{R}\tsum\limits_{i}\left\langle \widehat{S}%
_{z}^{i}\right\rangle _{R}^{2}+\tsum\limits_{R}P_{R}\left\langle \widehat{S}%
_{z}\right\rangle _{R}^{2}-\left\langle \widehat{S}_{z}\right\rangle ^{2} 
\nonumber \\
&\geq &\frac{N}{4}-\tsum\limits_{R}P_{R}\tsum\limits_{i}\left\langle 
\widehat{S}_{z}^{i}\right\rangle _{R}^{2}  \nonumber \\
&\geq &\frac{N}{4}-\frac{1}{4}N+\sum_{R}P_{R}\,\tsum\limits_{i}\left(
\left\langle \widehat{S}_{x}^{i}\right\rangle _{R}^{2}+\left\langle \widehat{%
S}_{y}^{i}\right\rangle _{R}^{2}\right)  \nonumber \\
&\geq &\frac{1}{N}\left( \left\langle \widehat{S}\,_{x}\right\rangle
^{2}+\left\langle \widehat{S}\,_{y}\right\rangle ^{2}\right)
\label{Eq.FinalIequal}
\end{eqnarray}%
for the case of a non-entangled state. This result is that in Sorensen et al.%
\cite{Sorensen01a}. \pagebreak

\section{Appendix E - Heisenberg Uncertainty Principle Results}

\label{Appendix - Heisenberg Uncertainty Principle Results}

Here we derive the results in SubSection \ref{SubSection - Sorensen and
Molmer 2001} leading to inequalities for the variance $\left\langle \Delta 
\widehat{J}\,_{x}^{2}\right\rangle $ considered as a function of $%
|\left\langle \widehat{J}\,_{z}\right\rangle |$ for states where the spin
operators are chosen such that $\left\langle \widehat{J}\,_{x}\right\rangle
=\left\langle \widehat{J}\,_{y}\right\rangle =0$.

From the Schwarz inequality $\left\langle \widehat{J}\,_{z}\right\rangle
^{2}\leq \left\langle \widehat{J}\,_{z}^{2}\right\rangle $ so that 
\begin{equation}
\left\langle \widehat{J}\,_{x}^{2}\right\rangle +\left\langle \widehat{J}%
\,_{y}^{2}\right\rangle +\left\langle \widehat{J}\,_{z}\right\rangle
^{2}\leq \left\langle \widehat{J}\,_{x}^{2}\right\rangle +\left\langle 
\widehat{J}\,_{y}^{2}\right\rangle +\left\langle \widehat{J}%
\,_{z}^{2}\right\rangle =J(J+1)
\end{equation}%
giving Eq. (\ref{Eq.SchwarzResult}). Subtracting $\left\langle \widehat{J}%
\,_{x}\right\rangle ^{2}=\left\langle \widehat{J}\,_{y}\right\rangle ^{2}=0$
from each side gives

\begin{equation}
\left\langle \Delta \widehat{J}\,_{x}^{2}\right\rangle +\left\langle \Delta 
\widehat{J}\,_{y}^{2}\right\rangle +\left\langle \widehat{J}%
\,_{z}\right\rangle ^{2}\leq J(J+1)
\end{equation}

Substituting for $\left\langle \Delta \widehat{J}\,_{y}^{2}\right\rangle $
from the Heisenberg uncertainty principle result in Eq. (\ref{Eq.HUP}) gives 
\begin{equation}
\left\langle \Delta \widehat{J}\,_{x}^{2}\right\rangle ^{2}-\left(
J(J+1)-\left\langle \widehat{J}\,_{z}\right\rangle ^{2}\right) \left\langle
\Delta \widehat{J}\,_{x}^{2}\right\rangle +\frac{1}{4}\xi \left\langle 
\widehat{J}\,_{z}\right\rangle ^{2}\leq 0
\end{equation}%
The left side is a parabolic function of $\left\langle \Delta \widehat{J}%
\,_{x}^{2}\right\rangle $ and for this to be negative requires $\left\langle
\Delta \widehat{J}\,_{x}^{2}\right\rangle $ to lie between the two roots of
this function, giving 
\begin{eqnarray}
\left\langle \Delta \widehat{J}\,_{x}^{2}\right\rangle &\geq &\frac{1}{2}%
\left\{ \left( J(J+1)-\left\langle \widehat{J}\,_{z}\right\rangle
^{2}\right) -\sqrt{\left( J(J+1)-\left\langle \widehat{J}\,_{z}\right\rangle
^{2}\right) ^{2}-\xi \left\langle \widehat{J}\,_{z}\right\rangle ^{2}}%
\right\}  \nonumber \\
&& \\
\left\langle \Delta \widehat{J}\,_{x}^{2}\right\rangle &\leq &\frac{1}{2}%
\left\{ \left( J(J+1)-\left\langle \widehat{J}\,_{z}\right\rangle
^{2}\right) +\sqrt{\left( J(J+1)-\left\langle \widehat{J}\,_{z}\right\rangle
^{2}\right) ^{2}-\xi \left\langle \widehat{J}\,_{z}\right\rangle ^{2}}%
\right\}  \nonumber \\
&&
\end{eqnarray}%
which are the required inequalities in Eq. (\ref{Eq.HUPRestriction1}) and (%
\ref{Eq.HUPRestriction2}).

\pagebreak

\section{Appendix F - "Separable but Non-Local" States}

\label{Appendix - Separable but Non Local States} \ref{Section - Criteria
for Spin Squeezing Based on Non-Physical States}.

It is instructive to apply the various entanglement tests to the so-called
separable but non-local states considered in Refs. \cite{Verstraete03a}, 
\cite{Schuch04a}, for which the sub-system states are definitely \emph{not}
SSR compliant. These states should not pass the the Hillery tests \cite%
{Hillery06a}, \cite{Hillery09a} for SSR neglected entanglement, but they may
pass the entanglement tests in this paper and in Ref. \cite{Dalton14a} since
these states would be regarded as SSR compliant entangled. Note that these
states are consistent with the global particle number SSR, so there is no
dispute about whether they are possible two mode quantum states. The issue
is rather whether they should be categorised as separable or entangled, and
that depends on how separable (and hence entangled) states are first
defined. As discussed previously, the interferometric measurements discussed
here do not enable us to choose one definition over the other - that is an
issue involved swhat types of quantum states would be allowed in the
separate sub-systems.

The first example of such states is the \emph{mixture} of \emph{two mode
coherent states }is represented by the two mode density operator%
\begin{eqnarray}
\widehat{\rho } &=&\tint \frac{d\theta }{2\pi }\,\left\vert \alpha ,\alpha
\right\rangle \left\langle \alpha ,\alpha \right\vert  \nonumber \\
&=&\tint \frac{d\theta }{2\pi }\,\left( \left\vert \alpha \right\rangle
\left\langle \alpha \right\vert \right) _{a}\otimes \left( \left\vert \alpha
\right\rangle \left\langle \alpha \right\vert \right) _{b}
\label{Eq.TwoModeCoherentStateMixture}
\end{eqnarray}%
where $\left\vert \alpha \right\rangle _{C}$ is a one mode coherent state
for mode $c=a,b$ with $\alpha =|\alpha |\,\exp (-i\theta )$, and modes $a,b$
are associated with bosonic annihilation operators $\widehat{a}$, $\widehat{b%
}$. The magnitude $|\alpha |$ is fixed. This state gobally but not locally
SSR compliant.

Now 
\begin{eqnarray}
\left\langle \widehat{a}^{\dag }\widehat{b}\right\rangle &=&Tr\tint \frac{%
d\theta }{2\pi }\,\widehat{a}^{\dag }\widehat{b}\,\left( \left\vert \alpha
\right\rangle \left\langle \alpha \right\vert \right) _{a}\otimes \left(
\left\vert \alpha \right\rangle \left\langle \alpha \right\vert \right) _{b}
\nonumber \\
&=&Tr\tint \frac{d\theta }{2\pi }\,\left( \left\vert \alpha \right\rangle
\left\langle \alpha \right\vert \,\widehat{a}^{\dag }\right) _{a}\otimes
\left( \widehat{b}\left\vert \alpha \right\rangle \left\langle \alpha
\right\vert \right) _{b}  \nonumber \\
&=&|\alpha |^{2}  \label{Eq.SimpleCorrTestTwoModeCohState1}
\end{eqnarray}%
But 
\begin{eqnarray}
\left\langle \widehat{a}^{\dag }\widehat{a}\,\widehat{b}^{\dag }\widehat{b}%
\right\rangle &=&Tr\tint \frac{d\theta }{2\pi }\,\left( \widehat{a}^{\dag }%
\widehat{a}\left\vert \alpha \right\rangle \left\langle \alpha \right\vert
\right) _{a}\otimes \left( \widehat{b}^{\dag }\widehat{b}\left\vert \alpha
\right\rangle \left\langle \alpha \right\vert \right) _{b}  \nonumber \\
&=&\tint \frac{d\theta }{2\pi }\,\left( \left\langle \alpha \right\vert 
\widehat{a}^{\dag }\widehat{a}\left\vert \alpha \right\rangle \right)
_{a}\otimes \left( \left\langle \alpha \right\vert \widehat{b}^{\dag }%
\widehat{b}\left\vert \alpha \right\rangle \right) _{b}  \nonumber \\
&=&|\alpha |^{4}  \label{Eq.SimpleCorrTestTwoModeCohState2}
\end{eqnarray}%
Hence we have $|\left\langle \widehat{a}^{\dag }\widehat{b}\right\rangle
|^{2}>0$ and $|\left\langle \widehat{a}^{\dag }\widehat{b}\right\rangle
|^{2}=\left\langle \widehat{a}^{\dag }\widehat{a}\,\widehat{b}^{\dag }%
\widehat{b}\right\rangle $. This shows the state is SSR\ compliant
entangled. However it fails the Hillery test for SSR neglected entanglement
which is consistent with being a SSR\ neglected separable state from the 
\cite{Verstraete03a}, \cite{Schuch04a} viewpoint.

The second example of such states has an overall density operator which is a
statistical mixture given by 
\begin{eqnarray}
\widehat{\rho } &=&\frac{1}{4}(\left\vert \psi _{1}\right\rangle
\left\langle \psi _{1}\right\vert )_{a}\otimes \left\vert \psi
_{1}\right\rangle \left\langle \psi _{1}\right\vert )_{b}+\frac{1}{4}%
(\left\vert \psi _{i}\right\rangle \left\langle \psi _{i}\right\vert
)_{a}\otimes \left\vert \psi _{i}\right\rangle \left\langle \psi
_{i}\right\vert )_{b}  \nonumber \\
&&+\frac{1}{4}(\left\vert \psi _{-1}\right\rangle \left\langle \psi
_{-1}\right\vert )_{a}\otimes \left\vert \psi _{-1}\right\rangle
\left\langle \psi _{-1}\right\vert )_{b}+\frac{1}{4}(\left\vert \psi
_{-i}\right\rangle \left\langle \psi _{-i}\right\vert )_{a}\otimes
\left\vert \psi _{-i}\right\rangle \left\langle \psi _{-i}\right\vert )_{b} 
\nonumber \\
&&  \label{Eq.VerstraeteState}
\end{eqnarray}%
where $\left\vert \psi _{\omega }\right\rangle =(\left\vert 0\right\rangle
+\omega \left\vert 1\right\rangle )/\sqrt{2}$, with $\omega =1,i,-,-i$. The $%
\left\vert \psi _{\omega }\right\rangle $ are superpositions of zero and one
boson states and consequently the local particle number SSR is violated by
each of the sub-system density operators $\left\vert \psi _{\omega
}\right\rangle \left\langle \psi _{\omega }\right\vert )_{a}$ and $%
\left\vert \psi _{\omega }\right\rangle \left\langle \psi _{\omega
}\right\vert )_{b}$.

Now using $\widehat{b}\left\vert \psi _{\omega }\right\rangle =(\omega
\left\vert 0\right\rangle )/\sqrt{2}$, $\left\langle \psi _{\omega
}\right\vert \widehat{a}^{\dag }=(\left\langle 0\right\vert \omega ^{\ast })/%
\sqrt{2}$ and $|\omega |^{2}=1$ 
\begin{eqnarray}
\left\langle \widehat{a}^{\dag }\widehat{b}\right\rangle &=&Tr\frac{1}{4}%
\dsum\limits_{\omega }(\widehat{a}^{\dag }\left\vert \psi _{\omega
}\right\rangle \left\langle \psi _{\omega }\right\vert _{a})\otimes (%
\widehat{b}\left\vert \psi _{\omega }\right\rangle \left\langle \psi
_{\omega }\right\vert _{b})  \nonumber \\
&=&\frac{1}{4}\dsum\limits_{\omega }\left\langle \psi _{\omega }\right\vert 
\widehat{a}^{\dag }\left\vert \psi _{\omega }\right\rangle _{a}\left\langle
\psi _{\omega }\right\vert \widehat{b}\left\vert \psi _{\omega
}\right\rangle _{b}  \nonumber \\
&=&\frac{1}{4}\dsum\limits_{\omega }\frac{1}{2}\omega ^{\ast }\frac{1}{2}%
\omega  \nonumber \\
&=&\frac{1}{4}  \label{Eq.SimpleTestEntangVers1}
\end{eqnarray}%
But 
\begin{eqnarray}
\left\langle \widehat{a}^{\dag }\widehat{a}\,\widehat{b}^{\dag }\widehat{b}%
\right\rangle &=&Tr\frac{1}{4}\dsum\limits_{\omega }(\widehat{a}^{\dag }%
\widehat{a}\left\vert \psi _{\omega }\right\rangle \left\langle \psi
_{\omega }\right\vert _{a})\otimes (\widehat{b}^{\dag }\widehat{b}\left\vert
\psi _{\omega }\right\rangle \left\langle \psi _{\omega }\right\vert _{b}) 
\nonumber \\
&=&\frac{1}{4}\dsum\limits_{\omega }\left\langle \psi _{\omega }\right\vert 
\widehat{a}^{\dag }\widehat{a}\left\vert \psi _{\omega }\right\rangle
_{a}\left\langle \psi _{\omega }\right\vert \widehat{b}^{\dag }\widehat{b}%
\left\vert \psi _{\omega }\right\rangle _{b}  \nonumber \\
&=&\frac{1}{4}\dsum\limits_{\omega }\frac{1}{2}|\omega |^{2}\frac{1}{2}%
|\omega |^{2}  \nonumber \\
&=&\frac{1}{4}  \label{Eq.SimpleTestEntangVers2}
\end{eqnarray}%
Hence we have $|\left\langle \widehat{a}^{\dag }\widehat{b}\right\rangle
|^{2}>0$ and $|\left\langle \widehat{a}^{\dag }\widehat{b}\right\rangle
|^{2}<\left\langle \widehat{a}^{\dag }\widehat{a}\,\widehat{b}^{\dag }%
\widehat{b}\right\rangle $. This shows the state is SSR compliant entangled.
However it fails the Hillery test for entanglement, so is consistent with
being a SSR\ neglected separable state \cite{Verstraete03a}, \cite{Schuch04a}
viewpoint. It should be noted however that the density operator can also be
written as 
\begin{eqnarray}
\widehat{\rho } &=&\frac{1}{4}(\left\vert 0\right\rangle \left\langle
0\right\vert )_{A}\otimes \left\vert 0\right\rangle \left\langle
0\right\vert )_{B}+\frac{1}{4}(\left\vert 1\right\rangle \left\langle
1\right\vert )_{A}\otimes \left\vert 1\right\rangle \left\langle
1\right\vert )_{B}  \nonumber \\
&&+\frac{1}{2}(\left\vert \Psi _{+}\right\rangle \left\langle \Psi
_{+}\right\vert )_{AB}  \label{Eq.VerstraeteState2}
\end{eqnarray}%
where $\left\vert \Psi _{+}\right\rangle _{AB}=(\left\vert 0\right\rangle
_{A}\left\vert 1\right\rangle _{B}+\left\vert 1\right\rangle _{A}\left\vert
0\right\rangle _{B})/\sqrt{2}$. In this form the terms correspond to a
statistical mixture of states with $0,1,2$ bosons. The first two terms
correspond to separable states, in which the sub-system density operators
are SSR compliant. The final term however is a one boson Bell state which is
generally regarded as the paradigm of a two mode entangled state. Hence
regarding the overall state as separable is highly questionable.\medskip
\pagebreak

\section{Appendix G - Derivation of Interferometer Results}

\label{Appendix - Derivatio n of Interferometer Result}

\subsection{General Theory - Two Mode Interferometer}

\label{SubSection - General Thy}

Introducing the free and interaction evolution operators via 
\begin{eqnarray}
\widehat{U} &=&\widehat{U}_{0}\,\widehat{U}_{int}  \nonumber \\
\widehat{U}_{0} &=&\exp (-i\widehat{H}_{0}t/\hbar )
\end{eqnarray}%
it is straightforward to show that for 
\begin{equation}
\widehat{M}=\frac{1}{2}(\widehat{b}^{\dag }\widehat{b}-\widehat{a}^{\dag }%
\widehat{a})
\end{equation}%
we have 
\begin{eqnarray}
\left\langle \widehat{M}\right\rangle &=&Tr(\widehat{M}\,_{H}\,\widehat{\rho 
})  \nonumber \\
\left\langle \Delta \widehat{M}^{2}\right\rangle &=&Tr(\left\{ \widehat{M}%
_{H}-\left\langle \widehat{M}_{H}\right\rangle \right\} ^{2}\,\widehat{\rho }%
)  \label{Eq.MeasuredResults}
\end{eqnarray}%
giving the mean and variance in terms of the input density operator and
interaction picture Heisenberg operators 
\begin{eqnarray}
\widehat{M}_{H} &=&\frac{1}{2}(\widehat{b}_{H}^{\dag }\widehat{b}_{H}-%
\widehat{a}_{H}^{\dag }\widehat{a}_{H})  \nonumber \\
\widehat{b}_{H} &=&\widehat{U}_{int}^{-1}\,\widehat{b}\,\widehat{U}%
_{int}\qquad \widehat{a}_{H}=\widehat{U}_{int}^{-1}\,\widehat{a}\,\widehat{U}%
_{int}
\end{eqnarray}%
where we have used the results $\widehat{U}_{0}^{-1}\,\widehat{b}\,\widehat{U%
}_{0}=\exp (-i\omega _{b}t)\,\widehat{b}$ and $\widehat{U}_{0}^{-1}\,%
\widehat{a}\,\widehat{U}_{0}=\exp (-i\omega _{a}t)\,\widehat{a}$.

The interaction picture Heisenberg operators satisfy 
\begin{equation}
i\hbar \frac{\partial }{\partial t}\widehat{b}_{H}=[\widehat{b}_{H},\widehat{%
V}_{H}]\qquad i\hbar \frac{\partial }{\partial t}\widehat{a}_{H}=[\widehat{a}%
_{H},\widehat{V}_{H}]
\end{equation}%
where 
\begin{eqnarray}
\widehat{V}_{H} &=&\mathcal{A(}t)\exp (-i\omega _{0}t)\,\exp (i\phi )\,%
\widehat{b}_{H}^{\dag }\widehat{a}_{H}\,\exp (+i\omega _{ba}t)\,  \nonumber
\\
&&+\mathcal{A(}t)\exp (+i\omega _{0}t)\,\exp (-i\phi )\,\widehat{a}%
_{H}^{\dag }\widehat{b}_{H}\,\exp (-i\omega _{ba}t)  \nonumber \\
&=&\mathcal{A(}t)\,\exp (i\phi )\,\widehat{b}_{H}^{\dag }\widehat{a}_{H}\,+%
\mathcal{A(}t)\,\exp (-i\phi )\,\widehat{a}_{H}^{\dag }\widehat{b}_{H}\,
\end{eqnarray}%
for resonance.

We then find that the Heisenberg picture operators satisfy coupled linear
equations 
\begin{equation}
i\hbar \frac{\partial }{\partial t}\widehat{b}_{H}=\mathcal{A(}t)\,\exp
(+i\phi )\,\widehat{a}_{H}\qquad i\hbar \frac{\partial }{\partial t}\widehat{%
a}_{H}=\mathcal{A(}t)\,\exp (-i\phi )\,\widehat{b}_{H}
\end{equation}%
which after replacing the time $t$ by the area variable $s$ then involve
time independent coefficients%
\begin{equation}
i\frac{\partial }{\partial s}\widehat{b}_{H}(s)=\exp (+i\phi )\,\widehat{a}%
_{H}(s)\qquad i\frac{\partial }{\partial s}\widehat{a}_{H}(s)=\,\exp (-i\phi
)\,\widehat{b}_{H}(s)
\end{equation}%
The equations can then be solved via Laplace transforms giving 
\begin{equation}
\widehat{b}_{H}(s,\phi )=\cos s\,\widehat{b}-i\exp (i\phi )\,\sin s\,%
\widehat{a}\qquad \widehat{a}_{H}(s,\phi )=-i\exp (-i\phi )\,\sin s\,%
\widehat{b}+\,\cos s\,\widehat{a}
\end{equation}%
where now $2s$ is the area for the classical pulse.

Hence we have in general 
\begin{eqnarray}
\widehat{M}_{H}(2s,\phi ) &=&\frac{1}{2}(\widehat{b}_{H}^{\dag }(s,\phi )%
\widehat{b}_{H}(s,\phi )-\widehat{a}_{H}^{\dag }(s,\phi )\widehat{a}%
_{H}(s,\phi ))  \nonumber \\
&=&\sin 2s\,(\sin \phi \,\widehat{S}_{x}+\cos \phi \,\widehat{S}_{y})+\cos
2s\,\widehat{S}_{z}  \label{Eq.TransMeasureable}
\end{eqnarray}%
The versatility of the measurement follows from the range of possible
choices of the pulse area $2s$ and the phase $\phi $.

Writing $2s=\theta $ we can then substitute into Eq.(\ref{Eq.MeasuredResults}%
) to obtain results for $\left\langle \widehat{M}\right\rangle $ and $%
\left\langle \Delta \widehat{M}^{2}\right\rangle $. These are set out in
SubSection \ref{SubSection - Simple Two Mode Interferometer} in Eqs. (\ref%
{Eq.GeneralResultMean}) and (\ref{Eq.GeneralResultVariance}) in terms of the
mean values of the spin operators and the matrix elements of the covariance
matrix for the spin operators. all for the quantum state$\widehat{\rho }$.

\subsection{Beam Splitter and Phase Changer}

For the \emph{beam splitter} we have $2s=\pi /2$ and $\phi $ (variable) so
that 
\begin{equation}
\widehat{M}_{H}(\frac{\pi }{2},\phi )=\sin \phi \,\widehat{S}_{x}+\cos \phi
\,\widehat{S}_{y}  \label{Eq.TransMeasBeamSplitter}
\end{equation}%
whilst for the \emph{phase changer} we have $2s=\pi $ and $\phi $
(arbitrary) so that%
\begin{equation}
\widehat{M}_{H}(\pi ,\phi )=-\widehat{S}_{z}
\label{Eq.TransMeasPhaseChanger}
\end{equation}

\subsection{Other Measureables}

We can also consider other choices for the measureable, which then enable us
to determine other moments of the spin operators. A case of particular
interest is the square of the population difference%
\begin{equation}
\widehat{M}_{2}=\left( \frac{1}{2}(\widehat{b}^{\dag }\widehat{b}-\widehat{a}%
^{\dag }\widehat{a})\right) ^{2}
\end{equation}%
It is then straightforward to show for the beam splitter case with $2s=\pi
/2 $ and $\phi $ (variable) 
\begin{eqnarray}
\widehat{M}_{2H}(\frac{\pi }{2},\phi ) &=&\left( \sin \phi \,\widehat{S}%
_{x}+\cos \phi \,\widehat{S}_{y}\right) ^{2}  \nonumber \\
&=&\sin ^{2}\phi \,(\widehat{S}_{x})^{2}+\cos ^{2}\phi \,(\widehat{S}%
_{y})^{2}+\sin \phi \,\cos \phi \,(\widehat{S}_{x}\widehat{S}_{y}+\widehat{S}%
_{y}\widehat{S}_{x})  \nonumber \\
&&
\end{eqnarray}%
Hence 
\begin{equation}
\left\langle \widehat{M}_{2}\right\rangle =\sin ^{2}\phi \,\left\langle (%
\widehat{S}_{x})^{2}\right\rangle +\cos ^{2}\phi \,\left\langle (\widehat{S}%
_{y})^{2}\right\rangle +\sin \phi \,\cos \phi \,\left\langle (\widehat{S}_{x}%
\widehat{S}_{y}+\widehat{S}_{y}\widehat{S}_{x})\right\rangle
\end{equation}%
showing that the mean for the new observable $\widehat{M}_{2}$ is a
sinusoidal function of the BS interferometer variable $\phi $ with
coefficients that depend on the means of $\widehat{S}_{x}^{2}$, $\widehat{S}%
_{y}^{2}$ and $\widehat{S}_{x}\widehat{S}_{y}+\widehat{S}_{y}\widehat{S}_{x}$%
.

\subsection{General Theory - Multi-Mode Interferometer}

The derivation follows the same steps as in SubSection \ref{SubSection -
General Thy}. However here we have the results $\widehat{U}_{0}^{-1}\,%
\widehat{b}_{i}\,\widehat{U}_{0}=\exp (-i(\omega _{b}+\omega _{i})t)\,%
\widehat{b}_{i}$ and $\widehat{U}_{0}^{-1}\,\widehat{a}_{i}\,\widehat{U}%
_{0}=\exp (-i(\omega _{a}+\omega _{i})t)\,\widehat{a}_{i}$. The factors
involving $\exp (-i\omega _{i})t$ cancel out in the derivation of the
Heisenberg equations, which here are 
\begin{equation}
i\hbar \frac{\partial }{\partial t}\widehat{b}_{iH}=\mathcal{A(}t)\,\exp
(+i\phi )\,\widehat{a}_{iH}\qquad i\hbar \frac{\partial }{\partial t}%
\widehat{a}_{iH}=\mathcal{A(}t)\,\exp (-i\phi )\,\widehat{b}_{iH}
\end{equation}%
and the solutions are 
\begin{equation}
\widehat{b}_{iH}(s,\phi )=\cos s\,\widehat{b}_{i}-i\exp (i\phi )\,\sin s\,%
\widehat{a}_{i}\qquad \widehat{a}_{iH}(s,\phi )=-i\exp (-i\phi )\,\sin s\,%
\widehat{b}_{i}+\,\cos s\,\widehat{a}_{i}
\end{equation}%
where now $2s$ is the area for the classical pulse.

Hence we have in general 
\begin{eqnarray}
\widehat{M}_{H}(2s,\phi ) &=&\frac{1}{2}\sum_{i}(\widehat{b}_{iH}^{\dag
}(s,\phi )\widehat{b}_{iH}(s,\phi )-\widehat{a}_{iH}^{\dag }(s,\phi )%
\widehat{a}_{iH}(s,\phi ))  \nonumber \\
&=&\sin 2s\,(\sin \phi \,\widehat{S}_{x}+\cos \phi \,\widehat{S}_{y})+\cos
2s\,\widehat{S}_{z}
\end{eqnarray}%
This leads to the same formal results (\ref{Eq.MeanMultiMode}) and (\ref%
{Eq.VarianceMultiMode}) for the mean and variance. The versatility of the
measurement again follows from the range of possible choices of the pulse
area $2s$ and the phase $\phi $.\pagebreak

\section{Appendix H - Limits on Interferometry Tests}

\label{Appendix - Limits on Interferometry Tests}

The\emph{\ tests} for entanglement in a particular quantum state are given
in terms of the\emph{\ mean value} and \emph{variance} for certain physical
quantities. Interferometers are used to enable these means and variances to
be determined from measurements on \emph{another} physical quantity when
either the state being tested is acted upon by the interferometer or it is
being unaffected. Quantum theory enables us to predict two things. Firstly,
for any physical quantity $\widehat{M}$ we can predict the \emph{possible
values} that measurements could result in. Results from a succession of
measurements would confirm what these values are. Secondly, for any quantum
state, we can predict the \emph{probability} that measurement leads to a
specific value. A single measurement only yields one of the possible values,
so \emph{independent repetitions} of such measurements is needed to confirm
what the probabilities for measuring particular values are - ideally an
infinite number of repeated measurements would be required. If this was
possible, the computed mean $\left\langle \widehat{M}\right\rangle $ and
variance $\left\langle \Delta \widehat{M}^{2}\right\rangle $ of the
measurements for the physical quantity $\widehat{M}$ would confirm the
quantum theory predictions for any quantum state. A finite but large number
of independent measurements - each based on the \emph{same} probabiity
distribution for the possible results, would enable us to\emph{\ estimate}
the \emph{actual} mean and variance of the measured values from the \emph{%
sample} measurements. These estimates would not be precisely accurate. The
question is - how \emph{big} would the sample of repeated measurements need
to be for the purpose of using the estimated mean and variance in the tests
for \emph{entanglement} ?

Statistical theory in the form of the \emph{central limit theorem} \cite%
{Gupta13a} can be applied here. This tells us if the number $R$ of repeated
measurements is large, then the mean of the\emph{\ sample} measurements
approaches the \emph{true} mean and the \emph{variance} in the \emph{sample
estimation} of the \emph{mean} is given by the \emph{true variance} divided
by $R$%
\begin{eqnarray}
\left\langle \widehat{M}\right\rangle _{sample} &\rightarrow &\left\langle 
\widehat{M}\right\rangle  \nonumber \\
\left\langle \Delta \left\langle \widehat{M}\right\rangle ^{2}\right\rangle
_{sample} &\rightarrow &\frac{\left\langle \Delta \widehat{M}%
^{2}\right\rangle }{R}  \label{Eq.CentralLimitThm}
\end{eqnarray}%
We can use our theoretical estimate of the variance $\left\langle \Delta 
\widehat{M}^{2}\right\rangle $ to get an idea of how large the sample of
measurements must be in order that the standard deviation of the sample
estimate for the mean is small enough that the mean can confidently be
stated to exceed or be less than the quantity on the other side of the
inequality in the entanglement test.\pagebreak

\section{Appendix \ I - Relative Phase State}

\label{Appendix - Relative Phase State}

The\emph{\ relative phase eigenstate} (see \cite{Dalton12a}, \cite{Dalton14c}%
) for an $N$ boson two mode system has provided an important example of
different outcomes for the simple spin squeezing and Hillery spin squeezing
tests, so here its properties are set out in more detail. The results for
interferometric measurements on the relative phase state are also presented.

The relative phase state is a globally compliant entangled state of the
sub-systems $a$ and $b$ and is defined by%
\begin{equation}
\left\vert N,\theta _{p}\right\rangle =\frac{1}{\sqrt{N+1}}%
\sum_{k=-N/2}^{N/2}\exp (ik\theta _{p}^{N})\left\vert N/2-k\right\rangle
_{a}\left\vert N/2+k\right\rangle _{b}  \label{Eq.RelPhaseEigenstate}
\end{equation}%
where $\theta _{p}^{N}=p(2\pi /(N+1))$, $p=-N/2,-N/2+1,..,+N/2$ is a
quasi-continuum of $N+1$ equispaced phase eigenvalues, and $\left\vert
N/2-k\right\rangle _{a}$, $\left\vert N/2+k\right\rangle _{b}$ are Fock
states for sub-systems $a$ and $b$. The Hermitian relative phase operator $%
\widehat{\Theta }_{N}$ for $N$ boson states is then defined as 
\begin{equation}
\widehat{\Theta }_{N}=\sum_{p}\theta _{p}^{N}\,\left\vert N,\theta
_{p}\right\rangle \left\langle N,\theta _{p}\right\vert
\label{Eq.RelPhaseOpr}
\end{equation}%
and $\left\vert N,\theta _{p}\right\rangle $ is an eigenvector with
eigenvalue $\theta _{p}^{N}$.

Since these states are entangled with maximum \emph{mode entropy}, are \emph{%
spin squeezed} and are \emph{fragmented} BEC\ (two modes have macroscopic
occupancy) it is of some interest to examine their interferometric
properties for the simple beam splitter interfometer. As shown in \cite%
{Dalton12a} the relative phase state has the following mean values for the
spin operators when $\widehat{\rho }=\left\vert N,\theta _{p}\right\rangle
\left\langle N,\theta _{p}\right\vert $ 
\begin{equation}
\left\langle \widehat{S}_{x}\right\rangle _{\rho }=\frac{N\pi }{8}\cos
\theta _{p}\qquad \left\langle \widehat{S}_{y}\right\rangle _{\rho }=-\frac{%
N\pi }{8}\sin \theta _{p}\qquad \left\langle \widehat{S}_{z}\right\rangle
_{\rho }=0  \label{Eq.MeanSxSyRelativePhaseState}
\end{equation}%
so that for the measurable 
\begin{equation}
\left\langle \widehat{M}\right\rangle =\frac{N\pi }{8}\sin (\phi -\theta
_{p})  \label{Eq.MeanMeasurRelPhaseState}
\end{equation}%
We thus have a large amplitude - proportional to $N$ - sinusoidal dependence
for the mean value of the measureable on the interferometer phase detuning $%
(\phi -\theta _{p})$, and which goes to zero when $\phi =\theta _{p}$. Since
we never have both $\left\langle \widehat{S}_{x}\right\rangle _{\rho }$ and $%
\left\langle \widehat{S}_{y}\right\rangle _{\rho }$ equal to zero the simple
correlation test confirms that the relative phase eigenstate is entangled.

As mentioned above, the relative phase state is highly spin squeezed. To
describe this it is convenient to introduce rotated spin operators $\widehat{%
J}_{x}$, $\widehat{J}_{y}$ and $\widehat{J}_{z}$ given by (see Ref \cite%
{Dalton12a}, Eqn. 179)%
\begin{eqnarray}
\widehat{J}_{x} &=&\widehat{S}_{z}  \nonumber \\
\widehat{J}_{y} &=&\sin \theta _{p}\,\widehat{S}_{x}+\cos \theta _{p}\,%
\widehat{S}_{y}  \nonumber \\
\widehat{J}_{z} &=&-\cos \theta _{p}\,\widehat{S}_{x}+\sin \theta _{p}\,%
\widehat{S}_{y}  \label{Eq.NewSpinOprsPhaseState2}
\end{eqnarray}%
The new spin operators are Schwinger spin operators for \emph{new modes} $%
c,d $ where%
\begin{equation}
\widehat{a}=-\exp (\frac{1}{2}i\theta _{p})\left( \widehat{c}-\widehat{d}%
\right) /\sqrt{2}\quad \widehat{b}=-\exp (-\frac{1}{2}i\theta _{p})\left( 
\widehat{c}+\widehat{d}\right) /\sqrt{2}  \label{Eq.NewModeOprs2}
\end{equation}%
and the relative phase state also an \emph{entangled} state for \emph{new}
modes. This can be shown by substituting for the $\left\vert
N/2-k\right\rangle _{a}$ and $\left\vert N/2+k\right\rangle _{b}$ in terms
of Fock states for the new modes $c,d$.

These new angular momentum operators are \emph{principal spin operators} for
which the covariance matrix is diagonal. For the mean values 
\begin{equation}
\left\langle \widehat{J}_{x}\right\rangle _{\rho }=0\qquad \left\langle 
\widehat{J}_{y}\right\rangle _{\rho }=0\qquad \left\langle \widehat{J}%
_{z}\right\rangle _{\rho }=-\frac{N\pi }{8}  \label{Eq.MeanNewSpinOprs}
\end{equation}%
In terms of spin operators discussed above (see Eqs. (\ref{Eq.OutputOprSx})
and (\ref{Eq.OutputOprSy})) we have $\widehat{J}_{x}=\widehat{S}_{z}$, $%
\widehat{J}_{y}=\widehat{S}_{x}^{\#}(\frac{3\pi }{2}+\theta _{p})$ and $%
\widehat{J}_{z}=\widehat{S}_{y}^{\#}(\frac{3\pi }{2}+\theta _{p})$ so the
variances for $\widehat{J}_{y}$ and $\widehat{J}_{z}$ can be measured using
the simple BS interferometer, and the mean for $\widehat{J}_{x}$ is also
measureable by simply measuring the mean population difference without
subjecting the relative phase eigenstate to the BS interaction.

Inverting these expressions and substituting gives the measureable in terms
of the new spin operators%
\begin{equation}
\widehat{M}_{H}=\cos (\phi -\theta _{p})\,\widehat{J}_{y}-\sin (\phi -\theta
_{p})\,\widehat{J}_{z}  \label{Eq.Measurable}
\end{equation}%
Hence we find for the variance of the measureable%
\begin{eqnarray}
\left\langle \Delta \widehat{M}^{2}\right\rangle &=&\cos ^{2}(\phi -\theta
_{p})\,C(\widehat{J}_{y},\widehat{J}_{y})+\sin ^{2}(\phi -\theta _{p})\,C(%
\widehat{J}_{z},\widehat{J}_{z})  \nonumber \\
&&-2\sin (\phi -\theta _{p})\,\cos (\phi -\theta _{p})\,C(\widehat{J}_{y},%
\widehat{J}_{z})  \label{Eq.VarianceMeasurRelPhaseState}
\end{eqnarray}%
As $\widehat{J}_{x}$,$\widehat{J}_{y}$ and $\widehat{J}_{z}$ are principal
spin operators $C(\widehat{J}_{y},\widehat{J}_{z})=0$ and substituting for
the variances $C(\widehat{J}_{y},\widehat{J}_{y})=1/4+1/8\,\ln N$ and $C(%
\widehat{J}_{z},\widehat{J}_{z})=(1/6-\pi ^{2}/64)\,N^{2}$ (see \cite%
{Dalton12a}) we get for the variance of the measureable for an input
relative phase eigenstate%
\begin{eqnarray}
\left\langle \Delta \widehat{M}^{2}\right\rangle &=&\cos ^{2}(\phi -\theta
_{p})\,(\frac{1}{4}+\frac{1}{8}\ln N)+\sin ^{2}(\phi -\theta _{p})\,(\frac{1%
}{6}-\frac{\pi ^{2}}{64})\,N^{2}  \nonumber \\
&\approx &\frac{1}{4}+(\phi -\theta _{p})^{2}\,(\frac{1}{6}-\frac{\pi ^{2}}{%
64})\,N^{2}  \label{Eq.VarianceMeasurRelPhaseStateResult}
\end{eqnarray}%
for $\phi \approx $. $\theta _{p}.$ The other variance is $C(\widehat{J}_{x},%
\widehat{J}_{x})=(1/12)\,N^{2}$. The variance for the measurable depends
sinusoidally on $2(\phi -\theta _{p})$. Thus the quantum noise in the
measureable also goes to essentially zero at $\phi =$ $\theta _{p}$, when
the mean value $\left\langle \widehat{M}\right\rangle $ also goes to zero.
The width $\Delta \phi $ for this low noise window scales as $1/N$ - which
corresponds to the Heisenberg limit. At the zero of the mean value, the
relative fluctuation varies as $1/N$ as in the Heisenberg limit. Since for $%
\phi =\theta _{p}$ we have $\widehat{M}_{H}=\,\widehat{J}_{y}$ $=\widehat{S}%
_{x}^{\#}(\frac{3\pi }{2}+\theta _{p})$ and $\left\langle \Delta \widehat{M}%
^{2}\right\rangle =\,(\frac{1}{4}+\frac{1}{8}\ln N)$ whilst $\left\langle 
\widehat{S}_{z}\right\rangle _{\rho }=\left\langle \widehat{J}%
_{z}\right\rangle _{\rho }=-\frac{N\pi }{8}$. Thus the spin squeezing test
in Eq.(\ref{Eq.SpinSqgTestNewSx}) is satisfied, confirming again that the
relative phase eigenstate is an entangled state of modes $a$ and $b$.

In regard to \emph{particle entanglement} \cite{Wiseman03a}, \cite%
{Dowling06b} with $\widehat{\rho }=\left\vert N,\theta _{p}\right\rangle
\left\langle N,\theta _{p}\right\vert $ and with $n_{a}=(N/2-k)$, $%
n_{b}=(N/2+k)$, the quantities in Eqs. (\textbf{132}) and (\textbf{133}) 
\textbf{of paper 1 }are given by%
\begin{eqnarray}
\,\widehat{\rho }^{(n_{a}n_{b})} &=&\frac{1}{N+1}\left\vert
N/2-k\right\rangle _{a}\left\langle N/2-k\right\vert _{a}\otimes \left\vert
N/2+k\right\rangle _{b}\left\langle N/2+k\right\vert _{b}
\label{Eq.ProjStateRelPhase} \\
P_{n_{a}n_{b}} &=&\frac{1}{N+1}  \label{Eq.ProbRelPhase}
\end{eqnarray}%
and since $\widehat{\rho }^{(n_{a}n_{b})}$ is a separable state, it follows
that $E_{P}($ $\widehat{\rho })=0$. Thus the measure of particle
entanglement is zero for what is clearly an \emph{entangled} state. Hence
the particle entanglement measure has not detected entanglement in this
example.

The relative phase state is therefore a promising candidate for use as an
input state in two mode interferometry. More elaborate interferometers where
the interferometric variable is associated with other systems whose
parameters are to be measured might be developed. The main issue would be
whether such a relative phase state could be prepared. This is an issue
being dealt with elsewhere \cite{Dalton14c}.\pagebreak


\begin{thebibliography}{99}
\bibitem{Dalton14a} Dalton, B. J., Heaney, L., Goold, J. Garraway, B. M. and
Busch, Th., \textit{New J. Phys. }\textbf{2014}, \textit{16}, 013026.

\bibitem{Peres93a} Peres, A. , \textit{Quantum Theory: Concepts and Methods: 
}Kluwer: Dortrecht, \textbf{1993}.

\bibitem{Verstraete03a} Verstraete, F. and Cirac, J. I. ,\textit{Phys. Rev.
Letts. }\textbf{2003, }\textit{91, }010404.

\bibitem{Bartlett06a} Bartlett, S. D., Rudolph, T. and Spekkens, R. W. ,%
\textit{Int. J. Quant. Infn. }\textbf{2006}, \textit{4}, 17.

\bibitem{Leggett01a} Leggett, A. J. , \textit{Rev. Mod. Phys. }\textbf{2001, 
\textit{73}, }307.

\bibitem{Dalton12a} Dalton, B.J. and Ghanbari, S. \textit{J. Mod. Opt. }%
\textbf{2012}, \textit{59}, 287, \textit{ibid }\textbf{2013}, \textit{60},
602.

\bibitem{Caves81a} Caves, C. M., \textit{Phys. Rev. D }\textbf{1981}, 
\textit{23, }1693.

\bibitem{Kitagawa93a} Kitagawa, M. and Ueda, M., \textit{Phys. Rev. A }%
\textbf{1993}, \textit{47, }5138.

\bibitem{Pezze09a} Pezze, L. and Smerzi, A., \textit{Phys. Rev. Letts. }%
\textbf{2009, }\textit{102, }100401.

\bibitem{Berry00a} Berry, D. W. and Wiseman, H. W., \textit{Phys. Rev.
Letts. }\textbf{2000, }\textit{85, }5098.

\bibitem{Braunstein94a} Braunstein, S. L. and Caves, C. M., \textit{Phys.
Rev. Letts. }\textbf{1994, }\textit{72, }3439.

\bibitem{Helstrom76a} Helstrom, C. W., \textit{Quantum Detection and
Estimation Theory: }Academic Press: New York, \textbf{1976}.

\bibitem{Sorensen01a} Sorensen, A., Duan, L.-M., Cirac, J.I. and Zoller, P. 
\textit{Nature }\textbf{2001, }\textit{409, }63.

\bibitem{Jaaskelainen06a} Jaaskelainen, M. and Meystre, P., \textit{Phys.
Rev. A }\textbf{2006}, \textit{73, }013602.

\bibitem{Li09a} Li, Y., Treutlein, P., Reichel, J. and Sinatra, A., \textit{%
Eur. Phys. J. B }\textbf{2009}, \textit{68, }365.

\bibitem{Rose57a} Rose, M.E. \textit{Elementary Theory of Angular Momentum: }%
Wiley: New York, \textbf{1957}.

\bibitem{Wineland94a} Wineland, D. J., Bollinger, J. J., Itano, W. M. and
Heinzen, D. J., \textit{Phys. Rev. A }\textbf{1994}, \textit{50, }67.

\bibitem{Toth09a} Toth, G., Knapp, C., Guhne, O. and Briegel, H. J., \textit{%
Phys. Rev. A }\textbf{2009, }\textit{79, }042334.

\bibitem{He11b} He, Q. Y., Peng, S.-G., Drummond, P. D.and Reid, M. D., 
\textit{Phys. Rev. A }\textbf{2011, }\textit{84, }022107.

\bibitem{Hoffmann03a} Hoffmann, H.F. and Takeuchi, S., \textit{Phys. Rev. A }%
\textbf{2003, }\textit{68, }032103.

\bibitem{Amico08a} Amico, L., Fazio, R., Osterloh, A. and Vedral, V., 
\textit{Rev. Mod. Phys. }\textbf{2008, }\textit{80}\textbf{, }517.

\bibitem{Barnett89a} Barnett, S. M. and Pegg, D. T., \textit{Phys. Rev. A }%
\textbf{1989, }\textit{39, }1665.

\bibitem{Hillery06a} Hillery, M. and Zubairy, M.S., \textit{Phys. Rev.
Letts. }\textbf{2006, }\textit{96, }050503.

\bibitem{He12a} He, Q. Y., Drummond, P. D., Olsen, M. K. and Reid, M. D., 
\textit{Phys. Rev. A }\textbf{2012, }\textit{86, }023626.

\bibitem{He12b} He, Q. Y., Vaughan, T. G., Drummond, P. D.and Reid, M. D., 
\textit{New J. Phys. }\textbf{2012}, \textit{14}, 093012.

\bibitem{He11a} He, Q. Y., Reid, M. D., Vaughan, T. G., Gross, C.
Oberthaler, M., and Drummond, P. D., \textit{Phys. Rev. Letts. }\textbf{%
2011, }\textit{106, }120405.

\bibitem{Raymer03a} Raymer, M. G., Funk, A. C., Sanders, B. C. and de Guise,
H. \textit{Phys. Rev. A }\textbf{2003}, \textit{67, }052104.

\bibitem{Duan00a} Duan, L-M., Giedke, G. Cirac, J. I. and Zoller, P., 
\textit{Phys. Rev. Letts. }\textbf{2000, }\textit{84, }2722.

\bibitem{Hyllus12a} Hyllus, P., Pezze, L., Smerzi, A. and Toth, G. \textit{%
Phys. Rev. A }\textbf{2012}, \textit{86, }012237.

\bibitem{Benatti11a} Benatti, F., Floreanini, R. and Marzolino, U. \textit{%
J. Phys. B: At. Mol. Opt. Phys. }\textbf{2011}, \textit{44}, 091001.

\bibitem{Sorensen01b} Sorensen, A. and Molmer, K., \textit{Phys. Rev. Letts. 
}\textbf{2001, }\textit{86, }4431.

\bibitem{Dalton16a} Dalton, B. J., .... and Reid, M. D. \textit{Mesoscopic
Two-Mode Entangled and Steerable States in Bose-Einstein Condensates. To be
published }\textbf{2016.}

\bibitem{Hillery09a} Hillery, M., Dung, H. T. and Niset, J. \textit{Phys.
Rev. A }\textbf{2009, }\textit{80, }052335.

\bibitem{Toth03a} Toth, G., Simon, C. and Cirac, J. I., \textit{Phys. Rev. A 
}\textbf{2003, }\textit{68, }062310.

\bibitem{Reid89a} Reid, M. D., \textit{Phys. Rev. A }\textbf{1989, }\textit{%
40, }913.

\bibitem{Yurke86a} Yurke, B., McCall, S. L. and Klauder, J. R., \textit{%
Phys. Rev. A }\textbf{1986, }\textit{33, }4033.

\bibitem{Ferris08a} Ferris, A. J., Olsen, M. K., Cavalcanti, E. and Davis,
M. J., \textit{Phys. Rev. A }\textbf{2008, }\textit{78, }060104.

\bibitem{Chiaca15a} Chianca, C. V. and Olsen, M. K. , \textit{Phys. Rev. A }%
\textbf{2015, }\textit{92, }043626.

\bibitem{Gross11a} Gross, C., Strobel, H., Nicklas, E. , Zibold, T.,
Bar-Gill, N. Kurizki, G. and Oberthaler, M. K., \textit{Nature }\textbf{2011}%
, \textit{480, }219.

\bibitem{Gross10a} Gross, C., Zibold, T., Nicklas, E., Esteve, J. and
Oberthaler, M. K., \textit{Nature }\textbf{2010}, \textit{464, }1165.

\bibitem{Esteve08a} Esteve, J., Gross, C., Weller, A., Giovanazzi, S. and
Oberthaler, M. K., \textit{Nature }\textbf{2008}, \textit{455, }1216.

\bibitem{Riedel10a} Riedel, M..F., Bohl, P., Li, Y., Hansch, T. W., Sinatra,
A. and Treutlein, P., \textit{Nature }\textbf{2010}, \textit{464, }1170.

\bibitem{Schuch04a} Schuch, N., Verstraete, F. and Cirac, J. I., \textit{%
Phys. Rev. A }\textbf{2004}, \textit{70, }042310.

\bibitem{Gupta13a} Gupta, B. C. and Guttmann, I., \textit{Statistics and
Probability with Applications: }Wiley, New Jersey, USA \textbf{2013.}

\bibitem{Wiseman03a} Wiseman, H. M. and Vaccaro, J. A. , \textit{Phys. Rev.
Letts. }\textbf{2003, }\textit{91, }097902.

\bibitem{Dowling06b} Dowling, M. R., Bartlett, S. D., Rudolph, T. and
Spekkens, R. W., \textit{Phys. Rev. A }\textbf{2006, }\textit{74, }052113.

\bibitem{Dalton14c} Dalton, B. J., \textbf{2015 }\textit{Relative Phase
States in Quantum Optics }In preparation.\bigskip
\end{thebibliography}
\end{document}